\documentclass[longauth]{aa}  

\usepackage{graphicx}
\usepackage{txfonts}

\usepackage{natbib}

\usepackage{enumitem}
\usepackage{multicol}

\newcommand{\Lsun}{{\hbox {$L_\odot$}}}
\newcommand{\Msun}{{\hbox {$M_\odot$}}}

\newcommand{\hto}{{\hbox {H\textsubscript{2}O}}}
\def\t#1#2#3#4#5#6{{\hbox {$#1_{#2#3}\text{--}#4_{#5#6}$}}}
\def\htot#1#2#3#4#5#6{\hbox {\hto(\t#1#2#3#4#5#6)}}

\mathchardef\mhyphen="2D

\begin{document} 

\title{Close-up view of a luminous star-forming galaxy at $z=2.95$}

\author{S. Berta \inst{1}           \and
        A.~J. Young\inst{2}         \and
        P. Cox\inst{3}              \and  
        R. Neri \inst{1}            \and
        B.~M. Jones\inst{4}         \and
        A.~J. Baker\inst{2}         \and  
        A. Omont\inst{3}            \and 
        L. Dunne\inst{5}            \and 
        A. Carnero~Rosell\inst{6,7} \and
        L. Marchetti\inst{8,9,10}   \and         
        M. Negrello\inst{5}         \and 
        C. Yang\inst{11}            \and
        D.~A. Riechers\inst{12,13}  \and 
        H. Dannerbauer\inst{6,7}    \and 
        I. Perez-Fournon\inst{6,7}  \and 
        P. van der Werf\inst{14}    \and 
        T. Bakx\inst{15,16}         \and
        R.~J. Ivison\inst{17}       \and   
        A. Beelen \inst{18}         \and      
        V. Buat\inst{18}            \and
        A. Cooray\inst{19}          \and 
        I. Cortzen\inst{1}          \and 
        S. Dye\inst{20}             \and
        S. Eales\inst{5}            \and
        R. Gavazzi\inst{3}          \and 
        A.~I. Harris\inst{21}       \and 
        C.~N. Herrera\inst{1}       \and 
        D. Hughes\inst{22}          \and 
        S. Jin\inst{6,7}            \and
        M. Krips\inst{1}            \and 
        G. Lagache\inst{18}         \and 
        M. Lehnert\inst{3}          \and
        H. Messias\inst{23}         \and  
        S. Serjeant\inst{24}        \and 
        F. Stanley\inst{3}          \and 
        S. Urquhart\inst{24}        \and
        C. Vlahakis\inst{25}        \and 
        A. Wei{\ss}\inst{26}         
}

 \institute{Institut de Radioastronomie Millim\'etrique (IRAM), 300 rue de la Piscine, 38400 Saint-Martin-d'H{\`e}res, France\\
              \email{berta@iram.fr}
        \and
             Department of Physics and Astronomy, Rutgers, The State University of New Jersey, 
             136 Frelinghuysen Road, 
             Piscataway, NJ 08854-8019, USA
        \and
             Sorbonne Universit{\'e}, UPMC 
             Universit{\'e} Paris 6 \& CNRS, UMR 7095, 
             Institut d'Astrophysique de Paris, 98b boulevard Arago, 75014 Paris, France
        \and     
             Jodrell Bank Centre for Astrophysics, Department of Physics and Astronomy, The School of Natural Sciences, The University of Manchester, Manchester M13 9PL, UK
        \and 
             School of Physics and Astronomy, Cardiff University, The Parade, Cardiff CF24 3AA, UK
        \and  
             Instituto Astrof{\'i}sica de Canarias (IAC), E-38205 La Laguna, Tenerife, Spain  
        \and 
             Universidad de La Laguna, Dpto. Astrofísica, E-38206 La Laguna, Tenerife, Spain
        \and
             University of Cape Town, Department of Astronomy. Private Bag X3 Rondebosch, 7701 Cape Town, South Africa
        \and 
             Department of Physics and Astronomy, University of the Western Cape, Private Bag X17, Bellville 7535, Cape Town, South Africa
        \and  
             Istituto Nazionale di Astrofisica, Istituto di Radioastronomia, via Gobetti 101, 40129 Bologna, Italy
        \and 
            European Southern Observatory, Alonso de C{\'o}rdova 3107, Casilla 19001, Vitacura, Santiago, Chile
        \and
            Department of Astronomy, Cornell University, Space Sciences Building, Ithaca, New York (NY) 14853, USA
        \and
            Max-Planck-Institut f\"ur Astronomie, K\"onigstuhl 17, D-69117 Heidelberg, Germany
        \and     
             Leiden University, Leiden Observatory, PO Box 9513, 2300 RA Leiden, The Netherlands
        \and
             Division of Particle and Astrophysical Science, Graduate School of Science, Nagoya University, Aichi 464-8602, Japan
        \and  
             National Astronomical Observatory of Japan, 2-21-1, Osawa, Mitaka, Tokyo 181-8588, Japan
        \and
             European Southern Observatory, Karl-Schwarzschild-Strasse 2, D-85748 Garching, Germany
        \and 
             Aix-Marseille Universit\'{e}, CNRS \& CNES, Laboratoire d'Astrophysique de Marseille, 
             38, rue Frédéric Joliot-Curie 
             13388 Marseille, France
        \and
             University of California Irvine, Physics \& Astronomy, FRH 2174, Irvine CA 92697, USA
        \and 
             School of Physics and Astronomy, University of Nottingham, University Park, Nottingham NG7 2RD, UK
        \and
             Department of Astronomy, University of Maryland, College Park, MD 20742, USA
        \and
             Instituto Nacional de Astrofísica, \'Optica y Electr\'onica, Astrophysics Department, 
             Apdo 51 y 216, 
             Tonantzintla, Puebla 72000 Mexico
        \and
             Instituto de Astrofísica e Ciências do Espaço, Tapada da Ajuda, Edifício Leste, 1349-018 Lisboa, Portugal 
       \and
             Department of Physical Sciences, The Open University, Milton Keynes MK7 6AA, UK
        \and 
            National Radio Astronomy Observatory, 520 Edgemont Road, Charlottesville VA 22903, USA
        \and  
            Max-Planck-Institut f{\"u}r Radioastronomie, Auf dem H{\"u}gel 69, 53121 Bonn, Germany.
    }

  \date{Received October 22nd, 2020 / accepted December 2nd, 2020}


\abstract{Exploiting the sensitivity of the IRAM NOrthern Extended Millimeter Array (NOEMA) and its ability to process 
large instantaneous bandwidths, we have studied the morphology and other properties of the molecular gas and dust in 
the starburst galaxy, H-ATLAS~J131611.5+281219 (HerBS-89a), at {\it z}=2.95. High angular resolution ($0\farcs3$) 
images reveal a partial $1\farcs0$ diameter Einstein ring in the dust continuum emission and 
the molecular emission lines of $\rm ^{12}CO$\,(9-8) and $\rm H_2O\,(2_{02}-1_{11})$. Together with lower angular resolution
($0\farcs6$) images, we report the detection of a series of molecular lines including the three fundamental transitions
of the molecular ion $\rm OH^+$, namely $\rm (1_1-0_1)$, $\rm (1_2-0_1)$ and 
$\rm (1_0-0_1)$, seen in absorption; the molecular ion $\rm CH^+(1-0)$ seen in absorption (and tentatively in emission);   
two transitions of amidogen ($\rm NH_2$), namely $(2_{02}-1_{11})$ and $(2_{20}-2_{11})$ seen in emission; 
and $\rm HCN(11-10)$ and/or $\rm NH(1_2-0_1)$ seen in absorption. The NOEMA data are 
complemented with Very Large Array data tracing 
the $\rm ^{12} CO(1-0)$ emission line, which provides a measurement of 
the total mass of molecular gas and an anchor for a CO excitation analysis. 
In addition, we present \textit{Hubble Space Telescope} imaging that reveals the
foreground lensing galaxy in the near-infrared ($\rm 1.15 \mu m$). Together with photometric data from
the Gran Telescopio Canarias, we derive a photometric redshift of 
$z_\textrm{phot}=0.9^{+0.3}_{-0.5}$ for the foreground lensing  galaxy. 
Modelling the lensing of HerBS-89a, we reconstruct
the dust continuum (magnified by a factor $\mu \simeq 5.0$) and
molecular emission lines (magnified by $\mu \sim 4 - 5$) in the
source plane, which probe scales 
of $\sim 0\farcs1$ (or 800~pc). The  $\rm ^{12}CO(9-8)$ and $\rm H_2O\,(2_{02}-1_{11})$ emission lines
have comparable spatial and kinematic distributions; the source-plane reconstructions do not 
clearly distinguish between a one-component and a two-component scenario, but the latter, 
which reveals two compact rotating components with sizes of $\rm \approx 1 \, kpc$, that are 
likely merging, more naturally accounts for the broad line widths observed in HerBS-89a.
In the core of HerBS-89a, very dense gas with $n_{\rm H_2}\rm \sim 10^{7-9} \, cm^{-3}$ is revealed by the 
$\rm NH_2$ emission lines and the possible $\rm HCN(11-10)$ absorption line.  
HerBS-89a is a powerful star forming galaxy with a molecular gas mass of 
$M_{\textrm{mol}} = (2.1 \pm 0.4) \, \times 10^{11} \, M_\odot$, an 
infrared luminosity of  $L_{\textrm{IR}} = (4.6 \pm 0.4) \, \times 10^{12} \, L_\odot$, 
and a dust mass of  $M_{\textrm{dust}} = (2.6 \pm 0.2) \, \times 10^{9} \, M_\odot$, yielding
a dust-to-gas ratio $\delta_{\rm GDR} \approx 80$. We derive a star formation rate $\rm SFR = 614\pm59 \, M_\odot \, yr^{-1}$ 
and a depletion timescale $\tau_\textrm{depl} = (3.4 \pm 1.0) \times 10^8$ years. 
The $\rm OH^+$ and $\rm CH^+$ absorption lines, which trace low ($\rm \sim 100 \, cm^{-3}$) 
density molecular gas, all have their main velocity component red-shifted by $\rm \Delta V \sim 100 \, km\, s^{-1}$ relative 
to the global CO reservoir. We argue that these absorption lines trace a rare example of gas inflow 
towards the center of a starburst galaxy, indicating that HerBS-89a is accreting gas from its surroundings.}

\authorrunning{S. Berta et al.}

\titlerunning{Close-up View of a $z=2.95$ Star-Forming Galaxy}

\maketitle


\section{Introduction}\label{sect:intro}

In the last two decades, surveys in the far-infrared and sub-millimeter wavebands have opened up a new window for our 
understanding of the formation and evolution of galaxies, revealing a population of massive, dust-enshrouded galaxies 
forming stars at enormous rates in the early Universe \citep[see, e.g.,][]{Blain2002, Carilli-Walter2013, Casey2014, Hodge2020}. 
In particular, the extragalactic imaging surveys done with the \textit{Herschel Space Observatory} \citep{Pilbratt2010}, such as
Herschel-ATLAS \citep{Eales2010}, HerMES \citep{Oliver2012}, and PEP \citep{lutz2011}, have increased the number of dust-obscured star-forming galaxies from
hundreds to several hundred thousand. Together with other large-area surveys, like the all-sky \textit{Planck-HFI} \citep{Planck2015}, the South Pole Telescope \citep[SPT][]{Carlstrom2011} cosmological survey \citep{Staniszewski2009,Vieira2010} and the Atacama Cosmology Telescope (ACT) \citep{Marsden2014,Gralla2020}, we have today vast samples of luminous dusty star-forming 
galaxies (DSFGs) that are amongst the brightest galaxies in the Universe, including numerous examples of strongly lensed systems 
\citep[e.g.,][]{Negrello2010, Negrello2017, Cox2011, Bussmann2013, Spilker2014, Canameras2015, Reuter2020} 
and rare cases of galaxies with intrinsic infrared 
luminosities, $L_{\rm FIR} \gtrapprox 10^{13} \, \Lsun$, and star formation rates (SFRs) in excess 
of $\rm 1000 \, \Msun \, yr^{-1}$, known as Hyper-Luminous Infrared Galaxies
\citep[HyLIRGs, see, e.g.,][]{Ivison1998, Ivison2013, Ivison2019, Fu2013, Oteo2016, Riechers2013, Riechers2017}.

Detailed follow-up studies of these galaxies to investigate their nature and physical properties 
require robust estimates 
of their distances. Due to the  dust obscuration in these objects, searching for CO emission 
lines via sub/millimeter spectroscopy has proved to be the most reliable method for measuring accurate redshifts, an approach
that has become more and more efficient thanks to the increased bandwidths of the receivers and backends, most notably at the NOrthern Extended Millimeter Array (NOEMA) and the Atacama Large Millimeter/submillimeter 
Array (ALMA) \citep[e.g.,][and references therein]{Weiss2013, Fudamoto2017, Neri2020, Reuter2020}.  

Using NOEMA, \cite{Neri2020} reported the results of a project whose aim was to measure robust spectroscopic redshifts for 
13 bright \textit{Herschel}-selected galaxies with $\rm S_{500 \mu m} > 80 \, mJy$, preferentially selecting lensed systems \citep{Negrello2010}. Reliable spectroscopic redshifts were derived 
for 12 individual sources, demonstrating the efficiency of deriving redshifts of high-$z$ galaxies using the new correlator 
and broadband receivers on NOEMA. Based on the success of this Pilot Programme, we started a comprehensive redshift survey 
of a sample of  125 of the brightest \textit{Herschel}-selected galaxies (using the same selection criteria as above), 
the NOEMA Large Program {\it z}-GAL, whose main scientific 
goal is to further characterize the properties of luminous DSFGs in the early Universe. 
Interestingly, half of the sources in the Pilot Programme sample display CO emission line widths
in excess of $\rm 800 \, km\, s^{-1}$. Based on their estimated locations relative to 
the $L^{\prime}_{\rm CO(1-0)}$ versus $\rm \Delta V(CO)$ relationship of \cite{Harris2012},
several of the sources are inferred 
to be gravitationally amplified, while a number of them appear to belong to the rare class of 
hyper-luminous infrared galaxies \citep{Neri2020}. 

One of these galaxies, H-ATLAS~J131611.5+281219 (hereafter HerBS-89a), at $z=2.9497$, 
displays a very strong 2-mm continuum (with a flux density 
$\rm S_{159 GHz} = 4.56 \pm 0.05 \, mJy$) and CO emission lines that are the broadest of the entire sample 
with a line width (FWHM) of $\rm \Delta V \sim 1100 \, km\, s^{-1}$. 
Both the 2-mm continuum and $\rm ^{12}CO\,(5-4)$ line emission 
were resolved by the $1\farcs2$ imaging, with an extension of $0\farcs9\pm0\farcs1$ in
the east-west direction. 
The corresponding infrared luminosity of HerBS-89a (between 8 and 1000 $\mu$m in the rest-frame; uncorrected for amplification) is estimated to 
be $(2.9 \pm 0.2) \times  10^{13} \, \Lsun$, which suggested that HerBS-89a could be a HyLIRG. 
 
In order to further explore the properties of HerBS-89a, we used NOEMA to perform high angular resolution observations at 1~mm. 
We observed, in addition to the underlying dust continuum, the emission lines of $\rm ^{12}CO(9-8)$ and $\rm H_2O\,(2_{02}-1_{11})$;
the three lines of the ground state of the molecular ion $\rm OH^+$, $\rm (1_1-0_1)$, 
$\rm (1_2-0_1)$ and $\rm (1_0-0_1)$, 
which are for the first time reported together in a high-$\it z$ galaxy; the $\rm CH^+(1-0)$ line seen in absorption 
(and tentatively in emission);
$\rm NH_2(2_{02}-1_{11})(5/2-3/2)$ seen in emission; and $\rm HCN(11-10)$ and/or $\rm NH(1_2-0_1)$ seen in absorption, 
here also reported for the first time in a high-$z$ galaxy.

In HerBS-89a, the $0\farcs3$ images of the molecular emission lines and the dust continuum 
reveal a partial $1\farcs0$ diameter Einstein ring.
All of the observed transitions of the molecular ions $\rm OH^+$ and $\rm CH^+$ are seen in 
absorption against the strong dust continuum. 
Together, the molecular emission and absorption lines 
allow us to probe simultaneously in HerBS-89a regions that have very different properties, ranging from the 
(very) dense molecular gas traced by  $\rm ^{12} CO(9-8)$, 
$\rm H_2O\,(2_{02}-1_{11})$, $\rm NH_2$ and $\rm HCN(11-10)$ (overlapping with $\rm NH(1_2-0_1)$),  
to the low $\rm H_2$ fraction, 
diffuse gas traced in $\rm OH^+$ and the reservoirs of turbulent, cold and low-density molecular gas 
traced by $\rm CH^+$, thereby providing a unique view at sub-kpc spatial resolution 
(in the source plane) of the physical properties and feedback activity in this high-$z$ system.

The NOEMA data are complemented by \textit{Hubble Space Telescope} (HST) imaging that traces the foreground massive lensing galaxy 
in the near-infrared (using the F110W filter around $\rm 1.1 \mu m$), and by optical/near-infrared data obtained with 
the Gran Telescopio Canarias (GTC) that provide constraints on the redshift of the lensing  galaxy. 
In addition, we also present data obtained with the Karl G. Jansky Very Large Array (VLA) on the $\rm ^{12}CO(1-0)$ 
emission line, which allow us to derive the mass of molecular gas in HerBS-89a and give an anchoring point 
for an analysis of its CO spectral line energy distribution. 

The structure of the paper is as follows. Section~\ref{section:obs} describes the NOEMA, VLA, \textit{HST}, and GTC observations, and 
the reduction of the respective data sets; Section~\ref{section:results} presents the main results, including 
the morphology of the source, the properties of the molecular emission and absorption lines and the underlying continuum; 
Section~\ref{sect:lensing_glx} describes the characteristics of the foreground lensing galaxy; 
Section~\ref{sect:lensing_model} presents the lensing model and the 
morphology of HerBS-89a in the source plane; and Section~\ref{sect:global-properties} outlines the global intrinsic properties of the dust 
and molecular gas (derived from the CO and water emission lines) corrected for amplification, 
including the CO excitation. 
The molecular gas kinematics and the dynamical mass are reported in Sect.~\ref{sect:kinematics}; the properties of 
the molecular absorption and emission lines other than CO and $\rm H_2O$ are outlined in Sect.~\ref{sect:other-lines}; 
and a discussion of the gas inflow suggested by the red-shifted molecular absorption lines 
of $\rm OH^+$ and $\rm CH^+$ is presented in Sect.~\ref{sect:inflow}. 
Finally, Section~\ref{section:conclusions} summarizes 
the main conclusions of this paper and outlines future prospects.

Throughout this paper we adopt a spatially flat $\Lambda$CDM cosmology 
with $H_{0}=67.4\,{\rm km\,s^{-1}\,Mpc^{-1}}$ and $\Omega_\mathrm{M}=0.315$ \citep{planck2020_2018_VI} and assume a \citet{chabrier2003} initial mass function (IMF). 
At the redshift $z=2.9497$ of HerBS-89a, one arc-second corresponds to 7.9 kpc and the luminosity distance to the source is $D_{\rm L} = 2.5 \times 10^4$~Mpc.


\section{Observations and Data Reduction}\label{section:obs}

\subsection{NOEMA}\label{sect:noema_data}

We used NOEMA to target high-frequency molecular lines in HerBS-89a, redshifted into the 1-mm band. The observations were carried 
out under two separate projects. 

The first project was a Discretionary Directorial Time (DDT) project, labelled E18AE (P.I.: R. Neri), observed 
on February 5, 2019 with ten antennas using the extended A-configuration, yielding an angular resolution of $0\farcs3$, for a total
observing time of 4.3 hours on-source time. This project, which was specifically tailored to measure the $\rm ^{12}CO(9-8)$ and $\rm H_2O(2_{02}-1_{11})$ 
emission lines, also enabled the detection of two strong absorption lines of the molecular ion $\rm OH^+$ 
corresponding to the redshifted ground state transitions of $\rm OH^+(1_1-0_1)$ and $\rm OH^+(1_2-0_1)$ 
($\rm \nu_{rest} = 1033.118$~GHz and 971.803~GHz, respectively)\footnote{The adopted frequencies correspond to 
the strongest of the hyperfine transitions for the specific $\rm \Delta \it J$ \citep[][and references therein]{Indriolo2015}.}.  

The second project, labelled W19DE (P.I.: S. Berta), was completed on March 30th, 2020 with ten antennas using the intermediate 
C-configuration, yielding an angular resolution of $1\farcs0 \times 0\farcs6$, for a total on-source observing time 
of 4.2 hours. This project was a follow-up of the DDT to measure the third ground state transition of $\rm OH^+(1_0-0_1)$ ($\rm \nu_{rest} = 909.158\, GHz$)\footnotemark[1] 
and the ground state transition of the molecular ion $\rm CH^+(1-0)$ ($\rm \nu_{rest} = 835.137 \, GHz$)\footnote{The rest frequency of the 
$\rm CH^+(1-0)$ transition is taken from \cite{Muller2010}.}. 

Observing conditions were excellent for both projects with an atmospheric phase stability of 20$\rm ^o$ RMS and 0.8~mm of 
precipitable water vapor. The correlator was operated in the low resolution mode to provide spectral channels with nominal 
resolution of 2~MHz. The NOEMA antennas were equipped with 2SB receivers that cover a spectral window of 7.744~GHz in each 
sideband and polarization. For the first series of observations, we covered the frequency range from 244.9 to 252.6~GHz and 
260.4 to 268.1~GHz; for the second series, the frequency ranges were from 209.4 to 217.1~GHz and 224.9 to 232.6~GHz. 

The strength of the continuum in HerBS-89a at 1-mm ($\rm \sim 20 \, mJy$ - see Table~\ref{tab:continuum}) 
enabled phase self-calibration, which significantly 
improved the image fidelity and dynamic range both in the continuum and molecular line emission 
(see Sects. \ref{sect:continuum} and \ref{sect:lines}). 
The flux calibrator used in both projects was LkH$\alpha$101. The phase calibrator for the high-angular resolution project (E18AE) was 1308+326. The data were calibrated, averaged in 
polarization, mapped, and analyzed in the GILDAS software package. The absolute flux calibration is accurate to within 
10\% and the $1\sigma$ error on the absolute position of HerBS-89a is estimated to be $<0\farcs2$.

\begin{figure*}[!ht]
\centering
\includegraphics[width=0.45\textwidth]{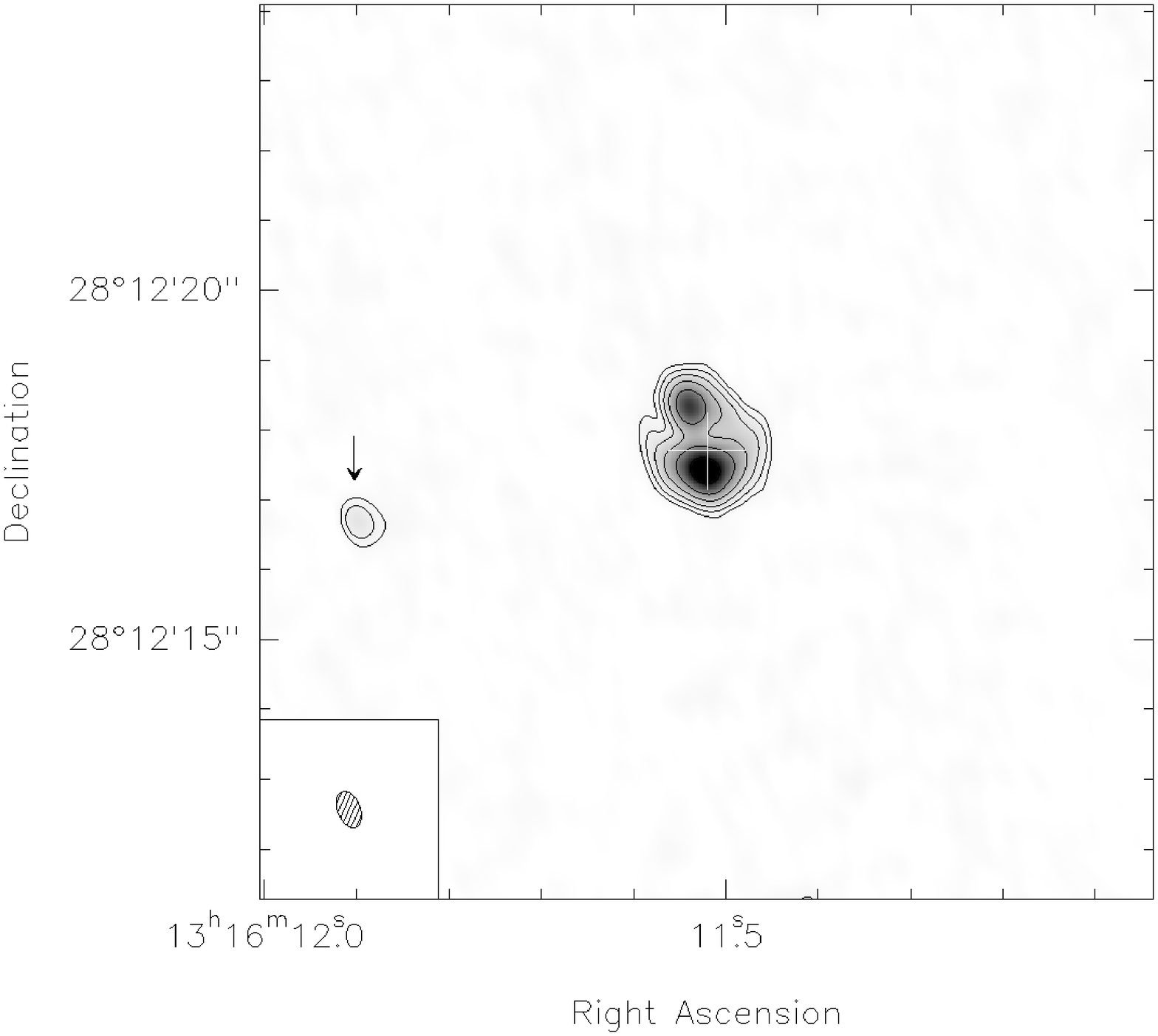}
\hspace{0.07\textwidth}
\includegraphics[width=0.45\textwidth]{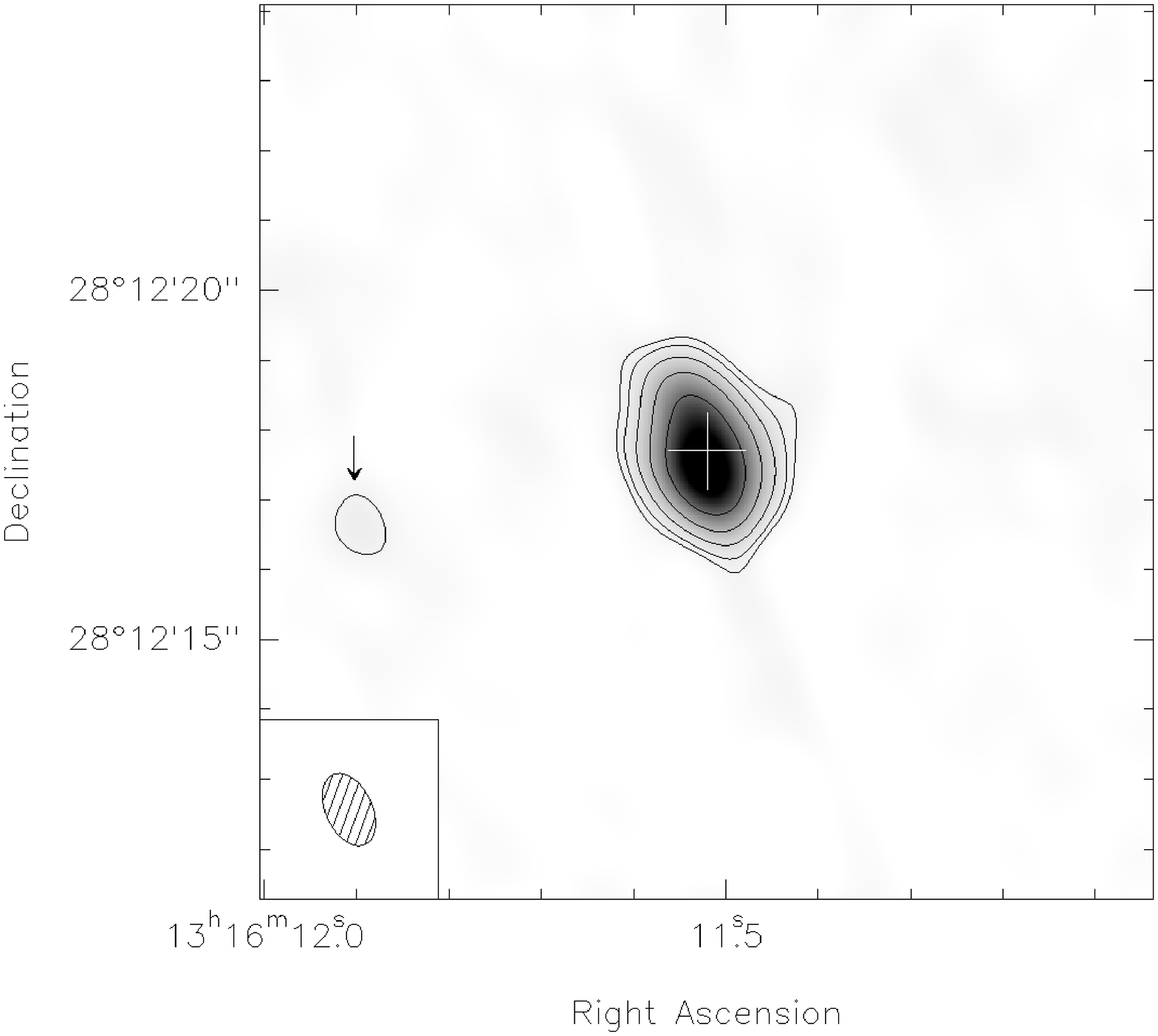}
\caption{Continuum images of HerBS-89a at 1.2~mm, resulting from merging the lower and upper side bands of the two 
sets of NOEMA observations obtained with natural weighting of the visibilities (see Sect.~\ref{sect:continuum} for further details). Contours are at 5, 10, 20, 40, 80$\sigma$ levels. North is up, East is left. {\em Left}: High-angular resolution continuum image at 254.6~GHz 
with a beam size of  $0\farcs43 \times 0\farcs22$. {\em Right}: Lower-angular resolution image at 220.8~GHz with a 
beam size of $1\farcs1 \times 0\farcs66$. The synthesized beam is shown in the lower left corner of
each image, and the cross indicates the position of the phase center (RA 13:16:11.52, DEC +28:12:17.7). 
The weak source HerBS-89b, about $6\farcs0$ to the east, is detected at both frequencies; the position of the source is 
indicated with an arrow and further details are provided in Sect.~\ref{sect:continuum}.}
\label{fig:2SB_continuum_maps}
\end{figure*}

\begin{figure}[!ht]
\centering
\includegraphics[width=0.48\textwidth]{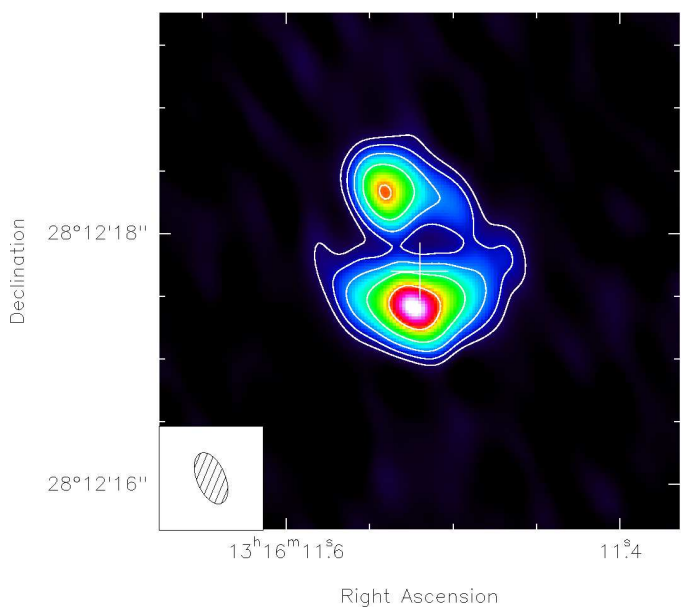}
\caption{Zoom-in of the continuum image of HerBS-89a at 254.6~GHz, obtained with uniform weighting, i.e. enhanced resolution, showing the details of the dust-continuum emission, in particular the faint extended emission of the $\sim 1\farcs0$ diameter partial 
Einstein ring. See caption of Fig.\ref{fig:2SB_continuum_maps} for further details.}
\label{fig:2SB_continuum_zoom}
\end{figure}

\subsection{VLA}\label{sect:vla_data}

The National Radio Astronomy Observatory's (NRAO) VLA was used to observe 
the $\rm ^{12}CO(1-0)$ emission line in HerBS-89a when the array was in the C configuration. The observations are part of a 
larger project (program I.D.:VLA/20A-083 - P.I.: D. Riechers) whose goal was to measure 
the $\rm ^{12}CO(1-0)$ emission lines in the sample of 
\textit{Herschel}-selected galaxies studied in \cite{Neri2020}; the complete results of that project will be presented 
in a forthcoming paper (Stanley et al. in prep.). 
The data were acquired on March 22, 2020 under stable atmospheric conditions. The 0.9~cm Ka band receivers were 
tuned (1~GHz bandwidth per IF) to the expected frequency of the CO emission line, i.e., 29.1848~GHz based on the redshift determined 
by \cite{Neri2020}, and to 38.499~GHz for the second baseband. In total, we observed for 1.6~hours (with 62~minutes on-source) 
recording $1024 \times 2$~MHz dual-polarization channels across a total bandwidth of 2~GHz, 
which was chosen to maximize line sensitivity while retaining as much bandwidth as possible. 
The 2~GHz bandwidth setup was used to maximize the potential for stacking of faint lines (for the entire VLA project), 
while at same time retaining 
sufficient spectral resolution (2~MHz, i.e. $\rm 16-21 \, km\, s^{-1}$) to finely sample the broad CO emission line of HerBS-89a. 
The 8-bit samplers were selected to maximize sensitivity. The source 3C286 was observed to determine accurate complex gain 
and bandpass correction solutions and was also observed to set the absolute scale flux density based on the \cite{Perley2017} models; 
the pointing accuracy was checked regularly. The data were calibrated, averaged in polarization, mapped used natural baseline weighting and 
analyzed in the CASA (Common Astronomy Software Applications) package.
The resulting line map has a spatial resolution of $1\farcs24 \times 0\farcs79$ 
(P.A. $\rm -62.3^o$) and a rms noise level of $\rm 32.1 \, \mu Jy/beam$ over a band width of 0.13 GHz. 
 The 
absolute flux scale is estimated to be accurate to within 10\%.

\begin{table}[!ht]
\caption{Far-infrared and sub/millimeter continuum flux densities of HerBS-89a (and HerBS-89b)} 
\centering
\begin{tabular}{c c c c c}
\hline\hline
     &   &  HerBS-89a      &  HerBS-89b       &            \\
$\lambda$ [mm]   & $\nu$ [GHz] &  \multicolumn{2}{c}{$S_{\nu}$ $[$mJy$]$} &  Ref.     \\
\hline            
\multicolumn{5}{c}{\textbf{\emph{Herschel}}}   \\
0.10  &              & $<$130.8    &       --       &  (1,2)   \\ 
0.16  &              & $<$136.2   &       --       &   "      \\ 
0.25  &              &  71.8$\pm$5.7          &       --       &   "      \\ 
0.35  &              & 103.4$\pm$5.7          &       --       &   "      \\ 
0.50  &              &  95.7$\pm$7.0          &       --       &   "      \\ 
\multicolumn{5}{c}{\bf SCUBA-2}           \\
0.85  &              &  52.8$\pm$4.3          &       --       &  (3)      \\ 
\multicolumn{5}{c}{\bf NOEMA}             \\
1.14  &   263.3       &  22.33$\pm$0.04        & 0.66$\pm$0.04  &  (4)    \\ 
1.21  &   248.3       &  19.22$\pm$0.04        & 0.55$\pm$0.04  &   "     \\ 
1.31  &   229.3       &  13.62$\pm$0.05        & 0.34$\pm$0.05  &   "     \\ 
1.41  &   212.4       &  11.31$\pm$0.04        & 0.25$\pm$0.04  &   "     \\ 
1.89  &   158.6       &  4.56$\pm$0.05         & 0.24$\pm$0.05  &  (5)    \\ 
2.01  &   149.0       &  3.40$\pm$0.30         &     <0.1       &   "       \\
2.09  &   143.2       &  3.02$\pm$0.04         &      --        &   "      \\
2.25  &   133.5       &  2.20$\pm$0.20         &      --        &   "      \\
2.69  &   111.5       &  1.10$\pm$0.08         &      --        &  (5)   \\
2.89  &   103.7       &  0.83$\pm$0.05         &      --        &   "     \\
3.12  &    96.0       &  0.56$\pm$0.06         &      --        &   "    \\
3.40  &    88.3       &  0.44$\pm$0.06         &      --        &   "    \\
\multicolumn{5}{c}{\bf VLA}          \\ 
7.89  &    38.0       &  $<$0.147          &      --        &  (4)    \\
10.34 &   29.0       &  $<$0.065            &      --        &   "        \\
\hline
\end{tabular}
\tablefoot{(1) \cite{Bakx2018}; (2) \citet{Valiante2016}; (3) \cite{Bakx2020}; (4) This paper; (5) \cite{Neri2020}. 
The widths of each of the NOEMA sidebands is 7.744~GHz and their frequency ranges are 
provided for the 1-mm data in Sect.~\ref{sect:noema_data} and in Table~2 of \cite{Neri2020} for the 2 and 3-mm data. 
The flux densities of HerBS-89b have been corrected for the primary beam.
The quoted statistical uncertainties do not account for the absolute flux calibration uncertainty. All the upper limits
are $\rm 3\sigma$.
}
\label{tab:continuum}
\end{table}

\subsection{{\it HST}}\label{sect:hst_data}
HerBS-89a was observed with the $\it HST$ in March 2012 as part of a cycle-19 SNAPshot proposal (program I.D.: 12488 - P.I.: M. Negrello), 
which aimed, amongst other goals, at characterising the nature of the putative lenses in a large sample of 
candidate lensing systems selected at 500\,$\mu$m in the \textit{Herschel} extragalactic surveys. 
Observations were obtained with the Wide Field Camera 3 (WFC3) 
using the wide-J filter F110W (peak wavelength 1.15\,$\mu$m). The total exposure time is 252\,seconds. 
Data were reduced using the Python {\sc AstroDrizzle} package, with parameters optimised to improve the final image quality. 
The pixel scale of the Infrared-Camera is $0\farcs128$, but the image was resampled to a finer pixel scale of 
$0\farcs064$ by exploiting the adopted sub-pixel dither pattern. The astrometry of the resulting image was calibrated by matching the positions of 11 stars from SDSS DR12 and 2MASS and is accurate to within $0\farcs1$.

\subsection{GTC}\label{sect:gtc_data}
HerBS-89a was observed with the GTC 10.4~m telescope in two observing runs (program ID: GTC03-19ADDT and GTC09-19ADDT; PI: H. Dannerbauer). First,  
optical imaging was obtained using the instrument OSIRIS on April 6th, 2019 .
The observations were conducted in service mode under clear skies (although with non-photometric conditions), integrating 
for 10 minutes in the Sloan r-band filter with a seeing of $0\farcs8$. The field of view is $7\farcm5 \times 6\farcm0$ 
and the pixel size $0\farcs254$. Standard procedures for data reduction and calibration of the raw images 
were performed with the IRAF data reduction package \citep{Tody1986}. 

The second series of observations was performed on June 7th, 2020 using the visitor instrument HiPERCAM for a total observing time of 1.25~hours under average seeing conditions ($0\farcs9$). HiPERCAM 
is a quintuple-beam CCD imager enabling the five Sloan filter \textit{ugriz} to be observed simultaneously over a field of view 
of $2\farcm8 \times 1\farcm4$ 
with a pixel size of $0\farcs081$ \citep[][and references therein]{Dhillon2018}. 
The HiPERCAM team designed a dedicated data reduction tool\footnote{http://deneb.astro.warwick.ac.uk/phsaap/hipercam/docs/html/} 
that applies standard procedures for the reduction and calibration of the raw images.


\begin{figure*}[!ht]
\centering
\includegraphics[width=0.75\textwidth]{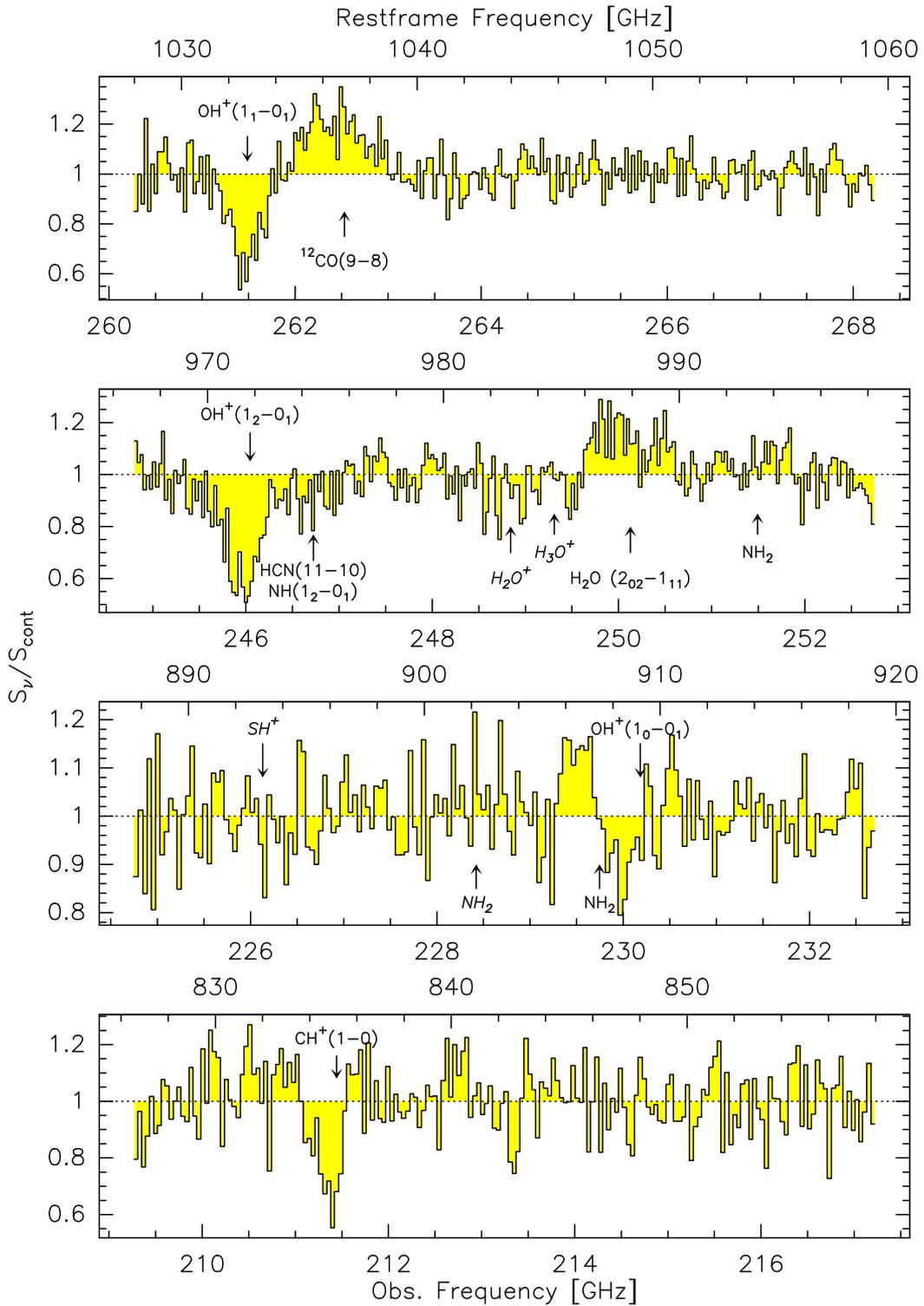} 
\caption{Spectra of HerBS-89a in the frequency ranges between 245 and 268~GHz and between
209 and 233~GHz. The spectra have been normalized by the continuum, which was modeled with a 
linear function (see text). The velocity channels were binned to $\rm 40 \, km\, s^{-1}$ for 
the 245--268~GHz data and $\rm 60 \, km\, s^{-1}$ for the 209--233~GHz data, 
reaching r.m.s. of 0.6 and 0.4 mJy per channel, respectively. 
The rest-frame and observed frequencies are given on the upper and lower horizontal axes, respectively. 
All the detected molecular emission and absorption lines are identified (with the arrows positioned at the line frequencies). 
In addition, the redshifted positions of molecular lines (at $z=2.9497$) that fall within the observed frequency range, 
but were not detected, are indicated (in italics), i.e., $\rm o\mhyphen H_3O^+ (0^-_0-1^+_0)$, 
$\rm H_2O^+ (3_{12}-3_{03})$, $\rm SH^+(2-1)$, and $\rm o\mhyphen NH_2 (2_{0,2}-1_{1,1}) (3/2-1/2)$.} 
\label{fig:spectra_cont_norm}
\end{figure*}

\section{NOEMA and VLA Results}\label{section:results}
In this section, we describe the new data obtained on HerBS-89a, outline the properties derived from the NOEMA 1-mm high-angular 
resolution observations for both the continuum and the molecular emission and absorption lines (Sect.~\ref{sect:NOEMA-results}), 
and present the results on the $\rm ^{12}CO(1-0)$ emission line measured with the VLA (Sect.~\ref{sect:CO(1-0)-VLA}).

\subsection{NOEMA Results}\label{sect:NOEMA-results}

\subsubsection{Continuum emission}\label{sect:continuum}
Figure \ref{fig:2SB_continuum_maps} presents the two continuum maps of HerBS-89a obtained by merging the lower and upper side-bands of 
each set of observations, resulting in a high-angular resolution image centered at $\rm \sim 254.6 \, GHz$ 
and a lower angular resolution image at $\rm \sim 220.8 \, GHz$. We reach a sensitivity of $\sigma=29$ and $33$ $\mu$Jy/beam in 
the two bands, respectively. Merging the two side-bands of the high-resolution continuum images improves the S/N ratio 
and resulted, by applying a uniform weight, 
in a final image with beam size of $0\farcs43 \times 0\farcs22$, to be compared to the $0\farcs55 \times 0\farcs33$ 
achieved by the A-configuration that was used for these observations with natural weighting.

Figure~\ref{fig:2SB_continuum_zoom} presents a zoom-in on this high-resolution continuum image of HerBS-89a. It displays 
an Einstein ring-like morphology, with a double source linked by weak arc structures,
indicating that HerBS-89a is lensed. The high quality of the image reveals the details of the morphology of the 
dust continuum, such as the differences between the northern and southern continuum peaks as well as 
the weaker emission extending between them.

For the lower frequency data obtained using the medium-compact C-configuration of NOEMA, 
the source is only marginally resolved. Combining the data from the upper and lower side-bands resulted 
in a beam of $1\farcs1 \times 0\farcs66$.

To the east of HerBS-89a, the weak unresolved source (labeled HerBS-89b), which was already detected in 2-mm continuum
by \cite{Neri2020}, is detected in the 1.2-mm continuum emission in both images (Fig.~\ref{fig:2SB_continuum_maps}), 
about $6"$ east of the phase center of our observations. The detection reported here confirms the authenticity of 
this source. However, there is no corresponding source in the SDSS catalogue at that position. 
The faintness of HerBS-89b precludes the extraction of  spectroscopic information from the available NOEMA data.  

Continuum aperture flux densities have been measured for each side-band separately. 
Table~\ref{tab:continuum} summarizes the values, both for 
HerBS-89a and the serendipitous source HerBS-89b, which is seen to be $\approx 40\times$  fainter 
than HerBS-89a. The effective frequencies of the adopted continuum bands and the statistical 
uncertainties on the flux densities are also provided. Note that the latter should be added 
in quadrature to the 10\% absolute flux calibration uncertainty (see Sect. \ref{sect:noema_data}).

\begin{figure*}[!ht]
\centering
\includegraphics[width=0.3\textwidth]{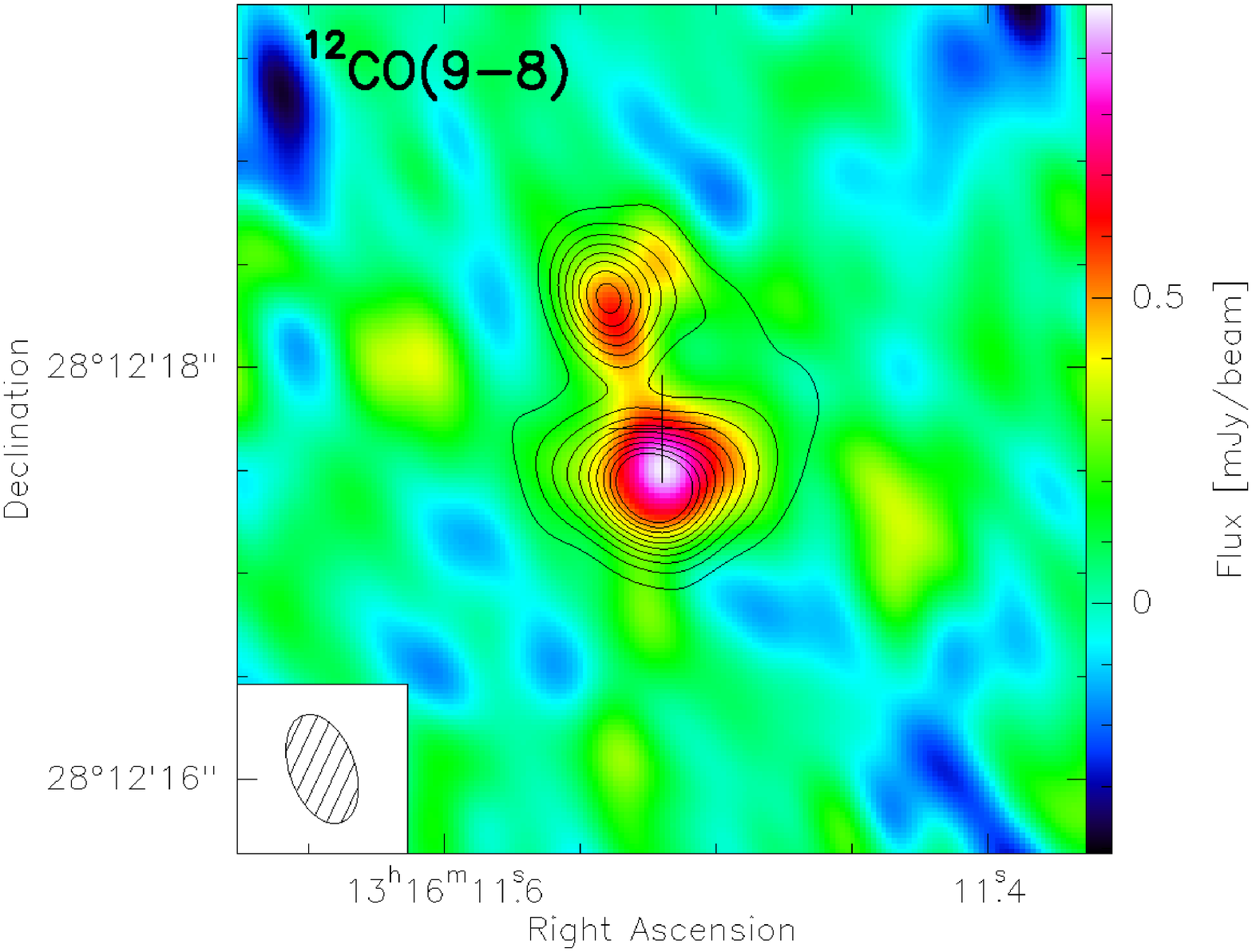}
\hspace{0.035\textwidth}
\includegraphics[width=0.3\textwidth]{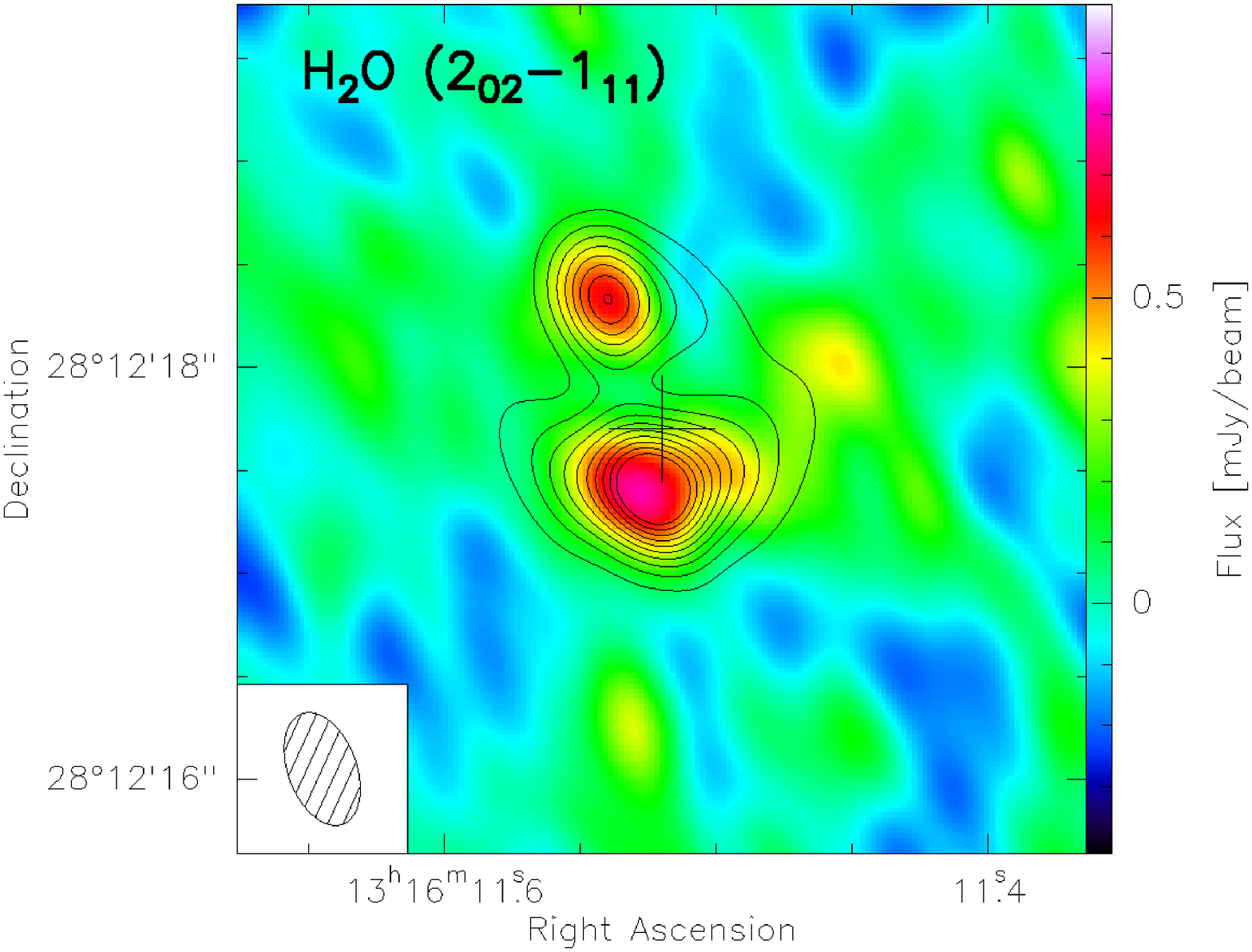}
\hspace{0.035\textwidth}
\includegraphics[width=0.3\textwidth]{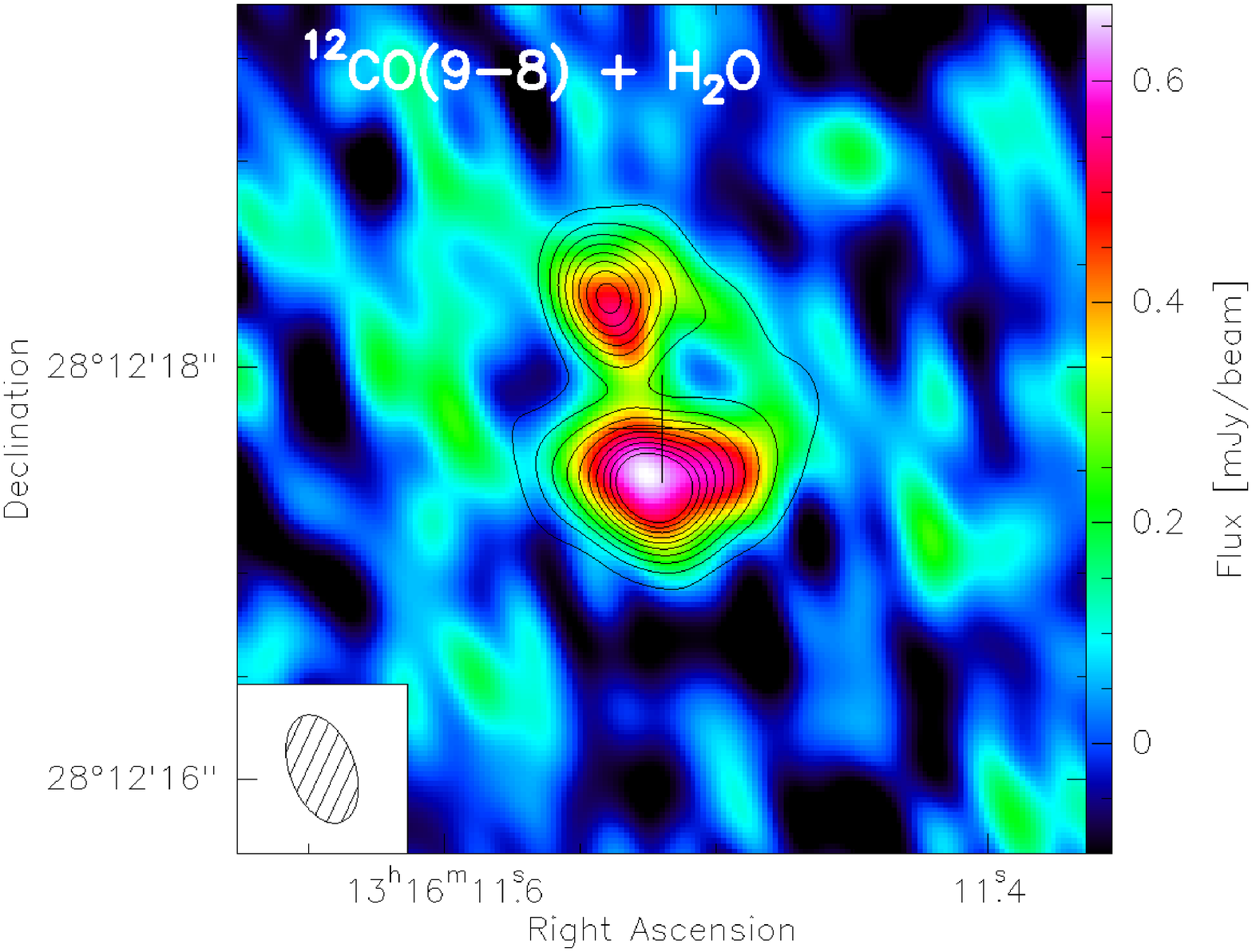}\\
\vspace{0.025\textwidth}
\includegraphics[width=0.3\textwidth]{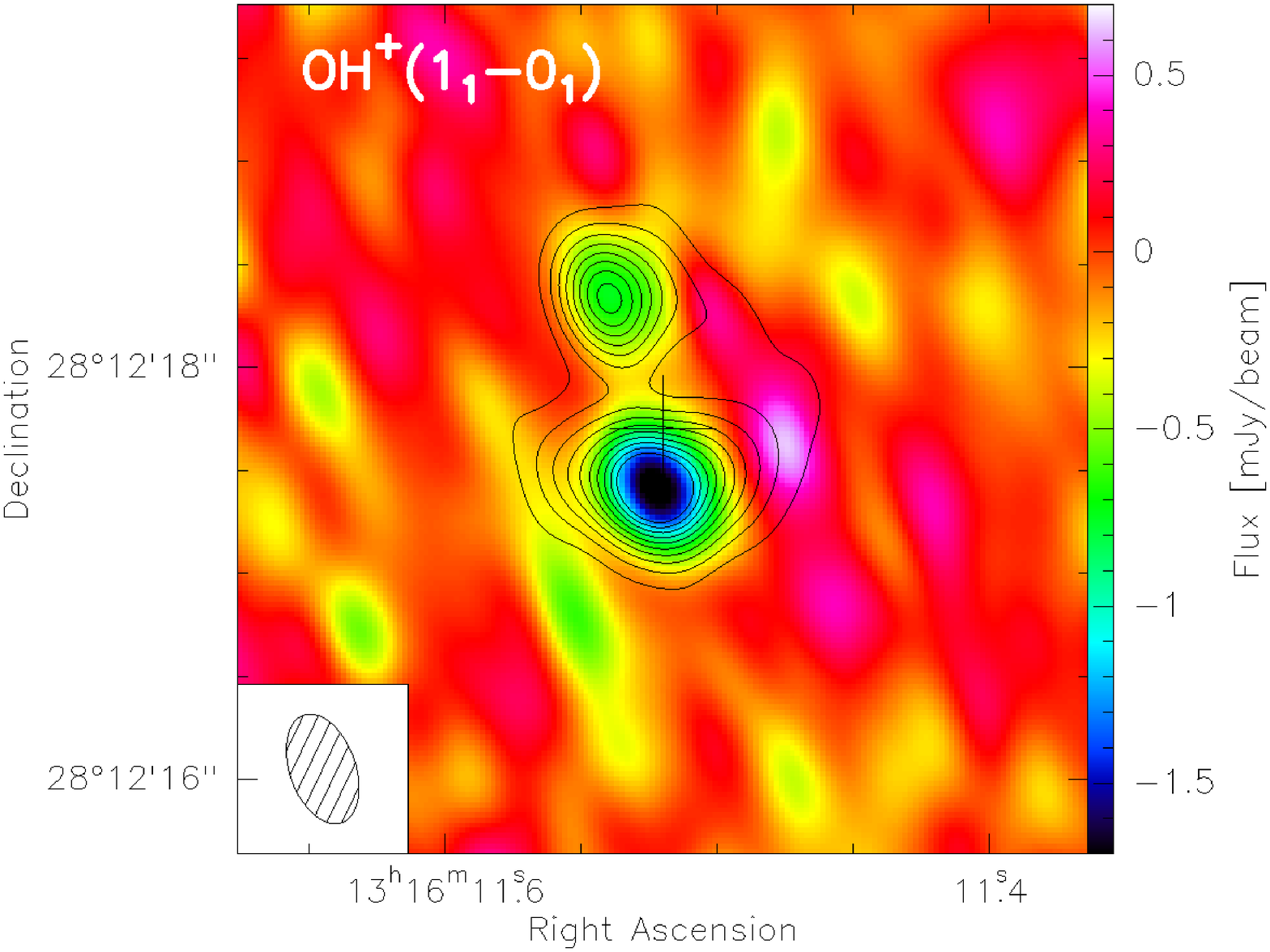}
\hspace{0.035\textwidth}
\includegraphics[width=0.3\textwidth]{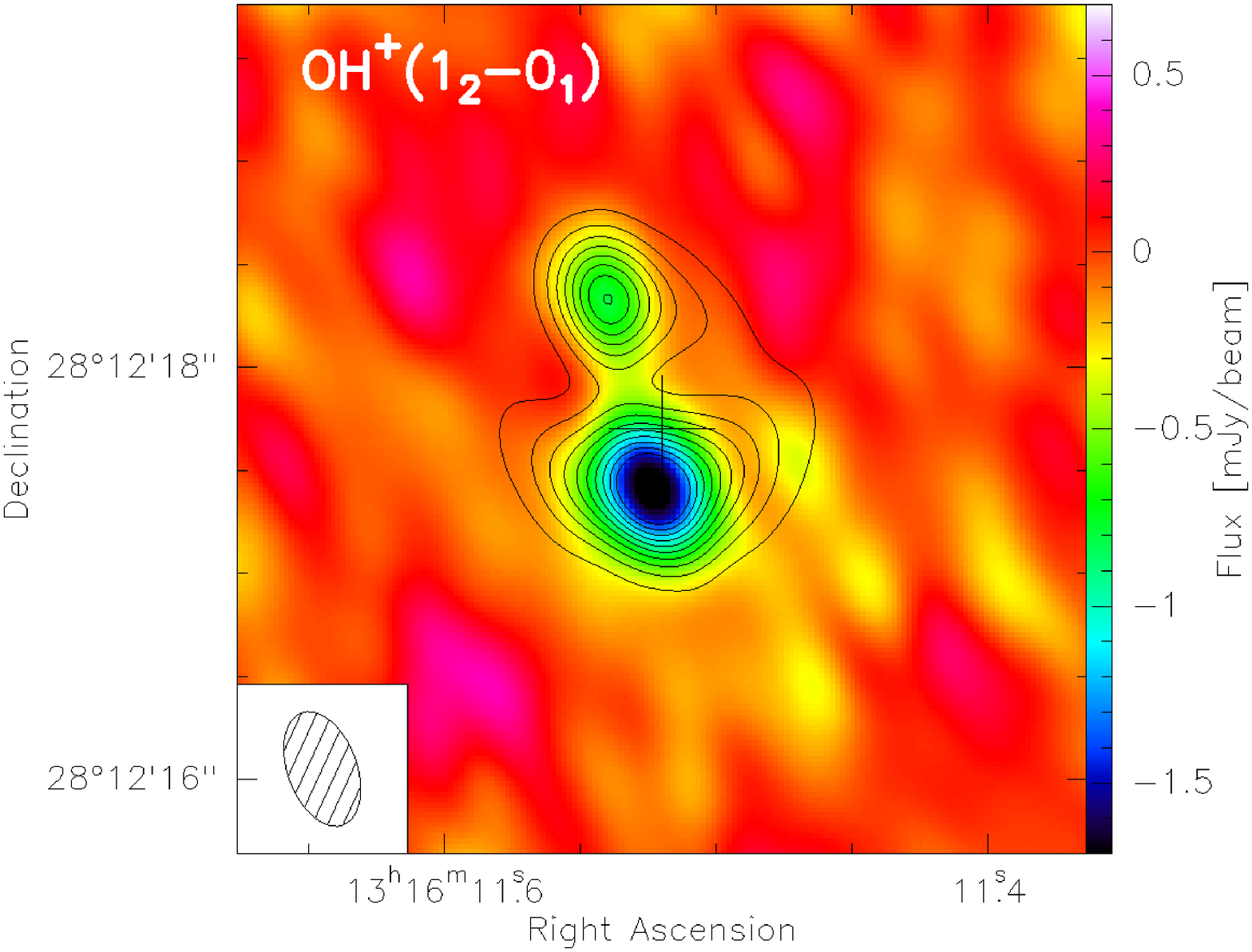}
\hspace{0.035\textwidth}
\includegraphics[width=0.3\textwidth]{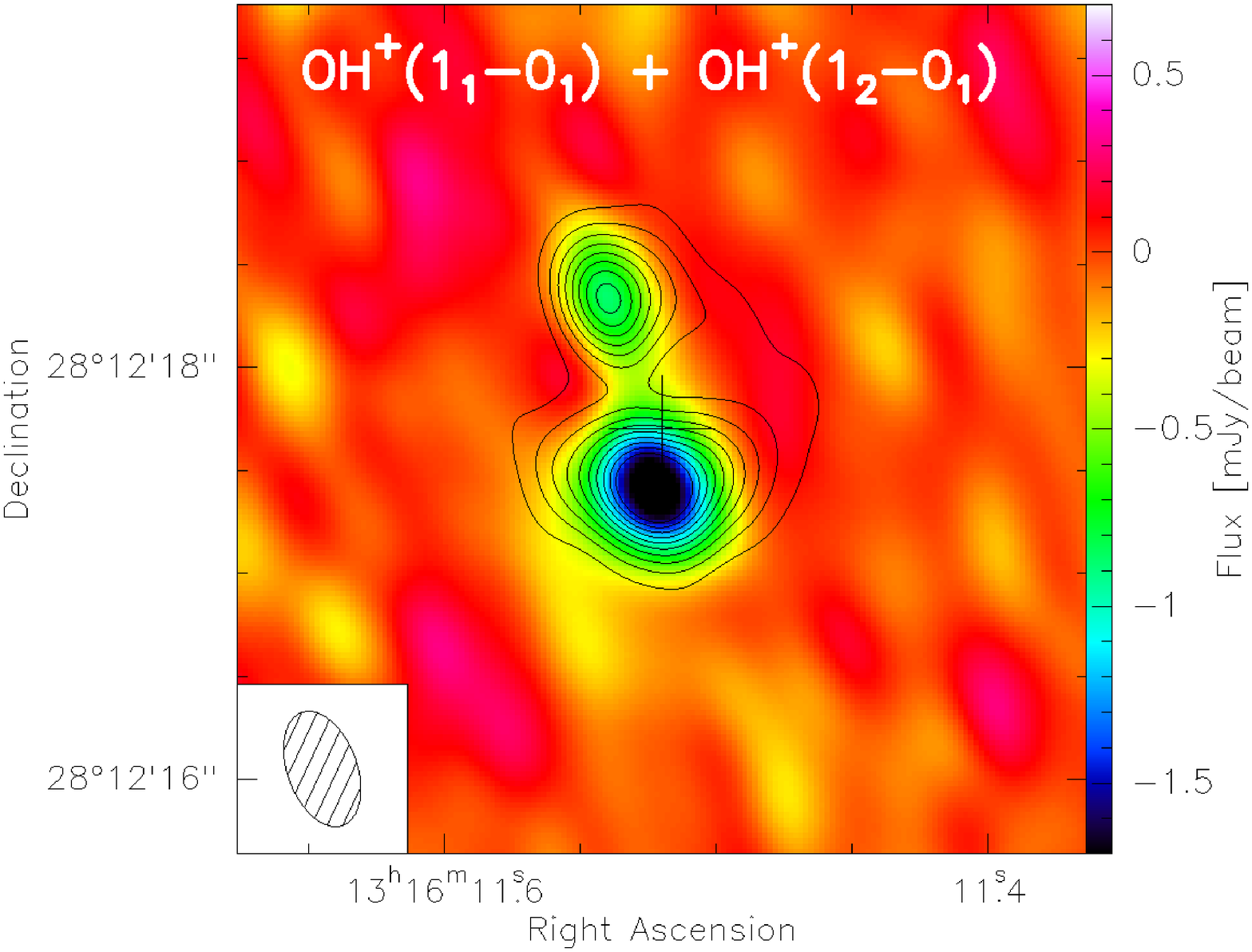}\\
\vspace{0.025\textwidth}
\includegraphics[width=0.3\textwidth]{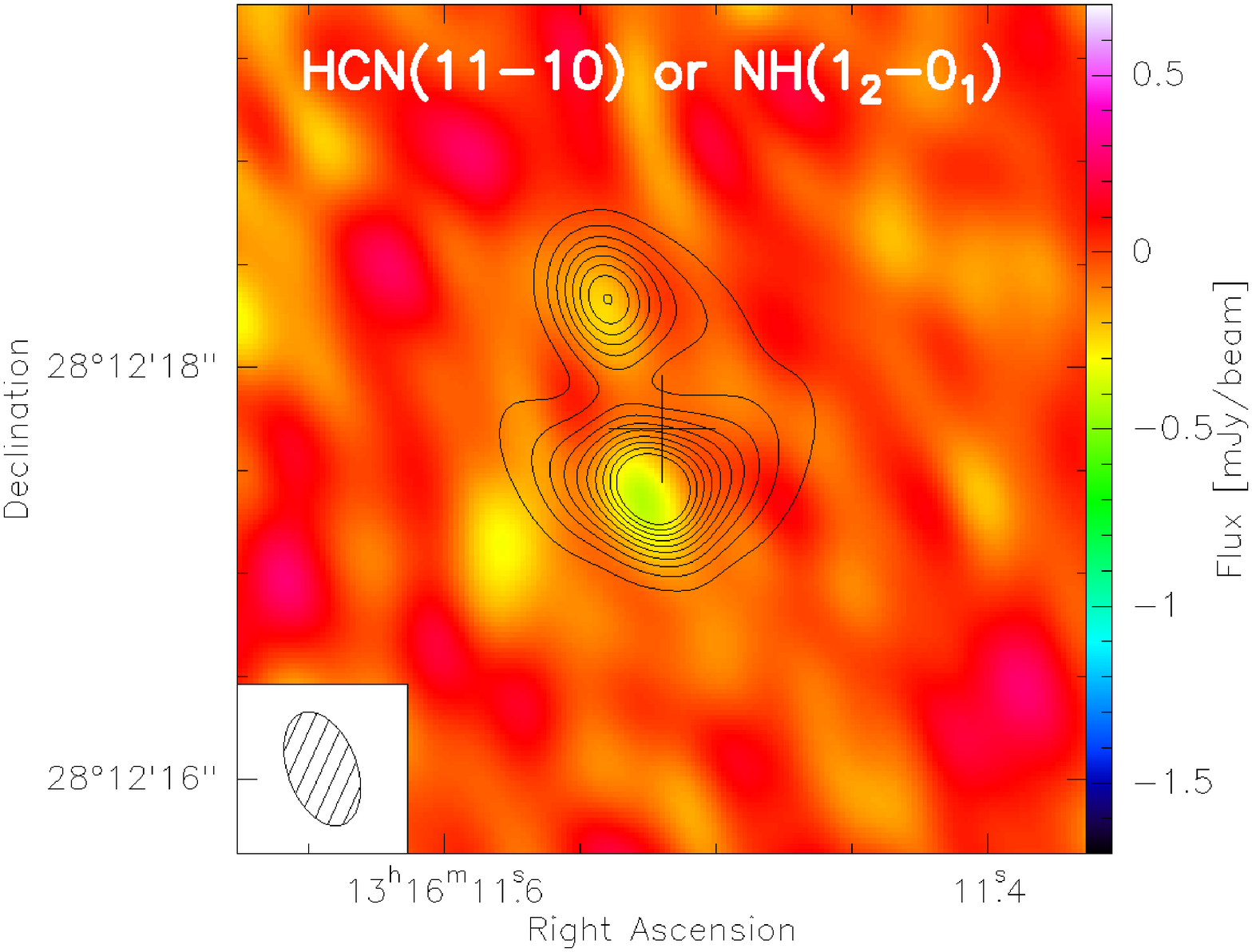}
\hspace{0.035\textwidth}
\includegraphics[width=0.3\textwidth]{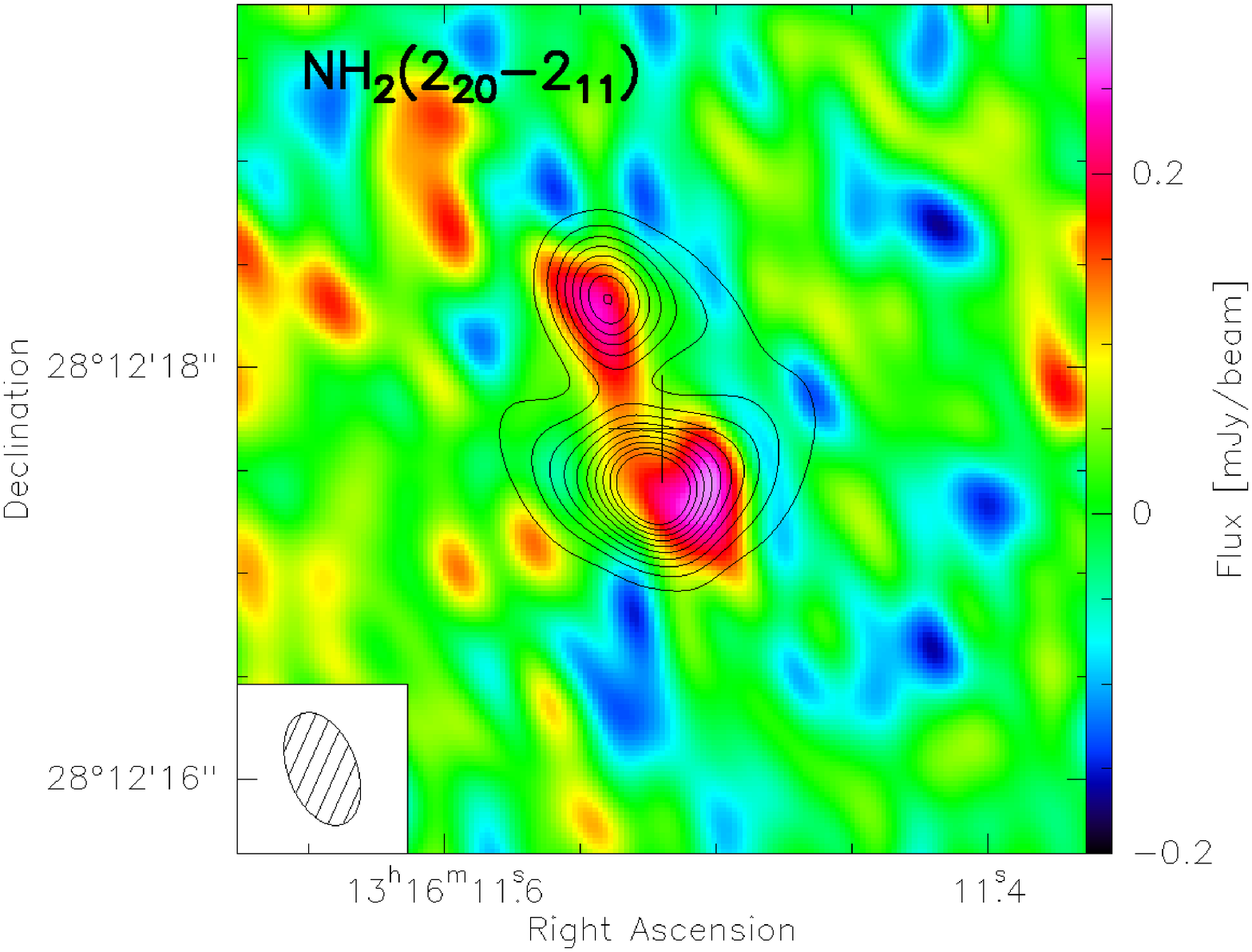}\\
\vspace{0.025\textwidth}
\includegraphics[width=0.3\textwidth]{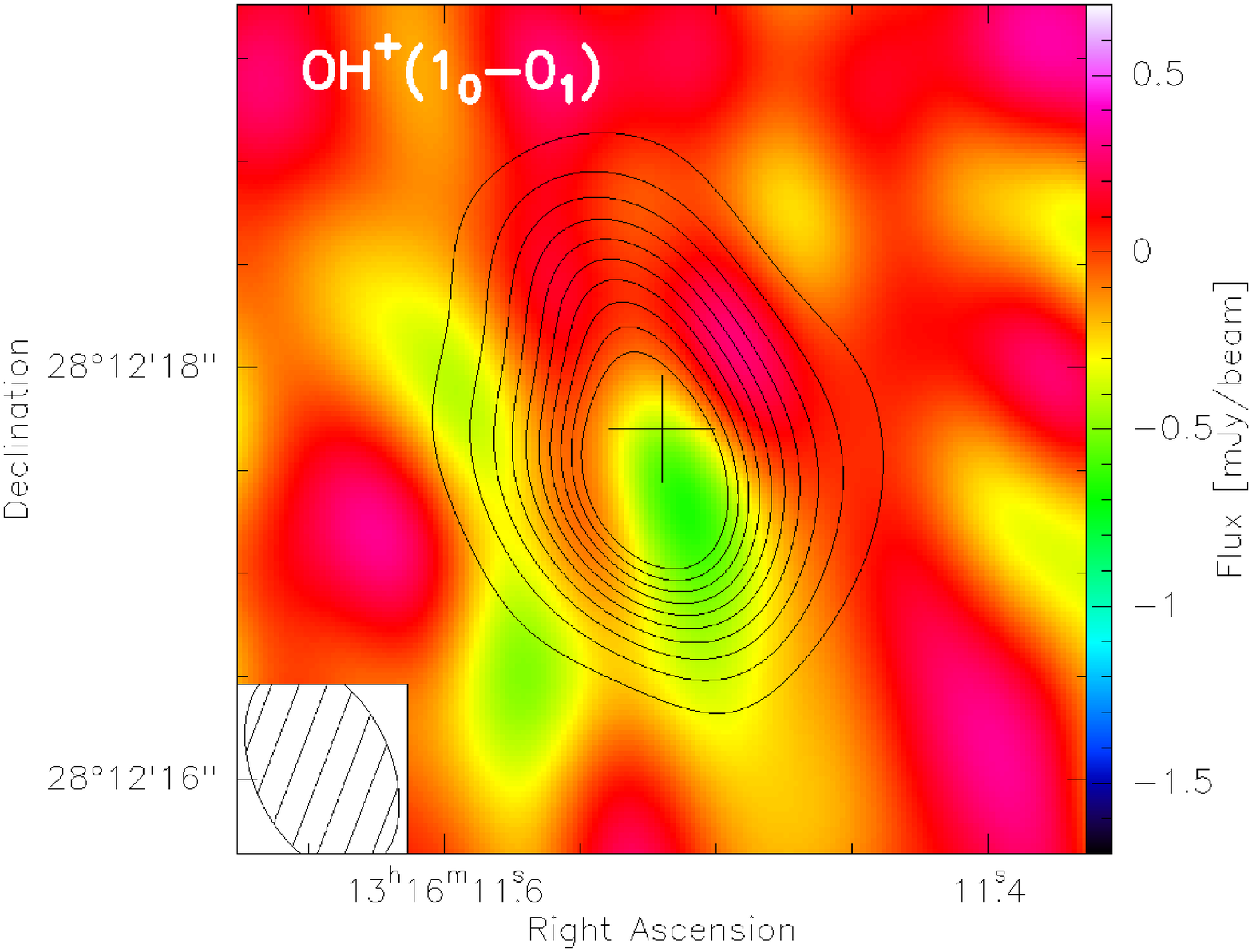}
\hspace{0.035\textwidth}
\includegraphics[width=0.3\textwidth]{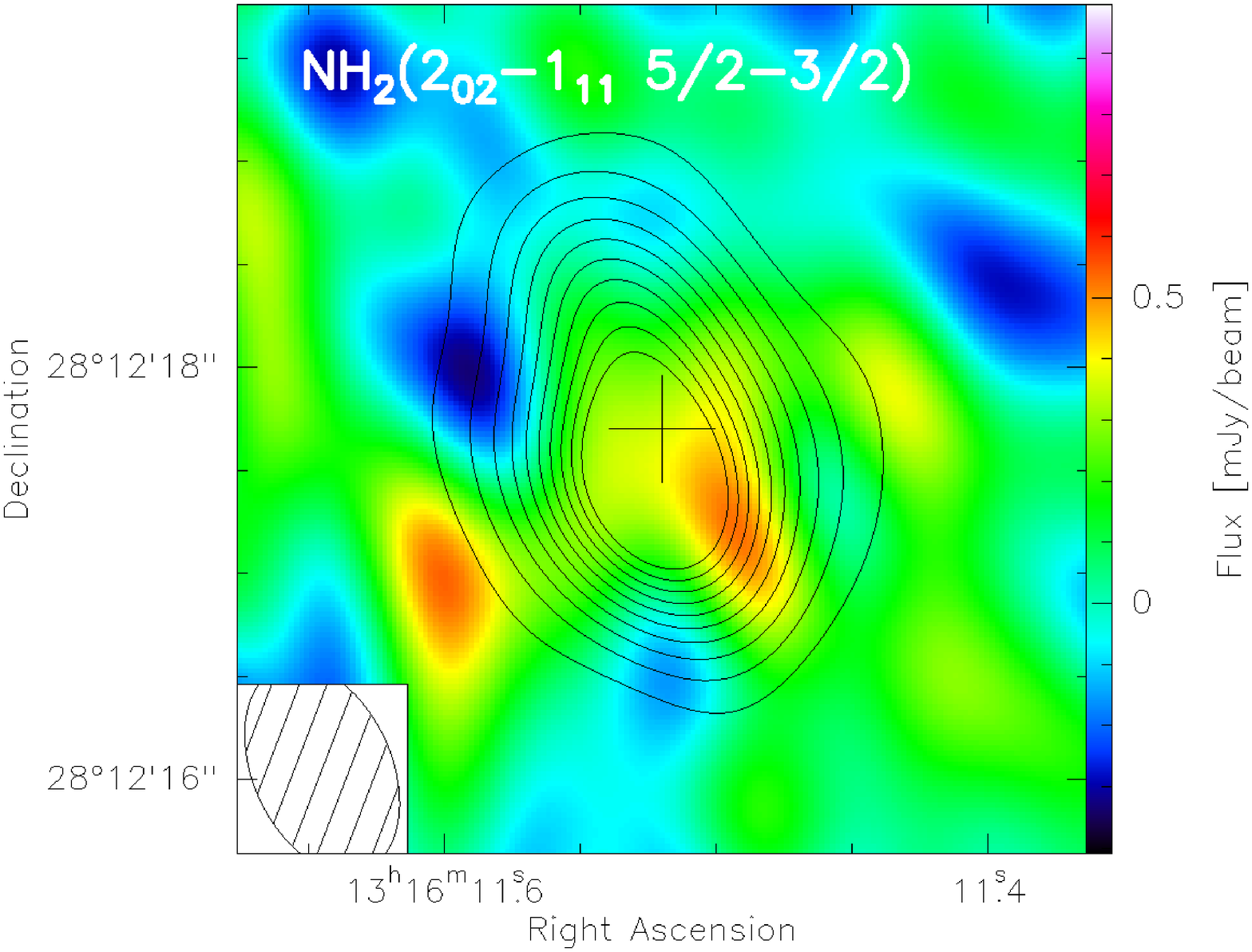}
\hspace{0.035\textwidth}
\includegraphics[width=0.3\textwidth]{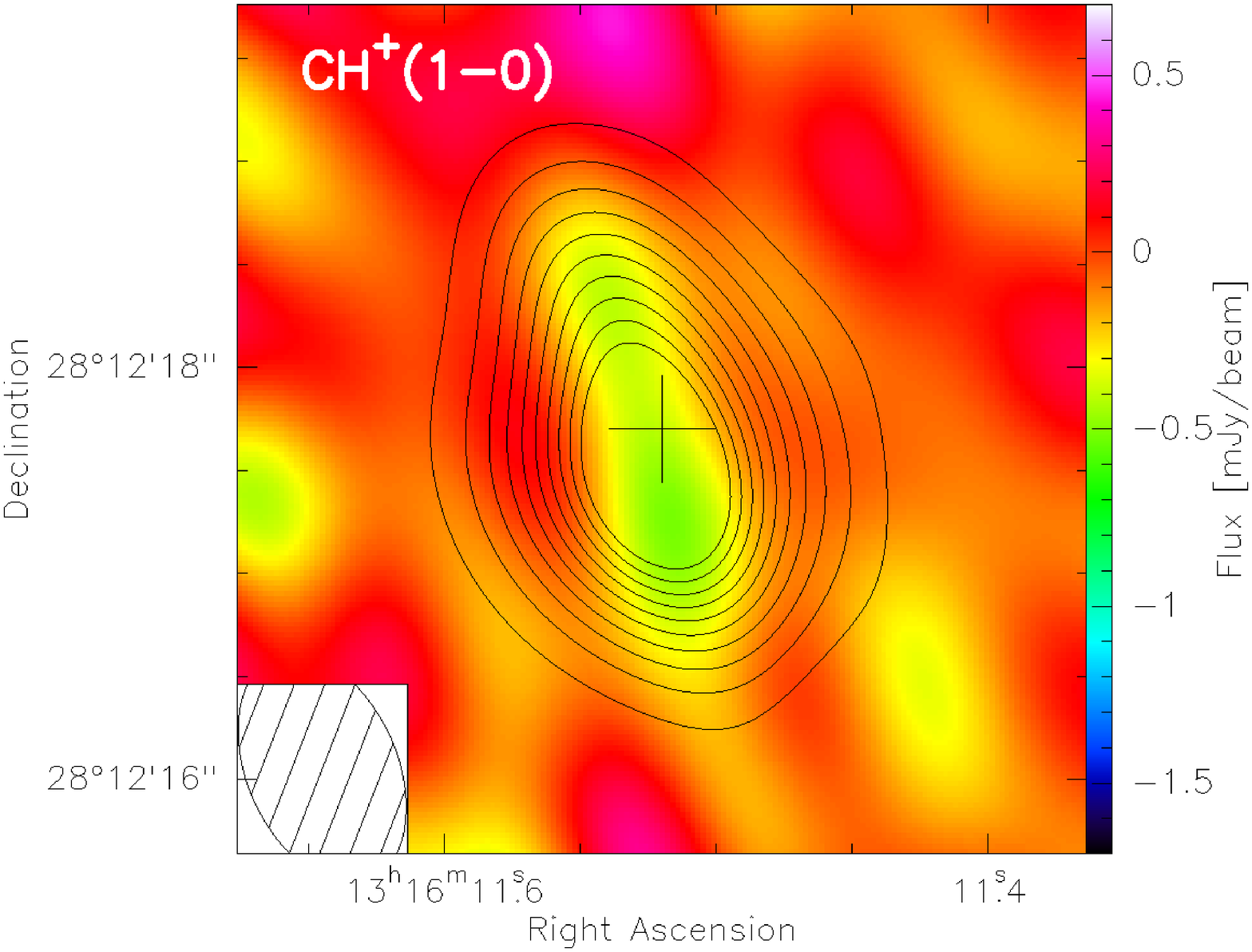}\\
\caption{Images of the molecular emission and absorption lines detected in
HerBS-89a, compared to the underlying continuum (natural weighting; shown as contours with 10$\sigma$ spacing) 
extracted in the respective side-band where each molecular line was detected (for each image the continuum has been 
subtracted from the line emission). The three upper rows display the high-angular 
resolution data, whereas the bottom row shows the lower-angular resolution images. The upper right panel presents 
the combined $\rm ^{12}CO(9-8)$ and $\rm H_2O(2_{02}-1_{11})$ image that enhances the low-level line emission, 
in particular along the western arc of the partial Einstein ring. The right panel in the second row displays the combined 
map of $\rm OH^+$ ($(1_1-0_1)$ and $(1_2-0_1)$) absorption lines. 
The synthesized beam is shown in the lower left corner of each of the images.}
\label{fig:overlay_lines_continuum}
\end{figure*}

\begin{figure*}[!ht]
\centering
\includegraphics[width=0.98\textwidth]{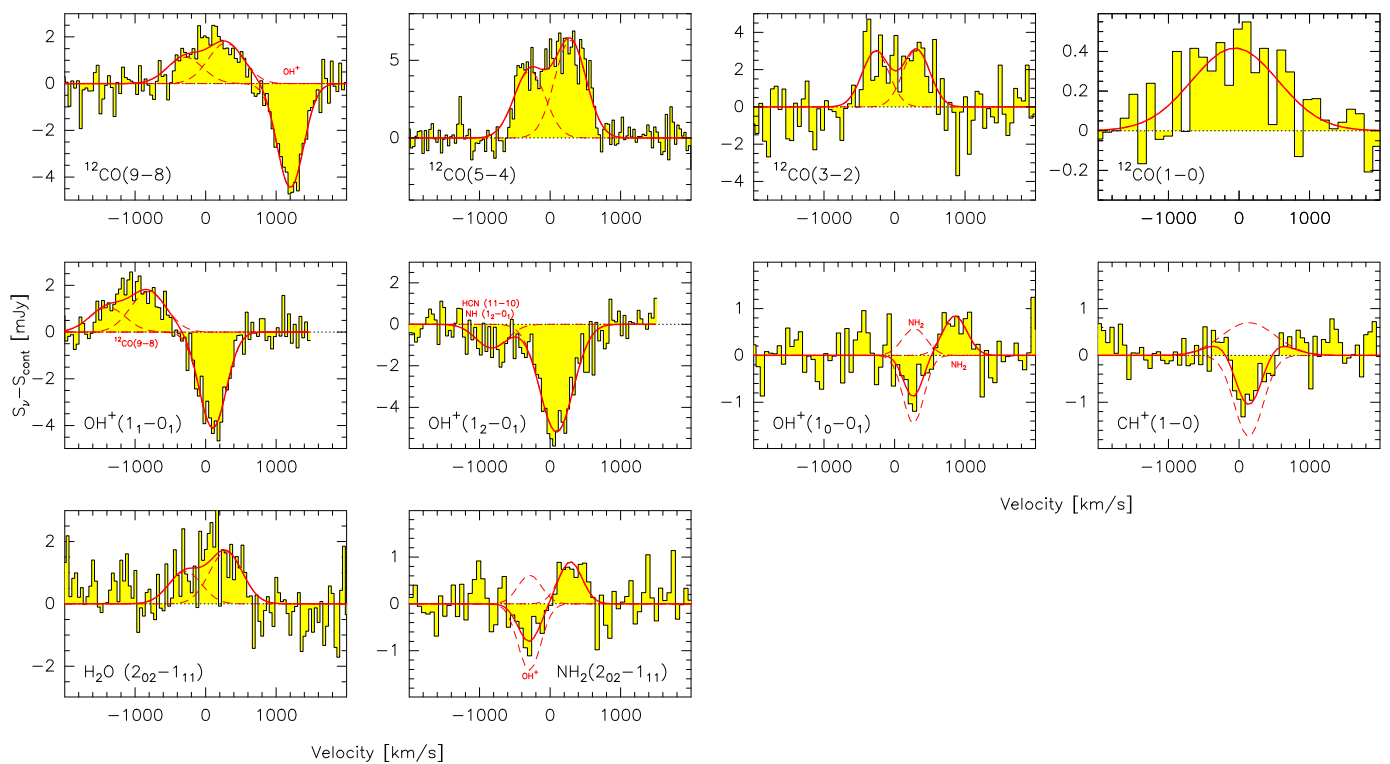}
\caption{The spectra of the molecular emission and absorption lines detected in HerBS-89a reported in this 
study, together with the $\rm ^{12}CO(5-4)$ and $\rm ^{12}CO(3-2)$ spectra from \citet{Neri2020}. 
Each molecular line is identified in the lower corner of each panel (labeled in black) and, in the case of a line 
overlap, i.e., $\rm ^{12}CO(9-8)$/$\rm OH^+(1_1-0_1)$, 
$\rm OH^+(1_2-0_1)$/$\rm HCN(11-10)$ and $\rm OH^+(1_0-0_1)$/$\rm o-NH_2 (2_{02}-1_{11})(5/2-3/2)$, the second molecular 
line is labelled in red. The spectra are displayed with the continuum subtracted and, in each panel, the molecular 
line (labeled in black) is plotted relative to the zero velocity corresponding to its rest frequency. 
Fits to the emission and absorption molecular lines are shown as solid lines, and, when multi-component and/or 
multiple lines have been fitted, they are displayed individually as dashed lines (see text for details).}
\label{fig:spectral_fit_GILDAS}
\end{figure*}

\subsubsection{Molecular emission and absorption lines}\label{sect:lines}
The large bandwidth of the NOEMA receivers allows us to search in HerBS-89a for 
redshifted emission and absorption lines 
over a wide range in frequency. The complete spectra of HerBS-89a, integrated over the areas subtended by the 
lines (see below) and normalized by the continuum, are displayed in 
Fig.~\ref{fig:spectra_cont_norm}. Each of the panels covers the frequency range of one sideband (LSB or USB) 
of one of the two observational projects discussed above, namely between 245 and 268~GHz (two upper panels) 
and between 209 and 233~GHz (two lower panels). Two strong molecular emission lines are present, 
$\rm ^{12}CO(9-8)$ and $\rm H_2O\,(2_{02}-1_{11})$, which both display wide profiles similar to those 
seen in the $\rm ^{12} CO(3-2)$ and $\rm ^{12}CO(5-4)$ emission lines \citep{Neri2020}. 
In addition, the three lines of the ground state of the molecular 
ion $\rm OH^+$ are all detected in absorption: the $\rm (1_1-0_1)$ line, which is adjacent in 
frequency to the $\rm ^{12}CO(9-8)$ emission line, and the 
$\rm (1_2-0_1)$ and $\rm (1_0-0_1)$ lines, which are detected for the first time in a high-$z$ galaxy. 
Blue-shifted from the strong $\rm OH^+(1_2-0_1)$ absorption line, another weaker absorption line is detected 
that is due to $\rm HCN(11-10)$ ($\rm \nu_{rest}=974.487$~GHz) and/or 
$\rm NH(1_2-0_1)$ ($\rm \nu_{rest}=974.471$~GHz) (see 
Sect.~\ref{sect:very-dense-gas} for a detailed discussion). 
Next to the $\rm OH^+(1_0-0_1)$ absorption line, another line is seen in emission, 
which corresponds to $\rm o\mhyphen NH_2 (2_{02}-1_{11}) (5/2-3/2)$ ($\rm \nu_{rest}=907.433$~GHz). 
A second $\rm NH_2$ line is detected with lower signal-to-noise ratio at higher frequency, 
the $\rm NH_2 (2_{20}-1_{11}) (5/2-3/2)$ transition 
at $\rm \nu_{rest}=993.3226$~GHz. Finally, the molecular ion $\rm CH^+$ is detected in the 
ground transition, $\rm CH^+(1-0)$; its profile is dominated by an absorption line that has a 
width similar to those of the $\rm OH^+$ absorption lines and at the same red-shifted velocity; 
in addition, a weak and broad ($\rm \sim 850 \, km\, s^{-1}$) emission component is likely present, 
of which the extreme red and blue wings are detected at low signal-to-noise at either side of the absorption line. 

Also identified in Fig.~\ref{fig:spectra_cont_norm} are the positions of red-shifted 
molecular lines that fall within the observed frequency range but remained undetected in HerBS-89a, including: 
$\rm SH^+(2-1)$ ($\rm \nu_{rest}=893.152$~GHz), $\rm o\mhyphen NH_2 (2_{02}-1_{11}) (3/2-1/2)$ 
($\rm \nu_{rest}=902.210$~GHz), $\rm H_2O^+ (3_{12}-3_{03})$ ($\rm \nu_{rest}=982.955$~GHz), and 
$\rm o\mhyphen H_3O^+ (0^-_0-1^+_0)$ ($\rm \nu_{rest}=984.708$~GHz), most of which are seen 
in the spectrum of the local merger Arp~220 \citep{Rangwala2011}.

Figure~\ref{fig:overlay_lines_continuum} shows the angular extents of the eight molecular emission and absorption lines 
detected with NOEMA in HerBS-89a, each compared to the continuum extracted in the respective 
sideband to which it belongs. The $\rm ^{12}CO(9-8)$ and $\rm H_2O(2_{02}-1_{11})$ emission lines
are detected in both lensed components of HerBS-89a, but are slightly shifted in position with respect to each other, 
with the water line being well centered on the continuum emission peaks. 
In order to recover more information on the distribution of the molecular emission lines, we combined the $\rm ^{12}CO(9-8)$ 
and $\rm H_2O(2_{02}-1_{11})$ velocity averaged images, under the assumption that they are probing roughly the same reservoir of 
high density and temperature molecular gas in HerBS-89a. The increased signal-to-noise ratio of the 
combined image (see upper right panel in Fig.~\ref{fig:overlay_lines_continuum}) reveals the molecular 
emission to the west, joining the southern and northern peaks, and resembling the 
image of the dust continuum emission. 

The two strong OH$^+$ absorption lines detected in our higher angular resolution data, namely the $\rm (1_1-0_1)$ and 
$\rm (1_2-0_1)$ transitions, also coincide with the southern and northern peaks of the continuum emission. 
The combined $\rm OH^+(1_1-0_1)$ and $\rm (1_2-0_1)$ map, integrated between -400 and +400~$\rm km \, s^{-1}$ 
from the line center in order to avoid contamination by neighboring lines (right panel in the second row of 
Fig.~\ref{fig:overlay_lines_continuum}), further highlights this spatial coincidence.

The faint absorption lines of our lower angular resolution data are detected mainly towards 
the southern component of HerBS-89a; the CH$^+$ absorption seems to be more extended than the $\rm OH^+$ absorption. 
The image of the $\rm HCN(11-10)$ and/or $\rm NH(1_2-0_1)$ 
absorption line shows two peaks exactly centered on the dust continuum emission peaks with no indication of emission 
extending further out. Finally, the $\rm NH_2 (2_{02}-1_{11}) (5/2-3/2)$ emission line peaks 
westward of the continuum peak, as expected for red-shifted gas (see below).

\begin{figure*}[!ht]
\centering
\includegraphics[width=0.45\textwidth]{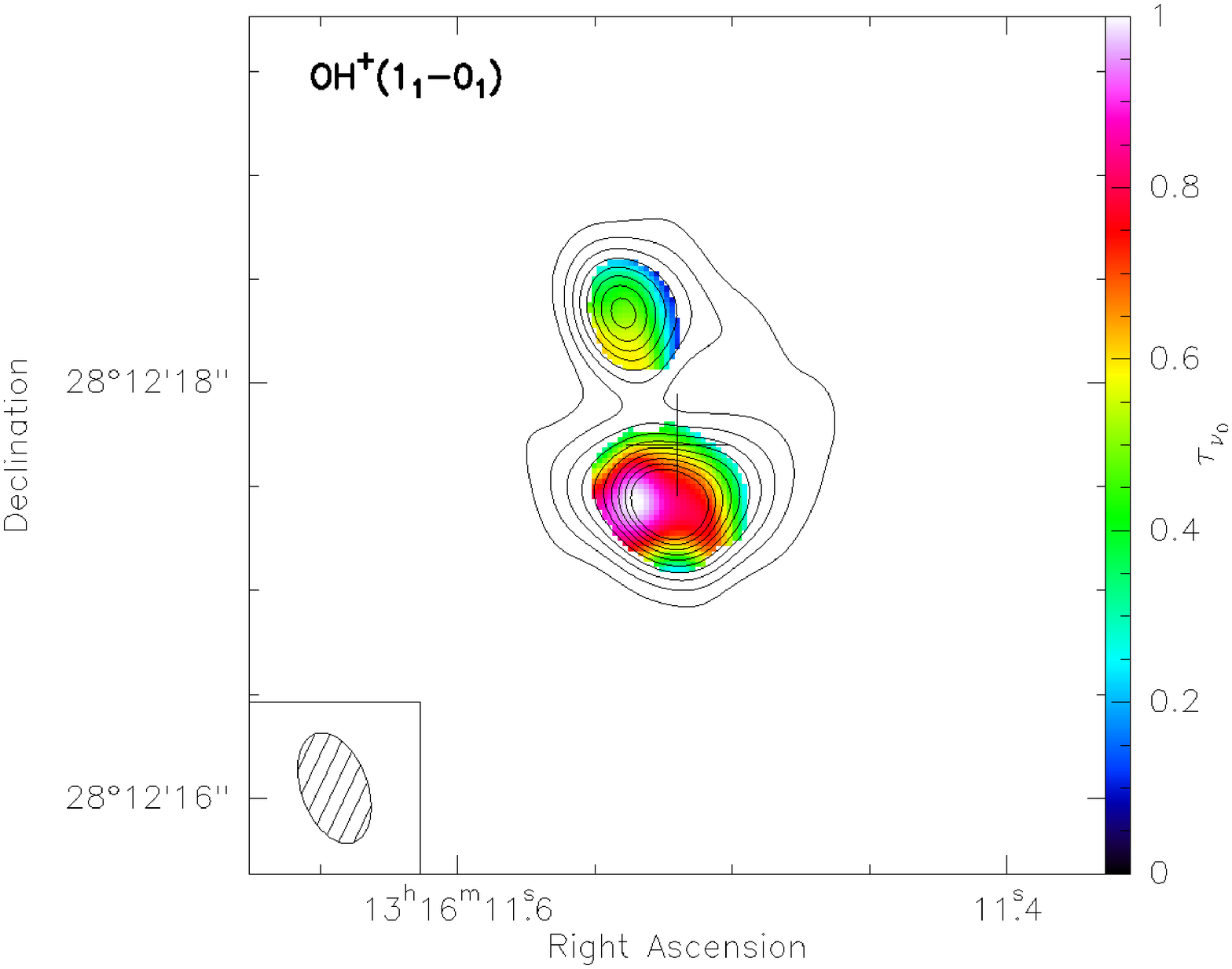}
\hspace{0.07\textwidth}
\includegraphics[width=0.45\textwidth]{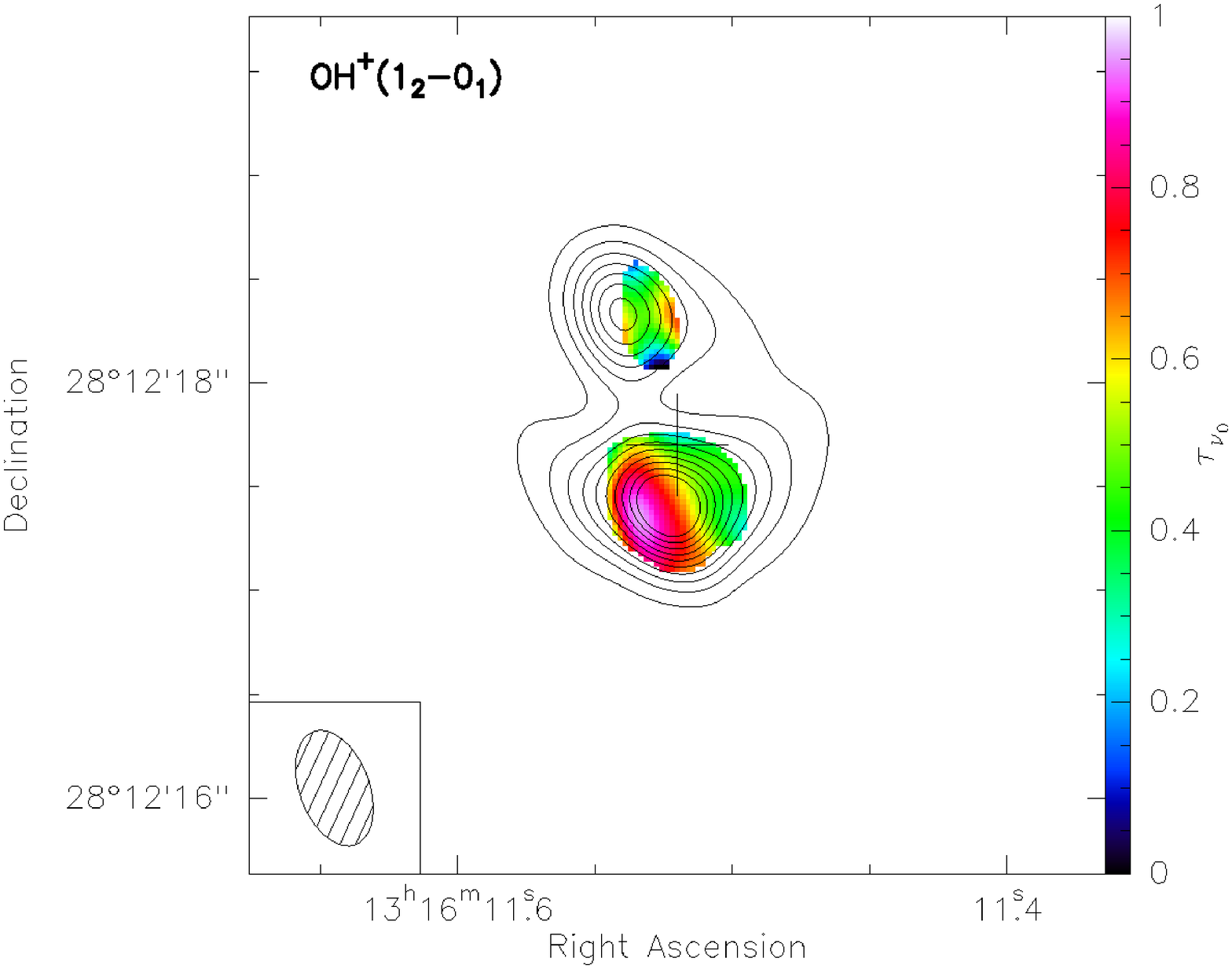}
\caption{Maps of the optical depth of the $\rm OH^+(1_1-0_1)$ and $\rm OH^+(1_2-0_1)$ absorption lines, evaluated as $F_\nu/F_\textrm{cont}=-e^{-\tau_\nu}$ at the central frequency $\nu_0$ of each line.}
\label{fig:abs_lines_tau_map}
\end{figure*}

\begin{figure*}[!ht]
\centering
\includegraphics[width=0.3\textwidth]{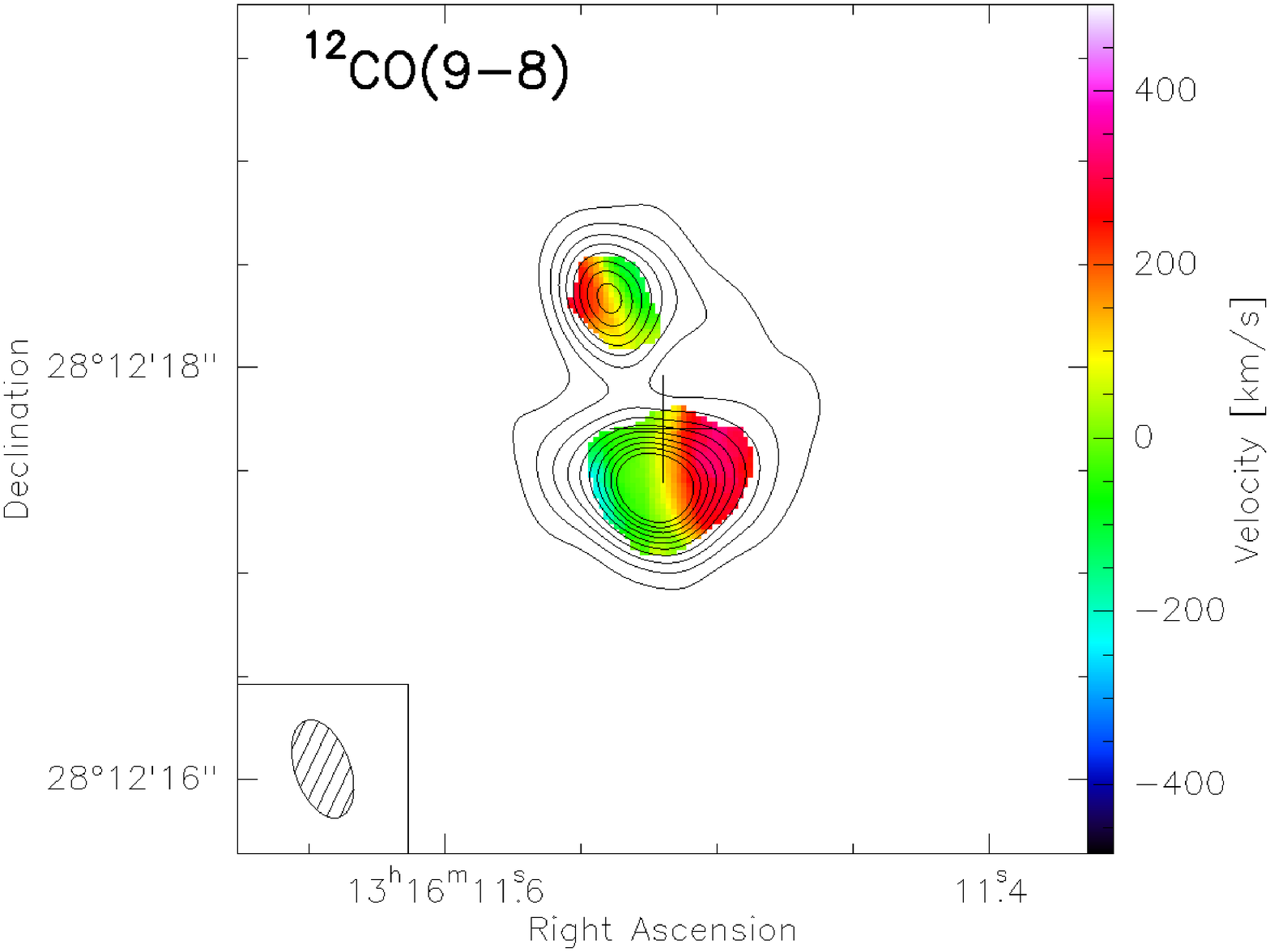}
\hspace{0.035\textwidth}
\includegraphics[width=0.3\textwidth]{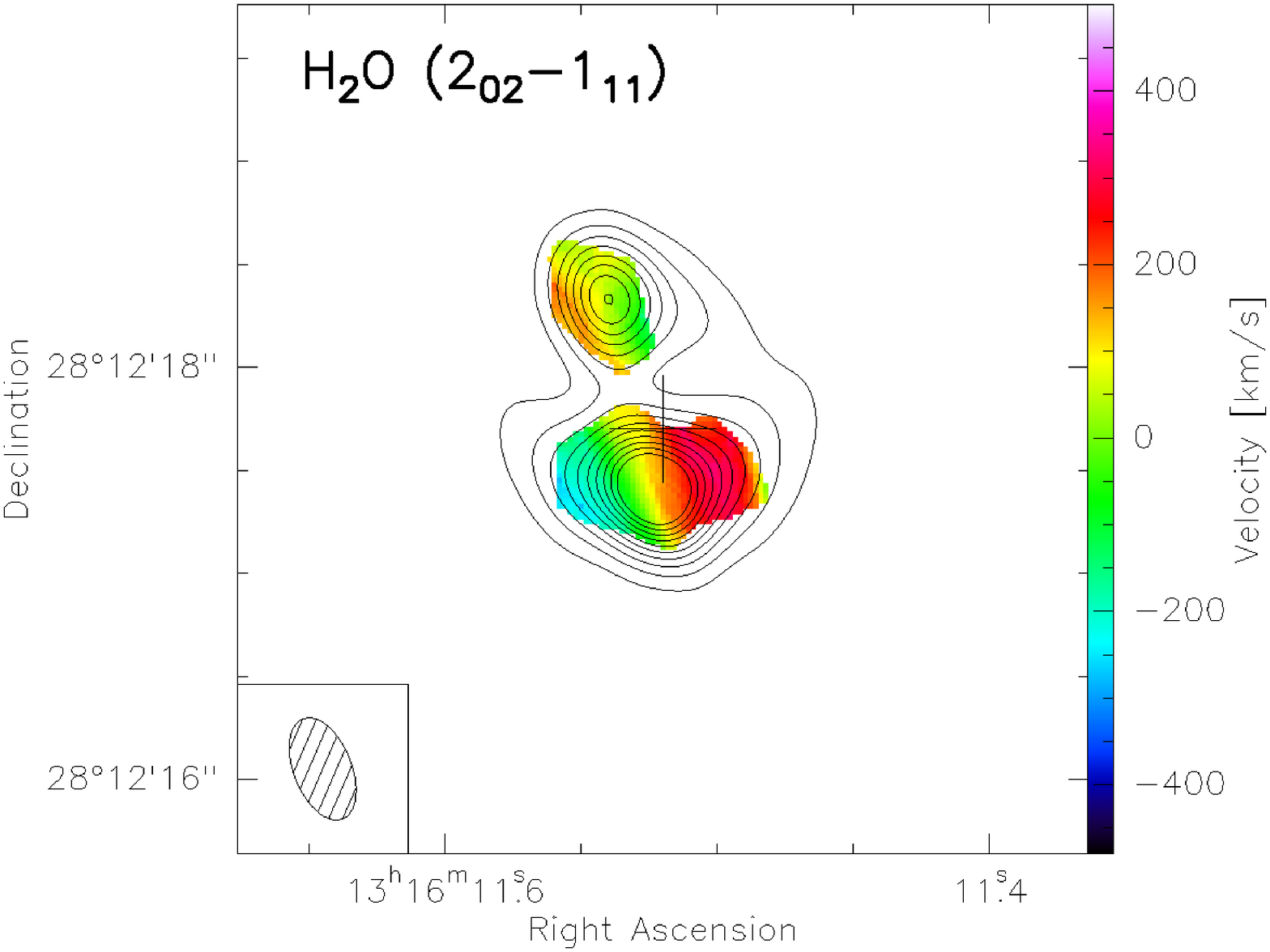}
\hspace{0.035\textwidth}
\includegraphics[width=0.3\textwidth]{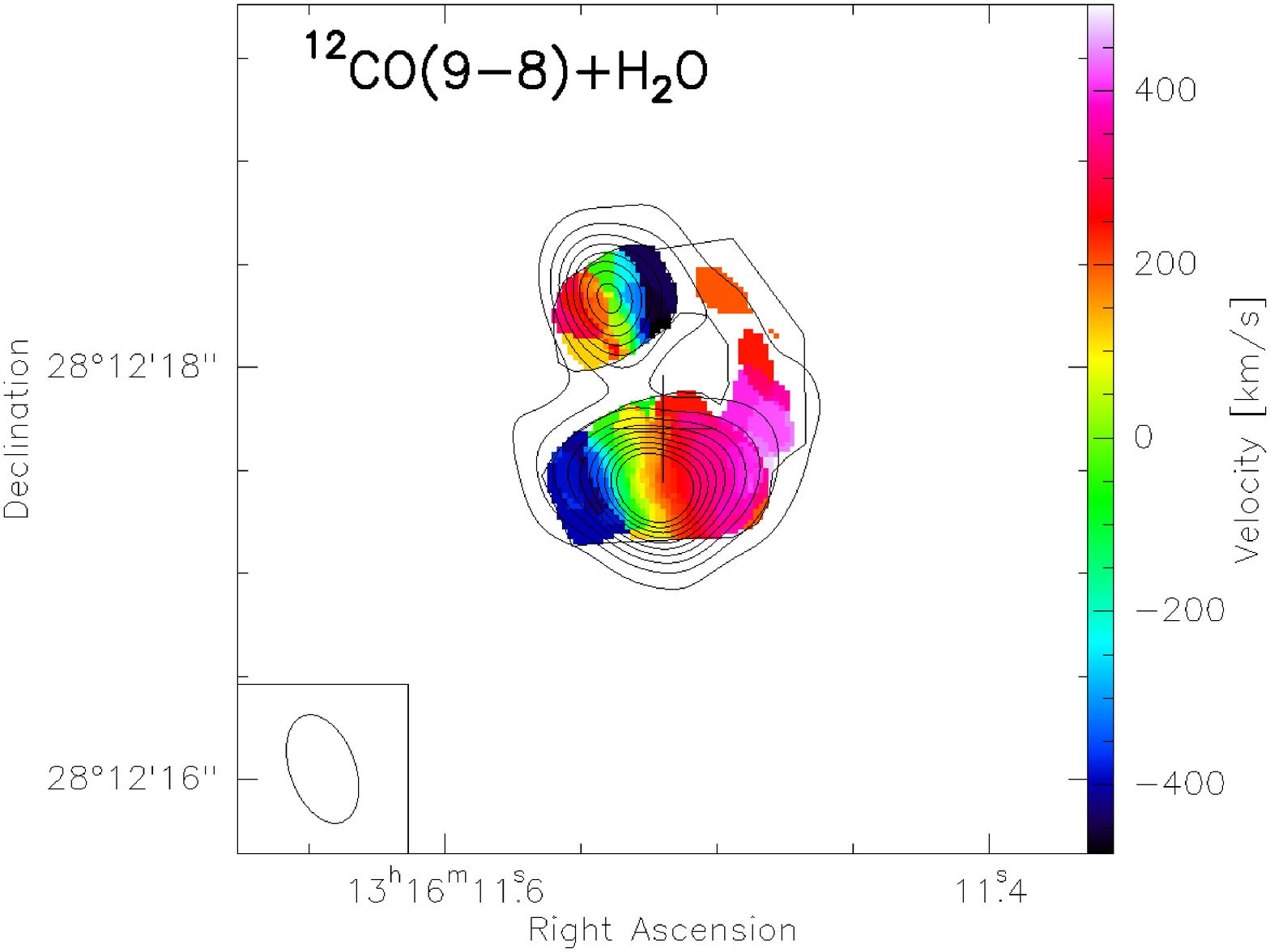}\\
\vspace{0.025\textwidth}
\includegraphics[width=0.3\textwidth]{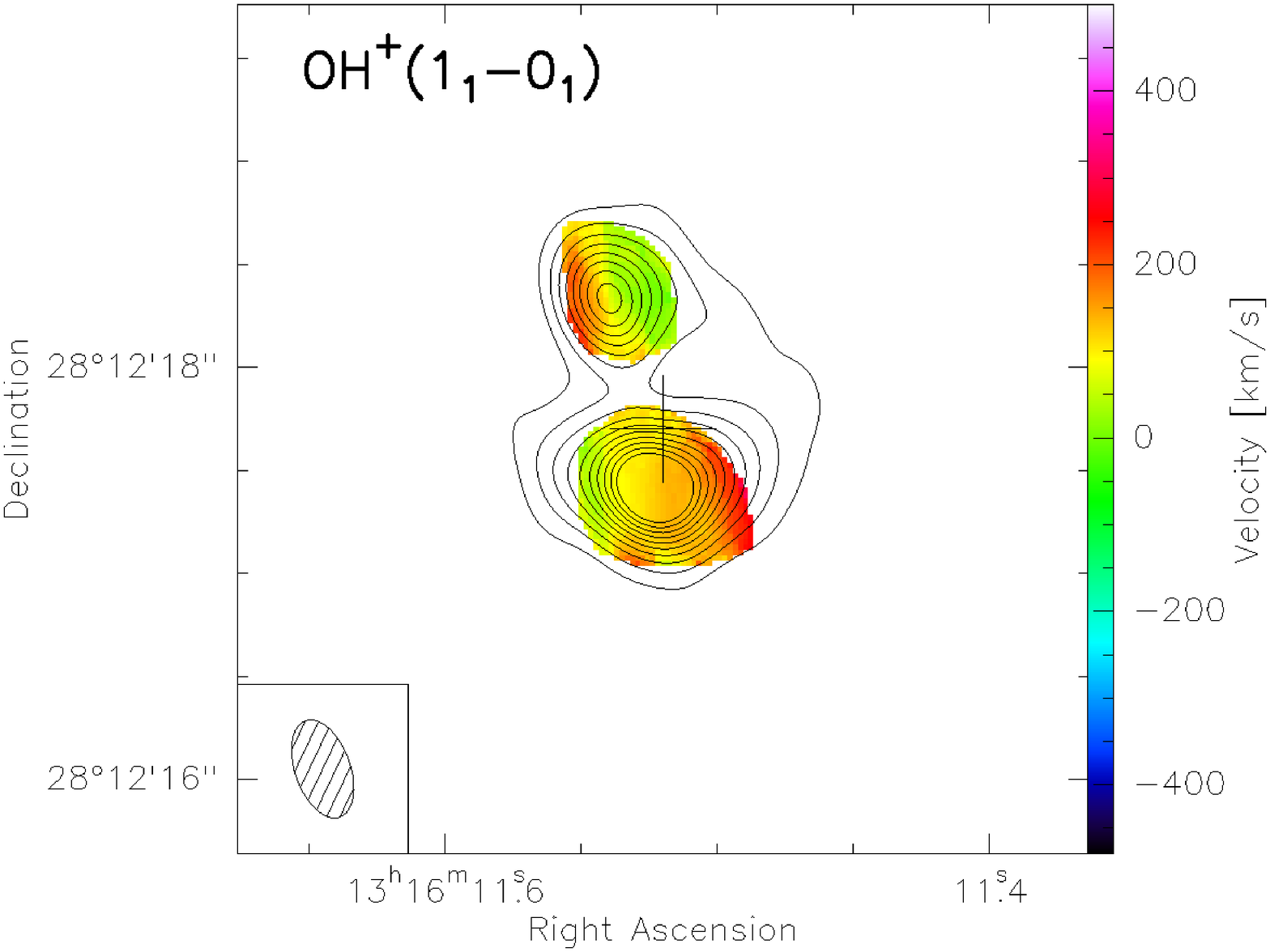}
\hspace{0.035\textwidth}
\includegraphics[width=0.3\textwidth]{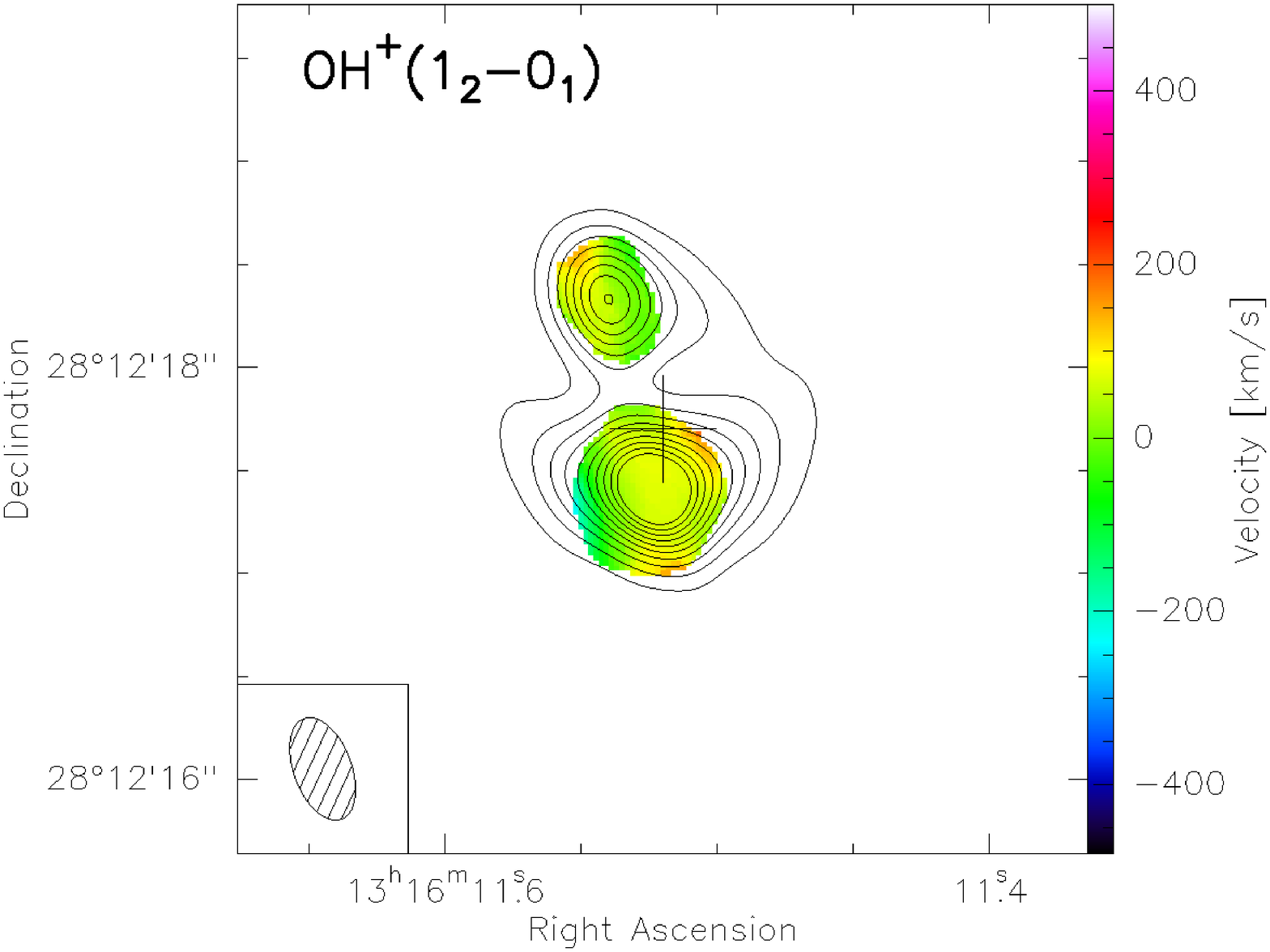}
\hspace{0.035\textwidth}
\includegraphics[width=0.3\textwidth]{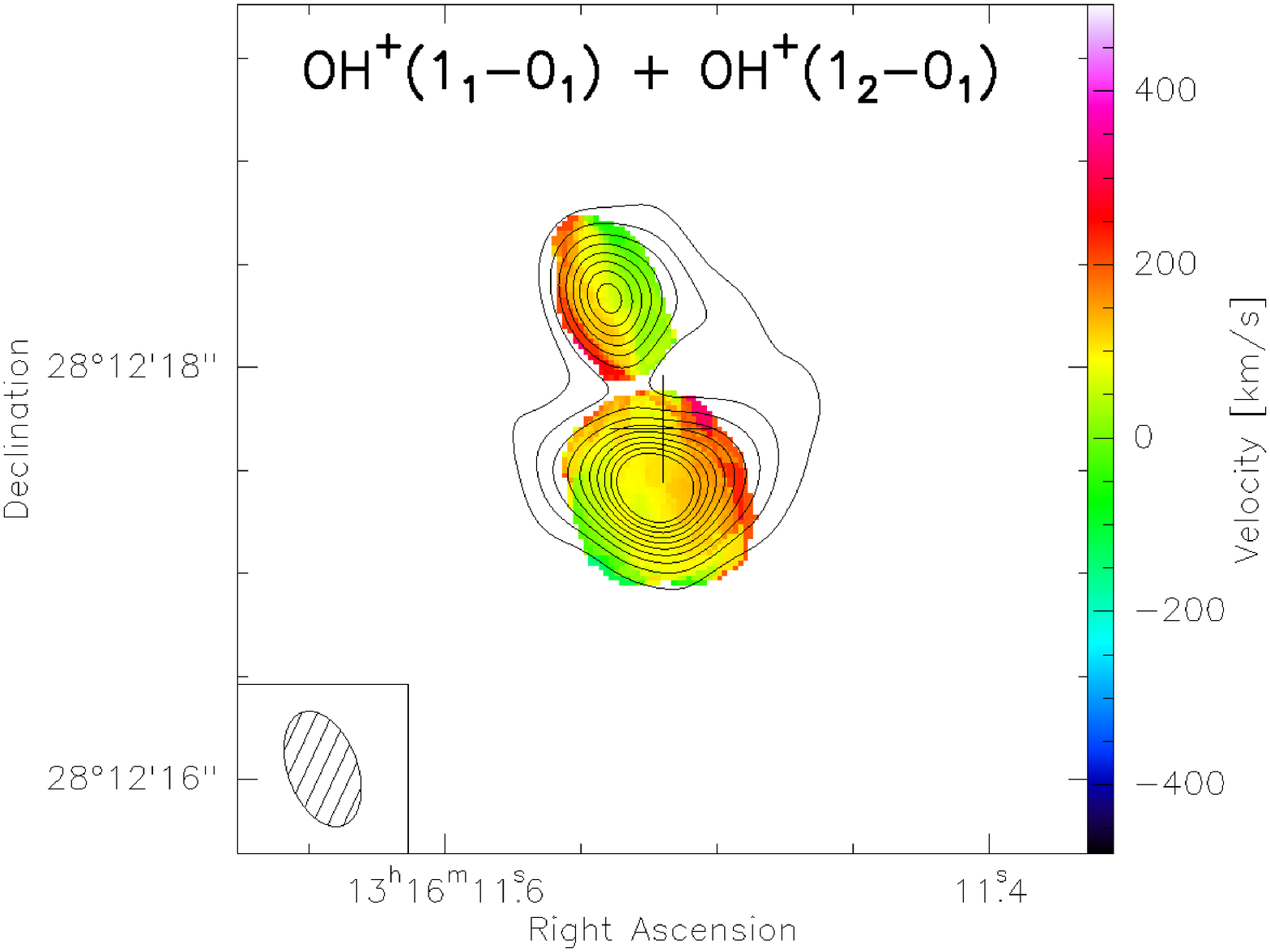}\\
\vspace{0.025\textwidth}
\includegraphics[width=0.3\textwidth]{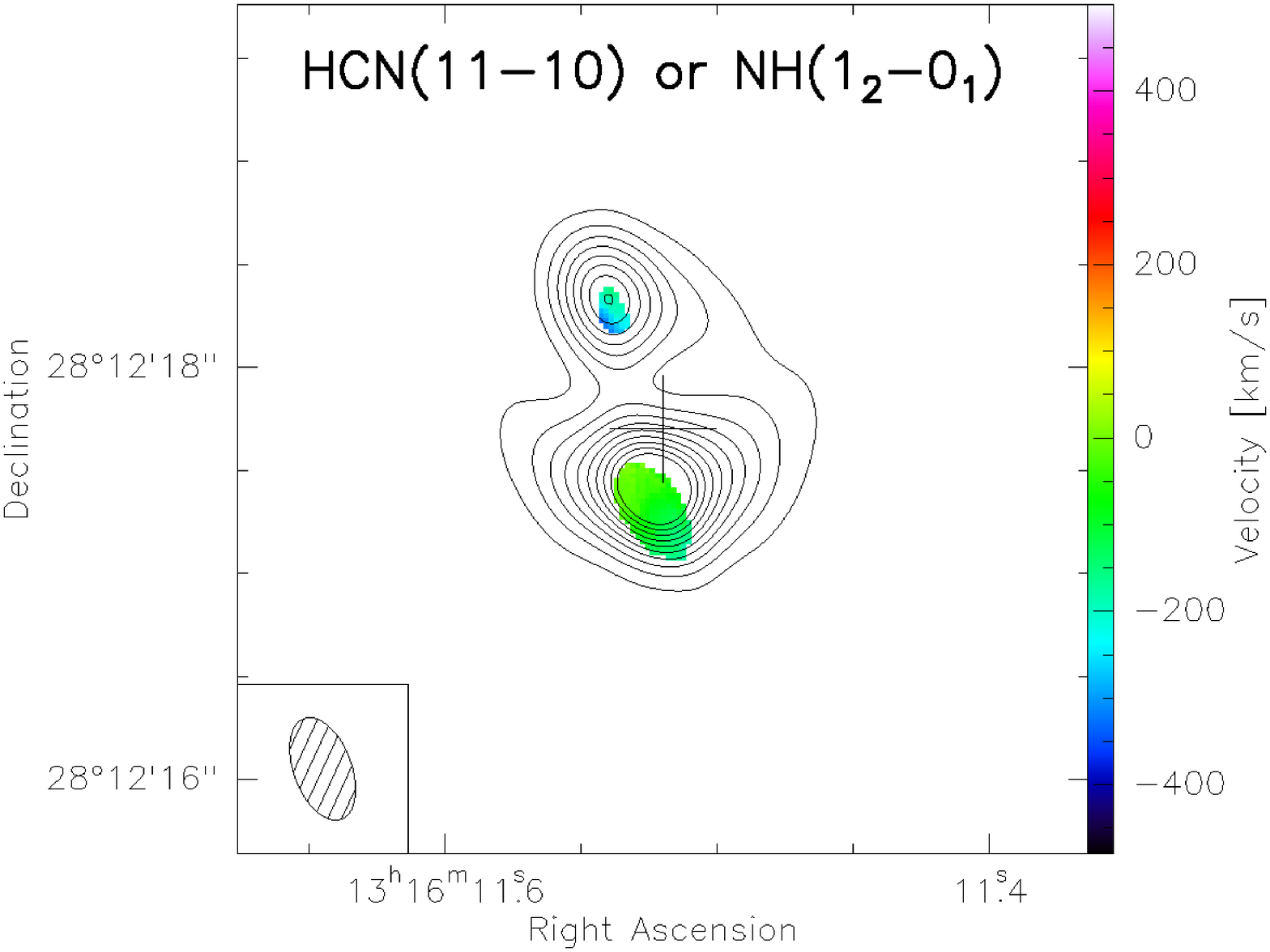}
\hspace{0.035\textwidth}
\includegraphics[width=0.3\textwidth]{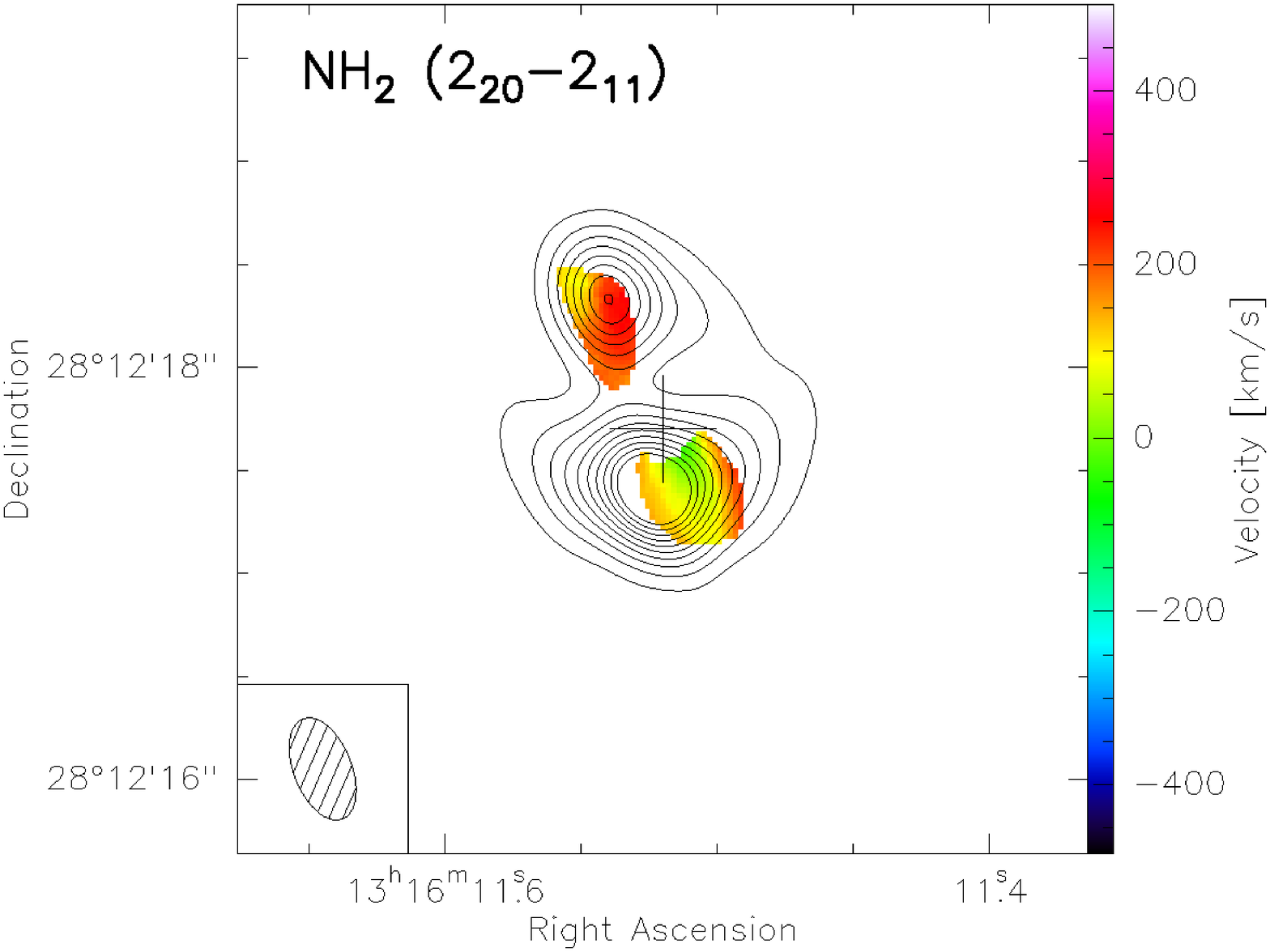}\\
\vspace{0.025\textwidth}
\includegraphics[width=0.3\textwidth]{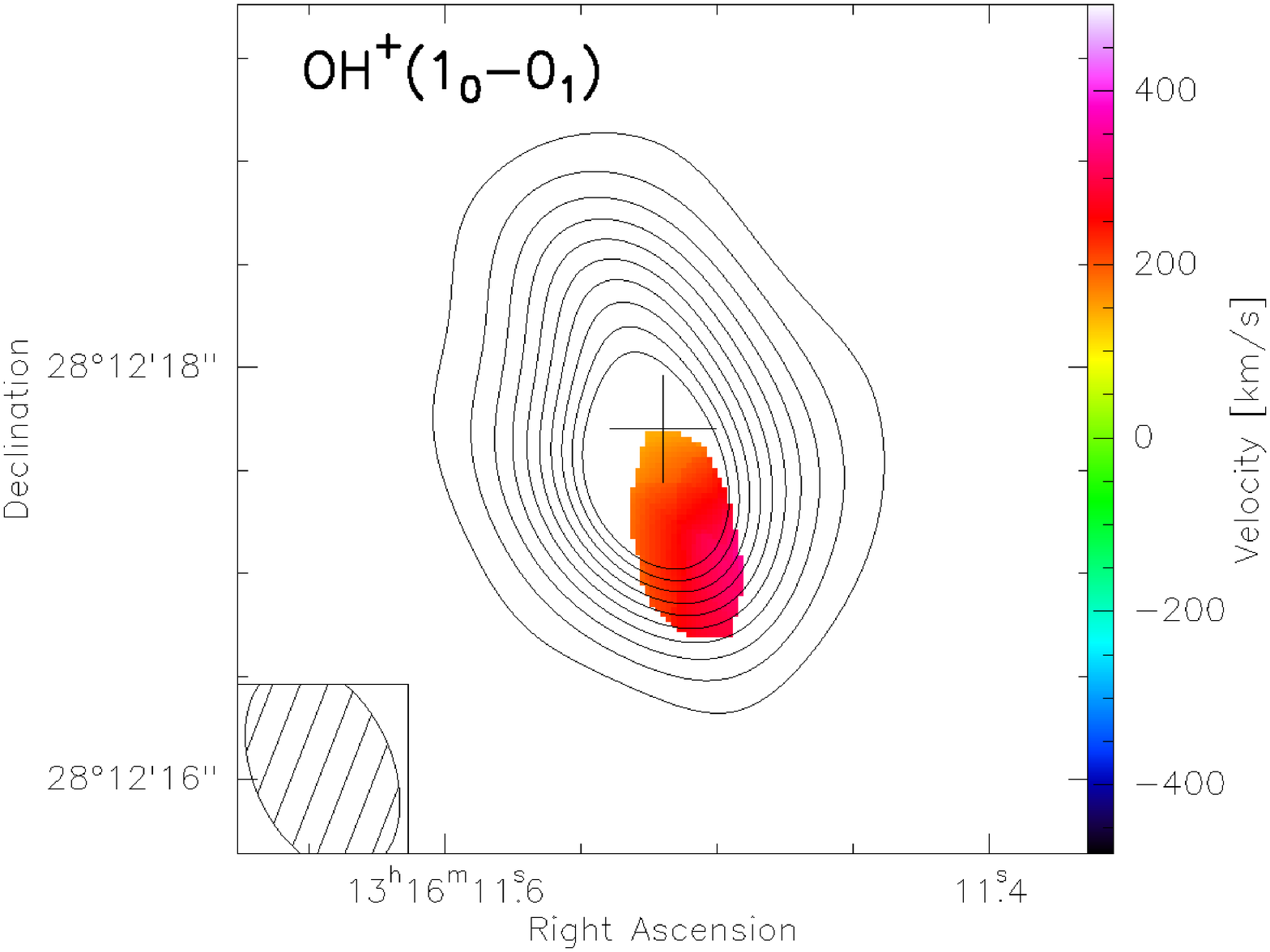}
\hspace{0.035\textwidth}
\includegraphics[width=0.3\textwidth]{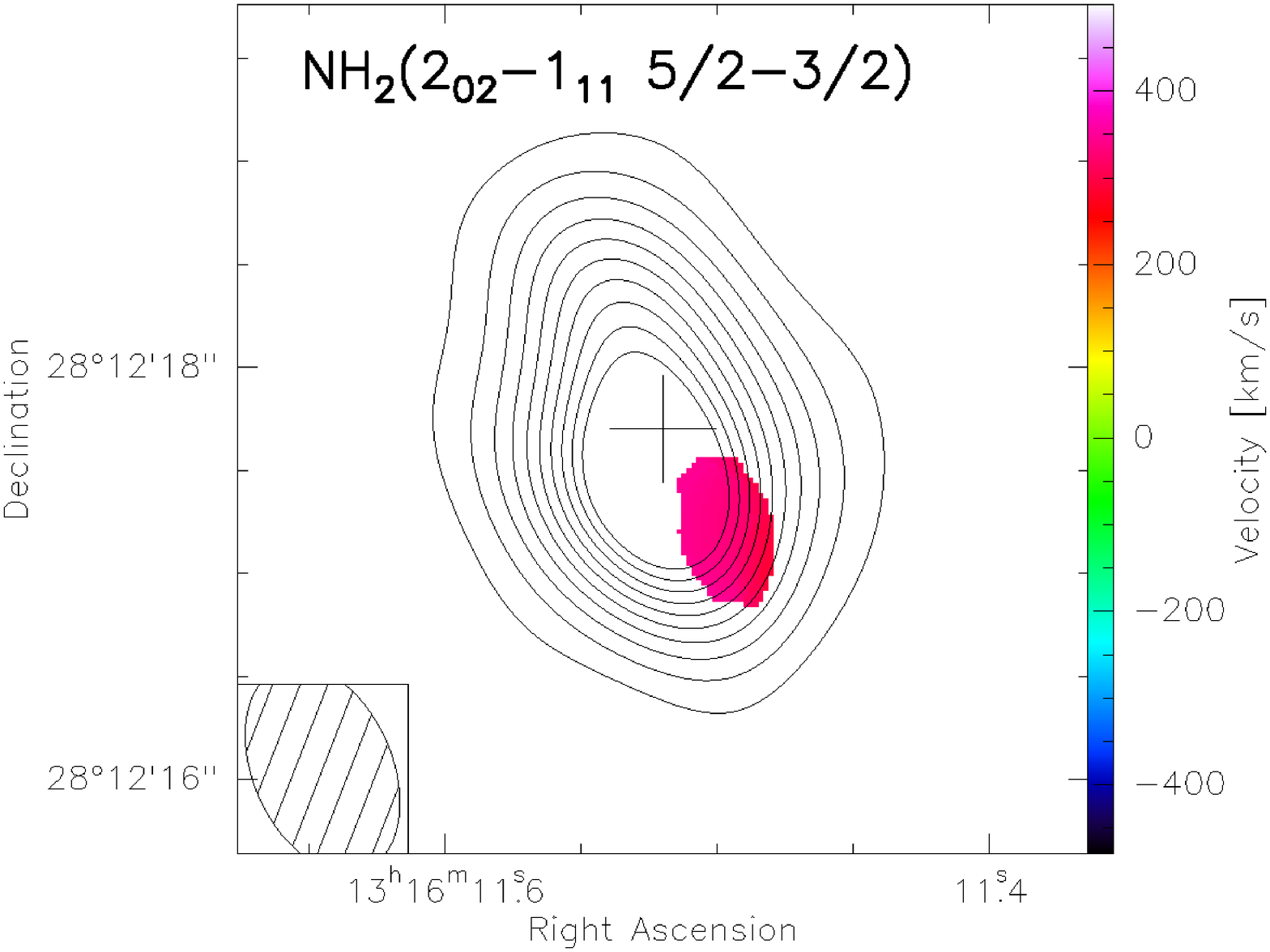}
\hspace{0.035\textwidth}
\includegraphics[width=0.3\textwidth]{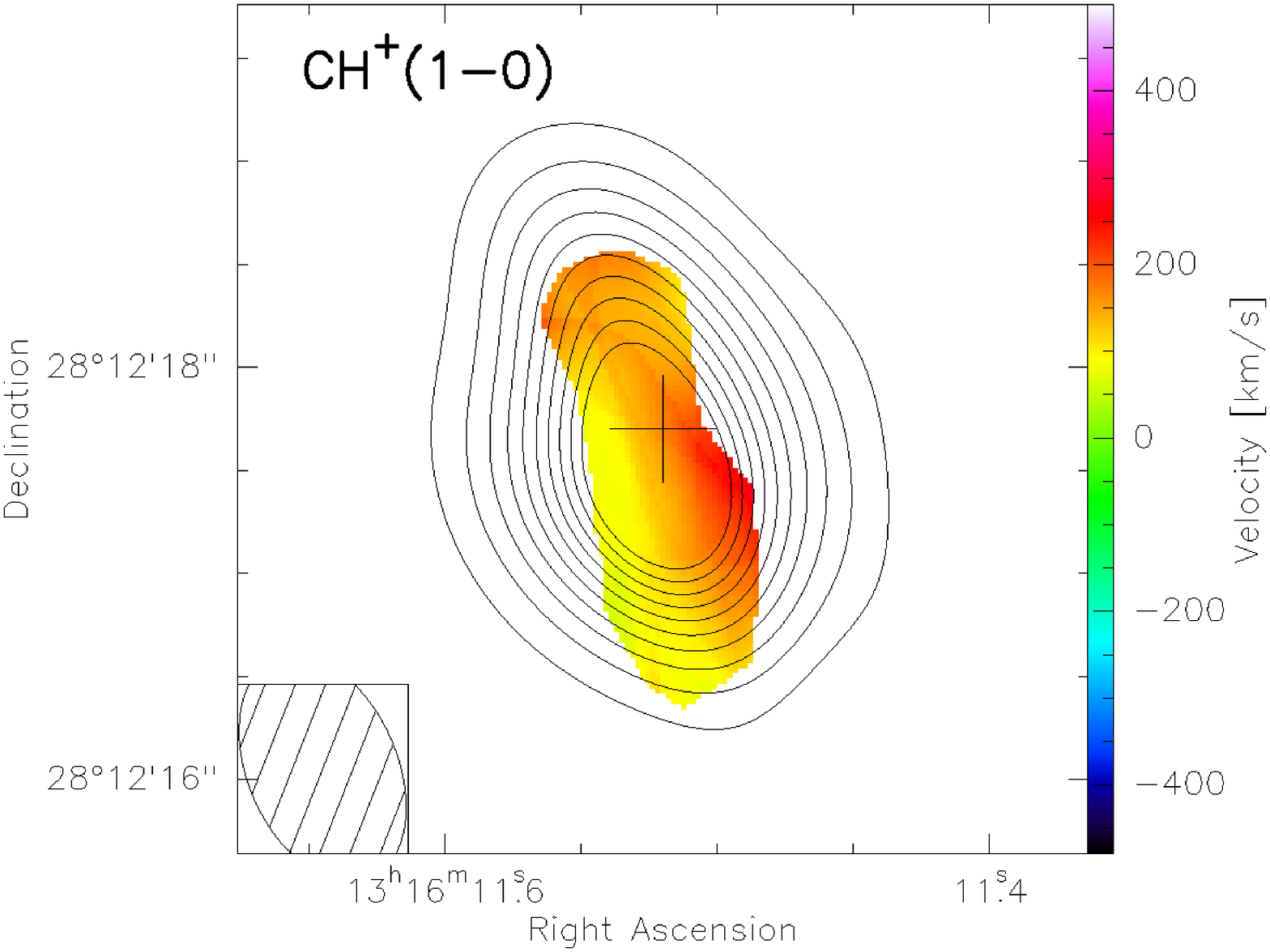}\\
\caption{Velocity field maps (shown in colors) in each of the molecular emission and absorption lines 
detected in HerBS-89a superimposed on the continuum at the individual frequencies of each molecular line 
(in contours of 10$\sigma$ spacing). The velocity maps were obtained above 3$\sigma$ thresholds in the 
zero-th moment map of the molecular lines. 
As in Fig.~\ref{fig:overlay_lines_continuum}, the three top rows display the high-angular resolution data, whereas the 
lower row shows the lower-angular resolution images. The upper right panel presents the combined $\rm ^{12}CO(9-8)$ 
and $\rm H_2O(2_{02}-1_{11})$ image that reveals the kinematics along the western arm of the partial Einstein ring. 
The right panel in the second row displays the combined velocity field of the two $\rm OH+$ absorption lines 
($(1_1-0_1)$ and $(1_2-0_1)$). The synthesized beam is shown in the lower left corner of each of the images.}
\label{fig:overlays_vel_cont}
\end{figure*}

In Fig.~\ref{fig:spectral_fit_GILDAS}, we show the integrated spectra of all the individual molecular emission and absorption lines detected in HerBS-89a, including the $\rm ^{12} CO(3-2)$ and $\rm ^{12} CO(5-4)$ emission 
lines reported in \citet{Neri2020}. 

For the two low-$J$ CO emission lines, \citet{Neri2020} fitted each of the double-peaked line profiles 
with two Gaussian components, with identical widths ($\rm FWHM = 631 \, km \, s^{-1}$)
and central velocities shifted by $\rm \pm 289 \,km\, s^{-1}$ from the nominal line frequency at the 
redshift of HerBS-89a ($z=2.9497$) (Table~\ref{tab:lines}). We adopt here the same approach for the 
$\rm ^{12}CO(9-8)$,  $\rm H_2O(2_{02}-1_{11})$ and $\rm NH_2 (2_{02}-1_{11}) (5/2-3/2)$ emission lines, 
fixing their component velocities to the values found for $\rm ^{12} CO(5-4)$ and $(3-2)$. We also force the two Gaussian 
components (of each line) to have the same width (see Table~\ref{tab:lines}). For the  
$\rm NH_2 (2_{20}-1_{11}) (5/2-3/2)$ emission line, the detection has too low S/N ratio to enable a 
reasonable fit and we therefore here report only upper limits to its line flux.

As mentioned above, the $\rm CH^+(1-0)$ line seems to display weak emission features on both sides 
of the absorption line. The fit to the profile shows the overlapping deep absorption feature and the less-well defined, broad component seen in emission. 

The $\rm OH^+ (1_2-0_1)$ absorption line displays strong, red-shifted absorption, whose 
profile is well reproduced by a narrow and deep absorption line, 
red-shifted by $\rm \sim 100 \, km\, s^{-1}$. Next to this $\rm OH^+$ absorption line, the broad weak 
absorption line, which corresponds to $\rm HCN(11-10)$ and/or $\rm NH(1_2-0_1)$, is centered at zero velocity and 
has a width of $\rm \sim 800 \, km\, s^{-1}$. As shown in Fig.~\ref{fig:overlay_lines_continuum}, 
this broad absorption line is very concentrated spatially, peaking on the northern and southern peaks of 
the dust continuum emission. 

The $\rm OH^+(1_0-0_1)$ and $\rm NH_2 (2_{02}-1_{11}) (5/2-3/2)$ lines are close to each other in frequency 
(i.e., 230.18 and 229.74~GHz, corresponding to a velocity difference of 
$\rm \Delta V \sim 150 \, km\, s^{-1}$). In the observed spectrum (Fig.~\ref{fig:spectral_fit_GILDAS})
there is an apparent shift for both species to the red by $\rm \sim 250 \, km\, s^{-1}$ 
with respect to their nominal frequencies. For the $\rm OH^+(1_0-0_1)$ absorption, this shift is larger than
for the two other $\rm OH^+$ transitions, and in the case of $\rm NH_2$, the profile seems 
to be inconsistent with those seen for other molecular emission lines tracing dense gas (CO and $\rm H_2O$). 
Since $\rm NH_2$ (amidogen) is a photodissociation product of $\rm NH_3$ (ammonia), it 
should probe the dense gas traced in the high-$J$ CO or water lines, and therefore display 
a similar double Gaussian profile. However, we only see the red-shifted component of the 
double-peaked line profile, most likely because the blue-shifted part of the $\rm NH_2$ emission 
line has been entirely suppressed by the $\rm OH^+$ absorption line. 
We have therefore fitted simultaneously the overlapping $\rm OH^+$ absorption and $\rm NH_2$ emission lines 
using three Gaussian profiles, namely: i) two components for $\rm NH_2$, where the velocities are 
derived from  the $\rm ^{12}CO(5-4)$ and $\rm ^{12}CO(3-2)$ emission lines and the ratio 
of the blue and red peaks similar to that of the
double Gaussian profile of the $\rm ^{12}CO(5-4)$ emission line; ii) one Gaussian for the $\rm OH^+$ 
absorption line whose width and the center velocity are derived from 
the narrow and deep $\rm OH^+(1_2-0_1)$ absorption line. 
The fit shown in Fig.~\ref{fig:spectral_fit_GILDAS} reproduces 
very well the observed spectrum including the velocity shift described above. 

Table \ref{tab:lines} summarizes the results of the spectral line fitting. For each molecular line, the main properties 
of the best Gaussian fits are listed, including the central velocity and frequency, 
the full width at half maximum (FWHM), the integrated intensity, and the intrinsic frequency. 
All the uncertainties are propagated into the derived quantities.

\begin{table*}
\caption{Summary of molecular emission and absorption line properties in HerBS-89a, obtained by the line fitting.} 
\centering
\begin{tabular}{l r@{$-$}l r c r@{.}l r@{$\pm$}l r@{.}l r@{.}l}
\hline	
\hline
Molecule	& \multicolumn{2}{c}{Transition}		& \multicolumn{1}{c}{$\nu_\textrm{rest}$}	& \multicolumn{1}{c}{$\nu_\textrm{obs}$}	& \multicolumn{2}{c}{$\rm V$}		& \multicolumn{2}{c}{FWHM}	& \multicolumn{2}{c}{$I_\textrm{line}$}	& \multicolumn{2}{c}{$\mu L'_\textrm{line}$}		\\
	&	&		& \multicolumn{1}{c}{$[$GHz$]$}		& \multicolumn{1}{c}{$[$GHz$]$}		& \multicolumn{2}{c}{$[$km s$^{-1}]$}	& \multicolumn{2}{c}{$[$km s$^{-1}]$}	& \multicolumn{2}{c}{$[$Jy km s$^{-1}]$} & 
	\multicolumn{2}{c}{[$\rm 10^{10} \, K \, km\, s^{-1} pc^{-2}$]}	\\
\hline	
\multicolumn{13}{c}{Emission Lines} \\
$^{12}$CO & $1$&$0$	     &  115.271 &  29.186	& -57&6$\pm$119.5 & 1443&293     & 0&64$\pm$0.13 & 25&76$\pm$5.23	 \\ 
       & $3$&$2$           &  345.796  &  87.549  &  0&0        & 1102&83      & 4&00$\pm$0.60 & 17&75$\pm$2.66    \\ 
       & $5$&$4$           &  576.268  & 145.902  &  0&0        & 1028&107     & 8&40$\pm$0.80 & 13&42$\pm$1.28    \\ 
       & $9$&$8$  	     & 1036.912 & 262.530	&  0&0 		  & 1168&94	     & 1&78$\pm$0.21 & 0&88$\pm$0.10	  \\ 
H$_2$O & $2_{02}$&$1_{11}$ &  987.927 & 250.127	&  0&0 		  & 1128&174	 & 1&59$\pm$0.37 & 0&86$\pm$0.20	  \\ 
NH$_2$ & $2_{02}$&$1_{11}^{\dagger}$ & 907.433 & 229.747 & 0&0 & 972&237$^\ast$ & 0&64$\pm$0.27$^\ast$ &	0&41$\pm$0.17 \\ 
       & $2_{20}$&$2_{11}^{\dagger}$ & 993.322 & 251.493 & 0&0 & \multicolumn{2}{c}{}    & $<$0&33   &	\multicolumn{2}{c}{} \\ 
\hline
\multicolumn{13}{c}{Absorption Lines} \\
OH$^+$ & $1_1$&$0_1$ 	     & 1033.118 & 261.546	& $+$102&2$\pm$13.0 & 447&32	& -1&96$\pm$0.14  \\	
       & $1_2$&$0_1$		 &  971.804 & 246.026	&  $+$91&6$\pm$20.1 & 558&51	& -3&08$\pm$0.28  \\	
       & $1_0$&$0_1$		 &  909.159 & 230.132	& $+$267&6$\pm$32.2 & 349&121	& -0&28$\pm$0.15$^{\ast\ast}$   \\	
CH$^+$ & $1$&$0$			 &  835.137 & 211.419	& $+$137&6$\pm$12.6 & 354&108$^{\ast\ast\ast}$	& -0&39$\pm$0.09$^{\ast\ast\ast}$	\\	
HCN     & $11$&$10^{\ddagger}$        &  974.487 & 246.724  & 0&0                 & 523&239 &  -0&63$\pm$0.28  \\	
NH      & $1_2$&$0_1^{\ddagger}$      &  974.471  & 246.720  & 0&0                 & 523&239 &  -0&63$\pm$0.28  \\	
\hline
\end{tabular}
\tablefoot{The data for the $\rm ^{12}CO(3-2)$ and $\rm ^{12}CO(5-4)$ emission lines are from \cite{Neri2020}. 
The fitting of each line is shown in Fig.~\ref{fig:spectral_fit_GILDAS}. Note that when one parameter is quoted without 
uncertainty, its value was fixed during the fitting procedure. The observed line luminosities $\mu L'_\textrm{line}$ 
have been derived using the standard relation given by \cite{Solomon-VandenBout2005} - 
see Sect.~\ref{sect:molecular-gas-mass-dust} and Eq.~\ref{eq:L'_line}. \\
$^\ast$ Combination of the two Gaussian components.\\
$^{\ast\ast}$ Only the visible red component. \\
$^{\ast\ast\ast}$ Net absorption only. \\
$^{\dagger}$ The two $\rm NH_2$ transitions are $(2_{02}-1_{11})\, (5/2-3/2)$ and $(2_{20}-2_{11})\, (5/2-3/2)$, respectively. In the case of the latter transition, we provide a $3\sigma$ upper limit.\\
$^{\ddagger}$ In the case of the absorption line observed at 246.72~GHz, we here provide the derived properties for HCN and NH separately, 
noting that both molecules likely do contribute to the observed absorption (see Sect.~\ref{sect:very-dense-gas}).
}
\label{tab:lines}
\end{table*}

\begin{table*}
\caption{Optical depths and column densities for the molecular absorption lines.}
\centering
\begin{tabular}{lcccc}
\hline
\hline
Line 			& $\tau(\nu_0)$	& $\rm \Delta$V        	& $\int \tau_\nu d\textrm{V}$	& $N_l$ \\
     			& 		& $[$km s$^{-1}]$  	& $[$km s$^{-1}]$   			& $[10^{14}$cm$^{-2}]$\\
\hline
$\rm OH^+ (1_1-0_1)$    &   $0.62\pm0.14$	&	$-280$ to $+530$	&   $191.4\pm17.2$		& $11.2\pm1.0$ \\ 
$\rm OH^+ (1_2-0_1)$    &   $0.61\pm0.16$	&	$-400$ to $+600$	&   $306.0\pm23.9$		& $9.6\pm0.8$ \\  
$\rm OH^+ (1_0-0_1)$    &   $0.24\pm0.09$	&	$-33$ to $+560$		&   $53.8\pm15.1$		& $7.2\pm2.0$ \\ 
$\rm CH^+ (1-0)$        &   $0.36\pm0.13$	&	$-145$ to $+500$	&   $91.4\pm22.8$		& $2.6\pm0.7$\\ 
$\rm HCN (11-10)$       &   $0.24\pm0.12$	&	$-570$ to $+430$	&   $66.1\pm18.3$		& $0.38\pm0.1$\\ 
$\rm NH (1_2-0_1)$       &   $0.24\pm0.12$	&	$-570$ to $+430$	&   $66.1\pm18.3$		& $6.7\pm1.9$\\ 
\hline
\hline
\end{tabular}
\tablefoot{
The frequency $\nu_0$ corresponds to the deepest dip of each lines. In the case of the absorption line observed at 246.72~GHz, we provide two possible column densities obtained assuming that the line is a transition of either  
HCN or NH separately, although we note that both lines likely contribute at the same time (see Sect.~\ref{sect:very-dense-gas}).}
\label{tab:tau_and_column_density}
\end{table*}

\subsection{Optical depths and column densities of lines seen in absorption}\label{sect:tau-and-column-densities}
The optical depth of an absorption line ($\tau_\nu$) can be evaluated from the observed spectrum normalized 
by the continuum as follows
\begin{equation}\label{eq:tau}
1-e^{-\tau_\nu} = \frac{S_\nu^\textrm{cont}-S_\nu}{S_\nu^\textrm{cont}} \textrm{,}
\end{equation}
where $S_{\nu}$ is the flux density of the spectrum and $S_\nu^\textrm{cont}$ that of the continuum only.
The optical depth is evaluated at the central frequency $\nu_0$ of the line. Table \ref{tab:tau_and_column_density} reports the values 
of $\tau(\nu_0)$ obtained from the integrated spectra shown in Fig.~\ref{fig:spectra_cont_norm}, i.e., averaged over the 
angles covered by the lines on the sky. 
In the case of the two strongest absorption lines (the $\rm OH^+(1_1-0_1)$ and $(1_2-0_1)$ transitions), 
maps of the opacity are displayed in Fig.~\ref{fig:abs_lines_tau_map}. 

Using Eq.~\ref{eq:tau} and the continuum-normalized spectra, we compute ``optical depth spectra'' 
that describe the variation of $\tau_\nu$ along the absorption lines, as a 
function of frequency or velocity.  The column density $N_l$ of the absorption line is derived 
by integrating over the velocity range of the line along the line of sight
(e.g., Comito et al. \citeyear{comito2003}; see also Indriolo et al. \citeyear{Indriolo2015,Indriolo2018}):
\begin{equation}\label{eq:diff_column_density}
N_l=\frac{8\pi\nu^3}{c^3}\frac{g_l}{A_{ul}g_u} \int \tau_\nu d\textrm{V} \textrm{.}  
\end{equation}
The quantity $\int\tau_\nu d\textrm{V}$ and the speed of light $c$ can be expressed in units of
$[$cm s$^{-1}]$, and the transition frequency $\nu$ in $[$Hz$]$ (see Table \ref{tab:lines} for the corresponding 
frequencies of each line), so that $N_l$ is directly computed in units of $[$cm$^{-2}]$. 
The quantities $g_l$ and $g_u$ are the lower and upper state statistical 
weights, and $A_{ul}$ is the spontaneous emission coefficient. We compute the column density for the strongest of the 
hyperfine transitions for the specific rotational $\Delta J$, and we assume that nearly all molecules are in the 
ground rotational state. The values of the statistical weights and the emission coefficients are taken  
from the Cologne Database for Molecular Spectroscopy \citep[CDMS;][]{mueller2005}.

Table~\ref{tab:tau_and_column_density} summarizes the results for the absorption lines observed in HerBS-89a, 
i.e, the three $\rm OH^+$ transitions, $\rm CH^+(1-0)$ and the absorption line observed at 246.72~GHz that is due to 
$\rm HCN(11-10)$ and/or $\rm NH(1_2-0_1)$,  listing the velocity integration ranges, the integrated optical depths, 
and the column densities. Uncertainties are computed from the dispersions of the spectra, via standard error propagation.

\subsubsection{Velocity fields}\label{sect:velocity}
Detailed views on the dynamics of HerBS-89a are presented in Fig.~\ref{fig:overlays_vel_cont}, 
which displays the velocity fields of all the molecular emission and absorption lines detected in this source.
The emission lines of $\rm ^{12}CO(9-8)$ and $\rm H_2O(2_{02}-1_{11})$ show similar, regular east-west velocity 
gradients, from blue to red along the southern arc, and reversed in the northern peak, as expected 
from gravitational lensing (see Sect.~\ref{sect:lensing_model}). 
The combined image of the $\rm ^{12}CO(9-8)$ and $\rm H_2O(2_{02}-1_{11})$ emission lines
reveals further details, in particular the velocity distribution along the western arc.
The velocity field of these dense gas tracers is consistent with the presence of kinematically distinct components 
that could indicate a merger system, or with a single rotating disc-like structure (see Sect.~\ref{sect:kinematics}). The main peaks 
of the three $\rm OH^+$ and one $\rm CH^+$ absorption lines are seen to be red-shifted across most of the 
continuum (Fig.~\ref{fig:overlays_vel_cont}). Finally, the red-shifted component of the $\rm NH_2 (2_{02}-1_{11}) (5/2-3/2)$ 
emission line is clearly seen west of the continuum peak, as expected from the velocity field revealed in 
the higher angular resolution data for the $\rm ^{12}CO(9-8)$ and $\rm H_2O(2_{02}-1_{11})$ emission lines.

\begin{figure}[!ht]
\centering
\includegraphics[width=0.48\textwidth]{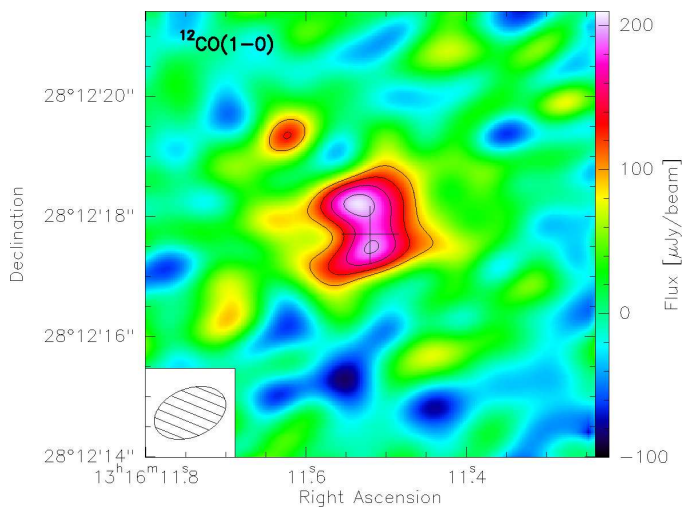}
\caption{$\rm ^{12}CO(1-0)$ emission in HerBS-89a. The 1$\sigma$ sensitivity is $\rm 32.1 \, \mu Jy/beam$ and the 
beam size (shown in the lower left corner) is $1\farcs24 \times 0\farcs79$ (P.A. $\rm -62.3^\circ$, E of N). 
The emission is resolved in the east-west direction. Contours are in $1\sigma$ steps starting from the $3\sigma$ level.}
\label{fig:CO10map}
\end{figure}

\subsection{VLA Results: The $^{12}$CO(1-0) emission line}\label{sect:CO(1-0)-VLA}

Figure~\ref{fig:spectral_fit_GILDAS} shows the VLA $\rm ^{12}CO (1-0)$ emission line spectrum, resampled to a velocity resolution of 
$\rm \sim 120 \, km\, s^{-1}$ to highlight the key properties of the line profile. 
Because of the limited S/N ratio, we fit the observed line with a single Gaussian profile. 
The position of the $\rm ^{12}CO(1-0)$ emission line is consistent with the nominal redshift of HerBS-89a.
Its FWHM is $\rm \Delta V \sim 1400\, km\, s^{-1}$, comparable to those of the three high-$J$ CO emission lines, 
and the observed total line intensity is $I_\nu=0.64\pm0.13 \, \textrm{Jy km s}^{-1}$ (Table \ref{tab:lines}). 

The emission line map, which is shown in Fig. \ref{fig:CO10map}, was obtained by integrating the VLA data cube between 
$-800$ and $\rm +500 \, km\, s^{-1}$ relative to the frequency of $\rm ^{12}CO (1-0)$ at $z=2.9497$. 
The $\rm ^{12} CO(1-0)$ line emission is slightly resolved and displays an east-west extension, similar 
to what is observed in the other CO lines for which the emission has been resolved 
(i.e., the $\rm ^{12}CO(9-8)$ and $(5-4)$ lines). It is noteworthy that, contrary to what is seen in the
$\rm ^{12}CO(9-8)$ and dust continuum emission, the northern component in $\rm ^{12}CO(1-0)$ appears 
slightly stronger than the southern component; however, the difference is at a $<1\sigma$ level. 
If confirmed, this inconsistency may be the result of differential magnification and indicate that the molecular gas reservoir 
traced in $\rm CO(1-0)$ is more extended than that traced by the high-$J$ CO and water transitions 
(see Sect.~\ref{sect:lensing_model} for a detailed discussion).

Due to the limited sensitivity of the VLA data, the underlying continuum of HerBS-89a was not detected. 
We measure an r.m.s. of $\sim 15.4$ and $\rm \sim 20.1 \, \mu Jy \, beam^{-1}$ in the continuum maps at 29 GHz (band width 0.74 GHz) and 38 GHz (band width 0.93 GHz) in the observed frame, with beam sizes of  $1\farcs23 \times 0\farcs79$ and $0\farcs96 \times 0\farcs59$, respectively.

\section{The foreground lensing galaxy}\label{sect:lensing_glx}

In this section, we derive the characteristics of the foreground lensing galaxy from the near-infrared \textit{HST} image 
and the GTC photometry, supplemented by NOEMA data.

\begin{figure}[!ht]
\centering
\includegraphics[width=0.48\textwidth]{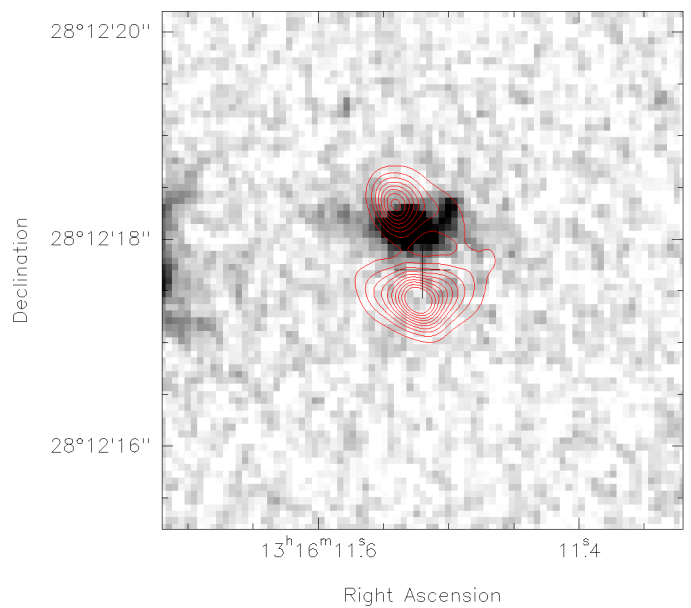}
\caption{
View of the foreground lensing galaxy in the \textit{HST} WFC3 F110W filter ($\rm 1.15 \, \mu m$) image, 
with the 254.6~GHz dust continuum measured with NOEMA superposed as red contours. 
}
\label{fig:NGP8_wide_field}
\end{figure}

\subsection{Imaging results (\textit{HST})}\label{sect:HST-results}

Figure~\ref{fig:NGP8_wide_field} displays the \textit{HST} wide-$J$ filter F110W image (at $\rm 1.15 \, \mu m$) towards 
HerBS-89a compared to the 1.2~mm dust continuum emission. The \textit{}{HST} image reveals a galaxy along the line of sight 
to HerBS-89a, which lies close to its northern component but is offset by $\sim 0\farcs2$ southwest from the northern dust 
continuum peak. The lensed galaxy itself remains undetected in the \textit{HST} image. The near-infrared emission of 
the foreground lensing galaxy is resolved in the \textit{HST} image. It is dominated by a bright bulge-like structure. It also displays a weak emission that extends in the east-west direction and that could resemble a disk. A bright spot to the North-West of the spheroidal component seems to wrap around the northern peak of the dust continuum, and could be a small satellite galaxy. 
From the \textit{HST} data, the foreground galaxy has an estimated flux density 
of $\rm \sim 4.36 \pm 0.88 \, \mu Jy$ at 1.15 $\mu$m.

\subsection{Photometric redshift (GTC)}\label{sect:GTC-results}

The foreground lensing galaxy is also detected in our GTC OSIRIS $r$-band imaging with a magnitude $r=24.5\pm 0.3$ AB mag. The deeper imaging with HiPERCAM detects the foreground galaxy in the \textit{griz} bands, while in the $u$ band the object remains undetected. The target lies very close to a bright ($r=19.5$) point-like object, $3\farcs3$ to the east, and the  photometric measurement needs to take properly into account the wings of the bright neighbor. 

The HiPERCAM maps were calibrated on SDSS DR12 stars present in the field. The code GALFIT \citep{peng2002} was used to subtract the profile of the bright point-like nearby object using a model PSF. Aperture photometry was performed on the residual image to measure the emission of the target lensing galaxy. Uncertainties were derived measuring the noise level in an annulus around the object, thus taking into account the presence of the nearby source. Finally, systematic zero-point uncertainties were propagated in the estimate of the magnitude uncertainties. Table~\ref{tab:optical} lists the derived magnitudes and the 3$\sigma$ $u$ band upper limit in the AB system \citep{Oke1983}. 

The foreground galaxy is also detected by \textit{WISE} \citep{wright2010} at 3.4 and 4.6~$\mu$m. However, the \textit{WISE} photometry likely includes contributions from the lensing galaxy, the lensed HerBS-89a galaxy, and the nearby bright point-like source. Because of the broad \textit{WISE} beam  ($6\farcs1$ at 3.4 $\mu$m), it is impossible to disentangle the different components. 
Therefore only the GTC and \textit{HST} data points have been retained in this analysis.

At $r=24.5$ it is far from easy to obtain an optical spectroscopic
redshift, but thanks to the sensitivity of HiPERCAM and the capability to observe simultaneously
the five $ugriz$ bands, together with the {\it HST} F110W imaging results, it is possible to derive an
photometric redshift.

\begin{figure}[!ht]
\centering
\rotatebox{-90}{\includegraphics[height=0.45\textwidth]{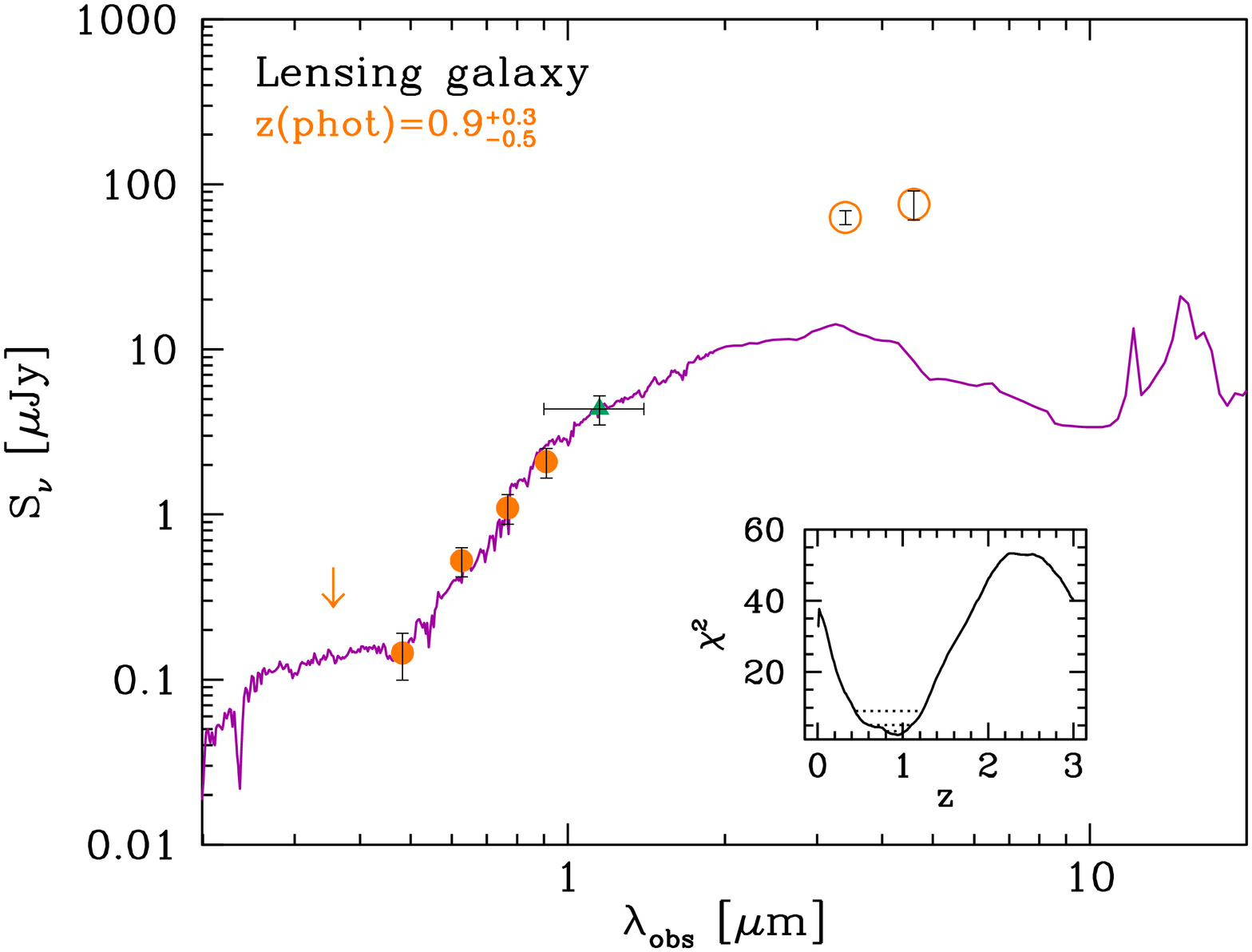}}
\caption{Best derivation of the photometric redshift of the foreground lensing galaxy, using the Le~PHARE 
code \citep{ilbert2006,arnouts1999}. The filled triangle shows the \textit{HST} F110W flux density and the filled circles 
represent the GTC measurements (see text). The $WISE$ data (open circles) have not been included in the fit. 
The curve shows the best fit model, a Sb spiral template \citep{polletta2007,ilbert2009}. The inset displays the distribution of $\chi^2$ 
as a function of redshift; the dotted horizontal lines correspond to the 1, 2, 3 $\sigma$ levels, defined as 
$\chi^2_\textrm{min}+1,\ 2.71,\, 6.63$. The best photometric redshift for the foreground galaxy is 
$z_\textrm{phot}=0.9^{+0.3}_{-0.5}$ where the errors are $3\sigma$.}
\label{fig:photoz_lens}
\end{figure}

To get the best possible result, and test systematics due to the chosen  method, we applied various photometric redshift codes \textit{EAZY} \citep{brammer2008}, Hyper-z \citep{Bolzonella2000}, 
Le~PHARE \citep{arnouts1999,ilbert2006}, and MAGPHYS-z \citep{dacunha2008,battisti2019} 
to the {\em ugriz} and F110W data, adopting different template 
libraries \citep{coleman1980,Bolzonella2000,berta2003,berta2013,babbedge2004,polletta2007,brammer2008,ilbert2009,dacunha2015}. 
Since it best accounts for the upper limits, the final choice is the Le~PHARE code. 

The best photometric redshift of the foreground lensing galaxy thus obtained 
is $z_\textrm{phot}=0.9$, with a 99\% confidence interval of $0.4< z_{\textrm{phot,}3\sigma}<1.2$. The best fit solution is obtained with a SB spiral template from the COSMOS library \citep{ilbert2009}, originally belonging to the SWIRE library \citep{polletta2007}.
Figure~\ref{fig:photoz_lens} shows the best fit model together with the $\chi^2$ distribution. 
The photometric redshift is best constrained by the $r-i$ color, tracing the D4000 break\footnote{The 4000~{\rm \AA} 
discontinuity is defined as the ratio between the average flux density between 4050 and 4250~{\rm \AA} and that between 3750 and 3950~{\rm \AA} 
\citep{Bruzual1983}.}. The lack of rest-frame near-infrared coverage and the upper limit in the $u$ band restrict the accuracy 
of the photo-$z$ measurement. Deeper data and a wider wavelength coverage would be necessary to improve its precision.

In addition, various attempts were made using templates that 
include emission lines \citep[e.g.][]{babbedge2004}, with the aim to verify if the $r$ and F110W bands could be contaminated by lines at $z\sim0.7$, namely $[${\sc Oii}$]$ ($\lambda3727$) and H$\alpha$ ($\lambda6563$), but none produced a better fit to the data.

\begin{table}[!ht]
\centering
\caption{Optical photometry of the foreground lensing galaxy.}
\begin{tabular}{c c c c c}
\hline
\hline
Band & $\lambda$ & mag & Flux Density & Instrument\\
    & $[\mu$m$]$ & $[$AB$]$ & [$\mu$Jy] & \\
\hline
$u$   & 0.36 & $>$25.0        & $<$0.36       & {\tiny GTC HiPERCAM} \\    
$g$   & 0.48 & 26.0$\pm$0.3   & 0.15$\pm$0.05 &     "        \\
$r$   & 0.64 & 24.5$\pm$0.3   & 0.63$\pm$0.06 & {\tiny GTC OSIRIS}   \\
$r$   & 0.63 & 24.6$\pm$0.2   & 0.53$\pm$0.11 & {\tiny GTC HiPERCAM} \\
$i$   & 0.77 & 23.8$\pm$0.2   & 1.10$\pm$0.22 &     "        \\
$z$   & 0.91 & 23.1$\pm$0.2   & 2.09$\pm$0.42 &     "        \\
F110W & 1.15 & 22.3$\pm$0.2   & 4.36$\pm$0.88 & \textit{\tiny HST} WFC3   \\
W1    & 3.35 & 19.4$\pm$0.1   & 63.1$\pm$6.1  & \textit{\tiny WISE} \\
W2    & 4.60 & 19.2$\pm$0.2   & 75.9$\pm$15.3 &     "        \\
\hline
\end{tabular}
\tablefoot{Due to the contamination of a nearby point-like source, the \textit{WISE} data 
were not considered in the derivation of the photometric redshift of the foreground 
lensing galaxy (see Sect.~\ref{sect:GTC-results}). The upper limit to the $u$-band 
flux density is given at a $3\sigma$ level. 
}
\label{tab:optical}
\end{table}

\begin{figure}[!ht]
\centering
\includegraphics[width=0.45\textwidth]{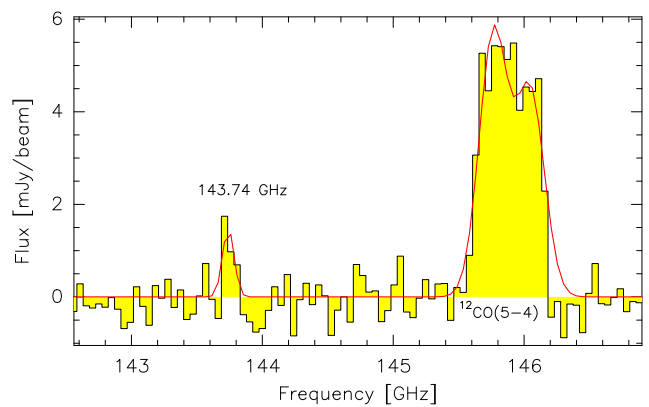}
\caption{Spectrum between 139.29 and 147.04~GHz towards HerBS-89a, from \cite{Neri2020}, showing 
the $\rm ^{12}CO(5-4)$ emission line of HerBS-89a (at 145.902~GHz) and the weaker, narrow ($\rm \sim 200 \, km \, s^{-1}$) 
emission line at 143.74~GHz, possibly belonging to the foreground lensing galaxy (see Sect.~\ref{sect:NOEMA-redshift-result} 
for further details).}  
\label{fig:possible_line_lens}
\end{figure}

\subsection{Spectroscopic redshift (NOEMA)}\label{sect:NOEMA-redshift-result}

An independent indication of the redshift of the lensing foreground can possibly be derived from the 2-mm spectral scan 
that was obtained to derive the spectroscopic resdhift of HerBS-89a in \cite{Neri2020}. 
In the frequency range 139.3 to 147.0~GHz, where the $\rm ^{12}CO(5-4)$ emission line of HerBS-89a 
was detected, a weaker narrow  ($\rm \sim 200 \, km \, s^{-1}$) emission line is seen at 143.74~GHz, 
close to the center position of HerBS-89a (see Fig.~\ref{fig:possible_line_lens}). 
The narrowness of this emission line excludes the possibility that it could be 
related to HerBS-89a. If it traces instead the foreground lensing galaxy, 
the line could correspond to a $\rm ^{12}CO$ emission line, either detected in the $J=2-1$ transition, 
in which case $z_{\rm lens} = 0.604$, or in the $J=3-2$ transition for which $z_{\rm lens} = 1.406$. The former is consistent within the $2 \sigma$ uncertainties with the photometric redshift derived from the GTC and \textit{HST} data.
A search for the counterpart of this emission line (at 1-mm) could allow us 
to verify the above identification, and eventually derive a precise spectroscopic redshift 
of the foreground galaxy gravitationally amplifying HerBS-89a.

\begin{table}[!ht]
\caption{Results of lens modelling: parameters of the best fit model for the foreground lens.}
\centering
\begin{tabular}{l c c c}
\hline
\hline
Parameter	& Value		&          \multicolumn{2}{c}{Uncertainties}            \\
            &           & 1$\sigma$             &  Systematic           \\
\hline
Ref. Point (RA)	& 13:16:09.79	& --		& -- \\ 
Ref. Point (Dec)& 28:12:18.17	& --		& -- \\ 
\hline
Position $\Delta$RA & -0$\farcs$0612	& 0.0005	& \\
Position $\Delta$Dec & -0$\farcs$0989	& 0.0008	& \\
Einstein Radius $b$ & 0$\farcs$4832	& 0.0006	& 0.0483 \\
Ellipticity $e$ & 0.177	& 0.002	& 0.018 \\
Position Angle $\theta_e$ & 32$^{\circ}$4 & 0.4	& 3.2 \\	
\hline
\end{tabular}
\tablefoot{The position angle's orientation is measured east of north.}
\label{tab:lensing_results_lens}
\end{table}

\begin{table*}[!ht]
\caption{Results of lens modelling: parameters of reconstructed source-plane emission (and absorption) in HerBS-89a. }
\centering
\begin{tabular}{l c c c c c c c c c}
\hline
\hline
Tracer & $I_0$	& $\Delta$RA	& $\Delta$Dec	& $e$ & $\theta_e$	& \multicolumn{2}{c}{$R_s$}	& $n$	& $\mu$	\\ 
	   & [Jy/beam] & \multicolumn{2}{c}{[mas]} &   & [deg]  & [mas] & [pc]					&			&	\\
\hline
Continuum$\rm ^a$ 	& 0.16$\pm$0.01  & -1.9$\pm$0.2 & -277.8$\pm$0.4 & 0.522$\pm$0.007	& -97.4$\pm$0.4	& 12$\pm$1	& 95$\pm$8	& 1.82$\pm$0.04 		& 5.05$\pm$0.03			\\	
\,\, {\em syst. unc.}		& \, (0.02) 	& (9.5) 	&  (9.5) 	& (0.052)	& (9.7) 	& (1.0)	& (10.0)	& (0.18) 	& (0.51)	\\	
\hline
$\rm CO(9-8)$			& 0.30 		& -9.6	& -254.5	& 0.408		& -92.8		& 0.4	& 3.0		& 3.21		& 5.23		\\	
\,\, {\em syst. unc.}		& (0.03)	& (11.2)	& (11.2)	& (0.041)	& (9.3)		& (0.1)	& (1.0)	& (0.32)	& (0.52)	\\	
\hline
$\rm CO(1-0)$			& 0.010		& 15.35		& -248.3	& 0.189		& 87.8		& 0.8		& 5.9		& 4.08		& 5.0	\\	
\,\, {\em syst. unc.}		& (0.001)	& (45.0)	& (45.0)	& (0.019)	& (8.8)		& (0.1)		& (0.6)		& (0.41)	& (1.0)		\\	
\hline
$\rm H_2O(2_{02}-1_{11})$			& 0.008		& -19.1	& -312.4	& 0.559		& -79.7		& 50.0		& 396.0		& 1.50		& 3.87		\\
\,\, {\em syst, unc.}		& (0.001)	& (22.5)	& (22.5)	& (0.056)	& (8.0)		& (5.0)	& (40.0)	& (0.15)	& (0.39)	\\	
\hline
$\rm OH^+(1_1-0_1)$ 		& 0.044 	& -3.3	& -287.1	& 0.045		& -23.9 	& 26.0		& 206.0		& 0.861		& 4.69		\\ 	
\,\, {\em syst. unc.}		& (0.004)	& (7.0)	& (7.0)	& (0.005)	& (2.4)		& (3.0)	& (24.0)	& (0.086)	& (0.47)	\\	
\hline
$\rm OH^+(1_2-0_1)$		& 0.042 	& -2.7 	& -314.1 	& 0.271		& -68.6		& 29.0		& 230.0		& 1.113		& 4.24		\\ 	
\,\, {\em syst. unc.}		& (0.004)	& (9.4)	& (9.4)	& (0.027)	& (6.9)		& (3.0)	& (24.0)	& (0.111)	& (0.42)	\\	
\hline
\end{tabular}
\tablefoot{(a) The dust continuum emission is measured at 254.5~GHz, and the uncertainties are $\pm 1\sigma$. For each parameter, the systematic uncertainties are tabulated separately in parentheses on a second line (see Sect.~\ref{sect:lensing_model}).
In the heading of the table, $e$ stands for the ellipticity, $\theta_e$ 
for the position angle (measured east of north), $R_s$ for the scale radius, $n$ for the S\'ersic index, 
and $\mu$ for the amplification factor. Surface brightness units
of Jy/beam are inherited from the surface brightness units of the
data. Strong covariance between $R_S$ and $n$ (see, e.g., Fig. \ref{fig:corner_lens}) means that the $\rm ^{12}CO(9-8)$ and $\rm ^{12}CO(1-0)$ data are formally compatible
with (larger) $R_S$ and (smaller) $n$ like those recovered for
the other reconstructions.}
\label{tab:lensing_results_source}
\end{table*}

\section{Lens Modelling}\label{sect:lensing_model}

In order to recover the source-plane morphology and the intrinsic properties of HerBS-89a, we derive a lens model from the 
high-angular resolution NOEMA continuum data as shown in Fig.~\ref{fig:2SB_continuum_zoom} 
(with initial lens parameters informed by \textit{HST} observations) 
and proceed to apply it to the molecular line data. We perform the lens modelling and source reconstruction using 
the \textit{lensmodel} package \citep{Keeton2001} with the \textit{pixsrc} extension 
\citep{Tagore&Keeton2014,Tagore&Jackson2016}.
For the sake of computational efficiency, all modelling has been done in the sky plane, 
although we recognize that fitting directly to visibility data can prevent specific choices in the interferometric 
imaging process (e.g., related to $uv$ weighting and deconvolution) from having undue impact on a model.
Although the quality of our NOEMA observations is excellent, 
the limited number of resolution elements across the source prevents us from leveraging
the full power of the non-parametric approach 
of \cite{Tagore&Keeton2014}, instead limiting us to the use of parametric (S\'ersic profile) components in 
reconstructing the sources. 

\begin{figure*}[!ht]
\centering
\includegraphics[width=0.98\textwidth]{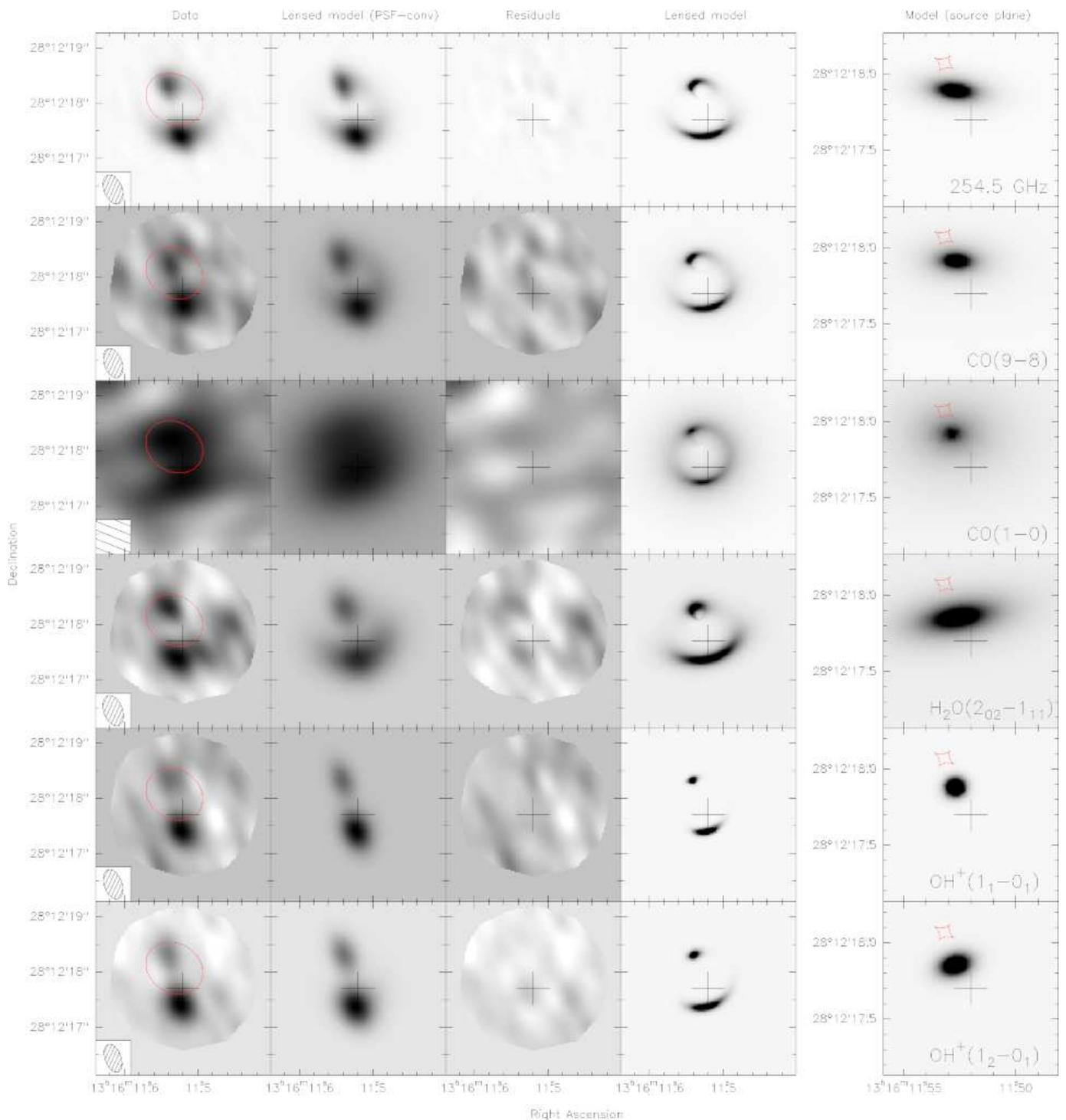}
\caption{Lens modelling results for the 1.2\,mm dust continuum emission and spectral lines in HerBS-89a (identified in the  
right most panels). In each row, from left to right:  observed image (with the synthesized beam shown in the lower left corner) and critical 
curve (red line); reconstructed model image convolved with the PSF; residuals; reconstructed model image at 
full angular resolution; and (separated and with a different angular scale) reconstructed source-plane image and caustic.
The residuals are within $\pm 2.5 \sigma$, showing that the model agrees well with the data. 
The cross shows the coordinates of the phase center of NOEMA observations (R.A. 13:16:11.52 and Dec. +28:12:17.7 in J2000). 
The $\rm OH^+$ absorption lines have been inverted to allow for a positive normalization of the S\'ersic profile reconstruction.}
\label{fig:lens_model_cont}
\end{figure*}

Because we are not using regularization to guide the
reconstruction of non-parametric sources, correlations in the
noise introduced by the finite $uv$ sampling do not require
special handling \`a la \citet{Riechers2008}. In all cases,
we convolve the lensed model images (fourth column of Fig. \ref{fig:lens_model_cont})
with the appropriate (two-dimensional Gaussian) synthesized beams
to generate the reconstructed model images (second column of Fig.
 \ref{fig:lens_model_cont}) that are directly compared to the data.

The foreground deflector is assumed to have a singular isothermal elliptical mass distribution (SIEMD) described by five 
parameters: Einstein radius ($b$), R.A., Dec., ellipticity ($e$),\footnote{Ellipticity is defined as $e = 1-q$, 
where $q$ is the axis ratio.} and ellipticity position angle ($\theta_e$; measured east of north). 
Table \ref{tab:lensing_results_lens} lists the best fit values of these parameters and their uncertainties as determined from a Markov chain Monte Carlo (MCMC) exploration of parameter space. 
The position of the lens is given relative to the reference position 
RA=13:16:09.79 and Dec=28:12:18.17.
Because of the very high S/N of the continuum data, we use the lens model derived from the continuum 
data when reconstructing the spectral line data.

The left panels of Fig.~\ref{fig:lens_model_cont} include the critical curve (i.e., the locus of infinite magnification 
in the image plane; red lines). In the case of a SIEMD with no external shear, this curve is also 
an iso-density contour of the deflector's 
mass distribution whose position and geometry can be compared to the \textit{HST} image of the lensing galaxy 
(Fig.~\ref{fig:NGP8_wide_field}). The corresponding source-plane caustic is plotted in the 
right panels of Fig.~\ref{fig:lens_model_cont}.

The continuum and (almost) all of the spectral line zeroth moment maps are fit 
well by single S\'ersic components in the source plane (see Table~\ref{tab:lensing_results_source} 
and  Fig.~\ref{fig:lens_model_cont} for fitting results, and below for more discussion of the line data). 
The profile is parametrized in terms of its position, normalization $I_0$, ellipticity $e$, position angle  $\theta_e$,
scale radius $R_s$, and S\'ersic index $n$:
\begin{equation}\label{eq:sersic}
I(r) = I_0\, \exp\left[-\left(\frac{r}{R_s}\right)^{1/n}\right] \textrm{.}
\end{equation}
Using the estimated photometric redshift of the foreground galaxy and its uncertainty (see Sect.~\ref{sect:GTC-results}), we estimate that the deflector mass enclosed 
within the Einstein radius is 
$M_\textrm{lens}=9.8^{+3.2}_{-4.9} \times 10^{10}$ M$_\odot$ (where the upper and lower bounds are at 3$\sigma$).

Statistical uncertainties in the lens and continuum source model were evaluated with two MCMC cycles. 
The first fixed the source model and determined the uncertainties of the lens parameters; the second fixed the 
lens model to its optimal parameters and let the source parameters vary. This approach was chosen for 
computing speed and for ease of comparison between continuum and line results (because 
the latter assume the lens model derived from the continuum data, i.e., the lens model is fixed in this case). However, it should be noted that the MCMC approach underestimates the lens$+$source errors.
Appendix~\ref{app:lens} presents the conditional posterior distributions of the lens and source model parameters.
 
To these statistical uncertainties, we add fiducial systematic uncertainties of 10\% associated 
with the choices of lensing potential, source brightness profile, interpolation errors, etc. 
The only exceptions are the source model positional errors, for which we adopt systematic uncertainties 
given by 
\begin{equation}\label{eq:systematic-errors} 
\frac{\sqrt{\textrm{HPBW}_\textrm{min} \, \textrm{HPBW}_\textrm{maj}}}{2} \frac{\sigma_I}{I_\textrm{line}}\frac{1}{\sqrt{\mu}} \textrm{,} 
\end{equation}
in terms of the synthesized beam dimensions $\rm HPBW_{min}$ and $\rm HPBW_{max}$, the S/N of 
the observed detection $\sigma_I/I_{\rm line}$, and the magnification $\mu$ \citep[see, e.g.,][]{Downes1999}.
For the lens model, we estimate systematic positional errors as the astrometric errors from the continuum imaging.
Because we expect systematic errors to dominate and because of computational constraints, we do not 
perform MCMC on the line spectral source parameters.

As mentioned above, nearly all of the spectral line data are fit well by single S\'ersic components; the exception 
is the H$_2$O($2_{02} - 1_{11}$) zeroth-moment map, which shows faint ($\sim 5\sigma$, 
compared to the $\sim 10\sigma$ peak) emission to the west and northwest that motivated some exploration 
of a two-component model. We adopt here a single-component fit for this line, due to numerical instability 
in the two-component fit, but note that there are some systematic features in the residuals associated 
with the peak of the emission. 
The results are listed in Table~\ref{tab:lensing_results_source}, along with the corresponding 
magnification factors $\mu$, for the high resolution data and for $\rm ^{12}CO(1-0)$. Because of the 
lower resolutions and S/N ratios,  we do not model the spectral lines and continuum that were measured at 
lower frequencies (from 209 to 232~GHz).

The magnification factor $\mu$ for the continuum and line data was computed  using samples from the 
lens and source parameter MCMC runs.  We estimate a magnification of $5.05 \pm 0.03 (\pm 0.51)$ for the 254.6 GHz 
dust continuum emission, where the term in parentheses represents the 10\% systematic uncertainty.

In the case of the $\rm ^{12}CO(1-0)$ line, we infer a lensing magnification $\mu=5.0$, with a large systematic uncertainty mainly due to possible different choices in how the zeroth-moment map is derived (also reflecting the limited S/N
of the current data set). Deeper and higher angular resolution observations 
of the $\rm ^{12}CO(1-0)$ emission would be helpful to better constrain the amplification factor as well as to study in greater detail the morphology (extent and kinematics) of the cold gas reservoir of HerBS-89a.

\section{Global properties of HerBS-89a: dust and molecular gas}\label{sect:global-properties}

In this section, we derive the intrinsic properties of HerBS-89a from the new millimeter, centimeter, and 
optical/near-infrared data presented in Sect.~\ref{section:results}, and the lensing model described 
in Sect.~\ref{sect:lensing_model}. The discussion is organized as follows: the analysis of the 
continuum spectral energy distribution and the derivation of the molecular gas mass and dust-to-gas ratio are 
described in Sections \ref{sect:sed} and \ref{sect:molecular-gas-mass-dust}; the CO line excitation is 
investigated in Sect.~\ref{sect:sled}, and the properties of the water emission line are examined in Sect.~\ref{sect:H2O-line}. 

\subsection{Continuum Spectral Energy Distribution}\label{sect:sed}
The new NOEMA continuum measurements (see Table~\ref{tab:continuum}), combined with the data from \citet{Neri2020} and 
\textit{Herschel} and SCUBA-2 photometry \citep{Bakx2018, Bakx2020}, provide an exquisite sampling of the far-infrared (FIR) 
and sub-mm spectral energy distribution (SED) of HerBS-89a. A total of 16 broad band measurements 
cover the range from 250~$\rm \mu m$ to 3 mm in the observed frame, 12 of which are spread along the Rayleigh-Jeans tail. 

\begin{figure}[!ht]
\centering
\includegraphics[width=0.45\textwidth]{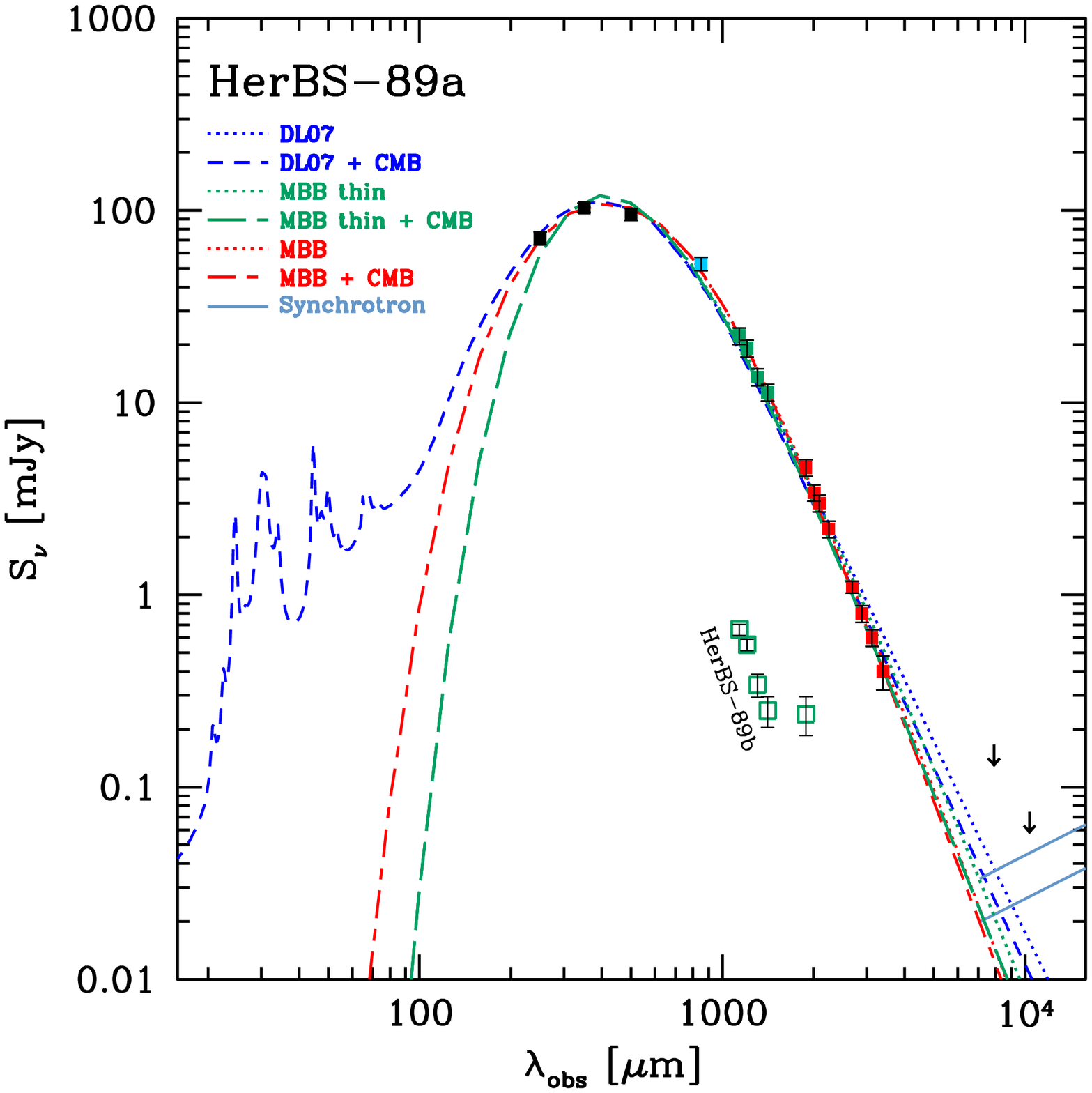}
\rotatebox{-90}{\includegraphics[height=0.45\textwidth]{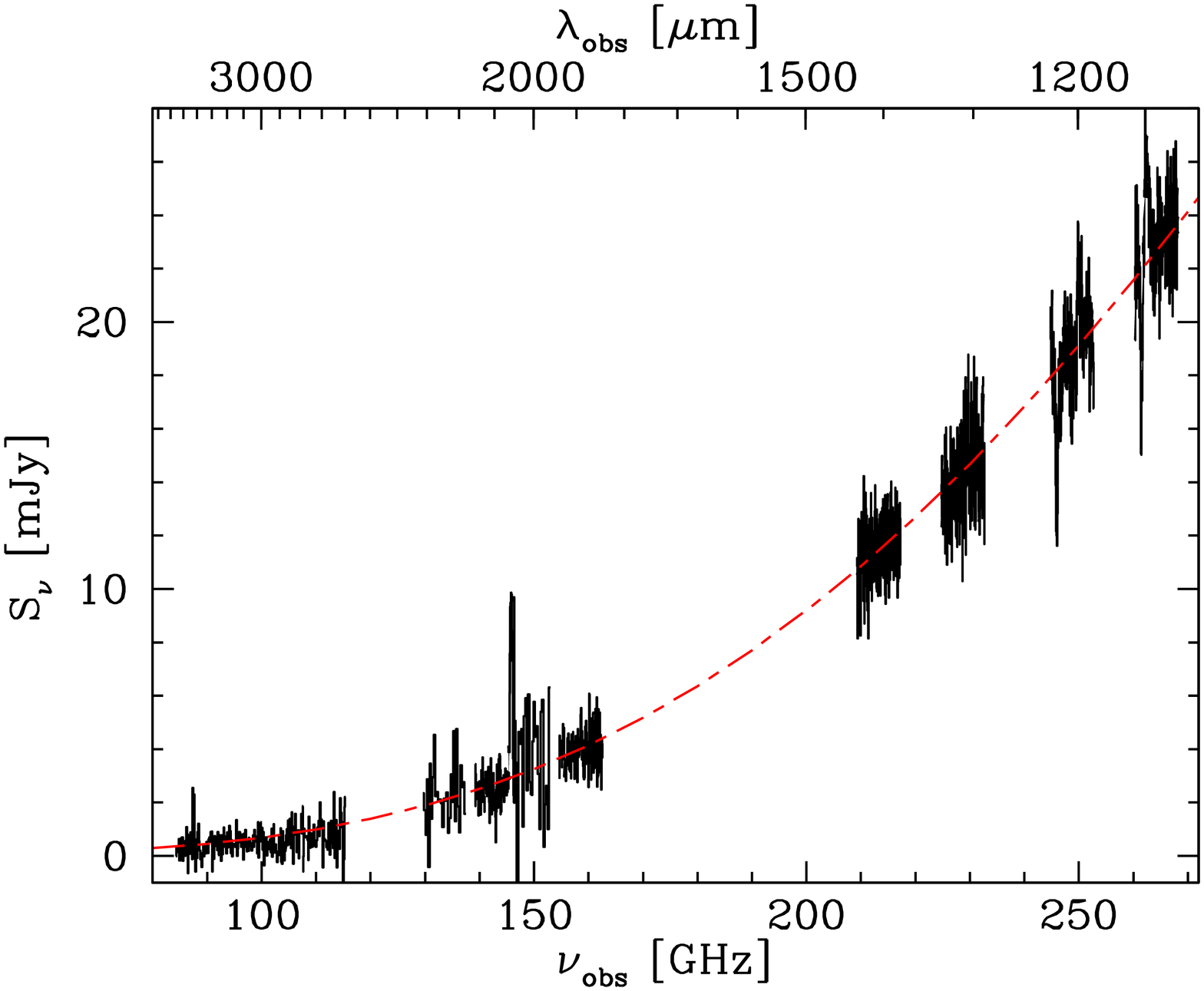}}
\caption{{\em Top panel}: Observed spectral energy distribution of HerBS-89a (filled symbols) and 89b (open symbols). 
The HerBS-89a data 
are fit with three different dust models: DL07 (blue lines), optically-thin MBB (green lines); MBB in general form (red lines). The dotted lines represent the models as they are, and the dashed lines include the correction for the effects of CMB (see text for details). 
The light-blue continuous lines represent the synchrotron emission with a spectral index $\alpha=-0.8$, normalized on the basis of the 
radio-FIR correlation subject to the redshift evolution found by \citet[][upper line]{magnelli2015} 
and \citet[][lower line]{delhaize2017}. The data include SPIRE flux densities \citep[black dots; from][]{Bakx2018}; 
SCUBA-2 photometry \citep[blue dot; from][]{Bakx2020}; the NOEMA derived continuum flux densities 
at 1~mm continuum (green dots) from this work and 2 and 3~mm (red dots) from \cite{Neri2020}; 
and 3$\sigma$ upper limits at 29 and 38~GHz (10 and 7.8~mm, respectively)) are derived from the VLA data (arrows). The data
for HerBS-89b are taken from this paper for 1~mm and \cite{Neri2020} for 2~mm flux densities.
{\em Lower panel}: HerBS-89a NOEMA spectra \citep[][and this work]{Neri2020} compared to the general form MBB model with CMB correction.}
\label{fig:SED_fit}
\end{figure}

Figure \ref{fig:SED_fit} shows the continuum data available for HerBS-89a as well as the 
continuum measurements extracted for the nearby source HerBS-89b. The latter source 
shows an increase of flux at wavelengths above $\rm \sim 2\,mm$ with respect to the dust emission, 
which could be due to non-thermal processes and rules out the possibility that HerBS-89b is 
a third lensed image of HerBS-89a.

Following and extending the work by \citet{Neri2020} and \citet{berta2016}, we model the SED of HerBS-89a using three different approaches: 
{\em i)} the  \citet[][hereafter DL07]{DL07} dust model;
{\em ii)} a single-temperature modified black body (MBB) in the optically-thin approximation; 
and {\em iii)} an MBB model in its general form.

In the case of the DL07 modelling, interstellar dust is described as a mixture of carbonaceous and amorphous silicate 
grains, whose size distributions are chosen to reproduce different observed extinction laws. 
We defer to \citet{dl01, DL07}, \citet{li2001}, and \citet{berta2016} for a detailed description of the 
model and its implementation. Despite the richness of the long-wavelength data, the SED is still lacking measurements in 
the mid-infrared, so that the fine details of the DL07 model, e.g., the possible contribution of the polycyclic aromatics hydrocarbons (PAHs), are not constrained. 
The dust mass estimate is nevertheless robust, because it is dominated by the colder dust components.

For the optically thin MBB case, the emergent luminosity from a given dust mass $M_\textrm{dust}$ is described as: 
\begin{equation}\label{eq:mbb}
L_\nu \sim M_\textrm{dust} \kappa_\nu B_\nu(T_\textrm{dust}) \textrm{,}
\end{equation}
where $B_{\nu}(T_\textrm{dust})$ is the Planck function, $T_\textrm{dust}$ is the dust temperature, 
and $\kappa_\nu=\kappa_0\left(\nu/\nu_0\right)^\beta$ is the mass absorption coefficient of dust at rest frequency $\nu$. 

The general form of the MBB differs from the optically thin approximation by the factor $\left(1-e^{-\tau_\nu} \right)$, 
with $\tau_\nu=\tau_0\left(\nu/\nu_0\right)^\beta$, instead of the $\nu^\beta$ term. Here $\nu_0$ is 
the frequency at which $\tau=\tau_0=1.0$ \citep[see][]{berta2016}. 
Therefore, the general form of the MBB tends to produce best fit solutions with higher dust temperature than 
the optically thin case. In the optically thin approximation, three free parameters are at play: $T$, $\beta$ and 
the model normalization. In the general form, $\nu_0$ is also a free parameter.

The MBB fits are limited to the data with rest-frame wavelength $\lambda_\textrm{rest}$\,>\,$50\,\mu$m, 
thus avoiding biases towards warmer temperatures. The effects of the cosmic microwave background (CMB) have been taken 
into account, following the prescription of \citet{dacunha2013}. A correction similar to that derived for the MBB has also been applied to the DL07 model.

\citet{berta2016} and \citet{Bianchi2013} present thorough 
discussions of the proper use of $\kappa_\nu$ and $\beta$. Appendix \ref{sect:kappa} summarizes the current $\kappa_\nu$ panorama.
Following the choice often found in the literature, we adopt the \citet{Draine2003} $\kappa_\nu$ and apply the correction prescribed by \citet{draine2014}:  $\kappa_\nu=0.047$ m$^2$ kg$^{-1}$ at 850 $\mu$m. The same correction is also applied to the dust masses derived with the DL07 approach.

The best fit models are shown in Fig. \ref{fig:SED_fit}, and the results are listed in Table~\ref{tab:sed_fit}. 
We include all the values obtained with our fits using the continuum flux densities not corrected for the lens magnification,
as well as our preferred source-plane values for the infrared luminosity and the dust mass 
based on the DL07 fit corrected for gravitational magnification 
(see Sect.~\ref{sect:lensing_model}). The CMB effect produces a steepening of the millimetric slope by 
$\sim+0.15$, comparable to the uncertainty on $\beta$. At the redshift $z=2.95$ and for the dust temperature of HerBS-89a, 
the effect is milder than what found at $z=3.6-5.8$ by \citet{jin2019}.

Finally, the VLA 3$\sigma$ upper limits for HerBS-89a that are shown in Fig.\ref{fig:SED_fit} have been
computed over an area equivalent to the NOEMA 263.3~GHz continuum (see Table \ref{tab:continuum}). 
For comparison, a power-law spectrum $S_\nu\propto\nu^{-0.8}$ is plotted, representing a synchrotron 
emission \citep[e.g.][]{ibar2010,thomson2014}, normalized on the basis of the 
radio-far-infrared correlation \citep{magnelli2015,delhaize2017}.

\begin{table}[!ht]      
\caption{Properties of HerBS-89a derived from the dust continuum spectrum.}
\centering
\begin{tabular}{lr@{$\pm$}ll}
\hline	
\hline
Quantity & \multicolumn{2}{c}{Value} & Units\\
\hline	
\multicolumn{4}{c}{DL07 model}\\
$\mu L_{\rm DL07}(8-1000 \, \mu \rm m)$ 	& (2.89&0.25) 10$^{13}$ & $L_\odot$ \\
$\mu M_{\rm DL07}\, \rm (dust)$		& (1.30&0.07) 10$^{10}$ & $M_\odot$ \\
\hline	
\multicolumn{4}{c}{Optically thin MBB}\\
$T_{\rm MBB, thin} \, \rm (dust)$		    & 28.0&1.3	  & K	\\		
$\beta_\textrm{MBB, thin}$		& 2.01&0.11	 	  & -- \\		
$\mu L_{\rm MBB, thin}(50-1000 \, \mu \rm m)$ 	& (1.94&0.08) 10$^{13}$ & $L_\odot$ \\
$\mu M_{\rm MBB, thin}\, \rm (dust)$		& (1.11&0.06) 10$^{10}$ & $M_\odot$ \\
\hline	
\multicolumn{4}{c}{MBB in general form}\\
$T_{\rm MBB, gen} \, \rm (dust)$		    & 49.2&2.7	  & K	\\		
$\beta_\textrm{MBB, gen}$		& 2.24&0.13	 	  & -- \\		
$\nu_\textrm{0, MBB, gen}$                         & 1214&121       & GHz \\
$\lambda_\textrm{0, MBB, gen}$                     & 247&24       & $\mu$m \\
$\mu L_{\rm MBB, gen}(50-1000 \, \mu \rm m)$ 	& (2.00&0.08) 10$^{13}$ & $L_\odot$ \\
$\mu M_{\rm MBB, gen}\, \rm (dust)$		& (6.06&0.71) 10$^{9}$ & $M_\odot$ \\
\hline
\multicolumn{4}{c}{De-magnified values from DL07}\\
$L_\textrm{IR}(8-1000 \, \mu \rm m)$ 	& (4.6&0.4) 10$^{12}$ & $L_\odot$ \\
$M_\textrm{dust}$		    & (2.6&0.2) 10$^{9}$ & $M_\odot$ \\
\hline
\end{tabular}
\tablefoot{The MBB results include the correction for the CMB effect on $T$ and $\beta$. The last two rows are based on the DL07 model and have been corrected for the gravitational lens magnification.}
\label{tab:sed_fit}
\end{table}

\subsection{Molecular gas mass and gas-to-dust ratio}\label{sect:molecular-gas-mass-dust}

The observed $\rm ^{12}CO(1-0)$ emission line intensity can be translated into a luminosity (in $\rm K \, km \, s^{-1} pc^2$), from 
which an estimate of the molecular gas mass can be made, using the following relation \citep{Solomon-VandenBout2005}:
\begin{equation}\label{eq:L'_line}
L^\prime_\textrm{Line}= 3.25\times10^7 \,  S_\textrm{Line} \Delta V \times \frac{D_L^2}{(1+z)} \frac{1}{\nu_\textrm{rest}^2} \textrm{,}
\end{equation}
where $S_\textrm{Line} \Delta V=I_\textrm{Line}$ is the integrated line flux in units of $\rm Jy \, km \, s^{-1}$; 
$\nu_\textrm{rest}$ is the line rest frequency in GHz; and
$D_L$ is the luminosity distance in Mpc.

The line luminosity, expressed in units of $L_\odot$, can be derived as:
\begin{equation}\label{eq:L_line}
L_\textrm{Line}= 3.2\times10^{-11} \, \nu_\textrm{rest}^3 L^\prime_\textrm{Line} \textrm{.}
\end{equation}

We define the molecular and total gas masses as
\begin{equation}
M_\textrm{mol}=M_{\textrm{H}_2}+M_\textrm{He} \textrm{,}
\end{equation}
\begin{equation}
M_\textrm{gas}=M_\textrm{mol}+M_\textrm{HI} \textrm{,}
\end{equation}
where $M_\textrm{He}$ is the mass of helium, $M_{\textrm{H}_2}$ is the mass of molecular hydrogen and 
$M_\textrm{HI}$ the mass of atomic hydrogen. 
Hereafter, we will assume that $M_\textrm{mol}\gg M_\textrm{HI}$ 
and thus that $M_{\rm gas} \approx M_{\rm mol}$.

The molecular hydrogen gas mass is computed from the $\rm ^{12}CO(1-0)$ luminosity applying a 
conversion factor $\alpha_\textrm{CO}$ in units of $M_\odot$ (K km s$^{-1}$ pc$^2$)$^{-1}$:
\begin{equation}\label{eq:alpha}
M_{\textrm{mol}} = \alpha_\textrm{CO} \, L^\prime_\textrm{CO(1-0)}\textrm{.}
\end{equation}

In the Milky Way and nearby star-forming galaxies with near-solar metallicity, the empirical conversion factor $\alpha_\textrm{CO}$ 
is $\alpha_\textrm{MW}= 4.4\pm 0.9$, already including a contribution due to helium  
\citep[see, e.g.,][]{magnelli2012b,bolatto2013,Carilli-Walter2013,tacconi2020}. For extreme local starbursts, there 
is good evidence from dynamical arguments that $\alpha_\textrm{CO}$ is 
0.8–1.5 \citep[e.g.,][]{downes1998,daddi2010,genzel2010}, as has also been suggested
that for DSFGs for which $\alpha_\textrm{CO}$=0.8  \citep{Carilli-Walter2013}. 
Applying a  $1.36\times$ correction 
factor to account for the helium contribution, the latter value becomes $\alpha_\textrm{CO}=1.09$. 
\citet{Dunne2020} used the far-infrared continuum of a small but statistically complete 
sample of 12 low redshift galaxies together with CO and C{\small I} measurements to determine the $\rm CO-H_2$ 
conversion factor $\alpha_\textrm{CO}$. In another paper, Dunne et al. (in prep.) combined this sample will others measurements 
found in the literature for a total of more than 70 far-infrared and sub-mm selected galaxies with CO and CI detection and 230 with CO only. 
In these studies, a value of $\kappa_{850}=0.065$ m$^2$ kg$^{-1}$ is adopted \citep[in agreement with the results by][see Appendix \ref{sect:kappa} for details.]{planck2011_XXI} and it is found 
that a value $\alpha_\textrm{CO}\simeq3$ is consistent with the behaviour of the three gas tracers in a diverse 
population of galaxies, ranging from high-$z$ powerful star-forming galaxies to local disks.

For the computation of the molecular gas mass of HerBS-89a,  we here adopt $\alpha_\textrm{CO}=3.0$ $M_\odot$ (K km s$^{-1}$ pc$^2$)$^{-1}$, and we multiply it by the usual factor 1.36 to account for the helium contribution. Possible consequences of using a different value are discussed at the end of this Section.

The resulting line luminosity and molecular gas mass, corrected for lensing magnification, are thus
\begin{equation}
L^\prime_\textrm{CO(1-0)}= (5.15 \pm 1.05) \times 10^{10} \, \textrm{K km s}^{-1}\, \textrm{ pc}^2 \textrm{,}
\end{equation}
\begin{equation}
M_\textrm{mol} = (2.1\pm 0.4) \, \times 10^{11} \, M_\odot\textrm{.}
\end{equation}

The molecular gas mass derived from the $\rm ^{12}CO(1-0)$ emission line 
and the SED-based estimate of the dust mass (see Sect.~\ref{sect:sed})
provide the basis for deriving the average gas-to-dust ratio ($\delta_{\rm GDR}$) in HerBS-89a:
\begin{equation}\label{eq:GDR}
\delta_\textrm{GDR}=\frac{M_\textrm{gas}}{M_\textrm{dust}} \textrm{,}
\end{equation}
where the assumption is made that the two estimates are representative for the whole galaxy, and that $M_\textrm{mol}\gg M_\textrm{HI}$. 

The data lead to a value of $\delta_\textrm{GDR}$ between 80 and 174, depending on whether the DL07 or the MBB (general form) dust mass estimate is used, with an uncertainty of $\pm$21 based on error propagation. Interestingly, this value is consistent within 
the errors with those seen in typical star-forming galaxies of solar metallicity, near the so-called 
``Main Sequence of star formation'' (MS), for which $\delta_\textrm{GDR}\sim 100$ \citep[e.g.,][]{leroy2011,magdis2012,remyruyer2014}. 
This result is nevertheless subject to the adopted value for $\alpha_\textrm{CO}$

The total molecular gas mass ($M_{\textrm{mol}}$) and SFR relate to each other via 
the so-called depletion timescale (in years):
\begin{equation}\label{eq:depl_timescale}
\tau_\textrm{depl} = \frac{M_\textrm{mol}}{\textrm{SFR}}\textrm{,}
\end{equation}
where the rate of star formation is computed as $\textrm{SFR}=1.09\, \times 10^{-10} \, L_{8-1000\mu\textrm{m}}$ with the infrared
luminosity expressed in units of $L_{\odot}$ (Kennicutt et al. \citeyear{kennicutt1998a}, after modification for 
a Chabrier et al. \citeyear{chabrier2003} IMF). 

Assuming that the observed infrared luminosity is entirely 
associated with the ongoing star formation activity of 
HerBS-89a, we derive an intrinsic star formation rate ${\rm SFR} = 614\pm59 \, M_\odot \, \rm yr^{-1}$  and 
a depletion timescale  $\tau_\textrm{depl} = (3.4\pm\,1.0) \times 10^8$ years. 
For its current formation rate, HerBS-89a would exhaust its molecular gas reservoir in only 340 million years, not 
considering any mass return to the interstellar medium or further gas inflow (see Sect. \ref{sect:inflow}).

\begin{figure*}[!ht]
\centering
\rotatebox{-90}{\includegraphics[height=0.75\textwidth]{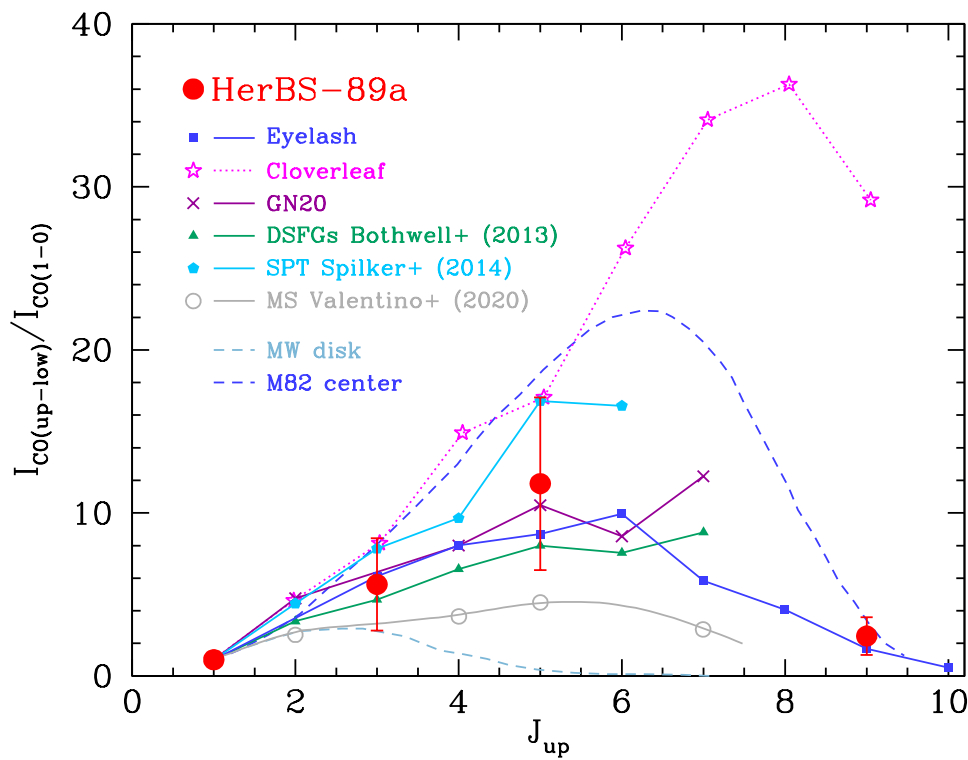}}
\caption{$^{12}$CO Spectral Line Energy Distribution (SLED) for HerBS-89a (red dots), compared to other well-studied 
local and high-$z$ galaxies (or samples of galaxies).  The CO SLEDs are normalized to the $\rm J = 1-0$ transition, after correcting for
the magnification,
and the line fluxes of HerBS-89a have been corrected for amplification (see Sect.~\ref{sect:lensing_model} and Table~\ref{tab:lensing_results_source}) before normalizing. 
The three samples of galaxies are from \citet{Bothwell2013} including 32 DSFGs at $1.2<z<4.1$,  \citet{Spilker2014}, 
including 22 SPT-selected DFSGs at $2.0<z<5.7$, and \citet{valentino2020}, including 78 ``main sequence'' star-forming galaxies. References to individual sources are: 
Cosmic Eyelash \citep{Swinbank2011,Danielson2011}; Cloverleaf \citep{barvainis1997,weiss2003,bradford2009,Riechers2011}; 
GN20 \cite[][and Cortzen et al. in prep.]{carilli2010, Cortzen2020}; 
M82 \citep{weiss2005b}; 
and the Milky Way disk \citep{fixsen1999}.}
\label{fig:sled}
\end{figure*}

The value for M$_\textrm{mol}$ was derived from the $\rm ^{12}CO(1-0)$ luminosity 
using $\alpha_\textrm{CO} = 3.0\times1.36$, a value that is close to 
the Milky Way value $\alpha_\textrm{MW}=4.4$ $M_\odot$ (K km s$^{-1}$ pc$^2$)$^{-1}$, commonly used for MS galaxies, 
rather than $\alpha_\textrm{SB}=1.09$, often associated to starbursts or outliers in the main sequence. This choice leads 
to $\delta_\textrm{GDR}$ and $\tau_\textrm{depl}$ values that would position HerBS-89a close to or on the MS star-forming 
galaxies on the integrated Kennicutt-Schmidt relation of star formation \citep[e.g.,][]{lada2015,liu2019}, which suggests that 
HerBS-89a is dominated by a ``secular'' star-forming mode.   

However, despite the evolution of the MS normalization 
\citep[increasing as a function of redshift, e.g.,][]{elbaz2011,whitaker2014,schreiber2015}, 
and because of the bending of the MS at the high stellar mass end, 
even at $z=3$, a $\textrm{SFR}\sim600$ $M_\odot$ yr$^{-1}$ would hardly be associated with a galaxy on the 
main sequence. It is therefore more likely that HerBS-89a is located in between the MS and the starbursts region.
Using $\alpha_\textrm{SB}$, the gas-to-dust ratio and the depletion time scale of HerBS-89a would both decrease 
by a factor of 3.74, yielding $\delta_\textrm{GDR}\simeq22-47$ and $\tau_\textrm{depl}=9.1\pm2.7\times 10^7$ yr.
Realistically, the actual value of $\alpha_\textrm{CO}$, on which these results depend, is likely between the two above mentioned extremes.

Further insights into the position of HerBS-89a with respect to the MS would require the knowledge of 
its stellar mass $M^\ast$. Detecting HerBS-89a and distinguishing it from its deflector seem out of reach for currently existing optical/NIR facilities, but the \textit{James Webb Space Telescope} (JWST) could provide the needed sensitivity.

\begin{figure}[!ht]
\centering
\includegraphics[width=0.48\textwidth]{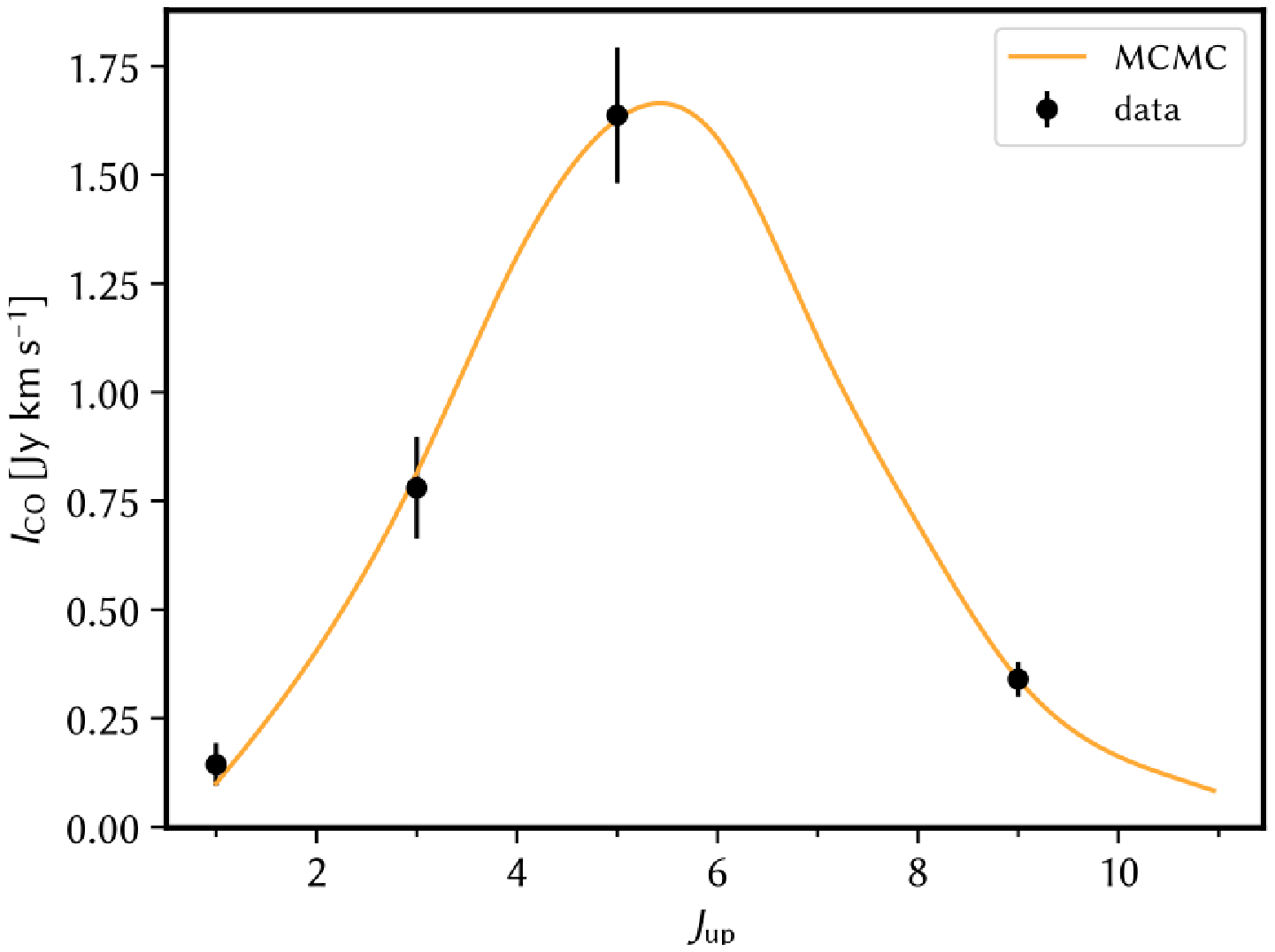}
\includegraphics[width=0.525\textwidth]{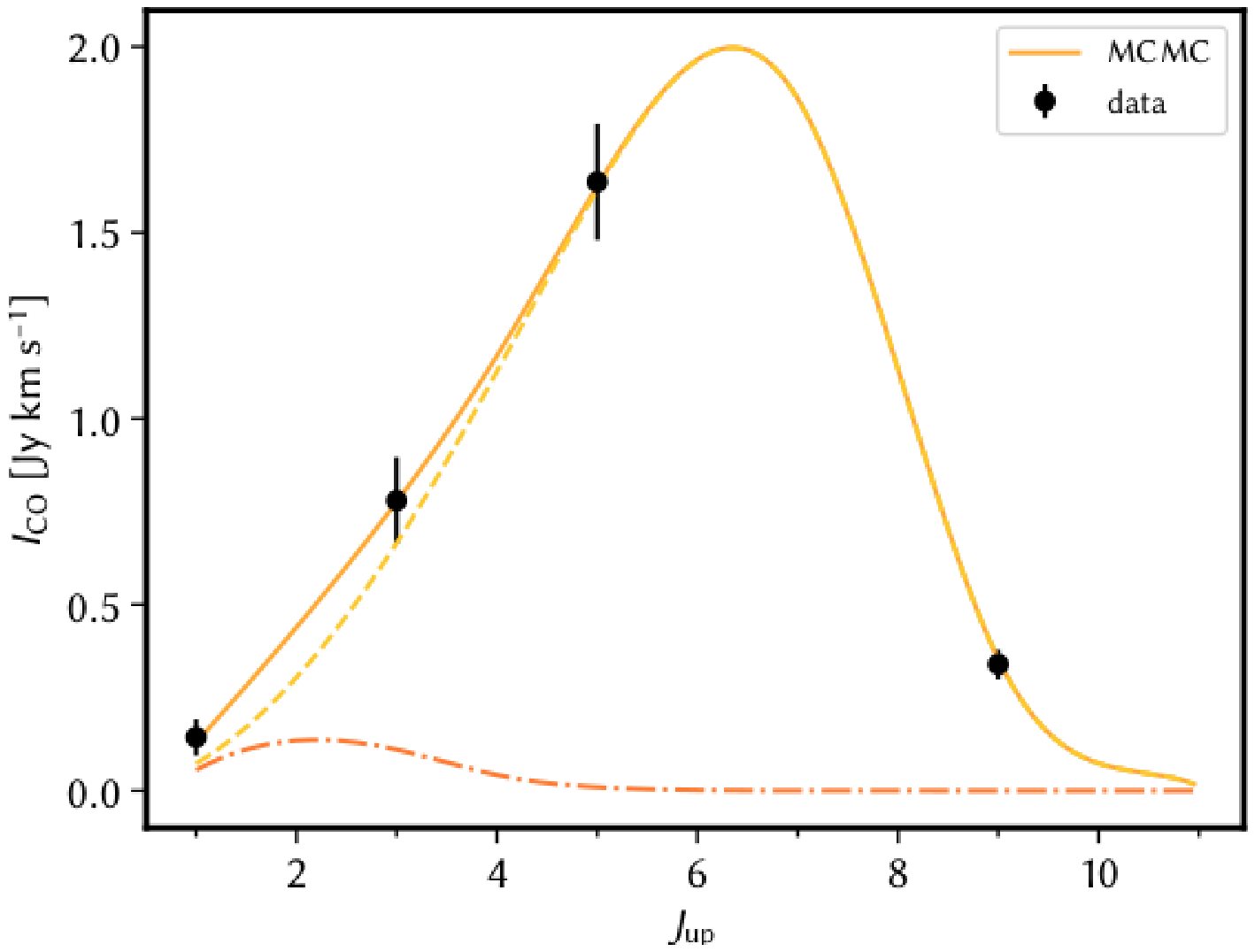}
\caption{Large Velocity Gradient (LVG) model fit to the $\rm ^{12}CO$ SLED of HerBS-89a (after correcting 
for lens magnification). {\it Top Panel}: The data are shown as black dots and the solid curve shows the best fit from the 
single component (MCMC) model corresponding to the maximum posterior possibility 
(see Sect.~\ref{sect:sled} for details). {\it Bottom Panel}: Two-components fit, including an additional cold and extended component (darker, dot-dashed line). }
\label{fig:SLED_fit}
\end{figure}

\subsection{CO Spectral Line Energy Distribution}\label{sect:sled}

Together with the measurements reported in \cite{Neri2020}, we have detected in total four $\rm ^{12}CO$ emission lines in HerBS-89a, 
namely the $J=$1-0, 3-2, 5-4 and 9-8 transitions. The $\rm ^{12}CO$ Spectral Line Energy Distribution (SLED) 
of HerBS-89a, normalized to the $J$=1-0 transition after correcting for the lens magnification (see Table~\ref{tab:lensing_results_source}), is shown in Fig.~\ref{fig:sled}. 
The velocity-integrated fluxes of the $\rm ^{12}CO$ lines increase monotonically with rotational 
quantum number up to $J$=5-4, with a turnover around that transition, before decreasing at higher $J$ 
values, with the $J$=9-8 emission line about a factor of 8 times weaker than the $J$=5-4 line. 
The peak of the CO SLED for HerBS-89a should therefore occur around the $J$=5-4 and 
$J$=6-5 transitions. This result makes HerBS-89a comparable to other starburst galaxies as shown in Fig.~\ref{fig:sled}, 
where various SLEDs of several individual DSFGs and one quasar (the Cloverleaf) are displayed together with the center 
of M82 and the Milky Way disk. In addition, the average SLEDs of three samples are shown: the (mostly) unlensed sample of 32 DSFGs 
selected at $\rm 850 \, \mu m$, at $1.2<z<4.1$ from \cite{Bothwell2013}, 
the sample of 22 (mostly lensed) high-$z$ ($2.0<z<5.7$) DSFGs selected from the SPT survey 
\citep{Spilker2014}, and the sample of 78 ``main sequence'' star-forming galaxies from \citet{valentino2020}.
The CO SLED of HerBS-89a most closely resembles that
of the Cosmic Eyelash, which has a similar peak around $J_{\rm up} \sim$5-6 and a falloff towards higher $J_{\rm up}$. 
It is also remarkably similar to the CO SLED of the starburst GN20 \citep{carilli2010}, up to 
$J_\textrm{up}=6$; however, recent observations by \citet[][and in prep.]{Cortzen2020} report an increase of the GN20 CO SLED at $J_\textrm{up}=7$.  

To investigate the CO line excitation further and constrain the physical conditions of the molecular gas in HerBS-89a, we model 
the CO line fluxes using a large velocity gradient (LVG) statistical equilibrium method 
\citep[following the approach described in][and references therein]{Yang2017}. 
The free parameters are the kinetic temperature of the molecular gas ($T_{\rm kin}$), the volume density 
($n_{\rm H_2}$), the column density of CO per unit velocity gradient ($N_{\rm CO}/{\rm dV}$), and the solid angle ($\rm \Omega_{app}$) of the source. The CO SLED only 
depends on $T_{\rm kin}$, $n_{\rm H_2}$ and $N_{\rm CO}/{\rm dV}$. A Bayesian approach is used to fit the line fluxes generated from the model 
and the code $emcee$ is adopted to perform a MCMC calculation. The resulting single component 
fit to the CO SLED of HerBS-89a is shown
in Fig.~\ref{fig:SLED_fit}. The SLED 
peaks around $J_{\rm up}$=6-5, and, from the single excitation component fitting, the molecular gas density is found to 
be $\rm log_{10}({\it n}_{H_2}/cm^{-3})=2.79^{+0.72}_{-0.54}$, the kinetic temperature 
$T_{\rm kin}=174^{+157}_{-92} \, {\rm K}$ and the CO column density 
${\rm log_{10}}(N_{\rm CO/dV}/(\rm cm^{-2} \, km \, s^{-1}))=17.59^{+0.55}_{-0.59}$, where the errors are $\rm 3\sigma$. 

A second fit including an additional, colder and more extended component is shown in the lower panel of Fig.~\ref{fig:SLED_fit}. 
Because of the paucity of available data points, the cold component is poorly constrained, with 
${\rm log_{10}}(n_{\rm H_2})=2.30^{+1.06}_{-0.57} \, \rm{cm^{-3}}$, $T_{\rm kin}=20^{+20}_{-7} \, {\rm K}$, and  
${\rm log_{10}}(N_{\rm CO/dV})=15.63^{+1.14}_{-0.79} \, {\rm cm^{-2} \, km \, s^{-1}}$, where the errors are $\rm 3\sigma$. The warm component is in 
broad agreement with the previous single component fitting results.

\begin{figure}[!ht]
\centering
\includegraphics[width=0.45\textwidth]{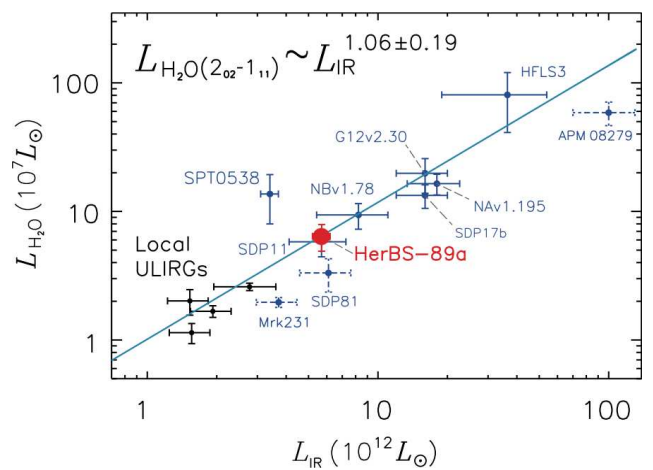}
\caption{
Correlation between $L_{\rm IR}$ and $L_{\rm H_2O(2_{02}-1_{11})}$ for local (black symbols) and high-redshift 
(blue symbols) galaxies. HerBS-89a is labeled with a red star and is located along the tight 
relationship between the infrared and water luminosity (shown as a light blue line). Figure is
adapted from \cite{Yang2016} and references therein.
}
\label{fig:water-infrared relation}
\end{figure}

\begin{figure*}[!ht]
\centering
\includegraphics[width=0.98\textwidth]{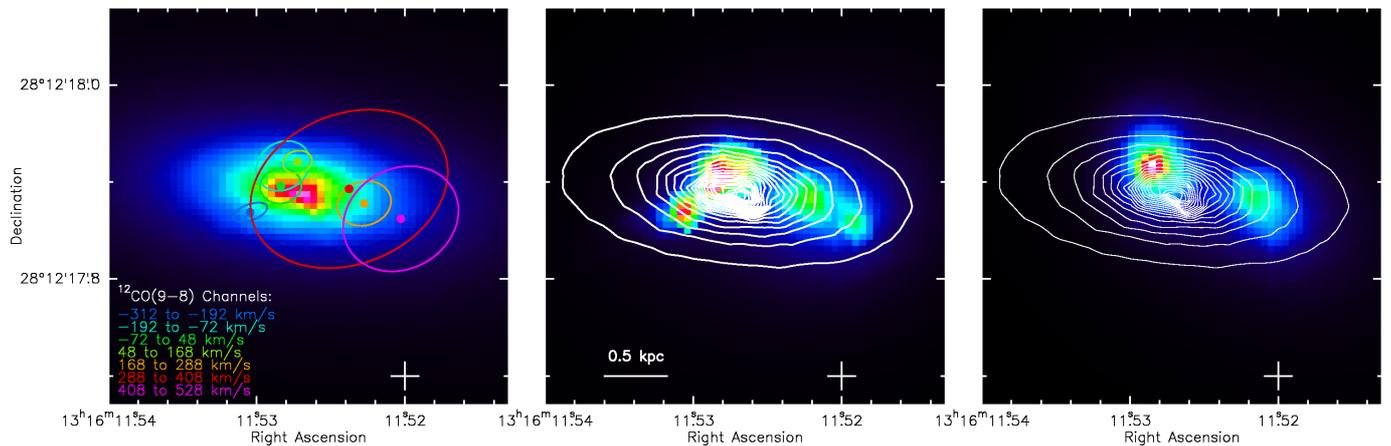}
\caption{{\em Left panel:} Positions (colored thick dots) and shapes (ellipses) of the reconstructed profiles of the individual 
$\rm ^{12}CO(9-8)$ channels, over-plotted on the map of the dust continuum emission in the 
source plane. The ellipses are plotted at 1/20$\rm ^{th}$ of the peak flux of each reconstructed channel map. 
{\em Central panel:} Zeroth moment map of the $\rm ^{12}CO(9-8)$ emission line derived from the reconstruction of the 
seven individual channels shown in the left panel. The contours show the dust continuum emission, starting at 1/20 of 
the peak flux in steps of 1/20$\rm ^{th}$ (see also Fig.~\ref{fig:co98_lens_model_ch}). 
{\em Right panel:} Zeroth moment map obtained splitting the $\rm ^{12}CO(9-8)$ in two halves, 
corresponding to the velocity ranges from -312 to $\rm 168 \, km \, s^{-1}$ and from 168 to $\rm 528 \, km \, s^{-1}$, and reconstructing them separately. 
In all panels, the cross at the lower right corner marks the phase center of the NOEMA observations.  }
\label{fig:co98_moment_0th_and_channels}
\end{figure*}

\begin{figure*}[!ht]
\centering
\includegraphics[width=0.97\textwidth]{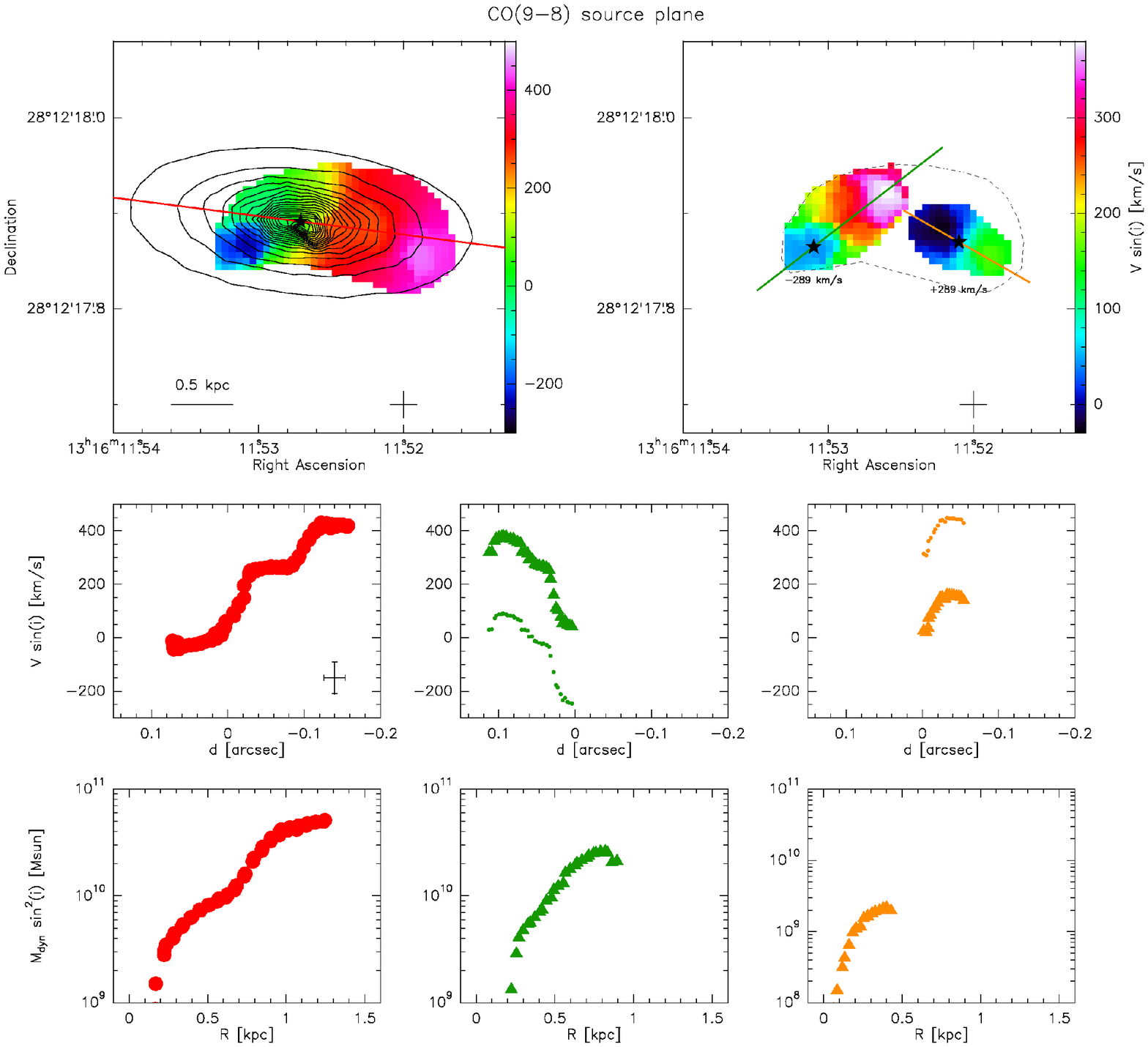}
\caption{Kinematic and dynamical analysis of the $\rm ^{12}CO(9-8)$ emission line in the source plane of HerBS-89a. 
{\em Top panels}: {\em Left} First moment map, compared to the S\'ersic profile of the dust continuum emission. The contours start 
from 1/20 of the peak intensity with steps of 1/20$\rm ^{th}$. The red line marks the continuum major axis.
The reference position is defined where the rest-velocity 
($V = 0 \, \rm km \, s^{-1})$ and is denoted with a star. From this position we calculate the radial distance.  
The cross at the lower right corner marks the phase center of the NOEMA observations. 
{\em Right}: Based on Figs.~\ref{fig:co98_lens_model_ch} and~\ref{fig:co98_moment_0th_and_channels}, 
two possible components are identified; the first moment maps of the two components are here shown, after shifting 
the velocity scale by $\rm \pm289\, km\, s^{-1}$ for clarity. The axes of the two components are shown as green and orange lines. 
The dashed line marks the contour of the velocity field shown in the top-left panel.
{\em Middle panels}: Velocity curves extracted along the axes marked in the top panels. Colors match those of the respective axes, with magenta points extracted from the top left moment map and the green and orange points extracted from the top right moment map.  In the two right-most panels, the data are shown both before (small symbols) and after (large symbols) shifting by $\rm \pm289\, km \, s^{-1}$. 
The amplitude of the error bars is shown in the left-most panel. {\em Bottom panels}: Dynamical mass profile as a function of radius, as derived from the velocity curves.}
\label{fig:kinematics_co98}
\end{figure*}

\subsection{The para-$\rm H_2O(2_{02}-1_{11})$ emission line}\label{sect:H2O-line}

Water is one of the most abundant molecules after $\rm H_2$ and CO in the 
gaseous interstellar medium (ISM). It serves as a unique diagnostic for probing the physical conditions of 
the ISM in both local \cite[e.g.,][]{Gonzalez-Alfonso2014} and high-redshift galaxies 
\cite[e.g.,][and references therein]{Omont2013, Riechers2013, Yang2016, Yang2020}. 
In particular, since the $E_\mathrm{upper}>200 \rm \, K$ levels of \hto\ are primarily excited through 
absorption of far-infrared photons from dust emission in warm dense regions, it is 
a useful diagnostic of the far-infrared radiation field independent of gas conditions. 
In addition, sub-millimeter water lines are the second strongest molecular emitter 
after the CO lines \cite[e.g.,][]{Yang2013}. The detection of the para-\htot202111\   
emission line ($\nu_{\rm rest} = 987.9268 \, \rm GHz$) in HerBS-89a is an illustration thereof: the line width is broad 
($\rm \Delta V \sim 1100 \, km\, s^{-1}$) and comparable to those 
of the mid/high-$J$ CO lines, 
consistent with the finding that the $J=2$ \hto\ lines are spatially 
co-located with the CO lines \cite[e.g.,][]{Omont2013, Yang2016}. The observed line 
flux is bright, with $I_{\rm H_2O} = 1.59\pm0.37 \rm \, Jy \, km\, s^{-1}$, and
comparable to the line flux of the $\rm ^{12}CO(9-8)$ emission line 
(Table~\ref{tab:lines}).  The CO and \hto\ lines have morphologies 
and velocity fields that are broadly comparable (see Figs.~\ref{fig:overlay_lines_continuum} 
and \ref{fig:overlays_vel_cont}) indicating that they both trace the same warm 
high-density gas in HerBS-89a. Using Eqs.~\ref{eq:L'_line} and \ref{eq:L_line}, 
we derive an intrinsic line luminosity (corrected for lensing magnification) 
of $L_{\rm H_2O(2_{02}-1_{11})} = (6.4\pm1.5) \times  10^7 L_\odot$, yielding a ratio 
$L_{\rm H_2O(2_{02}-1_{11})} / L_{\rm IR} = (1.12\pm0.36) \times 10^{-5}$, 
in terms of the 8 to 
$\rm 1000\,\mu m$ $L_{\rm IR}$ (Table~\ref{tab:sed_fit}). HerBS-89a lies 
on the correlation between $L_{\rm IR}$ and $L_{\rm H_2O}$, i.e. 
$L_{\rm H_2O(2_{02}-1_{11})} \propto L_{\rm IR}^{1.06\pm0.19}$ derived  
by \cite{Yang2016}, as shown in Fig.~\ref{fig:water-infrared relation}. This result is 
in line with the fact that far-infrared pumping is the 
dominant mechanism for the excitation of the sub-millimeter \hto\  
emission that originates in very dense, heavily obscured 
star-formation dominated regions \cite[see, e.g.,][]{Gonzalez-Alfonso2014, Yang2016}. 
The steeper-than-linear growth is likely the result of an increase of the optical 
depth in the dust continuum (at $\rm \sim 100 \, \mu m$) with increasing 
infrared luminosity. Finally, the ratio of $L_{\rm H_2O(2_{02}-1_{11})}  / L_{\rm IR}$ also 
indicates that an AGN component, if present in HerBS-89a, is not contributing 
much to the infrared luminosity, in contrast to a system like Mrk~231 
or APM~08279+5455 where the AGN is dominant \citep[and references therein][]{Yang2016}.

\section{Molecular gas kinematics}\label{sect:kinematics}

In an ideal situation, analysis of spectral line data cubes like those presented here for HerBS-89a can enable detailed analyses 
of source-plane kinematics. The lens model derived in Section 4 can in principle be applied to the emission 
in a particular line across many independent velocity channels, allowing them to be reconstructed in the source plane, 
used to compute moment maps, and fits to dynamical models
\citep[e.g.,][]{Sharon2019}. In practice, although the quality of our NOEMA observations 
is excellent, the S/N and the number of resolution elements (i.e., synthesized beams) 
across the observed size of HerBS-89a are not high enough to fully 
achieve this goal. Several factors are at play here.  
First, as noted in Sect.~\ref{sect:lensing_model} above, the combination of S/N and effective resolution limits us to the 
use of parametric (S\'ersic profile) sources in reconstructing individual channels.  Second, reconstruction of a given 
source-plane channel map tends to (a) favor sources that are very (sometimes implausibly) compact as a result of the use 
of a constrained emission profile, and (b) couple with the noise fluctuations in that channel in such a way 
that introduces uncertainties on the centroids of the reconstructed sources.  Finally, comparison of a 
delensed source-plane reconstruction with an observed channel map necessarily involves the convolution of the former 
to the (coarser) resolution of the latter, meaning that many possible source-plane reconstructions are formally compatible 
with the observed data.  

To provide a transparent (and possibly cautionary) picture of what properties can be derived from the current data,
we show in Figs.~\ref{fig:co98_moment_0th_and_channels} and \ref{fig:kinematics_co98} (as well as in
Figs.~\ref{fig:co98_lens_model_ch} and \ref{fig:h2o_lens_model_ch} in the Appendix) some of the results of our lens modelling efforts.   
Our main focus is the ${}^{12}$CO(9--8) line, which has higher S/N than 
the H$_2$O($2_{02} - 1_{11}$) line and is therefore more promising as a kinematic probe.  In order to improve 
the S/N, we rebin the $40\,{\rm km\,s^{-1}}$ data to a resolution of $120\,{\rm km\,s^{-1}}$, yielding a 
total of seven independent velocity channels that are suitable for modelling, although they do not span the line’s full 
velocity width.  Each of these channels is independently reconstructed with a S\'ersic profile in the source plane; these 
reconstructions are shown in the rightmost column of Fig.~\ref{fig:co98_lens_model_ch}, and are represented with the 
colored ellipses centered on dots in the left panel of Figure~\ref{fig:co98_moment_0th_and_channels} superposed on the 
dust continuum S\'ersic profile in the source plane. The modest (generally $< 2.5\sigma$ in the vicinity of the line emission) 
level of the data--model residuals in the third column of Fig.~\ref{fig:co98_lens_model_ch} indicates that these 
reconstructions are formally acceptable, although the five channels blue-ward of $168\,{\rm km\,s^{-1}}$ have scale radii 
that are notably smaller than those of the two most highly redshifted channels. The centroids of all seven channels fall 
within the envelope defined by the reconstructed dust continuum emission, and in turn by the reconstruction of 
the ${}^{12}$CO(9--8) zeroth moment map (second row of Fig.~\ref{fig:lens_model_cont}), giving confidence that 
the overall spatial extent of the warmest and densest material in HerBS-89a is understood.

The center panel of Fig.~\ref{fig:co98_moment_0th_and_channels} presents a source-plane zeroth moment map constructed by 
summing the source-plane reconstructions of the individual channel maps. This map seems to indicate an arc-like shaped 
morphology and the possibility of two separate components. However, we have here entered a regime in which the 
three limiting factors noted above are highly salient: because of the compactness of reconstructed S\'ersic-profile sources 
in most of the velocity channels, the moment map derived from them suggests that we have reliable information about the internal 
source structure on much finer scales than is actually the case.  It is also worth noting that when the moment map derived 
from the reconstructed channel maps is lensed forward into the image plane, it provides a poorer match to the observed 
${}^{12}$CO(9--8) moment map than does the single S\'ersic-component reconstruction in the second row of Fig.\ref{fig:lens_model_cont}. 
This apparent paradox occurs because small noise fluctuations in the observed channel maps have unavoidably affected 
the source centroids, shapes, and orientations of the corresponding source-plane reconstructions. Similar inconsistencies 
have been seen by other authors who have separately delensed both channel maps and integrated moment maps 
\cite[e.g.,][]{Dong2019}. Use of prior information on the expected relationships between emitting regions in adjacent 
velocity channels \citep[e.g.,][Young et al., in preparation]{Rizzo2018} would be one way to mitigate this tendency, 
but we have not used such an approach here.

Figure~\ref{fig:co98_moment_0th_and_channels} provides a further illustration of the degree to which small noise 
fluctuations in the observed channel maps can perturb the centroids of the source-plane reconstructions.  
Initially inspired by the apparent bent morphology (and nominal associated velocity gradients; see below) in the central panel of 
Fig.~\ref{fig:co98_moment_0th_and_channels}, we constructed new, roughly half-line moment maps from the bluest four 
and reddest three of our seven $120\,{\rm km\,s^{-1}}$ channels, and reconstructed these using the lens model optimized 
in Sec.~\ref{sect:lensing_model}. The sum of these two reconstructions (i.e., another zeroth moment map) is presented in 
the right panel of Fig.~\ref{fig:co98_moment_0th_and_channels}. The apparent double morphology might at first right 
suggest a pre-coalescence merging system; however, the “doubleness” of the morphology stems at least in part from the 
general compactness of our S\'ersic profile reconstructions.  Moreover, notable aspects of the internal source structure 
(e.g., the distinct southeastern offset of the reconstructed $-312$ to $-192\,{\rm km\,s^{-1}}$ channel map) in the center 
panel do not translate to the reconstructions used to derive the right panel. This level of inconsistency raises the 
question of whether detailed information on the intrinsic or the internal source structure (as distinct from the overall 
extent of emission) can be robustly recovered from this dataset, and suggests that the apparent small-scale structure 
in the resulting zeroth-moment maps has to be interpreted with caution. 

With the above discussion as preamble, we turn now to Fig.~\ref{fig:kinematics_co98}, which shows source-plane velocity 
fields derived from the seven-channel ($120\,{\rm km\,s^{-1}}$ resolution) reconstructions of Fig.~\ref{fig:co98_lens_model_ch} 
in two different ways. The left panel presents an intensity-weighted first moment map calculated in the usual way.  
We find that the dominant velocity gradient is in an east-west direction, roughly consistent with the east-west elongation of the 
reconstructed dust continuum emission. The comparison of the center and right panels in Fig.~\ref{fig:co98_moment_0th_and_channels} discussed above suggests that the perturbation from an \textit{exactly} east-west velocity gradient due to the most highly blue-shifted emission may not be robust.

Assuming a single rotationally supported disk, the radial profile of the dynamical mass of the system is given by
\begin{equation}\label{eq:Mdyn}
M_{\rm dyn}(R) = f \cdot {\frac {R\,V^2_{\rm obs}(R)}{G\,{\rm sin}^2\,i}} {\rm ,}
\end{equation}
where $i$ is the inclination angle of the rotating disk with respect to the line of sight, and $V_\textrm{obs}$ is 
the observed velocity as a function of radius. Although the inclination could be derived from 
the ellipticity of the light profile, as $i=\arccos{\left(1-e\right)}$, since the velocity gradient 
does not fully correspond to the continuum S\'ersic profile in the source plane, we opt to keep 
the $\sin(i)$ term explicit in our computations. The constant $f$ is a dimensionless scale factor that depends on the 
structure of the galaxy and has value $f \simeq 1.0$ for a thin disk embedded in a massive spheroid \citep{Lequeux1983}. 

Along the major axis of the continuum profile (red line in the top-left panel of Fig. \ref{fig:kinematics_co98}), we extract the velocity curve $V(R)$ shown in the middle-left panel. The corresponding bottom-left panel presents the radial dynamical mass profile, which leads to an estimate of  $M_{\rm dyn}\,{\rm sin}^2\,i \approx 5.1 \times 10^{10}\,M_\odot$ within $R = 1.25\,{\rm kpc}$. 

In contrast, the top-right panel of Fig.~\ref{fig:kinematics_co98} presents a first moment map in which masking has been used 
to highlight the two regions of strongest integrated emission (compare also to Figs. \ref{fig:co98_moment_0th_and_channels} 
and \ref{fig:co98_lens_model_ch}).  These two regions’ velocity scales have been artificially 
offset by $\pm 289\,{\rm km\,s^{-1}}$, corresponding to the separation
between the two peaks in the integrated CO spectrum (see Fig.\ref{fig:spectral_fit_GILDAS}) in order to better present them as
\textit{candidate} rotating disks. 
Velocity curves extracted along the axes of these two possible structures (green and orange lines) are shown in the middle row of Fig. \ref{fig:kinematics_co98},
both before (small symbols) 
and after (large symbols) applying the $\rm \pm289\, km \, s^{-1}$ shift.

The individual dynamical masses of these components, assuming rotational 
support as above, would be $M_{\rm dyn}\,{\rm sin}^2\,i \approx 2.6 \times 10^{10}\,M_\odot$ within $R = 0.85\,{\rm kpc}$ 
for the eastern component and $M_{\rm dyn}\,{\rm sin}^2\,i \approx 2.2 \times 10^{10}\,M_\odot$ within $R = 0.40\,{\rm kpc}$ 
for the western component, again under the assumption that $f = 1.0$ in Eq.~\ref{eq:Mdyn}. The bottom panels of Fig.~\ref{fig:kinematics_co98} 
show the dynamical mass radial distributions, limited to the extension of the first-moment maps. 

Assuming a baryon fraction $f_\textrm{baryon}$ \citep[e.g.,][]{tacconi2020}, and 
supposing that at high redshift the gas content of a galaxy is dominated by its molecular component and the atomic gas mass $M_\textrm{HI}$ can be neglected \citep[see, e.g.,][]{tacconi2018}, 
the dynamical mass could 
yield an estimate of the stellar mass $M^\ast=f_\textrm{baryon}M_\textrm{dyn}-M_\textrm{mol}-M_\textrm{HI}$
and of the molecular gas fraction $f_\textrm{mol}= M_\textrm{mol}/(f_\textrm{baryon}M_\textrm{dyn}-M_\textrm{mol})$, thereby solving the question of whether or not the galaxy lies on the MS. 
However, the dynamical mass profiles derived for HerBS-89a do not extend to large enough radii to allow for a determination of the total mass.

For the sake of completeness, we also performed the reconstructions of multiple velocity channels across 
the $\rm H_2O(2_{02}-1_{11})$ emission line, adopting the same approach as above, using a coarser velocity 
re-binning to boost the S/N ratio (see Appendix \ref{app:lens}). As already suggested by the
slightly different observed and reconstructed source-plane zeroth moment maps (see Fig.~\ref{fig:lens_model_cont}), 
the individual $\rm ^{12}CO(9-8)$ and $\rm H_2O(2_{02}-1_{11})$ velocity channels reconstruct to slightly different
locations in the source plane (see Figs.~\ref{fig:co98_lens_model_ch} and \ref{fig:h2o_lens_model_ch}). 
However, a first moment map derived from the reconstructed $\rm H_2O(2_{02}-1_{11})$ channel maps shows
a velocity gradient similar to the one shown in the top left panel of Fig.~\ref{fig:co98_lens_model_ch} for the $\rm ^{12}CO(9-8)$ emission line, albeit at a slightly different position angle, suggesting once more that 
this aspect of the system's kinematics has been robustly recovered. Due to the higher S/N of the $\rm ^{12}CO(9-8)$ emission
line, we have limited the kinematic modelling to this line, rather than adventuring further in the realm of water.

To summarize, given the uncertainties discussed above, ambiguity remains as to 
whether the broad, double-peaked spectral lines in HerBS-89a reflect a single rotating disk or a pair of galaxies 
on their way to merging, as in the case of previous studies of high-$z$ lensed starburst 
galaxies \citep[e.g.,][]{Genzel2003,Ivison2010,Sharon2015}. 
On balance, we favor the latter
scenario, which would provide a natural explanation of the very large widths of the $\rm ^{12}CO$ emission lines in this system. Observations with higher S/N and angular resolution that can support a more detailed lens model 
will be needed to resolve the one vs. two sources question in a definitive way for this system, shed further light 
on the properties of this galaxy, unveil the finer details of its structure and kinematics, and fully recover the secrets
of its starburst nature.

\section{Molecular lines other than CO and water}\label{sect:other-lines}
In the following sub-sections, we present and discuss the various molecular absorption and emission lines 
other than CO and $\rm H_2O$ that have been detected in the spectrum of HerBS-89a. 
The individual molecular lines are described starting with 
the very dense molecular gas tracers, namely: $\rm HCN$ and/or $\rm NH$, and $\rm NH_2$ (Sect.~\ref{sect:very-dense-gas}), followed by
the two molecular ions, $\rm OH^+$ and $\rm CH^+$, which probe the low density molecular gas with low $\rm H_2$ 
fractions (Sect.~\ref{sec:low-density-gas}). 

\subsection{Very dense gas tracers: $\rm NH_2$, and $\rm HCN$ and/or $\rm NH$}\label{sect:very-dense-gas}

\paragraph{$\bf NH_2$}
Amidogen ($\rm NH_2$) is a molecule that is an important reactant intermediate in the production 
and destruction of N-bearing 
molecules, in particular as a photodissociation product of ammonia ($\rm NH_3$). It is an asymmetrical molecule with a 
$\rm ^2B_1$ ground electronic state characterized by a complex rotational spectrum (similar to $\rm H_2O^+$). It was 
first detected in the interstellar medium (in SgrB2) by \cite{VanDishoeck1993} and \cite{Goicoechea2004}. 
$\rm NH_2$ was later detected in the \textit{Herschel} SPIRE-FTS 
spectra of Mrk~231 and Arp~220 \citep{Gonzalez-Alfonso2018, Rangwala2011}, where the $\rm NH_2(2_{02}-1_{11})(5/2-3/2)$ 
emission line is seen next to the $\rm OH^+(1_0-0_1)$ absorption line, and, possibly, the $\rm NH_2(2_{02}-1_{11})(3/2-1/2)$ 
emission line as well.

In the case of HerBS-89a, two emission lines of $\rm NH_2$ are detected. The first is the $\rm NH_2(2_{02}-1_{11})(5/2-3/2)$ emission line, which overlaps with the $\rm OH^+(1_0-0_1)$ absorption line, such that only the red-shifted part of the $\rm NH_2$ emission 
line is apparent in the spectrum. However, the nearby $\rm NH_2(2_{02}-1_{11})(3/2-1/2)$ remains undetected, which is 
in line with the expected ratio of $\approx 2$ between the $(5/3-3/2)$ and $(3/2-1/2)$ transitions. 
The second detected transition $\rm NH_2(2_{20}-2_{11})(5/2-3/2)$  is weaker but uncontaminated by nearby lines.

As the data are of low signal-to-noise, a detailed study of $\rm NH_2$, first reported here in a high-$z$ galaxy, 
remains out of reach. However, from the velocity-integrated emission map of the high-angular resolution data of the 
$\rm NH_2(2_{20}-2_{11})(5/2-3/2)$ emission line (Fig.\ref{fig:overlay_lines_continuum}), it is clear that 
the $\rm NH_2$ emission closely resembles the distributions of the $\rm ^{12}CO(9-8)$ and water emission lines, 
with the southern peak being stronger, in accordance with the fact that this molecule also traces the high-density 
gas. 
Dedicated future observations can further explore the properties of this molecule in HerBS-89a 
or in other high-$z$ galaxies to study the properties of their dense gas reservoirs.

\paragraph{$\bf HCN(11-10) \, and/or \, NH$} 
The absorption feature observed at 246.72~GHz in HerBS-89a may be due to redshifted $\rm HCN(11-10)$ and/or  
$\rm NH(1_2-0_1)$. With the current data, there is no way to distinguish between these molecular species, so we will
briefly describe each molecule separately. The absorption feature is relatively weak and detected at a 
low signal-to-noise ratio. As shown in Fig.~\ref{fig:overlay_lines_continuum}, it
is spatially well centered on both dust continuum emission peaks. This is the first detection 
of a high-$J$ HCN line and/or NH in a high-$z$ galaxy. Clearly, detecting other transitions of HCN and NH 
in HerBS-89a will help to determine the relative contributions of each of these species to the observed 
246.72~GHz absorption feature. 

\begin{itemize}

\item {\bf HCN} is known to be a reliable tracer of dense molecular gas ($n_{\rm H_2} \gtrapprox 3 \times 10^4 \, \rm cm^{-3}$), 
with critical densities 100 to 1000 times higher than of CO. Although HCN emission lines are typically one to two orders 
of magnitude fainter than CO emission lines, HCN has been observed in its low-$J$ transitions in 
local starburst galaxies, revealing a tight correlation between the HCN luminosity 
and the star formation rate \cite[e.g.,][]{Gao&Solomon2004, Zhang2014}. Emission lines of low-$J$ HCN have also been detected 
in a few high-$z$ galaxies \citep[][and references therein]{Weiss2007, Riechers2010a, Oteo2017, Canameras2020}, and HCN is seen in 
the stacked rest-frame 220-770~GHz spectrum of 22 high-$z$ DSFGs selected from the SPT survey \citep{Spilker2014}. 
  
Between $J$=5$-$4 and 11$-$10, HCN flips from emission to absorption as shown in the case of 
Arp~220 where \cite{Rangwala2011} detect the $J$=12$-$11 to 17$-$16 transitions in 
absorption\footnote{The $\rm HCN(11-10)$ transition was not discussed 
in \cite{Rangwala2011}}.
These transitions of HCN have critical densities around $\rm 10^9-10^{10} \, cm^{-3}$ and probe 
very dense ($n_{\rm H_2} \approx 3 \times 10^{6} \, \rm cm^{-3}$) gas. The high-$J$ lines of HCN are 
thought to be populated by radiative pumping of infrared photons (rather than by collisional excitation) 
in an intense high-temperature ($\rm > 350 \, K$) radiation field \cite[][]{Rangwala2011} .  

In HerBS-89a, the optical depth of the 246.72~GHz line is estimated to be $\tau(\nu_0) \sim 0.24\pm 0.12$ 
yielding, in the case of HCN, a column density of $\rm (0.38\pm0.1) \times 10^{14} \, cm^{-2}$ 
(Table~\ref{tab:tau_and_column_density}). Future observations of both high-$J$ HCN lines (in absorption) 
and low-$J$ lines (in emission) in HerBS-89a would allow us to confirm if this absorption line is indeed 
due (or partly due) to HCN, and, once verified, allow us to probe and constrain the properties 
of the reservoir of extremely dense gas in this high-$z$ starburst galaxy.

\item  {\bf NH} (imidogen) is a molecular radical that is linked (as for $\rm NH_2$) 
to the formation and destruction of $\rm NH_3$. It was first detected in the 
diffuse interstellar medium \citep{Meyer1991} and later in the dense molecular gas of SgrB2 \citep{Cernicharo2000, Goicoechea2004}. 
The \textit{Herschel} SPIRE-FTS spectra of Mrk~231 and Arp~220 \citep{Gonzalez-Alfonso2018, Rangwala2011} show the 
presence of $\rm NH$ with various transitions detected. $\rm NH$ has a spectrum similar to that of $\rm OH^+$, 
since these molecules are isoelectronic. Many of the transitions of $\rm NH$ overlap 
with $\rm OH^+$ transitions, in particular $\rm NH(1_2-1_0)$ with $\rm OH^+(1_0-0_1)$, with 
similar column densities derived for $\rm OH^+$ and $\rm NH$ in Mrk~231 
\citep{Gonzalez-Alfonso2018}. In the case of HerBS-89a, attributing all of 
the 246.72~GHz absorption line to $\rm NH$ implies
a column density of $\rm (6.7\pm1.9) \times 10^{14} \, cm^{-2}$, which is also
comparable to the column density of $\rm OH^+$ (Table~\ref{tab:tau_and_column_density}). 
Together with $\rm NH_2$, the likely detection of $\rm NH$ in HerBS-89a opens up the possibility of exploring 
the chemistry of N-bearing molecules in a high-$z$ galaxy by measuring other transitions of $\rm NH$ and searching for
$\rm NH_3$, and exploring whether (as is the case in SgrB2) the derived column densities are compatible with 
grain-surface chemistry and sputtering by shocks \citep[see, e.g.,][]{Goicoechea2004}.   

\end{itemize}

\subsection{Low density gas tracers: $\rm OH^+$ and $\rm CH^+$}\label{sec:low-density-gas} 

\paragraph{$\bf OH^+$}
The molecular ion $\rm OH^+$ is a reliable and powerful tracer of molecular outflows in DSFGs  
as has been demonstrated  with the \textit{Herschel} SPIRE-FTS studies in Arp~220 \citep{Rangwala2011} and 
Mrk~231 \citep{Gonzalez-Alfonso2018}, and in high-$z$ galaxies with NOEMA in HLFS~3 \citep{Riechers2013} 
and ALMA \citep[][who reported $\rm OH^+(1_1-0_1)$ absorption line in two lensed sources, i.e., the Cosmic Eyelash and SDP.17]{Indriolo2018}.  
As discussed in \cite{Indriolo2018}, $\rm OH^+$ absorption is thought to arise 
in the extended cool, diffuse, low $\rm H_2$ fraction gas that surrounds galaxies. $\rm OH^+$ has 
been found to be useful in constraining the cosmic-ray ionization rate of atomic hydrogen, $\rm \zeta_{H_2}$, 
particularly in gas with a low $\rm H_2$ fraction. The three ground 
state lines of $\rm OH^+$ 
(all are split into hyperfine components, which are blended due to the broad line widths 
measured in galaxies) are generally observed in absorption. In a few cases, they can be seen 
in emission as a red wing, as observed in Mrk~231 or Arp~220, or as recently reported for the $\rm OH^+(1_1-0_1)$ line 
in the $z=6.03$ quasar SDSS231038+1855 by \cite{Li2020} and the hot dust-obscured galaxy W0410$-$0913 at $z=3.63$ by 
\cite{Stanley2020}, where they can provide a measurement of the local density \citep[e.g.,][]{Rangwala2011}. 

In HerBS-89a, all three $\rm OH^+$ ground-state $1_J-0_1$ lines ($J = 0, 1, 2$) are seen in absorption (Fig.~\ref{fig:spectral_fit_GILDAS} 
middle panel). The $(1_1-0_1)$ and $(1_2-0_1)$ transitions are the strongest, while the $(1_0-0_1)$ transition is 
nearly five times weaker (Table~\ref{tab:lines}). None of the $\rm OH^+$ lines in HerBS-89a show any indication of emission above 
the continuum as in Mrk~231 or Arp~220 \citep{Gonzalez-Alfonso2018, Rangwala2011}. 
All three lines have comparable line widths with $\rm \Delta V \approx 500 \, km\, s^{-1}$ and the inferred $\rm OH^+$ column 
density is $N_{\rm OH^+} \sim 10^{15} \, \rm cm^{-2}$ (Table~\ref{tab:tau_and_column_density}), comparable 
to the value derived for the Cosmic Eyelash \citep{Indriolo2018}. The $\rm OH^+$ spectra show 
clear indications of red-shifted gas, with all three transitions peaking in opacity at $\rm \approx +100 \, km\, s^{-1}$ relative to the systemic velocity, 
including the $(1_0-0_1)$ transition after the  contamination of the $\rm NH_2$ emission 
line is accounted for (see Sec.~\ref{sect:lines}).

\paragraph{$\bf CH^+$}
The methylidyne cation, $\rm CH^+$, has been shown to be a sensitive tracer of feedback mechanisms 
in high-$z$ DSFGs \citep[]{Falgarone2017}, who report the detection of $\rm CH^+(1-0)$ in six lensed 
galaxies at redshifts $z \approx 2.5$, including the Cosmic Eyelash and SDP.17\footnote{The first detection of $\rm CH^+$ was 
made in the far-infrared spectrum of the planetary nebula NGC~7027 through  rotational lines 
from $J=2-1$ to $6-5$ seen in emission \cite{Cernicharo1997}.}. The $\rm CH^+(1-0)$ line is found 
both in emission and absorption, providing critical information on the mechanisms of mechanical energy 
dissipation and/or strong UV irradiation in these high-$z$ galaxies. Due to its high critical density for 
excitation ($\rm \sim 10^7 \, cm^{-3}$ for the $J=1-0$ transition), the $\rm CH^+$ line is seen in emission 
in dense ($\rm > 10^5 \, cm^{-3}$) gas, probing shocked regions powered 
by galactic winds, and in absorption, in large ($\rm > 10 \, kpc$) reservoirs of turbulent, cold ($\rm \sim 100\, K$), 
low-density ($\rm \sim 100 \, cm^{-3}$) gas \citep[see, e.g.,][]{Falgarone2017}. 

In HerBS-89a, the $\rm CH^+(1-0)$ is detected in absorption with a width of $\rm \approx 400 \, km\, s^{-1}$, 
comparable to that of the $\rm OH^+$ lines, and is also red-shifted to $\rm \approx +100 \, km \, s^{-1}$ 
(Fig.~\ref{fig:spectral_fit_GILDAS}). The inferred $\rm CH^+$ column density is estimated to 
be $N_{\rm CH^+} \sim 10^{14} \, \rm cm^{-2}$ (Table~\ref{sect:other-lines}), similar to what 
has been derived in the case of the Cosmic Eyelash \citep{Falgarone2017}. 

The low signal-to-noise, tentative detection of CH$^+$ emission could indicate that in HerBS-89a 
there is also dense gas ($\rm >10^5 \, cm^{-3}$) with a significant velocity dispersion 
(the line has a full width at zero intensity of $\rm \gtrapprox  1000 \, km\, s^{-1}$ and is centered at the systemic velocity), 
as was reported for three of the sources studied by \cite{Falgarone2017}. However, higher quality data 
are required to probe this dense gas component in greater detail.

\section{Inflowing gas}\label{sect:inflow}

In HerBS-89a, the three $\rm OH^+$ ground-state $1_J-0_1$ lines ($J = 0, 1, 2$) and the $\rm CH^+(1-0)$ line are all seen in 
absorption and red-shifted with respect to the systemic velocity (defined by the CO emission lines) by 
$\rm \approx +100 \, km\, s^{-1}$. As shown in the velocity maps of these absorption lines, the red-shifted gas is distributed 
over the southern and northern dust continuum peaks (in the image plane), uniformly covering the distribution 
of the blue- and red-shifted gas traced in the $\rm ^{12}CO(9-8)$ and water emission lines 
(Fig.~\ref{fig:overlay_lines_continuum}). The kinematics of the absorbing gas 
in HerBS-89a indicate that this red-shifted gas is therefore not kinematically related to the dense molecular gas.  
We argue here that we are in fact tracing, in both $\rm OH^+$ and $\rm CH^+$, 
low density gas that is flowing into the central regions of HerBS-89a, 

The number of high-$z$ sources for which measurements of absorption by molecular ions 
such as $\rm CH^+$ or $\rm OH^+$ are available is still low; however, it is noteworthy that the majority 
of the sources for which observations of these species have been made do show blue-shifted 
absorption lines, indicating vigorous outflow activity. Out of the five high-$z$ lensed galaxies where
$\rm CH^+(1-0)$ has been found in absorption, only one source
\footnote{In the case of the galaxy G09v1.40 reported in \cite{Falgarone2017}, adopting 
the precise redshift of $z=2.0924\pm0.0001$ \citep{Yang2017} shifts the $\rm CH^+(1-0)$ absorption line 
to $\rm \sim -200 \, km \, s^{-1}$ with respect to the systemic velocity of the 
source as defined by the CO emission lines; the $\rm CH^+(1-0)$ absorption line is therefore blue-shifted in this system  - see also
Butler et al. (in prep.).} 
shows a red-shifted line, namely the Cosmic Eyelash \citep{Falgarone2017}; for this source, the $\rm CH^+(1-0)$ absorption line 
is shifted by $\rm \Delta V \sim +150 \, km\, s^{-1}$ with respect to the systemic velocity as defined by the CO emission lines. 
A similar velocity shift in the Cosmic Eyelash is also observed in the $\rm OH^+(1_1-0_1)$ and $\rm H_2O^+(1_{11}-0_{00})$ absorption 
lines \citep{Indriolo2018}. Recently, Butler et al. (in prep.) report  $\rm OH^+(1_1-0_1)$ absorption lines towards ten
gravitationally lensed \textit{Herschel}-selected high-$z$ galaxies; for all of them, the line is invariably seen in absorption 
and blue-shifted with respect to the systemic velocity, with the exception of one source where the absorption line 
peaks at the systemic velocity. Together, the currently available observations of $\rm OH^+$ 
(and $\rm CH^+$) have established that these molecular ions are robust tracers of molecular gas outflows 
in high-$z$ galaxies, confirming the results of observations of $\rm OH^+$ or $\rm OH$ in local ultra-luminous 
galaxies \citep[e.g.,][and references therein]{Lu2017,Gonzalez-Alfonso2017}. 
In addition, recent studies of absorption lines in high-$z$ lensed 
DSFGs in the OH $\rm 119 \, \mu m$ doublet, in SPT-2319$-$55 at $z=5.29$ \citep{Spilker2018} and 
the quasar ULASJ131911+095051 at $z=6.13$ \citep{Herrera-Camus2020}, and in the $\rm H_2O(3_{30}-3_{21})$ 
and $\rm (4_{23}-4_{14})$ transitions in SPT-0346$-$52 at $z=5.656$ \citep{Jones2019}, have
revealed blue-shifted lines tracing powerful outflows (with rates of $\sim 200-500 \, M_\odot \rm yr^{-1}$).

In contrast to the above trends, the detection of red-shifted  molecular absorption lines in high-$z$ systems seems much less common. 
Only two cases are known, namely the Cosmic Eyelash \citep{Falgarone2017, Indriolo2018} and HerBS-89a (here described), 
out of the total of 18 high-$z$ sources observed to date for which $\rm CH^+(1-0)$ and/or $\rm OH^+(1_1-0_1)$ 
(or other tracers such as OH, $\rm H_2O$ or $\rm H_2O^+$) have been seen in absorption. 

Direct observational evidence of inflow is indeed scarce. Signatures of infalling gas are notoriously difficult 
to observe: the accreting material may have low metallicity, low density, and/or a small covering factor 
\citep[less than 10\% at $z \sim 1.5$; see, e.g.,][]{Fumagalli2011}, and it might be ionized or obscured by outflows. 
The need for a good alignment of an accretion structure with a background continuum source along the line of sight lowers the chance of detection even 
further.

The findings here described for molecular absorption lines in high-$z$ galaxies are in line with the scarcity of infall 
evidence gathered from previous studies. Using optical spectroscopy, \cite{Rubin2012} report evidence of cool, 
metal-enriched gas accretion onto galaxies at $z\sim0.5$. \cite{Wiseman2017} found a rare case of inflow of metal-poor 
gas from the intergalactic medium onto a $z=3.35$ galaxy. More recently, \cite{Ao2020} report evidence of infalling 
gas in a Lyman-$\alpha$ blob at $z=2.3$, and \cite{Daddi2020} presented observations revealing cold gas filaments 
accreting toward the center of a massive group galaxy at $z=2.91$. In the local Universe, a few examples of inflow 
activity have been found through \textit{Herschel} spectroscopic observations of compact luminous 
infrared galaxies. These include NGC~4418 and $\rm Zw \, 049.057$, where reversed P-Cygni profiles, a clear signature of 
infall, were revealed in the $\rm [OI] \,  63 \mu m$ fine-structure lines \citep{Gonzalez-Alfonso2012, Falstad2015}; 
Arp~299 where the absorption of the ground state of OH at 119~$\rm \mu m$ is found red-shifted by $\rm \sim +175 \, km \, s^{-1}$, 
suggesting a low excitation inflow in this source \citep{Falstad2017}; Circinus, which shows an unambiguous 
OH 119~$\rm \mu m$ inverted P-Cygni profile \citep{Stone2016}; and IRAS~10091+4704, the starburst-dominated system at $z=0.24$,
were a clear inverted P-Cygni OH profile indicates the presence of a fast inflow of molecular gas with a
rate of $\sim 100 \, M_\odot \rm yr^{-1}$ \citep{Herrera-Camus2020}.

\begin{figure}[!t]
\centering
\includegraphics[width=0.45\textwidth]{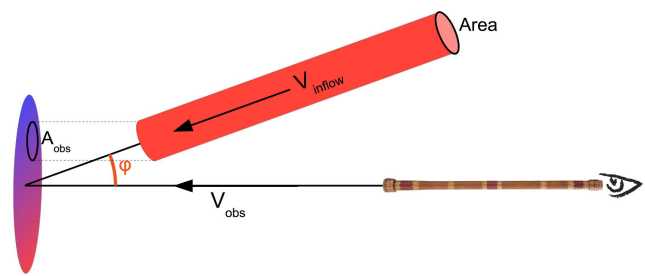}
\caption{Schematic of the simplified geometry of the inflow.}
\label{fig:geometry_infall}
\end{figure}

Theoretically, gas flowing into less massive ($M \leq 10^{12}\, M_\odot$) galaxies should be dynamically 
and thermally cold, while more massive halos receive most of their baryons as hotter ($T > 10^5 \, \rm K$) gas 
\citep[e.g.,][]{dekel2003,keres2005}. Thus cold, dense, metal-poor circum-galactic gas is often interpreted 
as direct evidence of accretion. Metal-poor Lyman-limit systems have been used as tracers of accretion 
in observations \citep[e.g.,][]{lehner2013} and simulations \citep[e.g.,][]{Fumagalli2011}.

Inflow and outflow rates of galaxies are constrained by cosmological simulations \citep[e.g.,][]{keres2005} and 
analytic models based on the mass assembly history of dark matter halos \citep{bouche2010,lilly2013}. 
\citet{erb2008} reproduces the gas mass fraction and metallicity distribution of galaxies with a 
simple chemical model involving gas inflow and outflow. Inverting their reasoning, \citet{yabe2015} 
constrain gas inflow and outflow rates based on the observed stellar masses, gas mass fractions, 
and metallicities of star-forming galaxies at $z\sim 1.4$.

Taken together, the above evidence strongly indicates that galaxies possess large reservoirs of circum-galactic 
gas eligible for accretion \citep{tumlinson2017}. Nevertheless, evidence for fuel does not necessarily 
imply the presence of fueling. 

We can describe the gas flow as a cylindrical channel that brings the gas onto the galaxy 
(see Fig.~\ref{fig:geometry_infall} for a schematic representation). Given the distance $d_\textrm{inflow}$ of the infalling clouds from the galaxy (i.e., given the line-of-sight distance that the gas has yet to travel), and assuming that the accreting gas 
travels at a constant speed $\textrm{V}_\textrm{inflow}$, the mass inflow rate can be computed as follows:
\begin{equation}\label{eq:inflow_rate}
\frac{dM_\textrm{X}}{dt}=N_l(\textrm{X}) m_\textrm{X} \frac{A_\textrm{obs}}{\sin{\phi}} \frac{\textrm{V}_\textrm{obs}}{d_\textrm{inflow}\cos{\phi}} {\rm .}
\end{equation}
Here X denotes either the $\rm OH^+$ or $\rm CH^+$ molecular ion; $N_l$ is the column density of the inflowing gas; $m_\textrm{X}$ is the mass of the molecular ion; $A_\textrm{obs}$ is the de-magnified projected area covered by the inflowing gas in the sky;  $\phi$ is the angle between the line of sight and the inflow direction; and $\textrm{V}_\textrm{obs}$ is the measured red-shifted velocity of the line. With these definitions, the term $V_\textrm{obs}/\cos{\phi}$ represents the de-projected inflow velocity $\rm V_\textrm{inflow}$, and the cross-sectional area of the cylinder perpendicular to the flow is given by $A_\textrm{inflow}=A_\textrm{obs}/\sin{\phi}$.

The $N_l(\textrm{OH}^+)$ value measured from the high angular resolution spectra for $\rm OH^+(1_1-0_1)$ and $(1_2-0_1)$ is $(9.6-11.2\times10^{14}) \, \rm cm^{-2}$ (Table~\ref{tab:tau_and_column_density}). The mass of the $\rm OH^+$ molecular ion is $m_{\textrm{OH}^+}=2.824\times10^{-23} \rm g$. The observed projected area can be computed from the S\'ersic radius $R_s$ of the $\rm OH^+$ source plane reconstruction (0.21-0.23~kpc; Table \ref{tab:lensing_results_source}) yielding $A_\textrm{obs}=0.13-0.17$~kpc$^2$. 
The observed velocity of the flow is $\rm 100 \, km\, s^{-1}$. The distance that the inflowing gas travels from the observed position to the center of the galaxy is one of the most uncertain quantities in Eq. \ref{eq:inflow_rate}. Typical distances found in the literature are of the order of one to a few kpc. Using the positions of lines and continuum in the source plane (Table~\ref{tab:lensing_results_source}), we can only compute the distances projected on the sky as between $\sim0.1$ and $\sim0.3$~kpc, which are of the same order than the systematic positional uncertainty computed via Eq.~\ref{eq:systematic-errors} ($\sim 0.08$~kpc). Assuming a geometry similar to that of $\rm OH^+$, because no source plane reconstruction has been attempted for the low angular resolution NOEMA data, a similar reasoning can be applied to the $\rm CH^+$ line. Based on these values and estimates and using Eq.~\ref{eq:inflow_rate}, we find
\begin{equation}
\frac{dM_{\textrm{OH}^+}}{dt} \sin{\phi}\cos{\phi} \sim 1.8 - 8.8 \times 10^{-6}\ M_\odot \textrm{yr}^{-1} \textrm{,}
\end{equation}
and
\begin{equation}
\frac{dM_{\textrm{CH}^+}}{dt} \sin{\phi}\cos{\phi} \sim 0.4 - 1.6 \times 10^{-6}\ M_\odot \textrm{yr}^{-1} \textrm{,}
\end{equation}
modulo the angle $\phi$ between the flow direction and the line of sight, which is unknown. 

In order to transform these results into the total inflow rate, we assume that the gas mass is dominated by hydrogen, and we adopt   
as abundances relative to hydrogen the quantities $f_{\textrm{OH}^+}=5.7\times 10^{-8}$ 
\citep[computed as the Galactic average value from Table 5 of][]{Indriolo2015} 
and $f_{\textrm{CH}^+}=7.6\times 10^{-9}$ \citep{godard2014,Indriolo2018}. 
By dividing Eq. \ref{eq:inflow_rate} by $f_\textrm{X}$ and multiplying by the ratio $m_\textrm{H}/m_\textrm{X}$ 
of the molecular ion's mass relative to H, we derive mass inflow rates of 
$\dot{M}_\textrm{inflow} \sin{\phi}\cos{\phi}  \sim 3 - 14  \ M_\odot \textrm{yr}^{-1}$ from $\rm OH^+$, 
and $\rm \sim 6 - 24 \ M_\odot \textrm{yr}^{-1}$ from $\rm CH^+$.\\

In the so-called ``bathtub'' or ``gas regulator'' model \citep[][among many others]{bouche2010,lilly2013,almeida2014,peng2014,dekel2014,somerville2015}, 
the evolution of the gas content of a galaxy can be formulated as follows:
\begin{equation}\label{eq:bathtub}
\dot{M}_\textrm{gas} = \Phi - \left(1-R+\left(\lambda-\eta_\textrm{rec}\right)\right)\epsilon M_\textrm{gas} + Y \textrm{,}
\end{equation}
where $\Phi$ is the inflow rate; $\textrm{SFR}=\epsilon M_\textrm{gas}$, as per the Schmidt-Kennicutt star formation law \citep{Schmidt1959,Kennicutt1998b} ; $R$ is the fraction of mass of newly formed stars quickly returned to the ISM through stellar winds and supernovae; $\Psi=\lambda\textrm{SFR}$ is the outflow rate, linked to the SFR by the mass loading factor $\lambda$; finally, $\eta_\textrm{rec}$ is the fraction of gas ejected by outflows falling back onto the galaxy. Additional factors, e.g., due to AGN feedback, possible mergers or other external causes, are here summarized by the term $Y$.

Different scenarios can be at play in the case of HerBS-89a. In the least likely scenario, we might be witnessing a flow channeling gas from the intergalactic medium onto the galaxy, i.e., the term $\Phi$ of Eq. \ref{eq:bathtub}. While such flow would in principle consist of pristine gas, the detected $\rm OH^+$ and $\rm CH^+$ transitions testify to the presence of enriched material. To produce these molecular ions, first of all, as the reaction rates go with $n^2$, pockets of dense gas should be present; second, the most important formation channel for $\rm OH^+$ (and $\rm CH^+$) requires high temperatures, which could be produced in an accretion shock in the infalling gas as it enters the halo of the galaxy, although this scenario remains speculative. 

\citet{yabe2015} use a simple analytic model of galaxy chemical evolution to constrain the inflow (and outflow) rates of three samples of star forming galaxies at $z=0,\ 1.4$ and 2.2 \citep[][and their work]{erb2006,peeples2011}. They show that the average gas inflow rate of the samples is 2 to 3$\times$ their SFRs. Turning into physical units, $\dot{M}_\textrm{inflow}$ is a few solar masses per year at $z=0$ growing to $\sim 100\, M_\odot$~yr$^{-1}$ at $z=2.2$. Model predictions reproduce these values for dark matter halos masses of $M_\textrm{DH}\sim 2\times 10^{12}\, M_\odot$ 
\citep[e.g.,][]{bouche2010,yabe2015}. 
If the SFR is fed by the inflow, then it is naturally proportional to the inflow rate.  the detailed process of gas accretion and its relation to the fuel of SFR is more complex and still poorly understood. The gas fraction and SFR efficiency (i.e., the reciprocal of $\tau_\textrm{depl}$) come into play. How they relate to $\dot{M}_\textrm{inflow}$ is yet to be fully investigated by observations, because of the rare occurrence of direct inflow measurements.

In this frame, HerBS-89a would require a very high inflow rate to sustain its large measured star formation rate. 
Although the angle between the line of sight and the inflow is unknown, the values derived above 
seem too low to support this scenario.

Alternatively, the flow could be part of a galactic fountain, in which a fraction of the gas expelled by the starburst winds falls back onto the galaxy (the term $\eta\epsilon M_\textrm{gas}$ in Eq.~\ref{eq:bathtub}). In this case, an outflow should also be detected, as described by $-\lambda\epsilon M_\textrm{gas}$. No evidence of any outflow feature was found in the spectra of HerBS-89a, neither in the form of blue-shifted emission (e.g., inverted P-Cygni profile) nor through absorption components of the detected lines. Nevertheless, it is worth noting that the absorption lines of $\rm OH^+$ here reported are not isolated, and have emission or absorption lines of other molecular species on their blue sides, perhaps preventing the identification of a blue-shifted component. Moreover --- despite the superb quality of the NOEMA spectra ---  the S/N reached in the spectra of the main emission lines is not high enough to detect any faint, very broad wings, that could indicate the presence of an outflow.

Finally, if HerBS-89a is in fact a merging pair or a multiple source 
(see Sect. \ref{sect:kinematics}), a third possibility is that the red-shifted $\rm OH^+$ and $\rm CH^+$ lines originate in a stream of matter between the two components. This case belongs to the term $Y$ in Eq.~\ref{eq:bathtub}, and would naturally explain the enriched composition of the flow. One possibility is that HerBS-89a is similar to a scaled-up version of the merger in the local Antennae galaxy (NGC 4038/9), in which an off-nuclear high concentration of dust and gas and a powerful starburst are found in the region between the two galaxies: a ``Hyper-Antennae''.

With the data in hand, it is however not possible to distinguish between these different scenarios and draw a definitive conclusion. Also in this case, higher angular resolution and deeper sensitivity observations of HerBS-89a will be required to shed further light on the nature of the mass inflow observed in this system.


\section{Summary and concluding remarks}
\label{section:conclusions}

We report new millimeter NOEMA and VLA observations of the lensed starburst galaxy HerBS-89a at $z=2.95$ 
that are complemented with near-infrared and optical data (obtained with the \textit{HST} and GTC) 
that reveal the foreground lensing galaxy. The ability to process large instantaneous bandwidths with NOEMA enables an 
in-depth exploration of a sizeable part of the 1-mm spectrum of HerBS-89a, covering in total 121~GHz in the rest-frame, 
leading to the detection of transitions of molecular ions and molecular species that have not been previously 
observed in a high-$z$ galaxy. In addition to the strong emission lines of 
$\rm ^{12}CO(9-8)$ and $\rm H_2O(2_{02}-1_{11})$, we detect the three $\rm OH^+$ ground-state $1_J-0_1$ lines 
($J = 0, 1, 2$) and the $\rm CH^+(1-0)$ line, all four seen in absorption and red-shifted with respect to the 
systemic velocity of HerBS-89a, two transitions of $\rm NH_2$ in emission, and $\rm HCN(11-10)$
and/or $\rm NH(1_2-0_1)$ lines, seen in absorption. Modelling the lensing of HerBS-89a, we reconstructed the 
dust continuum emission and the molecular emission lines in the source plane deriving the morphology and the 
kinematics of this high-$z$ galaxy. 

The main conclusions of this paper are as follows: 
\begin{itemize}
    \item The high-angular resolution ($0\farcs3$) images of HerBS-89a in  dust continuum and the emission lines of 
          $\rm ^{12}CO(9-8)$ and $\rm H_2O(2_{02}-1_{11})$ show two distinct components -  one to the north and a brighter
          slightly more extended component to the south, which are linked by an arc-like $1\farcs0$ diameter structure reminiscent 
          of an Einstein ring. HerBS-89a is gravitationally amplified by a foreground massive galaxy 
          at $z_\textrm{phot}=0.9^{+0.3}_{-0.5}$, with an estimated deflector mass of 
          $M_\textrm{lens}=9.8^{+3.2}_{-4.9} \times 10^{10} \, M_\odot$. Using lens modelling, we have reconstructed 
          the source-plane morphologies of HerBS-89a in dust continuum and molecular line emission. 
          The magnification factor is estimated to be $\mu \sim 5$ for the dust continuum and $\mu \sim 4-5$ for the emission lines. 
    \item HerBS-89a is a high-$z$ dusty star forming galaxy. From the $\rm ^{12}CO(1-0)$ 
          measurements, we derive a mass for the molecular gas reservoir of $M_{\textrm{mol}} = (2.1 \pm 0.4) \, \times 10^{11} \, M_\odot$. Modelling the available CO lines ($J=$1-0, 3-2, 5-4 and 9-8) with an LVG approach 
          indicates that the molecular gas in HerBS-89a has an average density of $n_{\rm H_2} \sim 500 \, \rm cm^{-3}$ 
          and a kinetic temperature $T_{\rm kin} \sim 200 \, \rm K$. Using the fine sampling of the dust continuum
          flux densities observed from $\rm 250 \, \mu m$ to 3.4~mm (or 63 to $\rm 860 \, \mu m$ in the rest-frame),  
          we derive an infrared luminosity of $L_{\textrm{IR}} = (4.6 \pm 0.4) \, \times 10^{12} \, L_\odot$, 
          and a dust mass of $M_{\textrm{dust}} = (2.6 \pm 0.2) \, \times 10^{9} \, M_\odot$, yielding a 
          dust-to-gas ratio $\delta_{\rm GDR} \approx 80$. The star formation rate is estimated to be 
          ${\rm SFR} = 614\pm59 \, M_\odot \, \rm yr^{-1}$ and the depletion timescale 
          $\tau_\textrm{depl} = (3.4\pm1.0) \times 10^8$ years. 
    \item Lens modelling suggests that the warmest and densest material in HerBS-89a (i.e., as traced by dust, 
          ${}^{12}$CO(9--8), and H$_2$O($2_{02}-1_{11}$) emission) is distributed across a region a few kpc across.  
          We cannot distinguish definitively between a single-disk and a two-component description of this distribution 
          on the basis of existing observations, but dynamical and molecular gas mass estimates are consistent in 
          either scenario, and the two-component scenario in particular provides a natural explanation for the very 
          large observed CO velocity widths.      
    \item A series of molecular emission and absorption lines are detected in HerBS-89a that trace
          very dense gas (up to $n_{\rm H_2} \sim 10^9 - 10^{10} \, \rm cm^{-3}$). 
          $\rm NH_2$ is seen in emission and detected in two transitions, while an absorption feature at 246.72~GHz 
          corresponds to HCN(11-10), $\rm NH(1_2-1_0)$ or a combination of both. Future observations of
          additional transitions of these species (particularly HCN and NH) would be needed to further explore 
          the densest regions of HerBS-89a.  
    \item The three fundamental transitions of $\rm OH^+$ and $\rm CH^+$(1-0) are all seen in absorption and all 
          red-shifted by $\rm \approx +100 \, km\, s^{-1}$ with respect to the systemic velocity of HerBS-89a. The molecular 
          ions $\rm OH^+$ and $\rm CH^+$ both trace low-density ($n_{\rm H_2} \sim 100 \, \rm cm^{-3}$) gas. The fact that
          the velocity red-shifted absorption lines cover substantially the distribution of the CO and water emission 
          lines in the image (and source) planes indicates that this low-density gas must be kinematically unrelated to 
          the dense gas. We here argue that we are tracing a rare case of 
          inflow of low-density gas, indicating that HerBS-89a is accreting matter from its surroundings. 
          Based on the available data, we have estimated that the mass inflow rate is of the order of
          $\approx 10-20 \, M_\odot \rm \, yr^{-1}$. Various scenarios are possible 
          to account for the observations including inflowing gas from the intergalactic medium, a galactic fountain 
          or a stream of matter between the two merging components.
\end{itemize}

The results here described  highlight the importance of processing large instantaneous bandwidth in the study 
of high-$z$ galaxies, as such observations allow us to efficiently measure multiple lines and 
explore in an unbiased way the entire available frequency range for the presence of molecular species. 
In the case of HerBS-89a, we were able to probe this system from the very highest to the lowest molecular gas 
densities with high-quality spectral data in the 830 to 1060~GHz rest-frame frequency range, which includes 
key molecular ionic transitions tracing feedback mechanisms. The findings here reported show that it is possible 
to unravel the complex physics at play within this starburst (likely to be a merger) system in the early Universe 
from the kinematics of its inner most regions 
to the low-density inflowing molecular gas. They provide the 
foundation for future follow-up observations to further study the nature of HerBS-89a and better constrain its
morphology and kinematics, which will need higher-angular resolution and higher signal-to-noise ratio data, as well as 
for future studies of larger samples of high-$z$ galaxies, with properties akin to those of HerBS-89a, 
selected from on-going large redshift surveys.


\begin{acknowledgements}
      This work is based on observations carried out under programs, E18AE (a Discretionary Directorial Time project) and W19DE,  
      using the IRAM NOrthern Extended Millimeter Array (NOEMA), and, program VLA/20A-083 using the 
      National Radio Astronomy Observatory's (NRAO) 
      Karl G. Jansky Very Large Array (VLA). IRAM is supported by INSU/CNRS (France), MPG (Germany) and IGN (Spain).
      The National Radio Astronomy Observatory is a facility of the National Science Foundation operated under cooperative 
      agreement by Associated Universities, Inc. The authors are grateful to the IRAM Director, K.F. Schuster, for 
      enabling the initial observations of this project. This paper makes also use of observations made with 
      the Gran Telescopio Canarias (GTC), installed at the Spanish Observatorio del Roque de los Muchachos of 
      the Instituto de Astrofísica de Canarias, in the island of La Palma, under Director’s Discretionary Time.
      The authors are grateful to the referee for constructive and insightful comments. 
      We would like to thank J. Cernicharo, F. Cox, and E. Falgarone for useful discussions about some aspects of this study 
      and E. Chapillon for her help with the JVLA data reduction. 
      A.J.Y. and A.J.B. acknowledge enlightening discussions with C.R. Keeton about the lens modelling. 
      The authors would also like to highlight interesting exchanges triggered by the thoughts of J. de Chabannes.
      This work benefited from the support of the project Z-GAL ANR-AAPG2019 of the French National Research Agency (ANR).
      A.J.B. and A.J.Y acknowledge support from the National Science Foundation grant AST-1716585. B.M.J. acknowledges 
      the support of a studentship grant from the UK Science and Technology Facilities Council (SFTC).  L.D. 
      acknowledges support from the ERC consolidator grant CosmicDust (PI: H. Gomez). T.B. acknowledges funding 
      from NAOJ ALMA Scientific Research Grant 2018-09B and JSPS KAKENHI No.~17H06130.
      C.Y. acknowledges support from an ESO Fellowship. S.D. is supported by a UK STFC Rutherford Fellowship. 
      H.D. acknowledges financial support from the  Spanish Ministry of Science, Innovation and Universities (MICIU) 
      under the 2014 Ramón y Cajal program RYC-2014-15686 and AYA2017-84061-P, the later one co-financed by FEDER 
      (European Regional Development Funds). D.A.R. acknowledges support from the National Science Foundation under 
      grant numbers AST-1614213 and AST-1910107 and from the Alexander von Humboldt Foundation through a Humboldt Research 
      Fellowship for Experienced Researchers. S.J. acknowledges financial support from the Spanish Ministry of Science, 
      Innovation and Universities (MICIU) under grant AYA2017-84061-P, co-financed by FEDER (European Regional Development Funds). A.C.R. acknowledges financial support from the Spanish Ministry of Science, Innovation and Universities (MICIU) under grant AYA2017-84061-P, co-financed by FEDER (European Regional Development Funds) and by the Spanish Space Research Program “Participation in the NISP instrument and preparation for the science of EUCLID” (ESP2017-84272-C2-1-R).
\end{acknowledgements}

\bibliographystyle{aa} 		
\bibliography{HerBS89a} 		

\begin{thebibliography}{184}
\expandafter\ifx\csname natexlab\endcsname\relax\def\natexlab#1{#1}\fi

\bibitem[{{Ao} {et~al.}(2020){Ao}, {Zheng}, {Henkel}, {Nie}, {Beelen}, {Cen},
  {Dijkstra}, {Francis}, {Geach}, {Kohno}, {Lehnert}, {Menten}, {Wang}, \&
  {Weiss}}]{Ao2020}
{Ao}, Y., {Zheng}, Z., {Henkel}, C., {et~al.} 2020, Nature Astronomy, 4, 670

\bibitem[{{Arnouts} {et~al.}(1999){Arnouts}, {Cristiani}, {Moscardini},
  {Matarrese}, {Lucchin}, {Fontana}, \& {Giallongo}}]{arnouts1999}
{Arnouts}, S., {Cristiani}, S., {Moscardini}, L., {et~al.} 1999, \mnras, 310,
  540

\bibitem[{{Babbedge} {et~al.}(2004){Babbedge}, {Rowan-Robinson},
  {Gonzalez-Solares}, {Polletta}, {Berta}, {P{\'e}rez-Fournon}, {Oliver},
  {Salaman}, {Irwin}, \& {Weatherley}}]{babbedge2004}
{Babbedge}, T.~S.~R., {Rowan-Robinson}, M., {Gonzalez-Solares}, E., {et~al.}
  2004, \mnras, 353, 654

\bibitem[{{Bakx} {et~al.}(2018){Bakx}, {Eales}, {Negrello}, {Smith},
  {Valiante}, {Holland}, {Baes}, {Bourne}, {Clements}, {Dannerbauer}, {De
  Zotti}, {Dunne}, {Dye}, {Furlanetto}, {Ivison}, {Maddox}, {Marchetti},
  {Micha{\l}owski}, {Omont}, {Oteo}, {Wardlow}, {van der Werf}, \&
  {Yang}}]{Bakx2018}
{Bakx}, T.~J.~L.~C., {Eales}, S.~A., {Negrello}, M., {et~al.} 2018, \mnras,
  473, 1751

\bibitem[{{Bakx} {et~al.}(2020){Bakx}, {Eales}, {Negrello}, {Smith},
  {Valiante}, {Holland}, {Baes}, {Bourne}, {Clements}, {Dannerbauer}, {De
  Zotti}, {Dunne}, {Dye}, {Furlanetto}, {Ivison}, {Maddox}, {Marchetti},
  {Micha{\l}owski}, {Omont}, {Oteo}, {Wardlow}, {van der Werf}, \&
  {Yang}}]{Bakx2020}
{Bakx}, T. J.~L.~C., {Eales}, S.~A., {Negrello}, M., {et~al.} 2020, \mnras,
  494, 10

\bibitem[{{Barvainis} {et~al.}(1997){Barvainis}, {Maloney}, {Antonucci}, \&
  {Alloin}}]{barvainis1997}
{Barvainis}, R., {Maloney}, P., {Antonucci}, R., \& {Alloin}, D. 1997, \apj,
  484, 695

\bibitem[{{Battisti} {et~al.}(2019){Battisti}, {da Cunha}, {Grasha}, {Salvato},
  {Daddi}, {Davies}, {Jin}, {Liu}, {Schinnerer}, {Vaccari}, \& {COSMOS
  Collaboration}}]{battisti2019}
{Battisti}, A.~J., {da Cunha}, E., {Grasha}, K., {et~al.} 2019, \apj, 882, 61

\bibitem[{{Berta} {et~al.}(2003){Berta}, {Fritz}, {Franceschini}, {Bressan}, \&
  {Pernechele}}]{berta2003}
{Berta}, S., {Fritz}, J., {Franceschini}, A., {Bressan}, A., \& {Pernechele},
  C. 2003, \aap, 403, 119

\bibitem[{{Berta} {et~al.}(2016){Berta}, {Lutz}, {Genzel},
  {F{\"o}rster-Schreiber}, \& {Tacconi}}]{berta2016}
{Berta}, S., {Lutz}, D., {Genzel}, R., {F{\"o}rster-Schreiber}, N.~M., \&
  {Tacconi}, L.~J. 2016, \aap, 587, A73

\bibitem[{{Berta} {et~al.}(2013){Berta}, {Lutz}, {Santini}, {Wuyts}, {Rosario},
  {Brisbin}, {Cooray}, {Franceschini}, {Gruppioni}, {Hatziminaoglou}, {Hwang},
  {Le Floc'h}, {Magnelli}, {Nordon}, {Oliver}, {Page}, {Popesso}, {Pozzetti},
  {Pozzi}, {Riguccini}, {Rodighiero}, {Roseboom}, {Scott}, {Symeonidis},
  {Valtchanov}, {Viero}, \& {Wang}}]{berta2013}
{Berta}, S., {Lutz}, D., {Santini}, P., {et~al.} 2013, \aap, 551, A100

\bibitem[{{Bianchi}(2013)}]{Bianchi2013}
{Bianchi}, S. 2013, \aap, 552, A89

\bibitem[{{Bianchi} {et~al.}(2019){Bianchi}, {Casasola}, {Baes}, {Clark},
  {Corbelli}, {Davies}, {De Looze}, {De Vis}, {Dobbels}, {Galametz},
  {Galliano}, {Jones}, {Madden}, {Magrini}, {Mosenkov}, {Nersesian}, {Viaene},
  {Xilouris}, \& {Ysard}}]{bianchi2019}
{Bianchi}, S., {Casasola}, V., {Baes}, M., {et~al.} 2019, \aap, 631, A102

\bibitem[{{Blain} {et~al.}(2002){Blain}, {Smail}, {Ivison}, {Kneib}, \&
  {Frayer}}]{Blain2002}
{Blain}, A.~W., {Smail}, I., {Ivison}, R.~J., {Kneib}, J.-P., \& {Frayer},
  D.~T. 2002, \physrep, 369, 111

\bibitem[{{Bolatto} {et~al.}(2013){Bolatto}, {Wolfire}, \&
  {Leroy}}]{bolatto2013}
{Bolatto}, A.~D., {Wolfire}, M., \& {Leroy}, A.~K. 2013, \araa, 51, 207

\bibitem[{{Bolzonella} {et~al.}(2000){Bolzonella}, {Miralles}, \&
  {Pell{\'o}}}]{Bolzonella2000}
{Bolzonella}, M., {Miralles}, J.~M., \& {Pell{\'o}}, R. 2000, \aap, 363, 476

\bibitem[{{Boselli} {et~al.}(2010){Boselli}, {Eales}, {Cortese}, {Bendo},
  {Chanial}, {Buat}, {Davies}, {Auld}, {Rigby}, {Baes}, {Barlow}, {Bock},
  {Bradford}, {Castro-Rodriguez}, {Charlot}, {Clements}, {Cormier}, {Dwek},
  {Elbaz}, {Galametz}, {Galliano}, {Gear}, {Glenn}, {Gomez}, {Griffin}, {Hony},
  {Isaak}, {Levenson}, {Lu}, {Madden}, {O'Halloran}, {Okamura}, {Oliver},
  {Page}, {Panuzzo}, {Papageorgiou}, {Parkin}, {Perez-Fournon}, {Pohlen},
  {Rangwala}, {Roussel}, {Rykala}, {Sacchi}, {Sauvage}, {Schulz}, {Schirm},
  {Smith}, {Spinoglio}, {Stevens}, {Symeonidis}, {Vaccari}, {Vigroux},
  {Wilson}, {Wozniak}, {Wright}, \& {Zeilinger}}]{boselli2010}
{Boselli}, A., {Eales}, S., {Cortese}, L., {et~al.} 2010, \pasp, 122, 261

\bibitem[{{Bothwell} {et~al.}(2013){Bothwell}, {Smail}, {Chapman}, {Genzel},
  {Ivison}, {Tacconi}, {Alaghband -Zadeh}, {Bertoldi}, {Blain}, \&
  {Casey}}]{Bothwell2013}
{Bothwell}, M.~S., {Smail}, I., {Chapman}, S.~C., {et~al.} 2013, \mnras, 429,
  3047

\bibitem[{{Bouch{\'e}} {et~al.}(2010){Bouch{\'e}}, {Dekel}, {Genzel}, {Genel},
  {Cresci}, {F{\"o}rster Schreiber}, {Shapiro}, {Davies}, \&
  {Tacconi}}]{bouche2010}
{Bouch{\'e}}, N., {Dekel}, A., {Genzel}, R., {et~al.} 2010, \apj, 718, 1001

\bibitem[{{Bradford} {et~al.}(2009){Bradford}, {Aguirre}, {Aikin}, {Bock},
  {Earle}, {Glenn}, {Inami}, {Maloney}, {Matsuhara}, {Naylor}, {Nguyen}, \&
  {Zmuidzinas}}]{bradford2009}
{Bradford}, C.~M., {Aguirre}, J.~E., {Aikin}, R., {et~al.} 2009, \apj, 705, 112

\bibitem[{{Brammer} {et~al.}(2008){Brammer}, {van Dokkum}, \&
  {Coppi}}]{brammer2008}
{Brammer}, G.~B., {van Dokkum}, P.~G., \& {Coppi}, P. 2008, \apj, 686, 1503

\bibitem[{{Bruzual A.}(1983)}]{Bruzual1983}
{Bruzual A.}, G. 1983, \apj, 273, 105

\bibitem[{{Bussmann} {et~al.}(2013){Bussmann}, {P{\'e}rez-Fournon}, {Amber},
  {Calanog}, {Gurwell}, {Dannerbauer}, {De Bernardis}, {Fu}, {Harris}, \&
  {Krips}}]{Bussmann2013}
{Bussmann}, R.~S., {P{\'e}rez-Fournon}, I., {Amber}, S., {et~al.} 2013, \apj,
  779, 25

\bibitem[{{Ca{\~n}ameras} {et~al.}(2015){Ca{\~n}ameras}, {Nesvadba}, {Guery},
  {McKenzie}, {K{\"o}nig}, {Petitpas}, {Dole}, {Frye}, {Flores-Cacho},
  {Montier}, {Negrello}, {Beelen}, {Boone}, {Dicken}, {Lagache}, {Le Floc'h},
  {Altieri}, {B{\'e}thermin}, {Chary}, {de Zotti}, {Giard}, {Kneissl}, {Krips},
  {Malhotra}, {Martinache}, {Omont}, {Pointecouteau}, {Puget}, {Scott},
  {Soucail}, {Valtchanov}, {Welikala}, \& {Yan}}]{Canameras2015}
{Ca{\~n}ameras}, R., {Nesvadba}, N.~P.~H., {Guery}, D., {et~al.} 2015, \aap,
  581, A105

\bibitem[{{Canameras} {et~al.}(2020){Canameras}, {Nesvadba}, {Kneissl},
  {Koenig}, {Yang}, {Beelen}, {Hill}, {Le Floc'h}, \& {Scott}}]{Canameras2020}
{Canameras}, R., {Nesvadba}, N.~P.~H., {Kneissl}, R., {et~al.} 2020, arXiv
  e-prints, arXiv:2009.13538

\bibitem[{{Carilli} {et~al.}(2010){Carilli}, {Daddi}, {Riechers}, {Walter},
  {Weiss}, {Dannerbauer}, {Morrison}, {Wagg}, {Dav{\'e}}, {Elbaz}, {Stern},
  {Dickinson}, {Krips}, \& {Aravena}}]{carilli2010}
{Carilli}, C.~L., {Daddi}, E., {Riechers}, D., {et~al.} 2010, \apj, 714, 1407

\bibitem[{{Carilli} \& {Walter}(2013)}]{Carilli-Walter2013}
{Carilli}, C.~L. \& {Walter}, F. 2013, \araa, 51, 105

\bibitem[{{Carlstrom} {et~al.}(2011){Carlstrom}, {Ade}, {Aird}, {Benson},
  {Bleem}, {Busetti}, {Chang}, {Chauvin}, {Cho}, {Crawford}, {Crites}, {Dobbs},
  {Halverson}, {Heimsath}, {Holzapfel}, {Hrubes}, {Joy}, {Keisler}, {Lanting},
  {Lee}, {Leitch}, {Leong}, {Lu}, {Lueker}, {Luong-Van}, {McMahon}, {Mehl},
  {Meyer}, {Mohr}, {Montroy}, {Padin}, {Plagge}, {Pryke}, {Ruhl}, {Schaffer},
  {Schwan}, {Shirokoff}, {Spieler}, {Staniszewski}, {Stark}, {Tucker}, {Vand
  erlinde}, {Vieira}, \& {Williamson}}]{Carlstrom2011}
{Carlstrom}, J.~E., {Ade}, P.~A.~R., {Aird}, K.~A., {et~al.} 2011, \pasp, 123,
  568

\bibitem[{{Casey} {et~al.}(2014){Casey}, {Narayanan}, \& {Cooray}}]{Casey2014}
{Casey}, C.~M., {Narayanan}, D., \& {Cooray}, A. 2014, \physrep, 541, 45

\bibitem[{{Cernicharo} {et~al.}(2000){Cernicharo}, {Goicoechea}, \&
  {Caux}}]{Cernicharo2000}
{Cernicharo}, J., {Goicoechea}, J.~R., \& {Caux}, E. 2000, \apjl, 534, L199

\bibitem[{{Cernicharo} {et~al.}(1997){Cernicharo}, {Liu},
  {Gonz{\'a}lez-Alfonso}, {Cox}, {Barlow}, {Lim}, \&
  {Swinyard}}]{Cernicharo1997}
{Cernicharo}, J., {Liu}, X.~W., {Gonz{\'a}lez-Alfonso}, E., {et~al.} 1997,
  \apjl, 483, L65

\bibitem[{{Chabrier}(2003)}]{chabrier2003}
{Chabrier}, G. 2003, \apjl, 586, L133

\bibitem[{{Clark} {et~al.}(2019){Clark}, {De Vis}, {Baes}, {Bianchi},
  {Casasola}, {Cassar{\`a}}, {Davies}, {Dobbels}, {Lianou}, {De Looze},
  {Evans}, {Galametz}, {Galliano}, {Jones}, {Madden}, {Mosenkov}, {Verstocken},
  {Viaene}, {Xilouris}, \& {Ysard}}]{clark2019}
{Clark}, C.~J.~R., {De Vis}, P., {Baes}, M., {et~al.} 2019, \mnras, 489, 5256

\bibitem[{{Clark} {et~al.}(2016){Clark}, {Schofield}, {Gomez}, \&
  {Davies}}]{clark2016}
{Clark}, C. J.~R., {Schofield}, S.~P., {Gomez}, H.~L., \& {Davies}, J.~I. 2016,
  \mnras, 459, 1646

\bibitem[{{Coleman} {et~al.}(1980){Coleman}, {Wu}, \& {Weedman}}]{coleman1980}
{Coleman}, G.~D., {Wu}, C.~C., \& {Weedman}, D.~W. 1980, \apjs, 43, 393

\bibitem[{{Comito} {et~al.}(2003){Comito}, {Schilke}, {Gerin}, {Phillips},
  {Zmuidzinas}, \& {Lis}}]{comito2003}
{Comito}, C., {Schilke}, P., {Gerin}, M., {et~al.} 2003, \aap, 402, 635

\bibitem[{{Cortzen} {et~al.}(2020){Cortzen}, {Magdis}, {Valentino}, {Daddi},
  {Liu}, {Rigopoulou}, {Sargent}, {Riechers}, {Cormier}, {Hodge}, {Walter},
  {Elbaz}, {B{\'e}thermin}, {Greve}, {Kokorev}, \& {Toft}}]{Cortzen2020}
{Cortzen}, I., {Magdis}, G.~E., {Valentino}, F., {et~al.} 2020, \aap, 634, L14

\bibitem[{{Cox} {et~al.}(2011){Cox}, {Krips}, {Neri}, {Omont}, {G{\"u}sten},
  {Menten}, {Wyrowski}, {Wei{\ss}}, {Beelen}, {Gurwell}, {Dannerbauer},
  {Ivison}, {Negrello}, {Aretxaga}, {Hughes}, {Auld}, {Baes}, {Blundell},
  {Buttiglione}, {Cava}, {Cooray}, {Dariush}, {Dunne}, {Dye}, {Eales},
  {Frayer}, {Fritz}, {Gavazzi}, {Hopwood}, {Ibar}, {Jarvis}, {Maddox},
  {Micha{\l}owski}, {Pascale}, {Pohlen}, {Rigby}, {Smith}, {Swinbank}, {Temi},
  {Valtchanov}, {van der Werf}, \& {de Zotti}}]{Cox2011}
{Cox}, P., {Krips}, M., {Neri}, R., {et~al.} 2011, \apj, 740, 63

\bibitem[{{da Cunha} {et~al.}(2008){da Cunha}, {Charlot}, \&
  {Elbaz}}]{dacunha2008}
{da Cunha}, E., {Charlot}, S., \& {Elbaz}, D. 2008, \mnras, 388, 1595

\bibitem[{da~Cunha {et~al.}(2013)da~Cunha, Groves, Walter, Decarli, Weiss,
  Bertoldi, Carilli, Daddi, Elbaz, Ivison, Maiolino, Riechers, Rix, Sargent, \&
  Smail}]{dacunha2013}
da~Cunha, E., Groves, B., Walter, F., {et~al.} 2013, \apj, 766, 12

\bibitem[{{da Cunha} {et~al.}(2015){da Cunha}, {Walter}, {Smail}, {Swinbank},
  {Simpson}, {Decarli}, {Hodge}, {Weiss}, {van der Werf}, {Bertoldi},
  {Chapman}, {Cox}, {Danielson}, {Dannerbauer}, {Greve}, {Ivison}, {Karim}, \&
  {Thomson}}]{dacunha2015}
{da Cunha}, E., {Walter}, F., {Smail}, I.~R., {et~al.} 2015, \apj, 806, 110

\bibitem[{{Daddi} {et~al.}(2010){Daddi}, {Elbaz}, {Walter}, {Bournaud},
  {Salmi}, {Carilli}, {Dannerbauer}, {Dickinson}, {Monaco}, \&
  {Riechers}}]{daddi2010}
{Daddi}, E., {Elbaz}, D., {Walter}, F., {et~al.} 2010, \apjl, 714, L118

\bibitem[{{Daddi} {et~al.}(2020){Daddi}, {Valentino}, {Rich}, {Neill},
  {Gronke}, {O'Sullivan}, {Elbaz}, {Bournaud}, {Finoguenov}, {Marchal},
  {Delvecchio}, {Jin}, {Liu}, {Calabro}, {Coogan}, {D'Eugenio}, {Gobat},
  {Kalita}, {Laursen}, {Martin}, {Puglisi}, {Schinnerer}, {Strazzullo}, \&
  {Wang}}]{Daddi2020}
{Daddi}, E., {Valentino}, F., {Rich}, R.~M., {et~al.} 2020, arXiv e-prints,
  arXiv:2006.11089

\bibitem[{{Danielson} {et~al.}(2011){Danielson}, {Swinbank}, {Smail}, {Cox},
  {Edge}, {Weiss}, {Harris}, {Baker}, {De Breuck}, {Geach}, {Ivison}, {Krips},
  {Lundgren}, {Longmore}, {Neri}, \& {Flaquer}}]{Danielson2011}
{Danielson}, A.~L.~R., {Swinbank}, A.~M., {Smail}, I., {et~al.} 2011, \mnras,
  410, 1687

\bibitem[{{Dekel} \& {Mandelker}(2014)}]{dekel2014}
{Dekel}, A. \& {Mandelker}, N. 2014, \mnras, 444, 2071

\bibitem[{{Dekel} \& {Woo}(2003)}]{dekel2003}
{Dekel}, A. \& {Woo}, J. 2003, \mnras, 344, 1131

\bibitem[{{Delhaize} {et~al.}(2017){Delhaize}, {Smol{\v{c}}i{\'c}},
  {Delvecchio}, {Novak}, {Sargent}, {Baran}, {Magnelli}, {Zamorani},
  {Schinnerer}, {Murphy}, {Aravena}, {Berta}, {Bondi}, {Capak}, {Carilli},
  {Ciliegi}, {Civano}, {Ilbert}, {Karim}, {Laigle}, {Le F{\`e}vre}, {Marchesi},
  {McCracken}, {Salvato}, {Seymour}, \& {Tasca}}]{delhaize2017}
{Delhaize}, J., {Smol{\v{c}}i{\'c}}, V., {Delvecchio}, I., {et~al.} 2017, \aap,
  602, A4

\bibitem[{{Dhillon} {et~al.}(2018){Dhillon}, {Dixon}, {Gamble}, {Kerry},
  {Littlefair}, {Parsons}, {Marsh}, {Bezawada}, {Black}, {Gao}, {Henry},
  {Lunney}, {Miller}, {Dubbeldam}, {Morris}, {Osborn}, {Wilson}, {Casares},
  {Mu{\~n}oz-Darias}, {Pall{\'e}}, {Rodriguez-Gil}, {Shahbaz}, \& {de Ugarte
  Postigo}}]{Dhillon2018}
{Dhillon}, V., {Dixon}, S., {Gamble}, T., {et~al.} 2018, in Society of
  Photo-Optical Instrumentation Engineers (SPIE) Conference Series, Vol. 10702,
  \procspie, 107020L

\bibitem[{{Dong} {et~al.}(2019){Dong}, {Spilker}, {Gonzalez}, {Apostolovski},
  {Aravena}, {B{\'e}thermin}, {Chapman}, {Chen}, {Hayward}, {Hezaveh}, {Litke},
  {Ma}, {Marrone}, {Morningstar}, {Phadke}, {Reuter}, {Sreevani}, {Stark},
  {Vieira}, \& {Wei{\ss}}}]{Dong2019}
{Dong}, C., {Spilker}, J.~S., {Gonzalez}, A.~H., {et~al.} 2019, \apj, 873, 50

\bibitem[{{Downes} {et~al.}(1999){Downes}, {Neri}, {Greve}, {Guilloteau},
  {Casoli}, {Hughes}, {Lutz}, {Menten}, {Wilner}, {Andreani}, {Bertoldi},
  {Carilli}, {Dunlop}, {Genzel}, {Gueth}, {Ivison}, {Mann}, {Mellier},
  {Oliver}, {Peacock}, {Rigopoulou}, {Rowan-Robinson}, {Schilke}, {Serjeant},
  {Tacconi}, \& {Wright}}]{Downes1999}
{Downes}, D., {Neri}, R., {Greve}, A., {et~al.} 1999, \aap, 347, 809

\bibitem[{{Downes} \& {Solomon}(1998)}]{downes1998}
{Downes}, D. \& {Solomon}, P.~M. 1998, \apj, 507, 615

\bibitem[{{Draine}(2003)}]{Draine2003}
{Draine}, B.~T. 2003, \araa, 41, 241

\bibitem[{{Draine} {et~al.}(2014){Draine}, {Aniano}, {Krause}, {Groves},
  {Sandstrom}, {Braun}, {Leroy}, {Klaas}, {Linz}, {Rix}, {Schinnerer},
  {Schmiedeke}, \& {Walter}}]{draine2014}
{Draine}, B.~T., {Aniano}, G., {Krause}, O., {et~al.} 2014, \apj, 780, 172

\bibitem[{{Draine} \& {Li}(2001)}]{dl01}
{Draine}, B.~T. \& {Li}, A. 2001, \apj, 551, 807

\bibitem[{{Draine} \& {Li}(2007)}]{DL07}
{Draine}, B.~T. \& {Li}, A. 2007, \apj, 657, 810

\bibitem[{{Dunne} {et~al.}(2020){Dunne}, {Maddox}, {Vlahakis}, \&
  {Gomez}}]{Dunne2020}
{Dunne}, L., {Maddox}, S.~J., {Vlahakis}, C., \& {Gomez}, H.~L. 2020, \mnras

\bibitem[{{Eales} {et~al.}(2010{\natexlab{a}}){Eales}, {Dunne}, {Clements},
  {Cooray}, {De Zotti}, {Dye}, {Ivison}, {Jarvis}, {Lagache}, {Maddox},
  {Negrello}, {Serjeant}, {Thompson}, {Van Kampen}, {Amblard}, {Andreani},
  {Baes}, {Beelen}, {Bendo}, {Benford}, {Bertoldi}, {Bock}, {Bonfield},
  {Boselli}, {Bridge}, {Buat}, {Burgarella}, {Carlberg}, {Cava}, {Chanial},
  {Charlot}, {Christopher}, {Coles}, {Cortese}, {Dariush}, {da Cunha},
  {Dalton}, {Danese}, {Dannerbauer}, {Driver}, {Dunlop}, {Fan}, {Farrah},
  {Frayer}, {Frenk}, {Geach}, {Gardner}, {Gomez}, {Gonz{\'a}lez-Nuevo},
  {Gonz{\'a}lez-Solares}, {Griffin}, {Hardcastle}, {Hatziminaoglou}, {Herranz},
  {Hughes}, {Ibar}, {Jeong}, {Lacey}, {Lapi}, {Lawrence}, {Lee}, {Leeuw},
  {Liske}, {L{\'o}pez-Caniego}, {M{\"u}ller}, {Nandra}, {Panuzzo},
  {Papageorgiou}, {Patanchon}, {Peacock}, {Pearson}, {Phillipps}, {Pohlen},
  {Popescu}, {Rawlings}, {Rigby}, {Rigopoulou}, {Robotham}, {Rodighiero},
  {Sansom}, {Schulz}, {Scott}, {Smith}, {Sibthorpe}, {Smail}, {Stevens},
  {Sutherland}, {Takeuchi}, {Tedds}, {Temi}, {Tuffs}, {Trichas}, {Vaccari},
  {Valtchanov}, {van der Werf}, {Verma}, {Vieria}, {Vlahakis}, \&
  {White}}]{Eales2010}
{Eales}, S., {Dunne}, L., {Clements}, D., {et~al.} 2010{\natexlab{a}}, \pasp,
  122, 499

\bibitem[{{Eales} {et~al.}(2010{\natexlab{b}}){Eales}, {Smith}, {Wilson},
  {Bendo}, {Cortese}, {Pohlen}, {Boselli}, {Gomez}, {Auld}, {Baes}, {Barlow},
  {Bock}, {Bradford}, {Buat}, {Castro-Rodr{\'\i}guez}, {Chanial}, {Charlot},
  {Ciesla}, {Clements}, {Cooray}, {Cormier}, {Davies}, {Dwek}, {Elbaz},
  {Galametz}, {Galliano}, {Gear}, {Glenn}, {Griffin}, {Hony}, {Isaak},
  {Levenson}, {Lu}, {Madden}, {O'Halloran}, {Okumura}, {Oliver}, {Page},
  {Panuzzo}, {Papageorgiou}, {Parkin}, {P{\'e}rez-Fournon}, {Rangwala},
  {Rigby}, {Roussel}, {Rykala}, {Sacchi}, {Sauvage}, {Schulz}, {Schirm},
  {Spinoglio}, {Srinivasan}, {Stevens}, {Symeonidis}, {Trichas}, {Vaccari},
  {Vigroux}, {Wozniak}, {Wright}, \& {Zeilinger}}]{eales2010b}
{Eales}, S.~A., {Smith}, M.~W.~L., {Wilson}, C.~D., {et~al.}
  2010{\natexlab{b}}, \aap, 518, L62

\bibitem[{{Elbaz} {et~al.}(2011){Elbaz}, {Dickinson}, {Hwang},
  {D{\'\i}az-Santos}, {Magdis}, {Magnelli}, {Le Borgne}, {Galliano},
  {Pannella}, {Chanial}, {Armus}, {Charmandaris}, {Daddi}, {Aussel}, {Popesso},
  {Kartaltepe}, {Altieri}, {Valtchanov}, {Coia}, {Dannerbauer}, {Dasyra},
  {Leiton}, {Mazzarella}, {Alexander}, {Buat}, {Burgarella}, {Chary}, {Gilli},
  {Ivison}, {Juneau}, {Le Floc'h}, {Lutz}, {Morrison}, {Mullaney}, {Murphy},
  {Pope}, {Scott}, {Brodwin}, {Calzetti}, {Cesarsky}, {Charlot}, {Dole},
  {Eisenhardt}, {Ferguson}, {F{\"o}rster Schreiber}, {Frayer}, {Giavalisco},
  {Huynh}, {Koekemoer}, {Papovich}, {Reddy}, {Surace}, {Teplitz}, {Yun}, \&
  {Wilson}}]{elbaz2011}
{Elbaz}, D., {Dickinson}, M., {Hwang}, H.~S., {et~al.} 2011, \aap, 533, A119

\bibitem[{{Erb}(2008)}]{erb2008}
{Erb}, D.~K. 2008, \apj, 674, 151

\bibitem[{{Erb} {et~al.}(2006){Erb}, {Shapley}, {Pettini}, {Steidel}, {Reddy},
  \& {Adelberger}}]{erb2006}
{Erb}, D.~K., {Shapley}, A.~E., {Pettini}, M., {et~al.} 2006, \apj, 644, 813

\bibitem[{{Falgarone} {et~al.}(2017){Falgarone}, {Zwaan}, {Godard}, {Bergin},
  {Ivison}, {Andreani}, {Bournaud}, {Bussmann}, {Elbaz}, {Omont}, {Oteo}, \&
  {Walter}}]{Falgarone2017}
{Falgarone}, E., {Zwaan}, M.~A., {Godard}, B., {et~al.} 2017, \nat, 548, 430

\bibitem[{{Falstad} {et~al.}(2017){Falstad}, {Gonz{\'a}lez-Alfonso}, {Aalto},
  \& {Fischer}}]{Falstad2017}
{Falstad}, N., {Gonz{\'a}lez-Alfonso}, E., {Aalto}, S., \& {Fischer}, J. 2017,
  \aap, 597, A105

\bibitem[{{Falstad} {et~al.}(2015){Falstad}, {Gonz{\'a}lez-Alfonso}, {Aalto},
  {van der Werf}, {Fischer}, {Veilleux}, {Mel{\'e}ndez}, {Farrah}, \&
  {Smith}}]{Falstad2015}
{Falstad}, N., {Gonz{\'a}lez-Alfonso}, E., {Aalto}, S., {et~al.} 2015, \aap,
  580, A52

\bibitem[{{Fixsen} {et~al.}(1999){Fixsen}, {Bennett}, \& {Mather}}]{fixsen1999}
{Fixsen}, D.~J., {Bennett}, C.~L., \& {Mather}, J.~C. 1999, \apj, 526, 207

\bibitem[{{Fu} {et~al.}(2013){Fu}, {Cooray}, {Feruglio}, {Ivison}, {Riechers},
  {Gurwell}, {Bussmann}, {Harris}, {Altieri}, {Aussel}, {Baker}, {Bock},
  {Boylan-Kolchin}, {Bridge}, {Calanog}, {Casey}, {Cava}, {Chapman},
  {Clements}, {Conley}, {Cox}, {Farrah}, {Frayer}, {Hopwood}, {Jia}, {Magdis},
  {Marsden}, {Mart{\'{\i}}nez-Navajas}, {Negrello}, {Neri}, {Oliver}, {Omont},
  {Page}, {P{\'e}rez-Fournon}, {Schulz}, {Scott}, {Smith}, {Vaccari},
  {Valtchanov}, {Vieira}, {Viero}, {Wang}, {Wardlow}, \& {Zemcov}}]{Fu2013}
{Fu}, H., {Cooray}, A., {Feruglio}, C., {et~al.} 2013, \nat, 498, 338

\bibitem[{{Fudamoto} {et~al.}(2017){Fudamoto}, {Ivison}, {Oteo}, {Krips},
  {Zhang}, {Weiss}, {Dannerbauer}, {Omont}, {Chapman}, \&
  {Christensen}}]{Fudamoto2017}
{Fudamoto}, Y., {Ivison}, R.~J., {Oteo}, I., {et~al.} 2017, \mnras, 472, 2028

\bibitem[{{Fumagalli} {et~al.}(2011){Fumagalli}, {Prochaska}, {Kasen}, {Dekel},
  {Ceverino}, \& {Primack}}]{Fumagalli2011}
{Fumagalli}, M., {Prochaska}, J.~X., {Kasen}, D., {et~al.} 2011, \mnras, 418,
  1796

\bibitem[{{Gao} \& {Solomon}(2004)}]{Gao&Solomon2004}
{Gao}, Y. \& {Solomon}, P.~M. 2004, \apj, 606, 271

\bibitem[{{Genzel} {et~al.}(2003){Genzel}, {Baker}, {Tacconi}, {Lutz}, {Cox},
  {Guilloteau}, \& {Omont}}]{Genzel2003}
{Genzel}, R., {Baker}, A.~J., {Tacconi}, L.~J., {et~al.} 2003, \apj, 584, 633

\bibitem[{{Genzel} {et~al.}(2010){Genzel}, {Tacconi}, {Gracia-Carpio},
  {Sternberg}, {Cooper}, {Shapiro}, {Bolatto}, {Bouch{\'e}}, {Bournaud},
  {Burkert}, {Combes}, {Comerford}, {Cox}, {Davis}, {Schreiber},
  {Garcia-Burillo}, {Lutz}, {Naab}, {Neri}, {Omont}, {Shapley}, \&
  {Weiner}}]{genzel2010}
{Genzel}, R., {Tacconi}, L.~J., {Gracia-Carpio}, J., {et~al.} 2010, \mnras,
  407, 2091

\bibitem[{{Godard} {et~al.}(2014){Godard}, {Falgarone}, \& {Pineau des
  For{\^e}ts}}]{godard2014}
{Godard}, B., {Falgarone}, E., \& {Pineau des For{\^e}ts}, G. 2014, \aap, 570,
  A27

\bibitem[{{Goicoechea} {et~al.}(2004){Goicoechea},
  {Rodr{\'\i}guez-Fern{\'a}ndez}, \& {Cernicharo}}]{Goicoechea2004}
{Goicoechea}, J.~R., {Rodr{\'\i}guez-Fern{\'a}ndez}, N.~J., \& {Cernicharo}, J.
  2004, \apj, 600, 214

\bibitem[{{Gonz{\'a}lez-Alfonso} {et~al.}(2014){Gonz{\'a}lez-Alfonso},
  {Fischer}, {Aalto}, \& {Falstad}}]{Gonzalez-Alfonso2014}
{Gonz{\'a}lez-Alfonso}, E., {Fischer}, J., {Aalto}, S., \& {Falstad}, N. 2014,
  \aap, 567, A91

\bibitem[{{Gonz{\'a}lez-Alfonso} {et~al.}(2018){Gonz{\'a}lez-Alfonso},
  {Fischer}, {Bruderer}, {Ashby}, {Smith}, {Veilleux}, {M{\"u}ller}, {Stewart},
  \& {Sturm}}]{Gonzalez-Alfonso2018}
{Gonz{\'a}lez-Alfonso}, E., {Fischer}, J., {Bruderer}, S., {et~al.} 2018, \apj,
  857, 66

\bibitem[{{Gonz{\'a}lez-Alfonso} {et~al.}(2012){Gonz{\'a}lez-Alfonso},
  {Fischer}, {Graci{\'a}-Carpio}, {Sturm}, {Hailey-Dunsheath}, {Lutz},
  {Poglitsch}, {Contursi}, {Feuchtgruber}, {Veilleux}, {Spoon}, {Verma},
  {Christopher}, {Davies}, {Sternberg}, {Genzel}, \&
  {Tacconi}}]{Gonzalez-Alfonso2012}
{Gonz{\'a}lez-Alfonso}, E., {Fischer}, J., {Graci{\'a}-Carpio}, J., {et~al.}
  2012, \aap, 541, A4

\bibitem[{{Gonz{\'a}lez-Alfonso} {et~al.}(2017){Gonz{\'a}lez-Alfonso},
  {Fischer}, {Spoon}, {Stewart}, {Ashby}, {Veilleux}, {Smith}, {Sturm},
  {Farrah}, {Falstad}, {Mel{\'e}ndez}, {Graci{\'a}-Carpio}, {Janssen}, \&
  {Lebouteiller}}]{Gonzalez-Alfonso2017}
{Gonz{\'a}lez-Alfonso}, E., {Fischer}, J., {Spoon}, H.~W.~W., {et~al.} 2017,
  \apj, 836, 11

\bibitem[{{Gralla} {et~al.}(2020){Gralla}, {Marriage}, {Addison}, {Baker},
  {Bond}, {Crichton}, {Datta}, {Devlin}, {Dunkley}, {D{\"u}nner}, {Fowler},
  {Gallardo}, {Hall}, {Halpern}, {Hasselfield}, {Hilton}, {Hincks},
  {Huffenberger}, {Hughes}, {Kosowsky}, {L{\'o}pez-Caraballo}, {Louis},
  {Marsden}, {Moodley}, {Niemack}, {Page}, {Partridge}, {Rivera}, {Sievers},
  {Staggs}, {Su}, {Swetz}, \& {Wollack}}]{Gralla2020}
{Gralla}, M.~B., {Marriage}, T.~A., {Addison}, G., {et~al.} 2020, \apj, 893,
  104

\bibitem[{{Harris} {et~al.}(2012){Harris}, {Baker}, {Frayer}, {Smail},
  {Swinbank}, {Riechers}, {van der Werf}, {Auld}, {Baes}, {Bussmann},
  {Buttiglione}, {Cava}, {Clements}, {Cooray}, {Dannerbauer}, {Dariush}, {De
  Zotti}, {Dunne}, {Dye}, {Eales}, {Fritz}, {Gonz{\'a}lez-Nuevo}, {Hopwood},
  {Ibar}, {Ivison}, {Jarvis}, {Maddox}, {Negrello}, {Rigby}, {Smith}, {Temi},
  \& {Wardlow}}]{Harris2012}
{Harris}, A.~I., {Baker}, A.~J., {Frayer}, D.~T., {et~al.} 2012, \apj, 752, 152

\bibitem[{{Hensley} \& {Draine}(2020)}]{Hensley2020}
{Hensley}, B.~S. \& {Draine}, B.~T. 2020, arXiv e-prints, arXiv:2009.00018

\bibitem[{{Herrera-Camus} {et~al.}(2020){Herrera-Camus}, {Sturm},
  {Graci{\'a}-Carpio}, {Veilleux}, {Shimizu}, {Lutz}, {Stone},
  {Gonz{\'a}lez-Alfonso}, {Davies}, {Fischer}, {Genzel}, {Maiolino},
  {Sternberg}, {Tacconi}, \& {Verma}}]{Herrera-Camus2020}
{Herrera-Camus}, R., {Sturm}, E., {Graci{\'a}-Carpio}, J., {et~al.} 2020, \aap,
  633, L4

\bibitem[{{Hodge} \& {da Cunha}(2020)}]{Hodge2020}
{Hodge}, J.~A. \& {da Cunha}, E. 2020, arXiv e-prints, arXiv:2004.00934

\bibitem[{{Ibar} {et~al.}(2010){Ibar}, {Ivison}, {Best}, {Coppin}, {Pope},
  {Smail}, \& {Dunlop}}]{ibar2010}
{Ibar}, E., {Ivison}, R.~J., {Best}, P.~N., {et~al.} 2010, \mnras, 401, L53

\bibitem[{{Ilbert} {et~al.}(2006){Ilbert}, {Arnouts}, {McCracken},
  {Bolzonella}, {Bertin}, {Le F{\`e}vre}, {Mellier}, {Zamorani}, {Pell{\`o}},
  {Iovino}, {Tresse}, {Le Brun}, {Bottini}, {Garilli}, {Maccagni}, {Picat},
  {Scaramella}, {Scodeggio}, {Vettolani}, {Zanichelli}, {Adami}, {Bardelli},
  {Cappi}, {Charlot}, {Ciliegi}, {Contini}, {Cucciati}, {Foucaud}, {Franzetti},
  {Gavignaud}, {Guzzo}, {Marano}, {Marinoni}, {Mazure}, {Meneux}, {Merighi},
  {Paltani}, {Pollo}, {Pozzetti}, {Radovich}, {Zucca}, {Bondi}, {Bongiorno},
  {Busarello}, {de La Torre}, {Gregorini}, {Lamareille}, {Mathez}, {Merluzzi},
  {Ripepi}, {Rizzo}, \& {Vergani}}]{ilbert2006}
{Ilbert}, O., {Arnouts}, S., {McCracken}, H.~J., {et~al.} 2006, \aap, 457, 841

\bibitem[{{Ilbert} {et~al.}(2009){Ilbert}, {Capak}, {Salvato}, {Aussel},
  {McCracken}, {Sanders}, {Scoville}, {Kartaltepe}, {Arnouts}, {Le Floc'h},
  {Mobasher}, {Taniguchi}, {Lamareille}, {Leauthaud}, {Sasaki}, {Thompson},
  {Zamojski}, {Zamorani}, {Bardelli}, {Bolzonella}, {Bongiorno}, {Brusa},
  {Caputi}, {Carollo}, {Contini}, {Cook}, {Coppa}, {Cucciati}, {de la Torre},
  {de Ravel}, {Franzetti}, {Garilli}, {Hasinger}, {Iovino}, {Kampczyk},
  {Kneib}, {Knobel}, {Kovac}, {Le Borgne}, {Le Brun}, {Le F{\`e}vre}, {Lilly},
  {Looper}, {Maier}, {Mainieri}, {Mellier}, {Mignoli}, {Murayama}, {Pell{\`o}},
  {Peng}, {P{\'e}rez-Montero}, {Renzini}, {Ricciardelli}, {Schiminovich},
  {Scodeggio}, {Shioya}, {Silverman}, {Surace}, {Tanaka}, {Tasca}, {Tresse},
  {Vergani}, \& {Zucca}}]{ilbert2009}
{Ilbert}, O., {Capak}, P., {Salvato}, M., {et~al.} 2009, \apj, 690, 1236

\bibitem[{{Indriolo} {et~al.}(2018){Indriolo}, {Bergin}, {Falgarone}, {Godard},
  {Zwaan}, {Neufeld}, \& {Wolfire}}]{Indriolo2018}
{Indriolo}, N., {Bergin}, E.~A., {Falgarone}, E., {et~al.} 2018, \apj, 865, 127

\bibitem[{{Indriolo} {et~al.}(2015){Indriolo}, {Neufeld}, {Gerin}, {Schilke},
  {Benz}, {Winkel}, {Menten}, {Chambers}, {Black}, {Bruderer}, {Falgarone},
  {Godard}, {Goicoechea}, {Gupta}, {Lis}, {Ossenkopf}, {Persson},
  {Sonnentrucker}, {van der Tak}, {van Dishoeck}, {Wolfire}, \&
  {Wyrowski}}]{Indriolo2015}
{Indriolo}, N., {Neufeld}, D.~A., {Gerin}, M., {et~al.} 2015, \apj, 800, 40

\bibitem[{{Ivison} {et~al.}(2019){Ivison}, {Page}, {Cirasuolo}, {Harrison},
  {Mainieri}, {Arumugam}, \& {Dudzevi{\v{c}}i{\={u}}t{\.{e}}}}]{Ivison2019}
{Ivison}, R.~J., {Page}, M.~J., {Cirasuolo}, M., {et~al.} 2019, \mnras, 489,
  427

\bibitem[{{Ivison} {et~al.}(1998){Ivison}, {Smail}, {Le Borgne}, {Blain},
  {Kneib}, {Bezecourt}, {Kerr}, \& {Davies}}]{Ivison1998}
{Ivison}, R.~J., {Smail}, I., {Le Borgne}, J.~F., {et~al.} 1998, \mnras, 298,
  583

\bibitem[{{Ivison} {et~al.}(2010){Ivison}, {Smail}, {Papadopoulos}, {Wold},
  {Richard}, {Swinbank}, {Kneib}, \& {Owen}}]{Ivison2010}
{Ivison}, R.~J., {Smail}, I., {Papadopoulos}, P.~P., {et~al.} 2010, \mnras,
  404, 198

\bibitem[{{Ivison} {et~al.}(2013){Ivison}, {Swinbank}, {Smail}, {Harris},
  {Bussmann}, {Cooray}, {Cox}, {Fu}, {Kov{\'a}cs}, {Krips}, {Narayanan},
  {Negrello}, {Neri}, {Pe{\~n}arrubia}, {Richard}, {Riechers}, {Rowlands},
  {Staguhn}, {Targett}, {Amber}, {Baker}, {Bourne}, {Bertoldi}, {Bremer},
  {Calanog}, {Clements}, {Dannerbauer}, {Dariush}, {De Zotti}, {Dunne},
  {Eales}, {Farrah}, {Fleuren}, {Franceschini}, {Geach}, {George}, {Helly},
  {Hopwood}, {Ibar}, {Jarvis}, {Kneib}, {Maddox}, {Omont}, {Scott}, {Serjeant},
  {Smith}, {Thompson}, {Valiante}, {Valtchanov}, {Vieira}, \& {van der
  Werf}}]{Ivison2013}
{Ivison}, R.~J., {Swinbank}, A.~M., {Smail}, I., {et~al.} 2013, \apj, 772, 137

\bibitem[{{James} {et~al.}(2002){James}, {Dunne}, {Eales}, \&
  {Edmunds}}]{james2002}
{James}, A., {Dunne}, L., {Eales}, S., \& {Edmunds}, M.~G. 2002, \mnras, 335,
  753

\bibitem[{{Jin} {et~al.}(2019){Jin}, {Daddi}, {Magdis}, {Liu}, {Schinnerer},
  {Papadopoulos}, {Gu}, {Gao}, \& {Calabro}}]{jin2019}
{Jin}, S., {Daddi}, E., {Magdis}, G.~E., {et~al.} 2019, arXiv e-prints,
  arXiv:1906.00040

\bibitem[{{Jones} {et~al.}(2013){Jones}, {Fanciullo}, {K{\"o}hler},
  {Verstraete}, {Guillet}, {Bocchio}, \& {Ysard}}]{jones2013}
{Jones}, A.~P., {Fanciullo}, L., {K{\"o}hler}, M., {et~al.} 2013, \aap, 558,
  A62

\bibitem[{{Jones} {et~al.}(2017){Jones}, {K{\"o}hler}, {Ysard}, {Bocchio}, \&
  {Verstraete}}]{jones2017}
{Jones}, A.~P., {K{\"o}hler}, M., {Ysard}, N., {Bocchio}, M., \& {Verstraete},
  L. 2017, \aap, 602, A46

\bibitem[{{Jones} {et~al.}(2019){Jones}, {Maiolino}, {Caselli}, \&
  {Carniani}}]{Jones2019}
{Jones}, G.~C., {Maiolino}, R., {Caselli}, P., \& {Carniani}, S. 2019, \aap,
  632, L7

\bibitem[{{Keeton}(2001)}]{Keeton2001}
{Keeton}, C.~R. 2001, arXiv e-prints, astro

\bibitem[{{Kennicutt}(1998{\natexlab{a}})}]{kennicutt1998a}
{Kennicutt}, Robert~C., J. 1998{\natexlab{a}}, \araa, 36, 189

\bibitem[{{Kennicutt}(1998{\natexlab{b}})}]{Kennicutt1998b}
{Kennicutt}, Robert~C., J. 1998{\natexlab{b}}, \apj, 498, 541

\bibitem[{{Kere{\v{s}}} {et~al.}(2005){Kere{\v{s}}}, {Katz}, {Weinberg}, \&
  {Dav{\'e}}}]{keres2005}
{Kere{\v{s}}}, D., {Katz}, N., {Weinberg}, D.~H., \& {Dav{\'e}}, R. 2005,
  \mnras, 363, 2

\bibitem[{{Lada}(2015)}]{lada2015}
{Lada}, C.~J. 2015, in Galaxies in 3D across the Universe, ed. B.~L. {Ziegler},
  F.~{Combes}, H.~{Dannerbauer}, \& M.~{Verdugo}, Vol. 309, 31--38

\bibitem[{{Lehner} {et~al.}(2013){Lehner}, {Howk}, {Tripp}, {Tumlinson},
  {Prochaska}, {O'Meara}, {Thom}, {Werk}, {Fox}, \& {Ribaudo}}]{lehner2013}
{Lehner}, N., {Howk}, J.~C., {Tripp}, T.~M., {et~al.} 2013, \apj, 770, 138

\bibitem[{{Lequeux}(1983)}]{Lequeux1983}
{Lequeux}, J. 1983, \aap, 125, 394

\bibitem[{{Leroy} {et~al.}(2011){Leroy}, {Bolatto}, {Gordon}, {Sand strom},
  {Gratier}, {Rosolowsky}, {Engelbracht}, {Mizuno}, {Corbelli}, {Fukui}, \&
  {Kawamura}}]{leroy2011}
{Leroy}, A.~K., {Bolatto}, A., {Gordon}, K., {et~al.} 2011, \apj, 737, 12

\bibitem[{{Li} \& {Draine}(2001)}]{li2001}
{Li}, A. \& {Draine}, B.~T. 2001, \apj, 554, 778

\bibitem[{{Li} {et~al.}(2020){Li}, {Wang}, {Riechers}, {Walter}, {Decarli},
  {Venamans}, {Neri}, {Shao}, {Fan}, {Gao}, {Carilli}, {Omont}, {Cox},
  {Menten}, {Wagg}, {Bertoldi}, \& {Narayanan}}]{Li2020}
{Li}, J., {Wang}, R., {Riechers}, D., {et~al.} 2020, \apj, 889, 162

\bibitem[{{Lilly} {et~al.}(2013){Lilly}, {Carollo}, {Pipino}, {Renzini}, \&
  {Peng}}]{lilly2013}
{Lilly}, S.~J., {Carollo}, C.~M., {Pipino}, A., {Renzini}, A., \& {Peng}, Y.
  2013, \apj, 772, 119

\bibitem[{{Liu} {et~al.}(2019){Liu}, {Schinnerer}, {Groves}, {Magnelli},
  {Lang}, {Leslie}, {Jim{\'e}nez-Andrade}, {Riechers}, {Popping}, {Magdis},
  {Daddi}, {Sargent}, {Gao}, {Fudamoto}, {Oesch}, \& {Bertoldi}}]{liu2019}
{Liu}, D., {Schinnerer}, E., {Groves}, B., {et~al.} 2019, \apj, 887, 235

\bibitem[{{Lu} {et~al.}(2017){Lu}, {Zhao}, {D{\'\i}az-Santos}, {Xu}, {Gao},
  {Armus}, {Isaak}, {Mazzarella}, {van der Werf}, {Appleton}, {Charmandaris},
  {Evans}, {Howell}, {Iwasawa}, {Leech}, {Lord}, {Petric}, {Privon}, {Sanders},
  {Schulz}, \& {Surace}}]{Lu2017}
{Lu}, N., {Zhao}, Y., {D{\'\i}az-Santos}, T., {et~al.} 2017, \apjs, 230, 1

\bibitem[{{Lutz} {et~al.}(2011){Lutz}, {Poglitsch}, {Altieri}, {Andreani},
  {Aussel}, {Berta}, {Bongiovanni}, {Brisbin}, {Cava}, {Cepa}, {Cimatti},
  {Daddi}, {Dominguez-Sanchez}, {Elbaz}, {F{\"o}rster Schreiber}, {Genzel},
  {Grazian}, {Gruppioni}, {Harwit}, {Le Floc'h}, {Magdis}, {Magnelli},
  {Maiolino}, {Nordon}, {P{\'e}rez Garc{\'\i}a}, {Popesso}, {Pozzi},
  {Riguccini}, {Rodighiero}, {Saintonge}, {Sanchez Portal}, {Santini}, {Shao},
  {Sturm}, {Tacconi}, {Valtchanov}, {Wetzstein}, \& {Wieprecht}}]{lutz2011}
{Lutz}, D., {Poglitsch}, A., {Altieri}, B., {et~al.} 2011, \aap, 532, A90

\bibitem[{{Magdis} {et~al.}(2012){Magdis}, {Daddi}, {B{\'e}thermin}, {Sargent},
  {Elbaz}, {Pannella}, {Dickinson}, {Dannerbauer}, {da Cunha}, {Walter},
  {Rigopoulou}, {Charmandaris}, {Hwang}, \& {Kartaltepe}}]{magdis2012}
{Magdis}, G.~E., {Daddi}, E., {B{\'e}thermin}, M., {et~al.} 2012, \apj, 760, 6

\bibitem[{{Magnelli} {et~al.}(2015){Magnelli}, {Ivison}, {Lutz}, {Valtchanov},
  {Farrah}, {Berta}, {Bertoldi}, {Bock}, {Cooray}, {Ibar}, {Karim}, {Le
  Floc'h}, {Nordon}, {Oliver}, {Page}, {Popesso}, {Pozzi}, {Rigopoulou},
  {Riguccini}, {Rodighiero}, {Rosario}, {Roseboom}, {Wang}, \&
  {Wuyts}}]{magnelli2015}
{Magnelli}, B., {Ivison}, R.~J., {Lutz}, D., {et~al.} 2015, \aap, 573, A45

\bibitem[{{Magnelli} {et~al.}(2012){Magnelli}, {Saintonge}, {Lutz}, {Tacconi},
  {Berta}, {Bournaud}, {Charmandaris}, {Dannerbauer}, {Elbaz},
  {F{\"o}rster-Schreiber}, {Graci{\'a}-Carpio}, {Ivison}, {Maiolino}, {Nordon},
  {Popesso}, {Rodighiero}, {Santini}, \& {Wuyts}}]{magnelli2012b}
{Magnelli}, B., {Saintonge}, A., {Lutz}, D., {et~al.} 2012, \aap, 548, A22

\bibitem[{{Marsden} {et~al.}(2014){Marsden}, {Gralla}, {Marriage}, {Switzer},
  {Partridge}, {Massardi}, {Morales}, {Addison}, {Bond}, {Crichton}, {Das},
  {Devlin}, {D{\"u}nner}, {Hajian}, {Hilton}, {Hincks}, {Hughes}, {Irwin},
  {Kosowsky}, {Menanteau}, {Moodley}, {Niemack}, {Page}, {Reese}, {Schmitt},
  {Sehgal}, {Sievers}, {Staggs}, {Swetz}, {Thornton}, \&
  {Wollack}}]{Marsden2014}
{Marsden}, D., {Gralla}, M., {Marriage}, T.~A., {et~al.} 2014, \mnras, 439,
  1556

\bibitem[{{Meyer} \& {Roth}(1991)}]{Meyer1991}
{Meyer}, D.~M. \& {Roth}, K.~C. 1991, \apjl, 376, L49

\bibitem[{{M{\"u}ller}(2010)}]{Muller2010}
{M{\"u}ller}, H.~S.~P. 2010, \aap, 514, L6

\bibitem[{{M{\"u}ller} {et~al.}(2005){M{\"u}ller}, {Schl{\"o}der}, {Stutzki},
  \& {Winnewisser}}]{mueller2005}
{M{\"u}ller}, H. S.~P., {Schl{\"o}der}, F., {Stutzki}, J., \& {Winnewisser}, G.
  2005, Journal of Molecular Structure, 742, 215

\bibitem[{{Negrello} {et~al.}(2017){Negrello}, {Amber}, {Amvrosiadis}, {Cai},
  {Lapi}, {Gonzalez-Nuevo}, {De Zotti}, {Furlanetto}, {Maddox}, {Allen},
  {Bakx}, {Bussmann}, {Cooray}, {Covone}, {Danese}, {Dannerbauer}, {Fu},
  {Greenslade}, {Gurwell}, {Hopwood}, {Koopmans}, {Napolitano}, {Nayyeri},
  {Omont}, {Petrillo}, {Riechers}, {Serjeant}, {Tortora}, {Valiante}, {Verdoes
  Kleijn}, {Vernardos}, {Wardlow}, {Baes}, {Baker}, {Bourne}, {Clements},
  {Crawford}, {Dye}, {Dunne}, {Eales}, {Ivison}, {Marchetti}, {Micha{\l}owski},
  {Smith}, {Vaccari}, \& {van der Werf}}]{Negrello2017}
{Negrello}, M., {Amber}, S., {Amvrosiadis}, A., {et~al.} 2017, \mnras, 465,
  3558

\bibitem[{{Negrello} {et~al.}(2010){Negrello}, {Hopwood}, {De Zotti}, {Cooray},
  {Verma}, {Bock}, {Frayer}, {Gurwell}, {Omont}, {Neri}, {Dannerbauer},
  {Leeuw}, {Barton}, {Cooke}, {Kim}, {da Cunha}, {Rodighiero}, {Cox},
  {Bonfield}, {Jarvis}, {Serjeant}, {Ivison}, {Dye}, {Aretxaga}, {Hughes},
  {Ibar}, {Bertoldi}, {Valtchanov}, {Eales}, {Dunne}, {Driver}, {Auld},
  {Buttiglione}, {Cava}, {Grady}, {Clements}, {Dariush}, {Fritz}, {Hill},
  {Hornbeck}, {Kelvin}, {Lagache}, {Lopez-Caniego}, {Gonzalez-Nuevo}, {Maddox},
  {Pascale}, {Pohlen}, {Rigby}, {Robotham}, {Simpson}, {Smith}, {Temi},
  {Thompson}, {Woodgate}, {York}, {Aguirre}, {Beelen}, {Blain}, {Baker},
  {Birkinshaw}, {Blundell}, {Bradford}, {Burgarella}, {Danese}, {Dunlop},
  {Fleuren}, {Glenn}, {Harris}, {Kamenetzky}, {Lupu}, {Maddalena}, {Madore},
  {Maloney}, {Matsuhara}, {Micha{\l}owski}, {Murphy}, {Naylor}, {Nguyen},
  {Popescu}, {Rawlings}, {Rigopoulou}, {Scott}, {Scott}, {Seibert}, {Smail},
  {Tuffs}, {Vieira}, {van der Werf}, \& {Zmuidzinas}}]{Negrello2010}
{Negrello}, M., {Hopwood}, R., {De Zotti}, G., {et~al.} 2010, Science, 330, 800

\bibitem[{{Neri} {et~al.}(2020){Neri}, {Cox}, {Omont}, {Beelen}, {Berta},
  {Bakx}, {Lehnert}, {Baker}, {Buat}, {Cooray}, {Dannerbauer}, {Dunne}, {Dye},
  {Eales}, {Gavazzi}, {Harris}, {Herrera}, {Hughes}, {Ivison}, {Jin}, {Krips},
  {Lagache}, {Marchetti}, {Messias}, {Negrello}, {Perez-Fournon}, {Riechers},
  {Serjeant}, {Urquhart}, {Vlahakis}, {Wei{\ss}}, {van der Werf}, {Yang}, \&
  {Young}}]{Neri2020}
{Neri}, R., {Cox}, P., {Omont}, A., {et~al.} 2020, \aap, 635, A7

\bibitem[{{Oke} \& {Gunn}(1983)}]{Oke1983}
{Oke}, J.~B. \& {Gunn}, J.~E. 1983, \apj, 266, 713

\bibitem[{{Oliver} {et~al.}(2012){Oliver}, {Bock}, {Altieri}, {Amblard},
  {Arumugam}, {Aussel}, {Babbedge}, {Beelen}, {B{\'e}thermin}, {Blain},
  {Boselli}, {Bridge}, {Brisbin}, {Buat}, {Burgarella},
  {Castro-Rodr{\'{\i}}guez}, {Cava}, {Chanial}, {Cirasuolo}, {Clements},
  {Conley}, {Conversi}, {Cooray}, {Dowell}, {Dubois}, {Dwek}, {Dye}, {Eales},
  {Elbaz}, {Farrah}, {Feltre}, {Ferrero}, {Fiolet}, {Fox}, {Franceschini},
  {Gear}, {Giovannoli}, {Glenn}, {Gong}, {Gonz{\'a}lez Solares}, {Griffin},
  {Halpern}, {Harwit}, {Hatziminaoglou}, {Heinis}, {Hurley}, {Hwang}, {Hyde},
  {Ibar}, {Ilbert}, {Isaak}, {Ivison}, {Lagache}, {Le Floc'h}, {Levenson},
  {Faro}, {Lu}, {Madden}, {Maffei}, {Magdis}, {Mainetti}, {Marchetti},
  {Marsden}, {Marshall}, {Mortier}, {Nguyen}, {O'Halloran}, {Omont}, {Page},
  {Panuzzo}, {Papageorgiou}, {Patel}, {Pearson}, {P{\'e}rez-Fournon}, {Pohlen},
  {Rawlings}, {Raymond}, {Rigopoulou}, {Riguccini}, {Rizzo}, {Rodighiero},
  {Roseboom}, {Rowan-Robinson}, {S{\'a}nchez Portal}, {Schulz}, {Scott},
  {Seymour}, {Shupe}, {Smith}, {Stevens}, {Symeonidis}, {Trichas}, {Tugwell},
  {Vaccari}, {Valtchanov}, {Vieira}, {Viero}, {Vigroux}, {Wang}, {Ward},
  {Wardlow}, {Wright}, {Xu}, \& {Zemcov}}]{Oliver2012}
{Oliver}, S.~J., {Bock}, J., {Altieri}, B., {et~al.} 2012, \mnras, 424, 1614

\bibitem[{{Omont} {et~al.}(2013){Omont}, {Yang}, {Cox}, {Neri}, {Beelen},
  {Bussmann}, {Gavazzi}, {van der Werf}, {Riechers}, {Downes}, {Krips}, {Dye},
  {Ivison}, {Vieira}, {Wei{\ss}}, {Aguirre}, {Baes}, {Baker}, {Bertoldi},
  {Cooray}, {Dannerbauer}, {De Zotti}, {Eales}, {Fu}, {Gao}, {Gu{\'e}lin},
  {Harris}, {Jarvis}, {Lehnert}, {Leeuw}, {Lupu}, {Menten}, {Micha{\l}owski},
  {Negrello}, {Serjeant}, {Temi}, {Auld}, {Dariush}, {Dunne}, {Fritz},
  {Hopwood}, {Hoyos}, {Ibar}, {Maddox}, {Smith}, {Valiante}, {Bock},
  {Bradford}, {Glenn}, \& {Scott}}]{Omont2013}
{Omont}, A., {Yang}, C., {Cox}, P., {et~al.} 2013, \aap, 551, A115

\bibitem[{{Oteo} {et~al.}(2016){Oteo}, {Ivison}, {Dunne}, {Smail}, {Swinbank},
  {Zhang}, {Lewis}, {Maddox}, {Riechers}, {Serjeant}, {Van der Werf}, {Biggs},
  {Bremer}, {Cigan}, {Clements}, {Cooray}, {Dannerbauer}, {Eales}, {Ibar},
  {Messias}, {Micha{\l}owski}, {P{\'e}rez-Fournon}, \& {van Kampen}}]{Oteo2016}
{Oteo}, I., {Ivison}, R.~J., {Dunne}, L., {et~al.} 2016, \apj, 827, 34

\bibitem[{{Oteo} {et~al.}(2017){Oteo}, {Zhang}, {Yang}, {Ivison}, {Omont},
  {Bremer}, {Bussmann}, {Cooray}, {Cox}, {Dannerbauer}, {Dunne}, {Eales},
  {Furlanetto}, {Gavazzi}, {Gao}, {Greve}, {Nayyeri}, {Negrello}, {Neri},
  {Riechers}, {Tunnard}, {Wagg}, \& {Van der Werf}}]{Oteo2017}
{Oteo}, I., {Zhang}, Z.~Y., {Yang}, C., {et~al.} 2017, \apj, 850, 170

\bibitem[{{Peeples} \& {Shankar}(2011)}]{peeples2011}
{Peeples}, M.~S. \& {Shankar}, F. 2011, \mnras, 417, 2962

\bibitem[{{Peng} {et~al.}(2002){Peng}, {Ho}, {Impey}, \& {Rix}}]{peng2002}
{Peng}, C.~Y., {Ho}, L.~C., {Impey}, C.~D., \& {Rix}, H.-W. 2002, \aj, 124, 266

\bibitem[{{Peng} \& {Maiolino}(2014)}]{peng2014}
{Peng}, Y.-j. \& {Maiolino}, R. 2014, \mnras, 443, 3643

\bibitem[{{Perley} \& {Butler}(2017)}]{Perley2017}
{Perley}, R.~A. \& {Butler}, B.~J. 2017, \apjs, 230, 7

\bibitem[{{Pilbratt} {et~al.}(2010){Pilbratt}, {Riedinger}, {Passvogel},
  {Crone}, {Doyle}, {Gageur}, {Heras}, {Jewell}, {Metcalfe}, {Ott}, \&
  {Schmidt}}]{Pilbratt2010}
{Pilbratt}, G.~L., {Riedinger}, J.~R., {Passvogel}, T., {et~al.} 2010, \aap,
  518, L1

\bibitem[{{Planck Collaboration} {et~al.}(2011){Planck Collaboration},
  {Abergel}, {Ade}, {Aghanim}, {Arnaud}, {Ashdown}, {Aumont}, {Baccigalupi},
  {Balbi}, {Banday}, {Barreiro}, {Bartlett}, {Battaner}, {Benabed},
  {Beno{\^\i}t}, {Bernard}, {Bersanelli}, {Bhatia}, {Bock}, {Bonaldi}, {Bond},
  {Borrill}, {Bouchet}, {Boulanger}, {Bucher}, {Burigana}, {Cabella},
  {Cardoso}, {Catalano}, {Cay{\'o}n}, {Challinor}, {Chamballu}, {Chiang},
  {Chiang}, {Christensen}, {Colombi}, {Couchot}, {Coulais}, {Crill}, {Cuttaia},
  {Dame}, {Danese}, {Davies}, {Davis}, {de Bernardis}, {de Gasperis}, {de
  Rosa}, {de Zotti}, {Delabrouille}, {Delouis}, {D{\'e}sert}, {Dickinson},
  {Donzelli}, {Dor{\'e}}, {D{\"o}rl}, {Douspis}, {Dupac}, {Efstathiou},
  {En{\ss}lin}, {Finelli}, {Forni}, {Frailis}, {Franceschi}, {Galeotta},
  {Ganga}, {Giard}, {Giardino}, {Giraud-H{\'e}raud}, {Gonz{\'a}lez-Nuevo},
  {G{\'o}rski}, {Gratton}, {Gregorio}, {Grenier}, {Gruppuso}, {Hansen},
  {Harrison}, {Henrot-Versill{\'e}}, {Herranz}, {Hildebrandt}, {Hivon},
  {Hobson}, {Holmes}, {Hovest}, {Hoyland}, {Huffenberger}, {Jaffe}, {Jaffe},
  {Jones}, {Juvela}, {Keih{\"a}nen}, {Keskitalo}, {Kisner}, {Kneissl}, {Knox},
  {Kurki-Suonio}, {Lagache}, {L{\"a}hteenm{\"a}ki}, {Lamarre}, {Lasenby},
  {Laureijs}, {Lawrence}, {Leach}, {Leonardi}, {Leroy}, {Lilje},
  {Linden-V{\o}rnle}, {L{\'o}pez-Caniego}, {Lubin}, {Mac{\'\i}as-P{\'e}rez},
  {MacTavish}, {Maffei}, {Mandolesi}, {Mann}, {Maris}, {Marshall},
  {Mart{\'\i}nez-Gonz{\'a}lez}, {Masi}, {Matarrese}, {Matthai}, {Mazzotta},
  {McGehee}, {Meinhold}, {Melchiorri}, {Mendes}, {Mennella},
  {Miville-Desch{\^e}nes}, {Moneti}, {Montier}, {Morgante}, {Mortlock},
  {Munshi}, {Murphy}, {Naselsky}, {Natoli}, {Netterfield},
  {N{\o}rgaard-Nielsen}, {Noviello}, {Novikov}, {Novikov}, {Osborne}, {Pajot},
  {Paladini}, {Pasian}, {Patanchon}, {Perdereau}, {Perotto}, {Perrotta},
  {Piacentini}, {Piat}, {Plaszczynski}, {Pointecouteau}, {Polenta}, {Ponthieu},
  {Poutanen}, {Pr{\'e}zeau}, {Prunet}, {Puget}, {Rachen}, {Reach}, {Rebolo},
  {Reich}, {Renault}, {Ricciardi}, {Riller}, {Ristorcelli}, {Rocha}, {Rosset},
  {Rubi{\~n}o-Mart{\'\i}n}, {Rusholme}, {Sandri}, {Santos}, {Savini}, {Scott},
  {Seiffert}, {Shellard}, {Smoot}, {Starck}, {Stivoli}, {Stolyarov}, {Stompor},
  {Sudiwala}, {Sygnet}, {Tauber}, {Terenzi}, {Toffolatti}, {Tomasi}, {Torre},
  {Tristram}, {Tuovinen}, {Umana}, {Valenziano}, {Varis}, {Vielva}, {Villa},
  {Vittorio}, {Wade}, {Wandelt}, {Wilkinson}, {Ysard}, {Yvon}, {Zacchei}, \&
  {Zonca}}]{planck2011_XXI}
{Planck Collaboration}, {Abergel}, A., {Ade}, P.~A.~R., {et~al.} 2011, \aap,
  536, A21

\bibitem[{{Planck Collaboration} {et~al.}(2016){Planck Collaboration}, {Ade},
  {Aghanim}, {Alves}, {Aniano}, {Arnaud}, {Ashdown}, {Aumont}, {Baccigalupi},
  {Banday}, {Barreiro}, {Bartolo}, {Battaner}, {Benabed}, {Benoit-L{\'e}vy},
  {Bernard}, {Bersanelli}, {Bielewicz}, {Bonaldi}, {Bonavera}, {Bond},
  {Borrill}, {Bouchet}, {Boulanger}, {Burigana}, {Butler}, {Calabrese},
  {Cardoso}, {Catalano}, {Chamballu}, {Chiang}, {Christensen}, {Clements},
  {Colombi}, {Colombo}, {Couchot}, {Crill}, {Curto}, {Cuttaia}, {Danese},
  {Davies}, {Davis}, {de Bernardis}, {de Rosa}, {de Zotti}, {Delabrouille},
  {Dickinson}, {Diego}, {Dole}, {Donzelli}, {Dor{\'e}}, {Douspis}, {Draine},
  {Ducout}, {Dupac}, {Efstathiou}, {Elsner}, {En{\ss}lin}, {Eriksen},
  {Falgarone}, {Finelli}, {Forni}, {Frailis}, {Fraisse}, {Franceschi},
  {Frejsel}, {Galeotta}, {Galli}, {Ganga}, {Ghosh}, {Giard}, {Gjerl{\o}w},
  {Gonz{\'a}lez-Nuevo}, {G{\'o}rski}, {Gregorio}, {Gruppuso}, {Guillet},
  {Hansen}, {Hanson}, {Harrison}, {Henrot-Versill{\'e}},
  {Hern{\'a}ndez-Monteagudo}, {Herranz}, {Hildebrand t}, {Hivon}, {Holmes},
  {Hovest}, {Huffenberger}, {Hurier}, {Jaffe}, {Jaffe}, {Jones},
  {Keih{\"a}nen}, {Keskitalo}, {Kisner}, {Kneissl}, {Knoche}, {Kunz},
  {Kurki-Suonio}, {Lagache}, {Lamarre}, {Lasenby}, {Lattanzi}, {Lawrence},
  {Leonardi}, {Levrier}, {Liguori}, {Lilje}, {Linden-V{\o}rnle},
  {L{\'o}pez-Caniego}, {Lubin}, {Mac{\'\i}as-P{\'e}rez}, {Maffei}, {Maino},
  {Mand olesi}, {Maris}, {Marshall}, {Martin}, {Mart{\'\i}nez-Gonz{\'a}lez},
  {Masi}, {Matarrese}, {Mazzotta}, {Melchiorri}, {Mendes}, {Mennella},
  {Migliaccio}, {Miville-Desch{\^e}nes}, {Moneti}, {Montier}, {Morgante},
  {Mortlock}, {Munshi}, {Murphy}, {Naselsky}, {Natoli}, {N{\o}rgaard-Nielsen},
  {Novikov}, {Novikov}, {Oxborrow}, {Pagano}, {Pajot}, {Paladini}, {Paoletti},
  {Pasian}, {Perdereau}, {Perotto}, {Perrotta}, {Pettorino}, {Piacentini},
  {Piat}, {Plaszczynski}, {Pointecouteau}, {Polenta}, {Ponthieu}, {Popa},
  {Pratt}, {Prunet}, {Puget}, {Rachen}, {Reach}, {Rebolo}, {Reinecke},
  {Remazeilles}, {Renault}, {Ristorcelli}, {Rocha}, {Roudier},
  {Rubi{\~n}o-Mart{\'\i}n}, {Rusholme}, {Sandri}, {Santos}, {Scott}, {Spencer},
  {Stolyarov}, {Sudiwala}, {Sunyaev}, {Sutton}, {Suur-Uski}, {Sygnet},
  {Tauber}, {Terenzi}, {Toffolatti}, {Tomasi}, {Tristram}, {Tucci}, {Umana},
  {Valenziano}, {Valiviita}, {Van Tent}, {Vielva}, {Villa}, {Wade}, {Wandelt},
  {Wehus}, {Ysard}, {Yvon}, {Zacchei}, \& {Zonca}}]{planck2016_xxix}
{Planck Collaboration}, {Ade}, P.~A.~R., {Aghanim}, N., {et~al.} 2016, \aap,
  586, A132

\bibitem[{{Planck Collaboration} {et~al.}(2020){Planck Collaboration},
  {Aghanim}, {Akrami}, {Ashdown}, {Aumont}, {Baccigalupi}, {Ballardini},
  {Banday}, {Barreiro}, {Bartolo}, {Basak}, {Battye}, {Benabed}, {Bernard},
  {Bersanelli}, {Bielewicz}, {Bock}, {Bond}, {Borrill}, {Bouchet}, {Boulanger},
  {Bucher}, {Burigana}, {Butler}, {Calabrese}, {Cardoso}, {Carron},
  {Challinor}, {Chiang}, {Chluba}, {Colombo}, {Combet}, {Contreras}, {Crill},
  {Cuttaia}, {de Bernardis}, {de Zotti}, {Delabrouille}, {Delouis}, {Di
  Valentino}, {Diego}, {Dor{\'e}}, {Douspis}, {Ducout}, {Dupac}, {Dusini},
  {Efstathiou}, {Elsner}, {En{\ss}lin}, {Eriksen}, {Fantaye}, {Farhang},
  {Fergusson}, {Fernandez-Cobos}, {Finelli}, {Forastieri}, {Frailis},
  {Fraisse}, {Franceschi}, {Frolov}, {Galeotta}, {Galli}, {Ganga},
  {G{\'e}nova-Santos}, {Gerbino}, {Ghosh}, {Gonz{\'a}lez-Nuevo}, {G{\'o}rski},
  {Gratton}, {Gruppuso}, {Gudmundsson}, {Hamann}, {Handley}, {Hansen},
  {Herranz}, {Hildebrandt}, {Hivon}, {Huang}, {Jaffe}, {Jones}, {Karakci},
  {Keih{\"a}nen}, {Keskitalo}, {Kiiveri}, {Kim}, {Kisner}, {Knox},
  {Krachmalnicoff}, {Kunz}, {Kurki-Suonio}, {Lagache}, {Lamarre}, {Lasenby},
  {Lattanzi}, {Lawrence}, {Le Jeune}, {Lemos}, {Lesgourgues}, {Levrier},
  {Lewis}, {Liguori}, {Lilje}, {Lilley}, {Lindholm}, {L{\'o}pez-Caniego},
  {Lubin}, {Ma}, {Mac{\'\i}as-P{\'e}rez}, {Maggio}, {Maino}, {Mandolesi},
  {Mangilli}, {Marcos-Caballero}, {Maris}, {Martin}, {Martinelli},
  {Mart{\'\i}nez-Gonz{\'a}lez}, {Matarrese}, {Mauri}, {McEwen}, {Meinhold},
  {Melchiorri}, {Mennella}, {Migliaccio}, {Millea}, {Mitra},
  {Miville-Desch{\^e}nes}, {Molinari}, {Montier}, {Morgante}, {Moss}, {Natoli},
  {N{\o}rgaard-Nielsen}, {Pagano}, {Paoletti}, {Partridge}, {Patanchon},
  {Peiris}, {Perrotta}, {Pettorino}, {Piacentini}, {Polastri}, {Polenta},
  {Puget}, {Rachen}, {Reinecke}, {Remazeilles}, {Renzi}, {Rocha}, {Rosset},
  {Roudier}, {Rubi{\~n}o-Mart{\'\i}n}, {Ruiz-Granados}, {Salvati}, {Sandri},
  {Savelainen}, {Scott}, {Shellard}, {Sirignano}, {Sirri}, {Spencer},
  {Sunyaev}, {Suur-Uski}, {Tauber}, {Tavagnacco}, {Tenti}, {Toffolatti},
  {Tomasi}, {Trombetti}, {Valenziano}, {Valiviita}, {Van Tent}, {Vibert},
  {Vielva}, {Villa}, {Vittorio}, {Wand elt}, {Wehus}, {White}, {White},
  {Zacchei}, \& {Zonca}}]{planck2020_2018_VI}
{Planck Collaboration}, {Aghanim}, N., {Akrami}, Y., {et~al.} 2020, \aap, 641,
  A6

\bibitem[{{Planck Collaboration} {et~al.}(2015{\natexlab{a}}){Planck
  Collaboration}, Aghanim, Altieri, Arnaud, Ashdown, Aumont, Baccigalupi,
  Banday, Barreiro, \& Bartolo}]{Planck2015}
{Planck Collaboration}, Aghanim, N., Altieri, B., {et~al.} 2015{\natexlab{a}},
  \aap, 582, 29

\bibitem[{{Planck Collaboration} {et~al.}(2015{\natexlab{b}}){Planck
  Collaboration}, {Fermi Collaboration}, {Ade}, {Aghanim}, {Aniano}, {Arnaud},
  {Ashdown}, {Aumont}, {Baccigalupi}, {Banday}, {Barreiro}, {Bartolo},
  {Battaner}, {Benabed}, {Benoit-L{\'e}vy}, {Bernard}, {Bersanelli},
  {Bielewicz}, {Bonaldi}, {Bonavera}, {Bond}, {Borrill}, {Bouchet},
  {Boulanger}, {Burigana}, {Butler}, {Calabrese}, {Cardoso}, {Casand jian},
  {Catalano}, {Chamballu}, {Chiang}, {Christensen}, {Colombo}, {Combet},
  {Couchot}, {Crill}, {Curto}, {Cuttaia}, {Danese}, {Davies}, {Davis}, {de
  Bernardis}, {de Rosa}, {de Zotti}, {Delabrouille}, {D{\'e}sert}, {Dickinson},
  {Diego}, {Digel}, {Dole}, {Donzelli}, {Dor{\'e}}, {Douspis}, {Ducout},
  {Dupac}, {Efstathiou}, {Elsner}, {En{\ss}lin}, {Eriksen}, {Falgarone},
  {Finelli}, {Forni}, {Frailis}, {Fraisse}, {Franceschi}, {Frejsel}, {Fukui},
  {Galeotta}, {Galli}, {Ganga}, {Ghosh}, {Giard}, {Gjerl{\o}w},
  {Gonz{\'a}lez-Nuevo}, {G{\'o}rski}, {Gregorio}, {Grenier}, {Gruppuso},
  {Hansen}, {Hanson}, {Harrison}, {Henrot-Versill{\'e}},
  {Hern{\'a}ndez-Monteagudo}, {Herranz}, {Hildebrand t}, {Hivon}, {Hobson},
  {Holmes}, {Hovest}, {Huffenberger}, {Hurier}, {Jaffe}, {Jaffe}, {Jones},
  {Juvela}, {Keih{\"a}nen}, {Keskitalo}, {Kisner}, {Kneissl}, {Knoche}, {Kunz},
  {Kurki-Suonio}, {Lagache}, {Lamarre}, {Lasenby}, {Lattanzi}, {Lawrence},
  {Leonardi}, {Levrier}, {Liguori}, {Lilje}, {Linden-V{\o}rnle},
  {L{\'o}pez-Caniego}, {Lubin}, {Mac{\'\i}as-P{\'e}rez}, {Maffei}, {Maino},
  {Mand olesi}, {Maris}, {Marshall}, {Martin}, {Mart{\'\i}nez-Gonz{\'a}lez},
  {Masi}, {Matarrese}, {Mazzotta}, {Melchiorri}, {Mendes}, {Mennella},
  {Migliaccio}, {Miville-Desch{\^e}nes}, {Moneti}, {Montier}, {Morgante},
  {Mortlock}, {Munshi}, {Murphy}, {Naselsky}, {Natoli}, {N{\o}rgaard-Nielsen},
  {Novikov}, {Novikov}, {Oxborrow}, {Pagano}, {Pajot}, {Paladini}, {Paoletti},
  {Pasian}, {Perdereau}, {Perotto}, {Perrotta}, {Pettorino}, {Piacentini},
  {Piat}, {Plaszczynski}, {Pointecouteau}, {Polenta}, {Popa}, {Pratt},
  {Prunet}, {Puget}, {Rachen}, {Reach}, {Rebolo}, {Reinecke}, {Remazeilles},
  {Renault}, {Ristorcelli}, {Rocha}, {Roudier}, {Rusholme}, {Sandri}, {Santos},
  {Scott}, {Spencer}, {Stolyarov}, {Strong}, {Sudiwala}, {Sunyaev}, {Sutton},
  {Suur-Uski}, {Sygnet}, {Tauber}, {Terenzi}, {Tibaldo}, {Toffolatti},
  {Tomasi}, {Tristram}, {Tucci}, {Umana}, {Valenziano}, {Valiviita}, {Van
  Tent}, {Vielva}, {Villa}, {Wade}, {Wandelt}, {Wehus}, {Yvon}, {Zacchei}, \&
  {Zonca}}]{planck2015_intXXVIII}
{Planck Collaboration}, {Fermi Collaboration}, {Ade}, P.~A.~R., {et~al.}
  2015{\natexlab{b}}, \aap, 582, A31

\bibitem[{{Polletta} {et~al.}(2007){Polletta}, {Tajer}, {Maraschi},
  {Trinchieri}, {Lonsdale}, {Chiappetti}, {Andreon}, {Pierre}, {Le F{\`e}vre},
  {Zamorani}, {Maccagni}, {Garcet}, {Surdej}, {Franceschini}, {Alloin},
  {Shupe}, {Surace}, {Fang}, {Rowan-Robinson}, {Smith}, \&
  {Tresse}}]{polletta2007}
{Polletta}, M., {Tajer}, M., {Maraschi}, L., {et~al.} 2007, \apj, 663, 81

\bibitem[{{Priestley} \& {Whitworth}(2020)}]{Priestley2020}
{Priestley}, F.~D. \& {Whitworth}, A.~P. 2020, \mnras, 494, L48

\bibitem[{{Rangwala} {et~al.}(2011){Rangwala}, {Maloney}, {Glenn}, {Wilson},
  {Rykala}, {Isaak}, {Baes}, {Bendo}, {Boselli}, {Bradford}, {Clements},
  {Cooray}, {Fulton}, {Imhof}, {Kamenetzky}, {Madden}, {Mentuch}, {Sacchi},
  {Sauvage}, {Schirm}, {Smith}, {Spinoglio}, \& {Wolfire}}]{Rangwala2011}
{Rangwala}, N., {Maloney}, P.~R., {Glenn}, J., {et~al.} 2011, \apj, 743, 94

\bibitem[{{R{\'e}my-Ruyer} {et~al.}(2014){R{\'e}my-Ruyer}, {Madden},
  {Galliano}, {Galametz}, {Takeuchi}, {Asano}, {Zhukovska}, {Lebouteiller},
  {Cormier}, {Jones}, {Bocchio}, {Baes}, {Bendo}, {Boquien}, {Boselli},
  {DeLooze}, {Doublier-Pritchard}, {Hughes}, {Karczewski}, \&
  {Spinoglio}}]{remyruyer2014}
{R{\'e}my-Ruyer}, A., {Madden}, S.~C., {Galliano}, F., {et~al.} 2014, \aap,
  563, A31

\bibitem[{{Reuter} {et~al.}(2020){Reuter}, {Vieira}, {Spilker}, {Weiss},
  {Aravena}, {Archipley}, {Bethermin}, {Chapman}, {De Breuck}, {Dong},
  {Everett}, {Fu}, {Greve}, {Hayward}, {Hill}, {Hezaveh}, {Jarugula}, {Litke},
  {Malkan}, {Marrone}, {Narayanan}, {Phadke}, {Stark}, \&
  {Strandet}}]{Reuter2020}
{Reuter}, C., {Vieira}, J.~D., {Spilker}, J.~S., {et~al.} 2020, arXiv e-prints,
  arXiv:2006.14060

\bibitem[{{Riechers} {et~al.}(2013){Riechers}, {Bradford}, {Clements},
  {Dowell}, {P{\'e}rez-Fournon}, {Ivison}, {Bridge}, {Conley}, {Fu}, {Vieira},
  {Wardlow}, {Calanog}, {Cooray}, {Hurley}, {Neri}, {Kamenetzky}, {Aguirre},
  {Altieri}, {Arumugam}, {Benford}, {B{\'e}thermin}, {Bock}, {Burgarella},
  {Cabrera-Lavers}, {Chapman}, {Cox}, {Dunlop}, {Earle}, {Farrah}, {Ferrero},
  {Franceschini}, {Gavazzi}, {Glenn}, {Solares}, {Gurwell}, {Halpern},
  {Hatziminaoglou}, {Hyde}, {Ibar}, {Kov{\'a}cs}, {Krips}, {Lupu}, {Maloney},
  {Martinez-Navajas}, {Matsuhara}, {Murphy}, {Naylor}, {Nguyen}, {Oliver},
  {Omont}, {Page}, {Petitpas}, {Rangwala}, {Roseboom}, {Scott}, {Smith},
  {Staguhn}, {Streblyanska}, {Thomson}, {Valtchanov}, {Viero}, {Wang},
  {Zemcov}, \& {Zmuidzinas}}]{Riechers2013}
{Riechers}, D.~A., {Bradford}, C.~M., {Clements}, D.~L., {et~al.} 2013, \nat,
  496, 329

\bibitem[{{Riechers} {et~al.}(2011){Riechers}, {Carilli}, {Maddalena}, {Hodge},
  {Harris}, {Baker}, {Walter}, {Wagg}, {Vand en Bout}, {Wei{\ss}}, \&
  {Sharon}}]{Riechers2011}
{Riechers}, D.~A., {Carilli}, C.~L., {Maddalena}, R.~J., {et~al.} 2011, \apjl,
  739, L32

\bibitem[{{Riechers} {et~al.}(2017){Riechers}, {Leung}, {Ivison},
  {P{\'e}rez-Fournon}, {Lewis}, {Marques-Chaves}, {Oteo}, {Clements}, {Cooray},
  \& {Greenslade}}]{Riechers2017}
{Riechers}, D.~A., {Leung}, T.~K.~D., {Ivison}, R.~J., {et~al.} 2017, \apj,
  850, 1

\bibitem[{{Riechers} {et~al.}(2008){Riechers}, {Walter}, {Brewer}, {Carilli},
  {Lewis}, {Bertoldi}, \& {Cox}}]{Riechers2008}
{Riechers}, D.~A., {Walter}, F., {Brewer}, B.~J., {et~al.} 2008, \apj, 686, 851

\bibitem[{{Riechers} {et~al.}(2010){Riechers}, {Wei{\ss}}, {Walter}, \&
  {Wagg}}]{Riechers2010a}
{Riechers}, D.~A., {Wei{\ss}}, A., {Walter}, F., \& {Wagg}, J. 2010, \apj, 725,
  1032

\bibitem[{{Rizzo} {et~al.}(2018){Rizzo}, {Vegetti}, {Fraternali}, \& {Di
  Teodoro}}]{Rizzo2018}
{Rizzo}, F., {Vegetti}, S., {Fraternali}, F., \& {Di Teodoro}, E. 2018, \mnras,
  481, 5606

\bibitem[{{Rubin} {et~al.}(2012){Rubin}, {Prochaska}, {Koo}, \&
  {Phillips}}]{Rubin2012}
{Rubin}, K. H.~R., {Prochaska}, J.~X., {Koo}, D.~C., \& {Phillips}, A.~C. 2012,
  \apjl, 747, L26

\bibitem[{{S{\'a}nchez Almeida} {et~al.}(2014){S{\'a}nchez Almeida},
  {Elmegreen}, {Mu{\~n}oz-Tu{\~n}{\'o}n}, \& {Elmegreen}}]{almeida2014}
{S{\'a}nchez Almeida}, J., {Elmegreen}, B.~G., {Mu{\~n}oz-Tu{\~n}{\'o}n}, C.,
  \& {Elmegreen}, D.~M. 2014, \aapr, 22, 71

\bibitem[{{Schmidt}(1959)}]{Schmidt1959}
{Schmidt}, M. 1959, \apj, 129, 243

\bibitem[{{Schreiber} {et~al.}(2015){Schreiber}, {Pannella}, {Elbaz},
  {B{\'e}thermin}, {Inami}, {Dickinson}, {Magnelli}, {Wang}, {Aussel}, {Daddi},
  {Juneau}, {Shu}, {Sargent}, {Buat}, {Faber}, {Ferguson}, {Giavalisco},
  {Koekemoer}, {Magdis}, {Morrison}, {Papovich}, {Santini}, \&
  {Scott}}]{schreiber2015}
{Schreiber}, C., {Pannella}, M., {Elbaz}, D., {et~al.} 2015, \aap, 575, A74

\bibitem[{{Sharon} {et~al.}(2015){Sharon}, {Baker}, {Harris}, {Tacconi},
  {Lutz}, \& {Longmore}}]{Sharon2015}
{Sharon}, C.~E., {Baker}, A.~J., {Harris}, A.~I., {et~al.} 2015, \apj, 798, 133

\bibitem[{{Sharon} {et~al.}(2019){Sharon}, {Tagore}, {Baker}, {Rivera},
  {Keeton}, {Lutz}, {Genzel}, {Wilner}, {Hicks}, {Allam}, \&
  {Tucker}}]{Sharon2019}
{Sharon}, C.~E., {Tagore}, A.~S., {Baker}, A.~J., {et~al.} 2019, \apj, 879, 52

\bibitem[{{Solomon} \& {Vanden Bout}(2005)}]{Solomon-VandenBout2005}
{Solomon}, P.~M. \& {Vanden Bout}, P.~A. 2005, \araa, 43, 677

\bibitem[{{Somerville} \& {Dav{\'e}}(2015)}]{somerville2015}
{Somerville}, R.~S. \& {Dav{\'e}}, R. 2015, \araa, 53, 51

\bibitem[{{Spilker} {et~al.}(2018){Spilker}, {Aravena}, {B{\'e}thermin},
  {Chapman}, {Chen}, {Cunningham}, {De Breuck}, {Dong}, {Gonzalez}, {Hayward},
  {Hezaveh}, {Litke}, {Ma}, {Malkan}, {Marrone}, {Miller}, {Morningstar},
  {Narayanan}, {Phadke}, {Sreevani}, {Stark}, {Vieira}, \&
  {Wei{\ss}}}]{Spilker2018}
{Spilker}, J.~S., {Aravena}, M., {B{\'e}thermin}, M., {et~al.} 2018, Science,
  361, 1016

\bibitem[{{Spilker} {et~al.}(2014){Spilker}, {Marrone}, {Aguirre}, {Aravena},
  {Ashby}, {B{\'e}thermin}, {Bradford}, {Bothwell}, {Brodwin}, {Carlstrom},
  {Chapman}, {Crawford}, {de Breuck}, {Fassnacht}, {Gonzalez}, {Greve},
  {Gullberg}, {Hezaveh}, {Holzapfel}, {Husband}, {Ma}, {Malkan}, {Murphy},
  {Reichardt}, {Rotermund}, {Stalder}, {Stark}, {Strandet}, {Vieira},
  {Wei{\ss}}, \& {Welikala}}]{Spilker2014}
{Spilker}, J.~S., {Marrone}, D.~P., {Aguirre}, J.~E., {et~al.} 2014, \apj, 785,
  149

\bibitem[{{Staniszewski} {et~al.}(2009){Staniszewski}, {Ade}, {Aird}, {Benson},
  {Bleem}, {Carlstrom}, {Chang}, {Cho}, {Crawford}, {Crites}, {de Haan},
  {Dobbs}, {Halverson}, {Holder}, {Holzapfel}, {Hrubes}, {Joy}, {Keisler},
  {Lanting}, {Lee}, {Leitch}, {Loehr}, {Lueker}, {McMahon}, {Mehl}, {Meyer},
  {Mohr}, {Montroy}, {Ngeow}, {Padin}, {Plagge}, {Pryke}, {Reichardt}, {Ruhl},
  {Schaffer}, {Shaw}, {Shirokoff}, {Spieler}, {Stalder}, {Stark}, {Vand
  erlinde}, {Vieira}, {Zahn}, \& {Zenteno}}]{Staniszewski2009}
{Staniszewski}, Z., {Ade}, P.~A.~R., {Aird}, K.~A., {et~al.} 2009, \apj, 701,
  32

\bibitem[{{Stanley} {et~al.}(2020){Stanley}, {Knudsen}, {Aalto}, {Fan},
  {Falstad}, \& {Humphreys}}]{Stanley2020}
{Stanley}, F., {Knudsen}, K.~K., {Aalto}, S., {et~al.} 2020, arXiv e-prints,
  arXiv:2011.09991

\bibitem[{{Stone} {et~al.}(2016){Stone}, {Veilleux}, {Mel{\'e}ndez}, {Sturm},
  {Graci{\'a}-Carpio}, \& {Gonz{\'a}lez-Alfonso}}]{Stone2016}
{Stone}, M., {Veilleux}, S., {Mel{\'e}ndez}, M., {et~al.} 2016, \apj, 826, 111

\bibitem[{{Swinbank} {et~al.}(2011){Swinbank}, {Papadopoulos}, {Cox}, {Krips},
  {Ivison}, {Smail}, {Thomson}, {Neri}, {Richard}, \& {Ebeling}}]{Swinbank2011}
{Swinbank}, A.~M., {Papadopoulos}, P.~P., {Cox}, P., {et~al.} 2011, \apj, 742,
  11

\bibitem[{{Tacconi} {et~al.}(2018){Tacconi}, {Genzel}, {Saintonge}, {Combes},
  {Garc{\'\i}a-Burillo}, {Neri}, {Bolatto}, {Contini}, {F{\"o}rster Schreiber},
  {Lilly}, {Lutz}, {Wuyts}, {Accurso}, {Boissier}, {Boone}, {Bouch{\'e}},
  {Bournaud}, {Burkert}, {Carollo}, {Cooper}, {Cox}, {Feruglio}, {Freundlich},
  {Herrera-Camus}, {Juneau}, {Lippa}, {Naab}, {Renzini}, {Salome}, {Sternberg},
  {Tadaki}, {{\"U}bler}, {Walter}, {Weiner}, \& {Weiss}}]{tacconi2018}
{Tacconi}, L.~J., {Genzel}, R., {Saintonge}, A., {et~al.} 2018, \apj, 853, 179

\bibitem[{{Tacconi} {et~al.}(2020){Tacconi}, {Genzel}, \&
  {Sternberg}}]{tacconi2020}
{Tacconi}, L.~J., {Genzel}, R., \& {Sternberg}, A. 2020, arXiv e-prints,
  arXiv:2003.06245

\bibitem[{{Tagore} \& {Jackson}(2016)}]{Tagore&Jackson2016}
{Tagore}, A.~S. \& {Jackson}, N. 2016, \mnras, 457, 3066

\bibitem[{{Tagore} \& {Keeton}(2014)}]{Tagore&Keeton2014}
{Tagore}, A.~S. \& {Keeton}, C.~R. 2014, \mnras, 445, 694

\bibitem[{{Thomson} {et~al.}(2014){Thomson}, {Ivison}, {Simpson}, {Swinbank},
  {Smail}, {Arumugam}, {Alexand er}, {Beelen}, {Brandt}, {Chandra},
  {Dannerbauer}, {Greve}, {Hodge}, {Ibar}, {Karim}, {Murphy}, {Schinnerer},
  {Sirothia}, {Walter}, {Wardlow}, \& {van der Werf}}]{thomson2014}
{Thomson}, A.~P., {Ivison}, R.~J., {Simpson}, J.~M., {et~al.} 2014, \mnras,
  442, 577

\bibitem[{{Tody}(1986)}]{Tody1986}
{Tody}, D. 1986, in Society of Photo-Optical Instrumentation Engineers (SPIE)
  Conference Series, Vol. 627, Instrumentation in astronomy VI, ed. D.~L.
  {Crawford}, 733

\bibitem[{{Tumlinson} {et~al.}(2017){Tumlinson}, {Peeples}, \&
  {Werk}}]{tumlinson2017}
{Tumlinson}, J., {Peeples}, M.~S., \& {Werk}, J.~K. 2017, \araa, 55, 389

\bibitem[{{Valentino} {et~al.}(2020){Valentino}, {Daddi}, {Puglisi}, {Magdis},
  {Liu}, {Kokorev}, {Cortzen}, {Madden}, {Aravena}, {G{\'o}mez-Guijarro},
  {Lee}, {Le Floc'h}, {Gao}, {Gobat}, {Bournaud}, {Dannerbauer}, {Jin},
  {Dickinson}, {Kartaltepe}, \& {Sanders}}]{valentino2020}
{Valentino}, F., {Daddi}, E., {Puglisi}, A., {et~al.} 2020, \aap, 641, A155

\bibitem[{{Valiante} {et~al.}(2016){Valiante}, {Smith}, {Eales}, {Maddox},
  {Ibar}, {Hopwood}, {Dunne}, {Cigan}, {Dye}, {Pascale}, {Rigby}, {Bourne},
  {Furlanetto}, \& {Ivison}}]{Valiante2016}
{Valiante}, E., {Smith}, M.~W.~L., {Eales}, S., {et~al.} 2016, \mnras, 462,
  3146

\bibitem[{{van Dishoeck} {et~al.}(1993){van Dishoeck}, {Jansen}, {Schilke}, \&
  {Phillips}}]{VanDishoeck1993}
{van Dishoeck}, E.~F., {Jansen}, D.~J., {Schilke}, P., \& {Phillips}, T.~G.
  1993, \apjl, 416, L83

\bibitem[{{Vieira} {et~al.}(2010){Vieira}, Crawford, Switzer, Ade, Aird, Ashby,
  Benson, Bleem, Brodwin, Carlstrom, Chang, Cho, Crites, de~Haan, Dobbs,
  Everett, George, Gladders, Hall, Halverson, Holder, Holzapfel, Hrubes, Joy,
  Keisler, Knox, Lee, Leitch, Lueker, Marrone, McIntyre, McMahon, Mehl, Meyer,
  Mohr, Montroy, Padin, Plagge, Pryke, Reichardt, Ruhl, Schaffer, Shaw,
  Shirokoff, Spieler, Stalder, Staniszewski, Stark, Vanderlinde, Walsh,
  Williamson, Yang, Zahn, \& Zenteno}]{Vieira2010}
{Vieira}, J.~D., Crawford, T.~M., Switzer, E.~R., {et~al.} 2010, \apj, 719, 763

\bibitem[{{Weingartner} \& {Draine}(2001)}]{wd01}
{Weingartner}, J.~C. \& {Draine}, B.~T. 2001, \apj, 548, 296

\bibitem[{{Wei{\ss}} {et~al.}(2013){Wei{\ss}}, {De Breuck}, {Marrone},
  {Vieira}, {Aguirre}, {Aird}, {Aravena}, {Ashby}, {Bayliss}, {Benson},
  {B{\'e}thermin}, {Biggs}, {Bleem}, {Bock}, {Bothwell}, {Bradford}, {Brodwin},
  {Carlstrom}, {Chang}, {Chapman}, {Crawford}, {Crites}, {de Haan}, {Dobbs},
  {Downes}, {Fassnacht}, {George}, {Gladders}, {Gonzalez}, {Greve},
  {Halverson}, {Hezaveh}, {High}, {Holder}, {Holzapfel}, {Hoover}, {Hrubes},
  {Husband}, {Keisler}, {Lee}, {Leitch}, {Lueker}, {Luong-Van}, {Malkan},
  {McIntyre}, {McMahon}, {Mehl}, {Menten}, {Meyer}, {Murphy}, {Padin},
  {Plagge}, {Reichardt}, {Rest}, {Rosenman}, {Ruel}, {Ruhl}, {Schaffer},
  {Shirokoff}, {Spilker}, {Stalder}, {Staniszewski}, {Stark}, {Story},
  {Vanderlinde}, {Welikala}, \& {Williamson}}]{Weiss2013}
{Wei{\ss}}, A., {De Breuck}, C., {Marrone}, D.~P., {et~al.} 2013, \apj, 767, 88

\bibitem[{{Wei{\ss}} {et~al.}(2007){Wei{\ss}}, {Downes}, {Neri}, {Walter},
  {Henkel}, {Wilner}, {Wagg}, \& {Wiklind}}]{Weiss2007}
{Wei{\ss}}, A., {Downes}, D., {Neri}, R., {et~al.} 2007, \aap, 467, 955

\bibitem[{{Wei{\ss}} {et~al.}(2003){Wei{\ss}}, {Henkel}, {Downes}, \&
  {Walter}}]{weiss2003}
{Wei{\ss}}, A., {Henkel}, C., {Downes}, D., \& {Walter}, F. 2003, \aap, 409,
  L41

\bibitem[{{Wei{\ss}} {et~al.}(2005){Wei{\ss}}, {Walter}, \&
  {Scoville}}]{weiss2005b}
{Wei{\ss}}, A., {Walter}, F., \& {Scoville}, N.~Z. 2005, \aap, 438, 533

\bibitem[{{Whitaker} {et~al.}(2014){Whitaker}, {Franx}, {Leja}, {van Dokkum},
  {Henry}, {Skelton}, {Fumagalli}, {Momcheva}, {Brammer}, {Labb{\'e}},
  {Nelson}, \& {Rigby}}]{whitaker2014}
{Whitaker}, K.~E., {Franx}, M., {Leja}, J., {et~al.} 2014, \apj, 795, 104

\bibitem[{{Wiseman} {et~al.}(2017){Wiseman}, {Perley}, {Schady}, {Prochaska},
  {de Ugarte Postigo}, {Kr{\"u}hler}, {Yates}, \& {Greiner}}]{Wiseman2017}
{Wiseman}, P., {Perley}, D.~A., {Schady}, P., {et~al.} 2017, \aap, 607, A107

\bibitem[{{Wright} {et~al.}(2010){Wright}, {Eisenhardt}, {Mainzer}, {Ressler},
  {Cutri}, {Jarrett}, {Kirkpatrick}, {Padgett}, {McMillan}, {Skrutskie},
  {Stanford}, {Cohen}, {Walker}, {Mather}, {Leisawitz}, {Gautier}, {McLean},
  {Benford}, {Lonsdale}, {Blain}, {Mendez}, {Irace}, {Duval}, {Liu}, {Royer},
  {Heinrichsen}, {Howard}, {Shannon}, {Kendall}, {Walsh}, {Larsen}, {Cardon},
  {Schick}, {Schwalm}, {Abid}, {Fabinsky}, {Naes}, \& {Tsai}}]{wright2010}
{Wright}, E.~L., {Eisenhardt}, P. R.~M., {Mainzer}, A.~K., {et~al.} 2010, \aj,
  140, 1868

\bibitem[{{Yabe} {et~al.}(2015){Yabe}, {Ohta}, {Akiyama}, {Iwamuro}, {Tamura},
  {Yuma}, {Dalton}, \& {Lewis}}]{yabe2015}
{Yabe}, K., {Ohta}, K., {Akiyama}, M., {et~al.} 2015, \apj, 798, 45

\bibitem[{{Yang} {et~al.}(2013){Yang}, {Gao}, {Omont}, {Liu}, {Isaak},
  {Downes}, {van der Werf}, \& {Lu}}]{Yang2013}
{Yang}, C., {Gao}, Y., {Omont}, A., {et~al.} 2013, \apjl, 771, L24

\bibitem[{{Yang} {et~al.}(2020){Yang}, {Gonz{\'a}lez-Alfonso}, {Omont},
  {Pereira-Santaella}, {Fischer}, {Beelen}, \& {Gavazzi}}]{Yang2020}
{Yang}, C., {Gonz{\'a}lez-Alfonso}, E., {Omont}, A., {et~al.} 2020, \aap, 634,
  L3

\bibitem[{{Yang} {et~al.}(2017){Yang}, {Omont}, {Beelen}, {Gao}, {van der
  Werf}, {Gavazzi}, {Zhang}, {Ivison}, {Lehnert}, \& {Liu}}]{Yang2017}
{Yang}, C., {Omont}, A., {Beelen}, A., {et~al.} 2017, \aap, 608, A144

\bibitem[{{Yang} {et~al.}(2016){Yang}, {Omont}, {Beelen},
  {Gonz{\'a}lez-Alfonso}, {Neri}, {Gao}, {van der Werf}, {Wei{\ss}}, {Gavazzi},
  {Falstad}, {Baker}, {Bussmann}, {Cooray}, {Cox}, {Dannerbauer}, {Dye},
  {Gu{\'e}lin}, {Ivison}, {Krips}, {Lehnert}, {Micha{\l}owski}, {Riechers},
  {Spaans}, \& {Valiante}}]{Yang2016}
{Yang}, C., {Omont}, A., {Beelen}, A., {et~al.} 2016, \aap, 595, A80

\bibitem[{{Zhang} {et~al.}(2014){Zhang}, {Gao}, {Henkel}, {Zhao}, {Wang},
  {Menten}, \& {G{\"u}sten}}]{Zhang2014}
{Zhang}, Z.-Y., {Gao}, Y., {Henkel}, C., {et~al.} 2014, \apjl, 784, L31

\end{thebibliography}

\begin{appendix}

\section{Details of lens model}\label{app:lens}

\begin{figure*}
\centering
\includegraphics[width=0.7\textwidth]{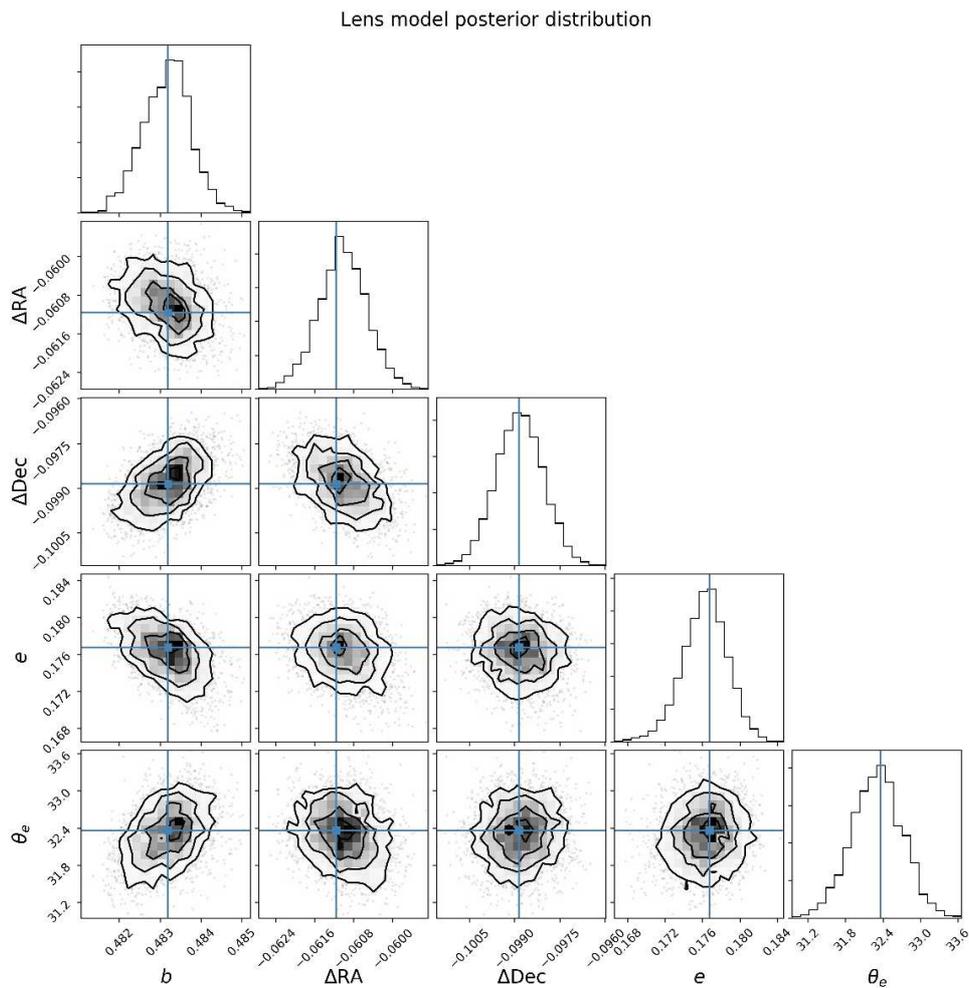}
\caption{Posterior distribution of the parameters of the foreground lens object, as obtained by the lens modelling 
and MCMC runs (see Sect.~\ref{sect:lensing_model} for details). The constraints are given on the following 
parameters: the position ($\rm \Delta$RA and $\rm \Delta$Dec in arcsec), the Einstein radius ($b$), 
the ellipticity ($e$), and the position angle ($\theta_e$) with the orientation measured east of north. 
The contours show the 0.5, 1, 1.5, and 2$\sigma$ of the integrated likelihood. The mean and the $\pm1\sigma$ interval 
around the mean are displayed as histograms along the diagonal of the plot.}
\label{fig:corner_lens}
\end{figure*}

\begin{figure*}
\centering
\includegraphics[width=0.7\textwidth]{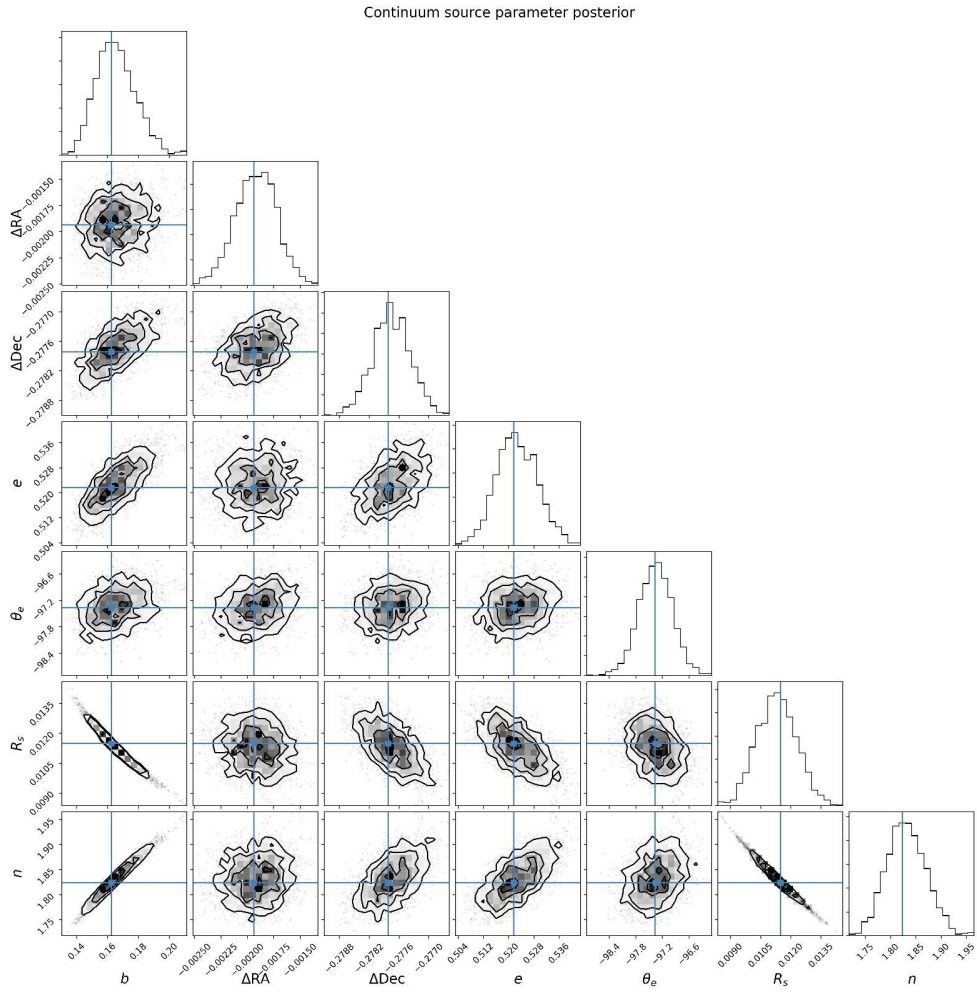}
\caption{Posterior distribution of the parameters of the HerBS-89a lensed source, as obtained by the lens modelling and MCMC 
runs (see Sect.~\ref{sect:lensing_model} for details). The constraints are given on the following parameters: 
the position ($\rm \Delta$RA and $\rm \Delta$Dec in arcsec), the Einstein radius ($b$), the ellipticity ($e$), 
the scale radius ($R_s$), the position angle ($\theta_e$) with the orientation measured east of north and the S\'ersic index ($n$). See Fig~\ref{fig:corner_lens} for further details.}
\label{fig:corner_source}
\end{figure*}

The statistical uncertainties on the lens model parameters, for both the foreground lens object and the HerBS-89a lensed source were computed via two MCMC run, focusing on the lens and the source, respectively (see Section~\ref{sect:lensing_model}). 
Figure~\ref{fig:corner_lens} and \ref{fig:corner_source} show the posterior distribution of the lens and source parameters, as obtained with MCMC.

In order to perform the kinematic analysis described in Sect. \ref{sect:kinematics}, we apply the lens model to high resolution emission line data preserving the velocity information by fitting S\'ersic profiles to  channel images in the data cubes. To obtain a sufficient S/N ratio, we bin the $\rm 40\,  km\, s^{-1}$ channels of $\rm ^{12}CO(9-8)$ and $\rm H_2O$ 
by a factor of 3 and 6, respectively. Despite this rebinning, the S/N ratios for 
some of the channels is still very low. 

\begin{figure*}
\centering
\includegraphics[width=0.9\textwidth]{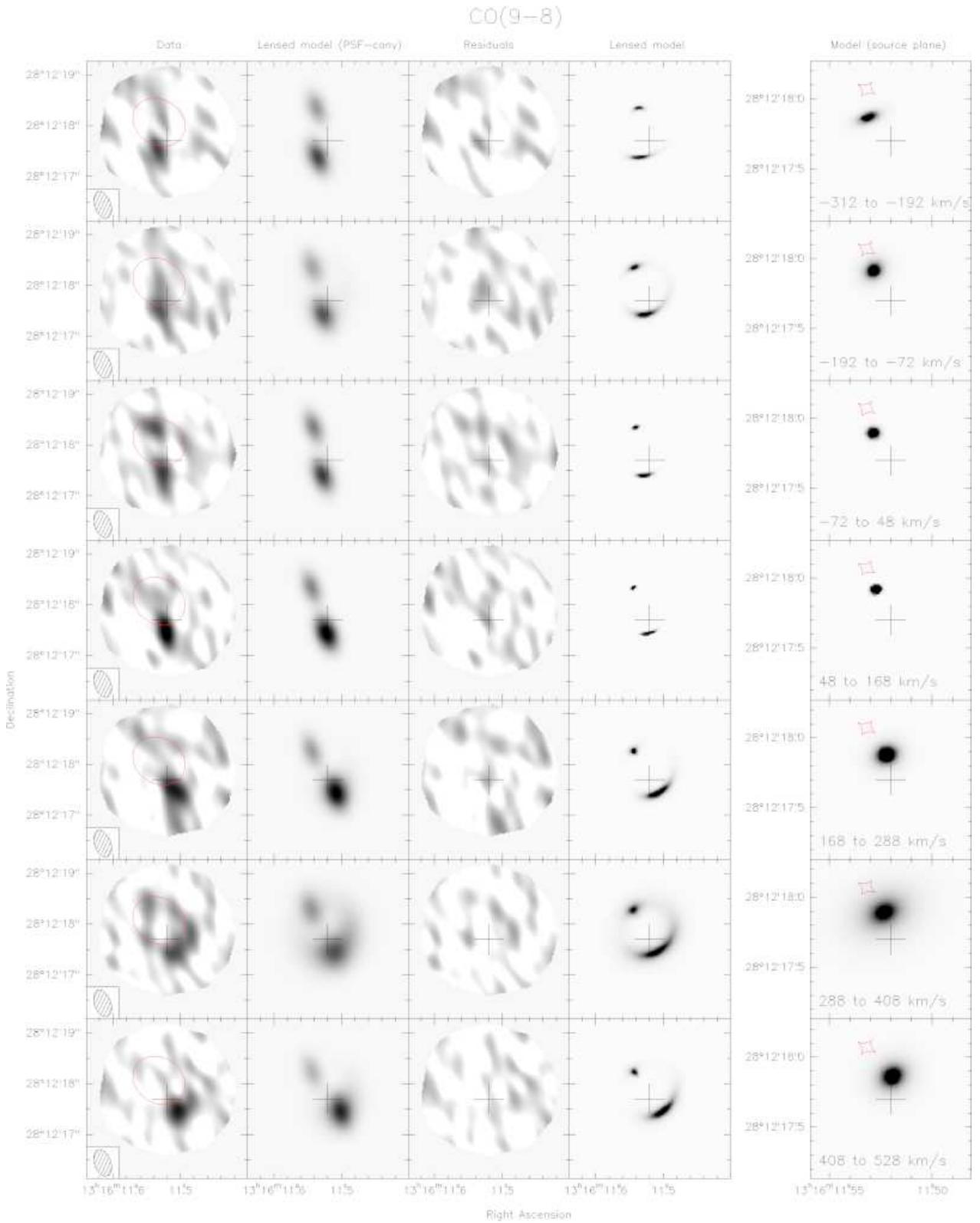}
\caption{Lens modelling of the $\rm ^{12}CO(9-8)$ spectral channels. The velocity ranges covered by each channel are marked in the right-hand panels. From left to right: observed image (with the synthesized beam shown in the lower left corner) and critical curve (red line); reconstructed model image convolved with the PSF; residuals, reconstructed model image at full angular resolution; and, separated and with a different angular scale, reconstructed source-plane images and caustic curve. The cross shows the central coordinate R.A. 13:16:11.52 and Dec. +28:12:17.7 (J2000).}
\label{fig:co98_lens_model_ch}
\end{figure*}

\begin{figure*}
\centering
\includegraphics[width=0.9\textwidth]{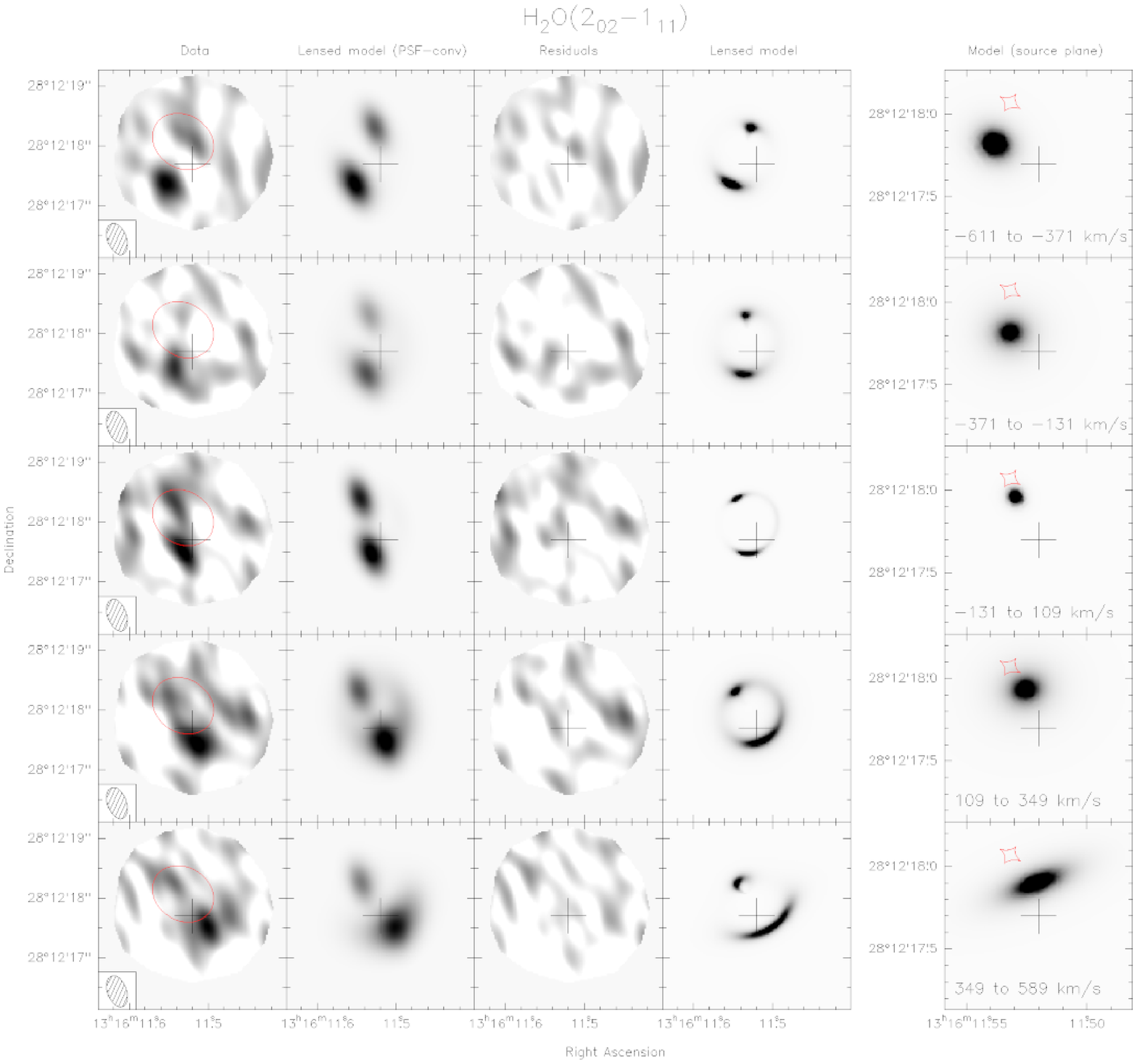}
\caption{Lens modelling of the $\rm H_2O(2_{02}-1_{11})$ spectral channels. The velocity ranges covered by each channel are marked in the right-hand panels. See Fig.~\ref{fig:co98_lens_model_ch} for further details.}
\label{fig:h2o_lens_model_ch}
\end{figure*}

Figures~\ref{fig:co98_lens_model_ch} and \ref{fig:h2o_lens_model_ch} present the results of this channel by channel modelling, for the $\rm ^{12}CO(9-8)$ and $\rm H_2O(2_{02}-1_{11}$ emission lines, respectively.

\section{Review of $\kappa_\nu$ values}\label{sect:kappa}

Although the use of the MBB model and Eq. \ref{eq:mbb} to derive dust masses is common and 
well-established in the literature, the actual value of $\kappa_\nu$ is far from being well defined. 
Moreover, it is not granted that a universal value shared by all galaxies does exist; 
evidence rather points to the opposite. 
We here review the current status on $\kappa_\nu$ taking into account recent studies 

The common preference found in the literature is to adopt the $\kappa_\nu$ values tabulated by \citet{li2001} 
or its updated version by \citet{Draine2003}. At 850~$\mu$m, the reference value is $\kappa_{850}=0.038$ m$^2$ kg$^{-1}$ 
with a dependence on frequency $\sim \nu^{2.07}$ at $\lambda=100$ and 600~$\mu$m \citep{berta2016}. 
To reconcile with a small conflict between the dust to hydrogen mass ratio predicted by their model and the elemental 
abundances in the solar neighborhood ISM, \citet{draine2014} re-normalized their dust properties such 
that $\kappa_{850}=0.047$ and $\kappa_{500}=0.116$ m$^2$ kg$^{-1}$. 
\citet{Hensley2020} discuss more recent finding, that could lead to further possible corrections towards larger $\kappa_\nu$. A larger value $\kappa_{850}=0.072$ m$^2$ kg$^{-1}$ come also from the THEMIS dust model \citep{jones2013,jones2017}, which is in agreement with the measured all sky extinction from SDSS quasars \citep{planck2016_xxix}. 

The tabulated $\kappa_\nu$ values mentioned above are theoretical \citep[see, e.g.,][]{wd01} and an alternative path is to 
adopt an empirical value of $\kappa_\nu$, measured at a reference frequency and then scale it to any other frequency 
using the power law with index $\beta$ as usual. Studying a sample of 22 local galaxies with known metallicities and 
available sub-mm data, \citet{james2002} derived $\kappa_{850}=0.07\pm0.02$ m$^2$ kg$^{-1}$. 

Using {\it Herschel} observations 
of local galaxies M~99 and M~100, \citet{eales2010b} derive a value of $\kappa_\nu\sim0.06$ m$^2$ kg$^{-1}$ at 350 $\mu$m, 
corresponding to $\kappa_{850}=0.01$ m$^2$ kg$^{-1}$ adopting a dust emissivity index $\beta=2$.
From early solar-neighborhood {\it Planck} results, a value  $\kappa_{850}=0.066$ m$^2$ kg$^{-1}$ is derived\footnote{\citet{planck2011_XXI} measure $\tau_{250}/N_\textrm{H} = 0.92 \times 10^{-25}$ cm$^2$ 
for the solar neighborhood. Using the relation $\tau_\nu=\kappa_0\left(\nu/\nu_0\right)^\beta R_\textrm{DG}\mu_\textrm{H}N_\textrm{H}$ \citep{bianchi2019,planck2015_intXXVIII}, this leads to  $\kappa_{850}=0.066$ m$^2$ kg$^{-1}$. Here $\beta=1.8$ and a dust to hydrogen mass ratio $R_\textrm{DG}=0.0091$ \citep{draine2014} were adopted. The quantity $\mu_\textrm{H}$ is the mass of the the hydrogen atom. It is worth to note the strong dependence of $\kappa_\nu$ on $\beta$ and $R_\textrm{DG}$: different choices lead to significantly different values of $\kappa_{850}$. Typical alternatives to the values mentioned above are, for example, $\beta=2$ (often found in the literature) and $R_\textrm{DG}=0.0074$ \citep{jones2017}. See \citet{bianchi2019} for an in depth description.},
roughly twice the commonly adopted value by \citet{Draine2003}. Recently, \citet{clark2016} assembled a sample of 22 
galaxies from the {\it Herschel Reference Survey} \citep{boselli2010} and derived a value of 
$\kappa_{500}=0.051^{+0.070}_{-0.026}$ m$^2$ kg$^{-1}$ at 500~$\mu$m, corresponding to 
$\kappa_{850}=0.018^{+0.023}_{-0.01}$. 
Significant variations of $\kappa_\nu$ at sub-galactic scales are shown by \citet{clark2019}.

Such very low values of $\kappa_\nu$ obtained for external galaxies are nevertheless subject to several caveats related to the derivation method, and thus need to be taken with caution. 
To mention a few: the sight-lines intercept a range of media, and possibly multiple temperature dust components; the assumptions made about dust-to-metals ratios need solid metallicity measurements and might depend on the adopted calibration; a variable $X_\textrm{CO}$ brings further uncertainties. For further discussions, see for example \citet{Priestley2020}. 

Finally, \citet{clark2016,clark2019} present a thorough compilation of the values of $\kappa_\nu$ found in the literature to date, 
and remarkably collect values spanning a range of more that three orders of magnitude (at their reference 500~$\mu$m wavelength; see the respective Figs.~1 in the above mentioned papers).

With these results in mind, it is clear that a proper comparison of dust masses to literature values must take into 
account the different assumptions underlying the SED modelling. In order to follow a popular choice in this field, 
we here adopted the \citet{Draine2003} $\kappa_\nu$, applying the correction prescribed by \citet{draine2014} 
to dust masses (as derived with both the MBB and DL07 modelling).

\end{appendix}


\end{document}